\documentclass[amsmath,amssymb,11pt]{article}
\usepackage{jheppub2}
\pdfoutput=1
\usepackage{graphicx,epsfig}
\usepackage{float}
\usepackage{subfig}
\usepackage{subfloat}
\usepackage[utf8]{inputenc}
\usepackage{hyperref}
\usepackage{caption}
\usepackage{color}

\newcommand*{\affmark}[1][*]{\textsuperscript{#1}}

\newcommand{\beq}{\begin{equation}}

\newcommand{\eeq}{\end{equation}}

 \newcommand{\be}{\begin{equation}}
 \newcommand{\ee}{\end{equation}}
 \newcommand{\bea}{\begin{eqnarray}}
 \newcommand{\eea}{\end{eqnarray}}

\usepackage{tikz}
\usetikzlibrary{positioning}
\usetikzlibrary{intersections}
\usetikzlibrary{fadings} 
\usetikzlibrary{arrows.meta} 
\usetikzlibrary{arrows}

\tikzfading[name=fade out,
inner color=transparent!0,
outer color=transparent!100]

\definecolor{cherryblossompink}{rgb}{1.0, 0.72, 0.77}
\definecolor{lightblue}{rgb}{0.68, 0.85, 0.9}

\usetikzlibrary{decorations.pathmorphing}
\usetikzlibrary{decorations.pathreplacing,decorations.markings}

\usetikzlibrary{backgrounds,automata}

\preprint{\texttt{IFT-UAM/CSIC-24-98}}

\title{Three-dimensional quantum black holes: a primer}

\author{Emanuele Panella,\affmark[a]}
\emailAdd{emanuele.panella.21@ucl.ac.uk}
\author{Juan F. Pedraza\affmark[b]}
\emailAdd{j.pedraza@csic.es}
\author{and Andrew Svesko\affmark[c]}
\emailAdd{andrew.svesko@kcl.ac.uk}

\affiliation{\affmark[a]Department of Physics and Astronomy, University College London,\\
Gower Street, London, WC1E 6BT, United Kingdom}
\affiliation{\affmark[b]Instituto de F\'isica Te\'orica UAM/CSIC,
Calle Nicol\'as Cabrera 13-15, 28049 Madrid, Spain}
\affiliation{\affmark[c]Department of Mathematics, King’s College London,
Strand, London, WC2R 2LS, United Kingdom}

\abstract{We review constructions of three-dimensional `quantum' black holes. Such spacetimes arise via holographic braneworlds and are exact solutions to an induced higher-derivative theory of gravity consistently coupled to a large$-c$ quantum field theory with an ultraviolet cutoff, accounting for all orders of semi-classical backreaction. Notably, such quantum-corrected black holes are much larger than the Planck length. We describe the geometry and horizon thermodynamics of a host of asymptotically (anti-) de Sitter and flat quantum black holes. A summary of higher-dimensional extensions is given. We survey multiple applications of quantum black holes and braneworld holography.
}

\begin{document}

\maketitle

\section{Overview} \label{sec:intro}

Semi-classical gravity remains a useful proxy to study quantum effects in gravity from the perspective of a macroscopic observer. In this context, quantum fields live in a classical dynamical spacetime where the combined system is characterized by the semi-classical Einstein equations \cite{Birrell:1982ix,Wald:1995yp}
\beq G_{ab}(g)+\Lambda g_{ab}=8\pi G_{N}\langle T_{ab}^{\text{QFT}}\rangle\;.\label{eq:semieineq}\eeq
On the left-hand side is the usual Einstein tensor for a classical spacetime $g_{ab}$ with cosmological constant $\Lambda$, while on the right-hand side $\langle T_{ab}^{\text{QFT}}\rangle$ is the expectation value of the (renormalized)  stress-energy tensor of the quantum field theory in some quantum state $|\Psi\rangle$. Semi-classical gravity should be viewed as an approximation and only valid in a certain regime. Indeed, the semi-classical approximation fails near the Planck scale as at this level quantum gravity effects become important, such that (\ref{eq:semieineq}) can no longer be trusted. Further, the semi-classical field equations (\ref{eq:semieineq}) are not expected to be valid for generic quantum states $|\Psi\rangle$, e.g., macroscopic superpositions \cite{Page:1981aj}. They are, however,  known to be valid when $|\Psi\rangle$ is approximately classical, i.e., a coherent state. 

Even in its regime of validity, solutions to semi-classical gravity, particularly black holes, are difficult to study consistently. Largely this is because solving  (\ref{eq:semieineq}) amounts to solving the problem of backreaction -- how quantum matter influences the classical geometry and vice versa -- a notoriously difficult and open problem as it requires  simultaneously solving a coupled system of geometry and quantum correlators. Often, in three spacetime dimensions and higher,\footnote{In two-dimensional dilaton gravity, the analog of the semi-classical Einstein equations can be solved exactly. This is because the quantum effective Polyakov action \cite{Polyakov:1981rd} capturing the two-dimensional conformal anomaly encodes nearly all backreaction effects \cite{Christensen:1977jc}.} the problem is examined perturbatively, offering limited insight, especially when backreaction effects become large. These difficulties only compound when there are a large number of quantum fields, or when the field theory is strongly coupled, as is the case for quantum chromodynamics and the standard model of particle physics.  

A context in which the physics of a large-$N$ number of strongly interacting quantum fields may be probed is the Anti-de Sitter/conformal field theory (AdS/CFT) correspondence \cite{Maldacena:1997re}. Born out of studies in string theory, AdS/CFT is a non-perturbative candidate model of
quantum gravity, where gravitational physics in a bulk $d+1$-dimensional asymptotically AdS spacetime has a dual description in terms of a CFT living on the $d$-dimensional conformal boundary of AdS. This duality is therefore a concrete realization of gravitational holography \cite{tHooft:1993dmi,Susskind:1994vu}. More specifically, in a large-$N$ expansion, the planar diagram limit of the CFT, the bulk is well-approximated by classical gravity (we will state this dictionary more precisely below). A powerful feature of the AdS/CFT correspondence is strong-weak
coupling duality: coupling constants between the bulk and boundary theories are inversely related, $G_{N}\sim N^{-1}$. Thus, computations of strongly coupled field theories may instead be performed via a classical gravity calculation. While the boundary geometry on which the CFT lives may be curved (and even contain black holes \cite{Hubeny:2009ru}), it is fixed.  Consequently, standard AdS/CFT holography alone is insufficient for addressing the problem of semi-classical backreaction.\footnote{There exist a set of AdS boundary conditions for which the boundary metric becomes dynamical \cite{Compere:2008us}.} 

Enter braneworld holography \cite{deHaro:2000wj}. Historically introduced as a possible solution to the hierarchy problem \cite{Arkani-Hamed:1998jmv,Randall:1999ee}, braneworld models treat the four-dimensional universe we experience as a membrane sitting in a spacetime with large extra-dimensions. When combined with holography, braneworlds function as a useful toolkit to address difficult problems in semi-classical gravity. In this framework, AdS/CFT duality is adapted to incorporate situations where a portion of the bulk, including its boundary, is removed by a $d$-dimensional Randall-Sundrum \cite{Randall:1999vf,Randall:1999ee} or Karch-Randall \cite{Karch:2000ct,Karch:2000gx} braneworld. Crucially, the geometry of the end-of-the-world (ETW) brane is dynamical, having an induced theory of gravity. More precisely, the brane serves as an infrared cutoff in the bulk, translating to a ultraviolet (UV) cutoff for the holographic CFT. As in holographic renormalization \cite{Kraus:1999di,Emparan:1999pm,deHaro:2000vlm,Skenderis:2002wp,Papadimitriou:2004ap}, a tower of higher-derivative corrections to the $d$-dimensional Einstein-Hilbert action are induced by the holographic cutoff $\text{CFT}_{d}$. From the brane perspective, the induced theory may thus be interpreted as a semi-classical theory of gravity \cite{Emparan:2002px}, where the higher-derivative corrections incorporate backreaction effects due to the CFT living on the brane, governed by
\beq G_{ij}+\Lambda_{d}g_{ij}+(\text{higher-curvature})=8\pi G_{d}\langle T^{\text{CFT}}_{ij}\rangle_{\text{planar}}\;.\label{eq:branesemieomintro}\eeq
Here $\Lambda_{d}$ and $G_{d}$ are induced cosmological and Newton constants on the brane, and the right-hand side indicates the holographic CFT is in its planar limit.\footnote{Going beyond the planar limit corresponds to including bulk quantum effects.}

At first glance it would appear the braneworld has only complicated the situation with its higher-derivative corrections: solving the induced field equations (\ref{eq:branesemieomintro}) requires solving the problem of backreaction in a complex higher-derivative theory of gravity. The computational advantage of braneworld holography, however, is that the semi-classical induced brane theory has an equivalent bulk description in terms of classical $\text{AdS}_{d+1}$ gravity coupled to a brane obeying Israel junction conditions. Thus,  exact spacetimes solving the classical bulk field equations with brane boundary conditions automatically correspond to exact solutions to the semi-classical brane equations of motion (\ref{eq:branesemieomintro}). Holographic braneworlds thus provide a means to exactly study the problem of backreaction without having to explicitly solve semi-classical field equations.  
%classical solutions to the bulk Einstein equations (with proper brane boundary conditions) exactly correspond to solutions to the highly complicated semi-classical equations of motion on the brane.
%\footnote{It is worth emphasizing that exact bulk solutions lead to exact solutions to the entire gravity theory on the brane, including the whole tower of higher-derivative terms. Importantly, while general higher derivative theories of gravity are pathological since they are typically accompanied by ghosts, one does not expect the brane theory to be pathological (assuming one does not truncate the series of terms) since the bulk theory and the procedure of integrating out the bulk are not pathological.}
In particular, classical $\text{AdS}_{d+1}$ black holes which localize on the ETW brane are conjectured to precisely map to  black holes in $d$-dimensions, including all orders of quantum backreaction \cite{Emparan:2002px}, i.e., `quantum' black holes. 

The primary purpose of this article is to review the state of the art regarding such holographic quantum black holes. Emphasis is given to a particular class of analytic black holes which localize on an $\text{AdS}_{4}$ braneworld, first uncovered by Emparan, Horowitz and Myers \cite{Emparan:1999wa,Emparan:1999fd}, corresponding to three-dimensional quantum black holes \cite{Emparan:2002px,Emparan:2020znc,Emparan:2022ijy,Panella:2023lsi}. These braneworld black holes lead to an important observation: backreaction can lead to the existence of black holes where there were none before. That is, famously, there are no black hole solutions in vacuum to classical Einstein gravity in three-dimensions with positive or vanishing cosmological constant. Rather, the geometry of a point mass in $\text{Mink}_{3}$ or $\text{dS}_{3}$ is described as a conical defect without a black hole horizon \cite{Deser:1983tn,Deser:1983nh}; Schwarzschild-$\text{dS}_{3}$, for example, is a conical defect with a single cosmological horizon but no black hole horizon. Quantum corrections due to backreaction alter the three-dimensional geometry in such a way that a black hole horizon is induced, leading to a type of (quantum) censorship of conical singularities. 
%Evidence for this can already be seen at the perturbative level, independent of the number or coupling strength of the quantum fields, however..
Meanwhile, classical black holes do exist in three-dimensional Einstein gravity with a negative cosmological constant \cite{Banados:1992wn,Banados:1992gq}, a consequence of the tendency for gravitational collapse afforded by the negatively curved geometry. Nonetheless, in such contexts backreaction yields behavior strikingly different from their classical counterparts. 

A secondary goal of this review is to advertise a host of applications of holographic quantum black holes. These include the exact study of semi-classical horizon thermodynamics, holographic entanglement entropy and complexity, and the prospect of probing black hole singularities. Combined, quantum black holes serve as a theoretical laboratory to exactly test ideas in semi-classical/quantum gravity, deserving of further exploration.

\subsection{Road map and high-level summary} 

\noindent \textbf{Backreaction without holography.} Section \ref{sec:perturbanalysis} sets the stage by exploring quantum corrections to three-dimensional geometries at the perturbative level. This demonstrates one need not appeal to AdS/CFT or holographic braneworlds to see how quantum backreaction alters classical three-dimensional geometry. %independent of the number or coupling strength of the backreacting quantum fields. 
For example, a conformally coupled scalar field in Schwarzschild-(A)dS$_{3}$ produces a Casimir effect with negative energy density such that, upon solving the semi-classical Einstein equations (\ref{eq:semieineq}), the linear order change to the $tt$-component of the metric in static patch coordinates goes like \cite{Souradeep:1992ia,Soleng:1993yh}
\beq \delta g_{tt}= \frac{2L_{\text{P}}F(M)}{r}\;,\label{eq:linearcorrection}\eeq
where $L_{\text{P}}=\hbar G_{3}$ is the Planck length in three-dimensions\footnote{We will always work in units where the speed of light $c=1$.} and $F(M)$ is a positive function of the mass of the point particle generating the conical defect. The $1/r$-correction -- which is solely a quantum effect -- modifies the original blackening factor and its root structure, implying a black horizon may arise due to backreaction. However, since the correction is on the order of the Planck length, this conclusion is tenuous since quantum gravity effects are expected to play a role at this scale. If there is a large-$c$ amount of such quantum fields, it is conceivable that their combined quantum effect is to produce a correction proportional to $cL_{\text{P}}\gg L_{\text{P}}$, for which quantum gravity effects can be neglected. Unfortunately, it is thus far unknown how to solve the backreaction problem with such a large number of fields via non-holographic methods. Thus, while the perturbative analysis is suggestive, it advocates for the holographic braneworld approach described above. 

\vspace{2mm}

\noindent \textbf{Holographic braneworlds and quantum black holes.} In Section \ref{sec:BWholo} we present a general portrait of braneworld holography and quantum black holes. Historically, this framework is a natural extension of holographic renormalization, a prescription where bulk infrared (IR) divergences are eliminated by introducing an IR cutoff hypersurface near the AdS boundary, adding local counterterms, and employing a minimal subtraction scheme. In standard braneworld holography, the IR cutoff surface is replaced by an end-of-the-world brane $\mathcal{B}$ of tension $\tau$. The would be divergent counterterms now 
combine with the brane action as seen from the bulk, leading to an induced theory of higher-derivative gravity on the brane (\ref{eq:Ibgravgen})
\beq 
\begin{split} 
I_{\text{Bgrav}}&=\frac{1}{16\pi G_{d}}\int_{\mathcal{B}} \hspace{-1mm}d^{d}x\sqrt{-h}\biggr[R-2\Lambda_{d}+\frac{L_{d+1}^{2}}{(d-4)(d-2)}\left(R_{ij}^{2}-\frac{dR^{2}}{4(d-1)}\right)+\cdots\biggr]\;,
\end{split}
\label{eq:Ibgravgenintro}\eeq
coupled to a CFT with an ultraviolet cutoff. Here $G_{d}$ and $\Lambda_{d}$ are induced scales, dependent on the bulk Newton and cosmological constant and brane tension, and the ellipsis denotes an infinite tower of higher-derivative terms, proportional to increasing positive powers of the bulk AdS length scale $L_{d+1}$. A metric variation of this action results in the semi-classical equations of motion (\ref{eq:branesemieomintro}), solutions of which are conjectured to constitute quantum-corrected geometries due to the backreaction of a large-$c$ holographic CFT with a UV cutoff. 
%Rather than explicitly solving said equations, we instead follow \cite{Emparan:2002px} and opt to look for classical bulk black hole solutions with horizons that localize on the brane. 

\vspace{2mm}

\noindent \textbf{Benchmarking quantum black holes.} Section \ref{sec:qbhtaxonomy} is devoted to constructing three-dimensional quantum black holes using holographic braneworlds, including static and rotating black holes in AdS$_{3}$ (section \ref{ssec:qbtzbhs}), dS$_{3}$ (section \ref{ssec:qdsbhs}), and Minkowski$_{3}$ (section \ref{ssec:qbhsmink}). In all cases, the bulk geometry is taken to be the AdS$_{4}$ C-metric, with either a Karch-Randall (asymptotically AdS$_{3}$) or Randall-Sundrum (asymptotically flat or dS$_{3}$) end-of-the-world brane embedded inside. The neutral, static geometries induced on the brane include corrections of the type (\ref{eq:linearcorrection}), supporting the intuition gained from the perturbative analysis. For example, the static quantum BTZ black hole is (\ref{eq:qBTZ})
\beq
\hspace{-2mm}ds^{2}_{\text{qBTZ}} = -\left(\frac{\bar{r}^{2}}{\ell_{3}^{2}}-8 \mathcal{G}_3 M-\frac{\ell F(M)}{\bar{r}} \right)d\bar{t}^2 +\left(\frac{\bar{r}^{2}}{\ell_{3}^{2}}-8 \mathcal{G}_3 M-\frac{\ell F(M)}{\bar{r}} \right)^{-1}  d\bar{r}^2 + \bar{r}^2 d\bar{\phi}^2
\label{eq:qBTZintro}\;,\eeq
where $M$ is the black hole mass, and $\mathcal{G}_{3}$ is a renormalized Newton's constant due to an all-order resummation of the higher-derivative terms appearing in the induced action (\ref{eq:Ibgravgenintro}). One notable difference between the black hole (\ref{eq:qBTZintro}) and the perturbative geometry (\ref{eq:linearcorrection}) is that $\ell\sim cL_{\text{P}}\gg L_{\text{P}}$, where $c$ is the central charge of the cutoff CFT$_{3}$ on the brane. Thus, the correction appears on a scale where quantum gravitational effects can be consistently ignored; the geometry (\ref{eq:qBTZintro}) is a quantum-corrected black hole much larger than the Planck length. Further, the geometry (\ref{eq:qBTZintro}) is an exact solution to the semi-classical brane theory.

\vspace{2mm}

\noindent \textbf{Quantum black hole thermodynamics.} Section \ref{sec:qbhthermo} focuses on the horizon thermodynamics of quantum black holes. As with the geometry, the thermodynamics of the braneworld black hole is induced from the thermodynamics of the bulk AdS$_{4}$ black hole. For example, the bulk and brane horizon temperatures coincide. Meanwhile, the bulk entropy, given by the Bekenstein-Hawking area relation, is reinterpreted as generalized entropy
\beq S_{\text{BH}}^{(4)}\Longleftrightarrow S_{\text{gen}}^{(3)}\;,\eeq
the sum of gravitational and matter entanglement entropies. Consequently, from the brane perspective, the first law of thermodynamics of quantum black holes is 
\beq dM=TdS_{\text{gen}}^{(3)}+...\;,\eeq
where the ellipsis refers to possible additional variations such as rotation or charge.
Thus, by accounting for semi-classical backreaction,  classical gravitational entropy is replaced by its semi-classical generalization. Since the bulk thermodynamic variables are exactly known, the first law holds to all orders of backreaction. This is highly non-trivial from the brane perspective. Indeed, the induced semi-classical theory includes an infinite tower of higher-derivative terms and the matter entropy of the cutoff CFT$_{3}$. To determine the entropy and mass non-perturbatively in backreaction would require a resummation of the infinite tower of terms and knowledge of how to compute the von Neumann entropy of the CFT$_{3}$, a supremely challenging task. Via holography, this resummation is performed by the bulk. 

Holographic braneworlds, moreover, provide a natural higher-dimensional origin of \emph{extended} black hole thermodynamics (section \ref{ssec:extbhthermo}), where the cosmological constant is treated as a dynamical pressure. This is because the cosmological constant on the brane is partly induced by the brane tension. Tuning the brane tension alone amounts to varying the induced brane cosmological constant, such that mechanical work performed by the brane is interpreted as extended thermodynamics of the black hole on the brane. Intriguingly, all AdS$_{3}$ quantum black holes obey a semi-classical generalization of the reverse isoperimetric inequality, indicating quantum black holes have a maximal entropy state at fixed volume. 

Quantum black holes can also have a wildly different thermal phase structure than their classical counterparts (section \ref{ssec:ptsqbhs}). Specifically, in the case of the static quantum BTZ black hole, large enough backreaction trigger \emph{reentrant} phase transitions, e.g., a transition from thermal AdS to the quantum black hole back to thermal AdS. The first of these transitions can be understood as a quantum analog of the familiar first-order Hawking-Page transition of AdS black holes, while the second transition back to thermal AdS is entirely a consequence of non-perturbative backreaction effects. 
 
%due to the cutoff CFT. 

\vspace{2mm}

\noindent \textbf{Puzzles in higher-dimensions.} Section \ref{app:evapbhshighd} briefly reviews the history and status of higher-dimensional braneworld black holes and their semi-classical interpretation. Indeed, while we focus on three-dimensional quantum black holes, the conjecture \cite{Emparan:2002px,Tanaka:2002rb} that classical bulk black holes correspond to quantum-corrected black holes on the brane, in principle, holds in arbitrary spacetime dimensions. After assessing arguments that imply four and higher-dimensional quantum black holes must be evaporating, we describe static four-dimensional braneworld black holes and their peculiar features when viewed as quantum black holes. We conclude with a short summary regarding the holographic duals of evaporating black holes.

\vspace{2mm}

\noindent \textbf{Applications.} In Section \ref{sec:applications} we present a non-exhaustive list of applications and possible avenues for future research using the quantum black hole constructions described in this review. Historically, braneworlds, motivated by string theory, were conceived as possible resolutions to problems in high energy phenomenology, e.g., the hierarchy problem of the standard model of particle physics \cite{Randall:1999ee,Arkani-Hamed:1998jmv}. Though there are yet to be any experimental signatures of extra dimensions or braneworlds, the view taken here is that braneworld holography provides an invaluable framework to test ideas in (non-perturbative) semi-classical gravity and beyond. For example, prescriptions in holographic information theory, i.e., entanglement entropy and complexity, can be tested in the context of quantum black holes. Moreover, classical ideas such as cosmic censorship can be revisited using quantum black holes as a guide.
%Indeed, since there are so few examples of exact quantum black hole solutions to non-perturbative semi-classical gravity,  it is worth revisiting 

\vspace{2mm}

\noindent \textbf{Appendices.} While unnecessary to follow the core narrative of this review, we include a number of Appendices to be self-contained and pedagogical. The conventions used in the majority of this review are presented in Appendix \ref{app:conventions}. In Appendix \ref{app:holoren} we provide further details about holographic regularization, including a detailed derivation of the local counterterms that give rise to the induced gravity theory on the brane. Appendix \ref{app:braneworldbasics} summarizes relevant history and physics of braneworlds. Geometric elements of the C-metric utilized in the main text are given in Appendix \ref{app:AdSCmetprops}. Appendix \ref{app:onshellaction} presents a derivation of the thermodynamics of the static, neutral quantum BTZ black hole using a (bulk) canonical partition function.

%%%%%%%%%%%%%%%%%%%%%%%%%%%%%%%%%%%%%%%%%%%%%%%%%%%%%%%%%%%%%%%%%%%

\section{Black holes and backreaction in 3D: a perturbative analysis} \label{sec:perturbanalysis}

\subsection*{Three-dimensional black holes and conical defects}

In vacuum general relativity, black holes tend to disappear when lowering the dimension of spacetime from four to three. This can be understood at the level of dimensional analysis. If the only dimensionful parameter is three-dimensional Newton's constant $G_{3}$, then introducing a massive object of mass $M$ does not introduce an additional length scale needed to characterize a black hole horizon solely in terms of its mass; indeed, $G_{3}M$ is dimensionless.\footnote{Since we have set $c=1$, mass has dimensions of inverse length while $G_{3}$ has dimensions of length.} In fact, a massive point particle in flat $(2+1$)-dimensional general relativity is a conical defect, with angular deficit $\delta=2\pi(1-\sqrt{1-8G_{3}M})$ and a conical singularity at the origin \cite{Deser:1983tn}.
Moreover, while a cosmological constant $\Lambda$ will introduce another length scale, this alone is not sufficient to have a black hole horizon. Gravitational attraction is also required. 

To elaborate, there are black holes in asymptotically $\text{AdS}_{3}$. Namely, the Ba\~nados-Teitelboim-Zanelli (BTZ) black hole \cite{Banados:1992wn,Banados:1992gq}
\beq ds^{2}=-N(r)dt^{2}+N^{-1}(r)dr^{2}+r^{2}(d\phi+N_{\phi}dt)^{2}\;,\label{eq:3dmetgen}\eeq
with lapse and shift metric functions
\beq N(r)\equiv-8G_{3}M+\frac{r^{2}}{L_{3}^{2}}+\frac{(4G_{3}J)^{2}}{r^{2}}\;,\quad N_{\phi}\equiv-\frac{4G_{3}J}{r^{2}}\;,\eeq
for mass $M$ and spin $J$. The roots of the lapse, 
\beq r_{\pm}^{2}=\frac{L_{3}^{2}}{2}\left[8G_{3}M\pm\sqrt{(8G_{3}M)^{2}-\left(\frac{8G_{3}J}{L_{3}}\right)^{2}}\right]\;,\eeq
characterize the outer ($r_{+}$) and inner/Cauchy ($r_{-}$) horizons, with $r_{+}\geq r_{-}\geq0$, assuming $ML_{3}\geq J>0$ to avoid naked singularities. The reason we can interpret the BTZ metric as a `black hole' is because the negatively curved geometry of $\text{AdS}_{3}$ provides an innate geometric tendency for gravitational collapse (see, e.g., \cite{Ross:1992ba}).\footnote{Unlike the higher-dimensional black holes, the BTZ black hole does not possess a curvature singularity at $r=0$; indeed, the curvature is constant everywhere. Rather, $r=0$ describes a timelike/causal singularity.} Alternatively, there are no black holes in $\text{dS}_{3}$; the positively curved $\text{dS}_{3}$ background leads to an inability for collapse.\footnote{Here we are considering vacuum general relativity. Black holes in dS$_{3}$ can arise in pure modified theories of gravity, e.g., `new massive gravity' \cite{deBuyl:2013ega} or topological massive gravity \cite{Nutku:1993eb,Anninos:2009jt}.} Consequently, a point mass in $\text{dS}_{3}$,  is described by a conical defect \cite{Deser:1983nh} with a single cosmological horizon. 

To see this latter point, consider the Kerr-$\text{dS}_{3}$ metric. The line element formally takes the same form as (\ref{eq:3dmetgen}) except now with lapse and shift functions
\beq N(r)\equiv 1-8G_{3}M-\frac{r^{2}}{R_{3}^{2}}+\frac{(4G_{3}J)^{2}}{r^{2}}\;,\quad N_{\phi}\equiv+\frac{4G_{3}J}{r^{2}}\;,\eeq
where $R_{3}$ denotes the $\text{dS}_{3}$ length scale and `$+$' sign in $N_{\phi}$ is convention. Next,  introduce dimensionless parameters $\gamma\equiv r_{+}/R_{3}$ and $\alpha\equiv-4G_{3}J/\gamma R_{3}=ir_{-}/R_{3}$, where $r_{\pm}$ are
\beq r^{2}_{\pm}=\frac{R_{3}^{2}}{2}\left[(1-8G_{3}M)\pm\sqrt{(1-8G_{3}M)^{2}-\left(\frac{8G_{3}J}{R_{3}}\right)^{2}}\right]\;,\label{eq:rcKerrdS3}\eeq
with only a single positive root, $r_{+}$, identified as the cosmological horizon.
%\beq r_{+}=\frac{R_{3}}{2}\left(\sqrt{m+i\frac{8G_{3}J}{R_{3}}}+\sqrt{m-i\frac{8G_{3}J}{R_{3}}}\right)\;.\eeq
Then, the coordinate transformation \cite{Deser:1983nh,Bousso:2001mw,Panella:2023lsi}
\beq 
\begin{split}
 \tilde{t}= \gamma t+\alpha R_{3}\phi\;,\quad \tilde{\phi}=\gamma\phi-\alpha t/R_{3}\;,\quad \tilde{r}/R_{3}=\sqrt{\frac{(r/R_{3})^{2}+\alpha^{2}}{\gamma^{2}+\alpha^{2}}}\; 
\end{split}
\label{eq:coordtranstocon}\eeq
brings the Kerr-$\text{dS}_{3}$ geometry into an empty $\text{dS}_{3}$ form, i.e.,
\beq ds^{2}=-\left(1-\frac{\tilde{r}^{2}}{R_{3}^{2}}\right)d\tilde{t}^{2}+\left(1-\frac{\tilde{r}^{2}}{R_{3}^{2}}\right)^{-1}d\tilde{r}^{2}+\tilde{r}^{2}d\tilde{\phi}^{2}\;.\label{eq:conicaldS3}\eeq
Here, however, the coordinates $(\tilde{t},\tilde{\phi})$ do not have the same periodicity as standard $\text{dS}_{3}$, where $(t,r,\phi)\sim(t,r,\phi+2\pi)$. Rather,
\beq (\tilde{t},\tilde{\phi})\sim(\tilde{t}+2\pi R_{3}\alpha,\tilde{\phi}+2\pi\gamma)\;.\label{eq:periodcondsKdS3}\eeq
 Thence, Kerr$-\text{dS}_{3}$ is a conical defect geometry with angular deficit $\delta=2\pi(1-\gamma)$.
 %and is a quotient of $\text{dS}_{3}$.
 %Moreover, the Schwarzschild-$\text{dS}_{3}$ geometry (where $J=0=\alpha$) is also a conical defect with deficit $\delta=\sqrt{1-8G_{3}M}$, or a particle of mass $m$ whose stress-energy tensor $T_{ab}=m\delta(r)\delta^{0}_{a}\delta^{0}_{b}$ sources the geometry. 
 %Hence, there are no black solutions to three-dimensional Einstein's equations with positive cosmological constant.

It is worth pointing out that the $\text{AdS}_{3}$ geometry (\ref{eq:3dmetgen}) is not always a black hole. For $8G_{3}M<0$, the geometry is a conical defect, taking the form of empty $\text{AdS}_{3}$ (the line element (\ref{eq:conicaldS3}) with Wick rotation $R_{3}=-iL_{3}$), with the same periodicity (\ref{eq:periodcondsKdS3}), where now $\alpha\equiv r_{+}/L_{3}$ and $\gamma\equiv4G_{3}iJ/L_{3}$.\footnote{The coordinate transformation (\ref{eq:coordtranstocon}) is now $\tilde{t}=\gamma t-\alpha L_{3}\phi$, $\tilde{\phi}=\gamma\phi+\alpha t/L_{3}$ and $\tilde{r}/L_{3}=\sqrt{\frac{(r/L_{3})^{2}+\alpha^{2}}{\gamma^{2}-\alpha^{2}}}$.} In particular, when $J=0$, the states with $-1<8G_{3}M<0$ correspond to conical defects with angular deficit $\delta =2\pi(1-\sqrt{-8G_{3}M})$, while for $8G_{3}M<-1$ the geometry has a conical excess; at $8G_{3}M=-1$, the BTZ geometry is exactly empty $\text{AdS}_{3}$. Further,  when $8G_{3}M<0$ (for arbitrary $J$) the metric components are well-defined everywhere, i.e., there is no horizon and the conical singularity at $r=0$ is `naked'.

\subsection*{Backreaction and quantum dressing}

Another way to introduce a dimensionful parameter is to allow for quantum effects. Namely, for $\hbar\neq0$, there exists the three-dimensional Planck length $L_{\text{P}}=\hbar G_{3}$ (though there is no notion of Planck mass in three-dimensions). The question then is whether such quantum effects can modify the classical three-dimensional geometry so as to induce a (black hole) horizon when there was none before. 

Evidence of this comes from perturbatively solving the semi-classical Einstein equations (\ref{eq:semieineq}) for a conformally coupled scalar field $\Phi$ \cite{Steif:1993zv,Shiraishi:1993qnr,Lifschytz:1993eb,Martinez:1996uv,Casals:2016ioo,Casals:2019jfo,Emparan:2022ijy,Panella:2023lsi}, characterized by the action
\beq I=\frac{1}{16\pi G_{3}}\int d^{3}x\sqrt{-g}\left[R-2\Lambda\right]-\frac{1}{2}\int d^{3}x\sqrt{-g}\left[(\nabla\Phi)^{2}+\frac{1}{8}R\Phi^{2}\right]\;.\eeq
In such a set-up, the first step is to determine the renormalized stress tensor $\langle T_{ab}\rangle$. This is accomplished by first constructing the Green function associated with the equation of motion for the scalar field $\Phi$ in $\text{(A)dS}_{3}$, $(\Box-\frac{1}{8}R)\Phi=0$. Generically, the Green function is 
\beq G(x,x')=\frac{1}{4\pi}\frac{1}{|x-x'|}+\frac{\lambda}{4\pi}\frac{1}{|x+x'|}\;,\eeq
where $\lambda=0,1,-1$ corresponds to the scalar field obeying transparent, Neumann, or Dirichlet boundary conditions, respectively. For non-rotating backgrounds\footnote{See, e.g., \cite{Steif:1993zv,Casals:2019jfo,Panella:2023lsi} for rotating backgrounds.} and assuming transparent boundary conditions, the renormalized stress-tensor has the form
\beq \langle T^{a}_{\;b}\rangle=\frac{\hbar F(M)}{8\pi r^{3}}\text{diag}(1,1,-2)\;,\label{eq:stresstensorconfcoupscalar}\eeq
where $F(M)$ is a positive function of the mass. The explicit expression for $F(M)$ is different depending on whether the background is conical (A)dS$_{3}$ or the BTZ geometry (see, e.g., \cite{Casals:2016ioo,Casals:2019jfo,Emparan:2022ijy,Panella:2023lsi} for details), while the radial dependence and diagonal tensorial structure will change when considering non-transparent boundary conditions \cite{Lifschytz:1993eb}. In any case, the Green function of the quantum scalar field in the BTZ background is in the Hartle-Hawking state, satisfying the Kubo-Martin-Schwinger (KMS) condition at the black hole temperature \cite{Lifschytz:1993eb}.

Having the stress-tensor (\ref{eq:stresstensorconfcoupscalar}) source the right-hand side of the semi-classical Einstein's equations (\ref{eq:semieineq}), one finds for the static geometry the $\mathcal{O}(L_{P})$ correction to the three-dimensional metric (\ref{eq:3dmetgen}) is
\beq
 N(r)=\frac{r^2}{L_3^2} - 8G_3M - \delta g_{tt} \ , \qquad \delta g_{tt}=\frac{2L_\text{P}F(M)}{r} \ ,
\label{eq:pertbackAdS3geom}\eeq
for the $\text{AdS}_{3}$ geometries, and 
\beq 
 N(r)=1-8G_{3}M-\frac{r^2}{R_3^2} - \delta g_{tt} \ , \qquad \delta g_{tt}=\frac{2L_\text{P}F(M)}{r} \ ,
\label{eq:pertbackdS3geom}\eeq
for conical $\text{dS}_{3}$. Notably, $\delta g_{tt}>0$, indicating gravitational attraction \cite{Souradeep:1992ia,Soleng:1993yh}.\footnote{For conical (A)dS$_{3}$ the attractive gravitational effect is a by-product of a negative Casimir energy density from (\ref{eq:stresstensorconfcoupscalar}), $\rho_{\text{Cas}}=-\langle T^{t}_{\;t}\rangle=-\hbar F(M)/8\pi r^{3}$. This follows because a region of localized negative energy has a repulsive effect on its
exterior, however, the further one enters the region, the repulsive effect is lessened. Thus, at finite $r$ there is an
effective attraction from the Casimir energy \cite{Emparan:2022ijy}.} This attractive effect suggests a horizon induced due to semi-classical backreaction appears to dress the naked AdS$_{3}$ conical singularity or, in the case of conical dS$_{3}$, lead to a black hole horizon in addition to its classical cosmological horizon. 

The `horizon' radius, however, is proportional to the Planck length, the scale when quantum gravitational effects are expected to become important. Consequently, the above perturbative semi-classical analysis cannot be trusted, and we are not able to conclude the backreacted geometry results in a genuine horizon. This is not to say semi-classical backreaction will always result in Planckian-sized horizons. Indeed, both gravitational and quantum effects are in play: a large $c\gg1$ number of conformally coupled scalars results in a combined quantum effect $\propto c\hbar$ which may gravitate to yield semi-classical black holes, i.e., those with horizon radius $\sim Gc\hbar=cL_{\text{P}}\gg L_{\text{P}}$, for which quantum gravity effects may be safely neglected. To verify this, the backreaction problem of
a large number of fields must be non-linearly accounted for, and, thus far, perturbative methods have been unable accomplish this consistently. Braneworld holography provides a framework for which the backreaction problem due to a large-$c$ holographic conformal field theory can be exactly solved.

%%%%%%%%%%%%%%%%%%%%%%%%%%%%%%%%%%%%%%%%%%%%%%%%%%%%%%%%%%%%%

\section{Braneworld holography and quantum black holes} \label{sec:BWholo}

The perturbative analysis above suggests semi-classical backreaction due to quantum fields can lead to the appearance of a black hole horizon when there was none before. Due to the limitations of the perturbative approach, however, the observation is cursory at best. Since the perturbative correction to the classical geometry is on the order of the Planck scale we cannot definitively argue a black horizon appears. Only if there are a large-$c$ number of quantum fields present would this conclusion be plausible. The only known framework where a solution with these requirements can be consistently achieved  is braneworld holography, where one innately works in a large-$c$ limit. Below we summarize the relevant aspects of holographic braneworlds. 

\subsection{AdS/CFT dictionary and holographic renormalization}

The AdS/CFT correspondence \cite{Maldacena:1997re}, in its strongest form, describes a duality between a theory of gravity and conformal field theory at the level of their partition functions, summarized by Gubser, Klebanov, Polyakov and Witten (GKPW) \cite{Gubser:1998bc,Witten:1998qj}
\beq \biggr\langle e^{-\int_{\partial\mathcal{M}}\mathcal{O}\phi_{(0)}}\biggr\rangle_{\hspace{-1mm}\text{CFT}}=Z_{\text{grav}}[\phi_{(0)}]|_{\mathcal{M}}\;.\label{eq:GKPWdiction}\eeq
On the right-hand side we have the gravitational partition function of a bulk field $\Phi$ over an asymptotically $d+1$-dimensional AdS spacetime $\mathcal{M}$, with conformal boundary $\partial\mathcal{M}$, and $\phi_{(0)}$ is the fixed boundary value of the bulk field $\Phi$. On the left-hand side is the generating functional for the dual $d$-dimensional CFT living on $\partial\mathcal{M}$, where $\mathcal{O}$ is the field theory operator dual to the bulk field. Taking variations with respect to $\phi_{(0)}$ and then setting $\phi_{(0)}=0$, one can obtain correlation functions of $\mathcal{O}$, sourced by $\phi_{(0)}$. This equivalence of partition functions (\ref{eq:GKPWdiction}) is often dubbed the standard AdS/CFT dictionary, and, at least formally, defines a model of non-perturbative quantum gravity. 

One of the essential features of the AdS/CFT correspondence is that it can probe strongly coupled field theories on a non-dynamical background using weakly coupled, classical (super)gravity. This is because, typically, the holographic field theories are non-abelian gauge theories with gauge group of rank $N$ and 't Hooft coupling $\lambda$, where, at large-$N$ and $\lambda\gg1$ (the planar-diagram limit) the dynamics are effectively classical, with $\mathcal{O}(1/N)$ corrections in the dual field theory corresponding to bulk gravity quantum corrections $\mathcal{O}(G)$.\footnote{A concrete realization of AdS/CFT duality is that of $\mathcal{N}=4$ super-Yang-Mills theory, a superconformal field theory, which is dual to type IIB string theory on $\text{AdS}_{5}\times S^{5}$, where the 't Hooft coupling $\lambda$ controls curvature scale of $\text{AdS}_{5}$ whilst the string coupling is $g_{s}\sim N^{-1}$. In the large-$N$ limit, stringy interactions are thus suppressed and $\lambda\gg1$ forces curvatures to be small, such that the string theory may be replaced by an effectively classical gravity.} More generally, the dual field theory degrees of freedom are encoded in the central charge $c$, which, for known holographic theories scale with $N$, i.e., $c\sim N^{\alpha}$ for positive, real $\alpha$. Thus, the large-$c$ limit coincides with the classical limit\footnote{In this section we have set $\hbar=1$.} in the bulk, and the right-hand side of the dictionary (\ref{eq:GKPWdiction}) may be approximately given by a sum over classical saddles $\{\Phi_{i}\}$
\beq \lim_{c\to\infty}\biggr\langle e^{-\int_{\partial\mathcal{M}}\mathcal{O}\phi_{(0)}}\biggr\rangle_{\hspace{-1mm}\text{CFT}}=\sum_{i}e^{-I^{\text{on-shell}}_{\text{grav}}[\Phi_{i}]}\;,\label{eq:treelevelgkpw}\eeq
where each field configuration $\Phi_{i}$ is a solution to the bulk classical gravity equations of motion subject to the prescribed boundary conditions.\footnote{For bulk $\text{AdS}_{d+1}$ $\mathcal{M}$ with conformal boundary $\partial\mathcal{M}$, the $d$-dimensional field theory lives on a $d$-dimensional manifold that belongs to the conformal class of $\partial\mathcal{M}$. By choosing an appropriate conformal frame, the field theory may be placed on $\partial\mathcal{M}$. The choice of boundary metric fixes the boundary condition the bulk saddle-point geometry must obey and amounts to fixing the non-normalizable mode of the bulk graviton.} A particular case of interest is to turn off all sources $\phi_{(0)}$ except those the boundary value of the bulk metric. In such an event, at large-$c$, it is consistent to turn off all bulk fields except the metric, such that the bulk is described by a pure theory of gravity, often taken to be the Einstein-Hilbert action.

\subsubsection*{Holographic renormalization}\label{sec:holoren}

The standard dictionary (\ref{eq:GKPWdiction}), however,  requires special care in regards to divergences. Indeed, even at tree-level (\ref{eq:treelevelgkpw}), the gravity partition function exhibits long distance infrared (IR) divergences, which correspond to ultraviolet (UV) divergences in the CFT correlation functions. These divergences may be removed via holographic renormalization, a prescription that adds appropriate local counterterms \cite{Kraus:1999di,Emparan:1999pm,deHaro:2000vlm,Skenderis:2002wp,Papadimitriou:2004ap} in a minimal subtraction scheme. Since they will become relevant momentarily, let us outline the holographic renormalization procedure, leaving further computational details for Appendix \ref{app:holoren}. 

Consider a bulk asymptotically $\text{AdS}_{d+1}$ spacetime $\mathcal{M}$ of curvature scale $L_{d+1}$ and cosmological constant $\Lambda_{d+1}=-d(d-1)/2L_{d+1}^{2}$, governed by classical Einstein gravity  
\beq
I_{\text{bulk}}=\frac{1}{16\pi G_{d+1}}\int_{\mathcal{M}} d^{d+1}x\sqrt{-\hat{g}}\left(\hat{R}-2\Lambda_{d+1}\right)+\frac{1}{8\pi G_{d+1}}\int_{\partial\mathcal{M}}d^{d}x\sqrt{-h}K\;.
\label{eq:BulkTheory}\eeq
Here $G_{d+1}$ is the $d+1$-dimensional Newton's constant, $\hat{g}_{ab}$ is the metric endowed on $\mathcal{M}$, and $K$ in the Gibbons-Hawking York (GHY) boundary term is the trace of the extrinsic curvature of the boundary submanifold $\partial\mathcal{M}$ endowed with induced metric $h_{ij}$. Working in the large-$c$, planar-diagram limit, the bulk gravity theory has a dual holographic description in terms of a $\text{CFT}_{d}$ living on the asymptotic conformal boundary $\partial\mathcal{M}$. 

Asymptotically, the bulk AdS spacetime can be cast in Fefferman-Graham gauge \cite{Fefferman1985,Fefferman:2007rka}, such that near the boundary
\beq ds^{2}=\hat{g}_{ab}dx^{a}dx^{b}=L_{d+1}^{2}\left(\frac{d\rho^{2}}{4\rho^{2}}+\frac{1}{\rho}g_{ij}(x,\rho)dx^{i}dx^{j}\right)\;,\label{eq:FGexpand}\eeq
where the $d$-dimensional metric  has the expansion $g_{ij}(x,\rho)=g^{(0)}_{ij}(x)+\rho g^{(2)}_{ij}(x)+...+\rho^{d/2}g^{(d)}_{ij}(x)$. The conformal boundary is located at $\rho=0$. By perturbatively solving the bulk Einstein's equations, the higher-order metric coefficients $g^{(k>0)}_{ij}(x)$ may be cast covariantly in terms of the metric $g_{ij}^{(0)}$ and derivatives thereof.

\begin{figure}[t!]
\centering
 \includegraphics[width=6cm]{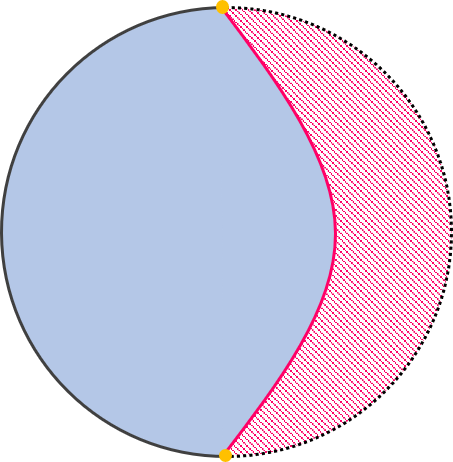}
\put(-63,35){ \large{$\rho=\epsilon$}}
\put(-120,80){ \Large{$\text{AdS}_{d+1}$}}
\caption{\small \textbf{Holographic regularization}. A constant timeslice of empty AdS$_{d+1}$. An IR cutoff surface is introduced at $\rho=\epsilon$ (thick, red line). The regulated action follows from integrating out the bulk radial coordinate from $\epsilon<\rho<\rho_{c}$. As $\epsilon\to0$, the cutoff surface recedes to the AdS boundary. }
\label{fig:holorenorm}\end{figure}

On-shell, the bulk action (\ref{eq:BulkTheory}) has IR divergences at $\rho=0$. To isolate and regulate these divergences, introduce an IR cutoff $\rho=\epsilon$, for $\epsilon\ll1$, near the asymptotic boundary, and integrate over bulk coordinate $\rho$ between $\epsilon<\rho<\rho_{c}$, where $\rho_{c}>\epsilon$ is some constant.
%\footnote{In principle, one would like to integrate $\epsilon<\rho<\infty$, where $\rho=\infty$ is the other side of the asymptotic boundary. However, for $d>2$, the entire analysis takes place near the $\rho=0$ region and breaks down far into the bulk to the $\rho=\infty$ region. Indeed, the expansion (\ref{eq:FGexpand}) is valid near the $\rho=0$ boundary. When $d=2$, however, the three-dimensional Weyl tensor is identically zero everywhere, such that the perturbative expansion truncates and the $\rho$-integration can be carried out explicitly, e.g., \cite{deHaro:2000vlm,Skenderis:1999nb}.} 
See Figure \ref{fig:holorenorm} for an illustration. This procedure produces a regulated bulk action, 
\beq I^{\text{reg}}_{\text{bulk}}=\frac{1}{16\pi G_{d+1}}\left[\int_{\rho>\epsilon}d^{d+1}x\sqrt{-\hat{g}}\left(\hat{R}-2\Lambda_{d+1}\right)+2\int_{\rho=\epsilon}d^{d}x\sqrt{-h}K\right]\;.\label{eq:regaction}\eeq
Using the perturbative expansion for $g_{ij}(x,\rho)$, the regulated action (\ref{eq:regaction}) may be divided into a contribution $I_{\text{div}}$ which diverges in the limit $\epsilon\to0$, and a finite contribution $I_{\text{fin}}$ 
\beq I^{\text{reg}}_{\text{bulk}}=I_{\text{div}}+I_{\text{fin}}\;.\eeq
Schematically, the IR divergent contribution is (see Appendix \ref{app:holoren} for details)
\beq I_{\text{div}}=\frac{L_{d+1}}{16\pi G_{d+1}}\int d^{d}x\sqrt{g_{(0)}}\left[\epsilon^{-d/2}a_{(0)}+\epsilon^{-d/2+1}a_{(2)}+...+\epsilon^{-1}a_{(d-2)}-\log\epsilon a_{(d)}\right]\;,\eeq
with coefficients $a_{(0)},a_{(2)},...$ that are covariant combinations of $g^{(0)}_{ij}$ and its derivatives. In terms of the boundary metric $h_{ij}$, it may be cast as
\beq \hspace{-5mm} I_{\text{div}}\hspace{-1mm}=\hspace{-1mm}\frac{L_{d+1}}{16\pi G_{d+1}}\int_{\partial\mathcal{M}} \hspace{-4mm} d^{d}x\sqrt{-h}\left[\frac{2(d-1)}{L_{d+1}^{2}}+\frac{R}{(d-2)}+\frac{L^{2}_{d+1}}{(d-2)^{2}(d-4)}\left(\hspace{-1mm} R_{ij}^{2}-\frac{dR^{2}}{4(d-1)}\right)\hspace{-1mm}+...\right]\label{eq:Idivv2}\eeq
where the ellipsis indicates higher-curvature and higher-derivative contributions (see, e.g., \cite{Kraus:1999di,Papadimitriou:2004ap,Elvang:2016tzz,Bueno:2022log}).  The finite contribution $I_{\text{fin}}\sim \mathcal{O}(\epsilon^{0})+\mathcal{O}(\epsilon)...$ survives the $\epsilon\to0$ limit, though it will also typically include higher-curvature terms. Its interpretation will be given momentarily.

At this stage the renormalized action is obtained by minimal subtraction, 
\beq I^{\text{ren}}_{\text{bulk}}=\lim_{\epsilon\to0}(I^{\text{reg}}_{\text{bulk}}+I_{\text{ct}})\;,\eeq
where a local counterterm action has been introduced, $I_{\text{ct}}=-I_{\text{div}}$, to precisely cancel the IR divergences. Then, via the standard AdS/CFT dictionary, variations with respect to the metric $h_{ij}$ of the renormalized action yields the quantum expectation value of the stress-tensor of the holographic CFT, 
\beq \langle T_{ij}^{\text{CFT}}\rangle=\lim_{\epsilon\to0}\left(-\frac{2}{\sqrt{\hat{g}(x,\rho)}}\frac{\delta I^{\text{ren}}_{\text{bulk}}}{\delta \hat{g}^{ij}(x,\epsilon)}\right)\equiv-\frac{2}{\sqrt{h}}\frac{\delta W_{\text{CFT}}[h]}{\delta h^{ij}}\;,\eeq
such that the renormalized bulk action is identified with the quantum effective action of the CFT, $W_{\text{CFT}}[h]$. Thus, at leading order, the finite action $I_{\text{fin}}$ characterizes the CFT.

\subsection{Braneworld holography}

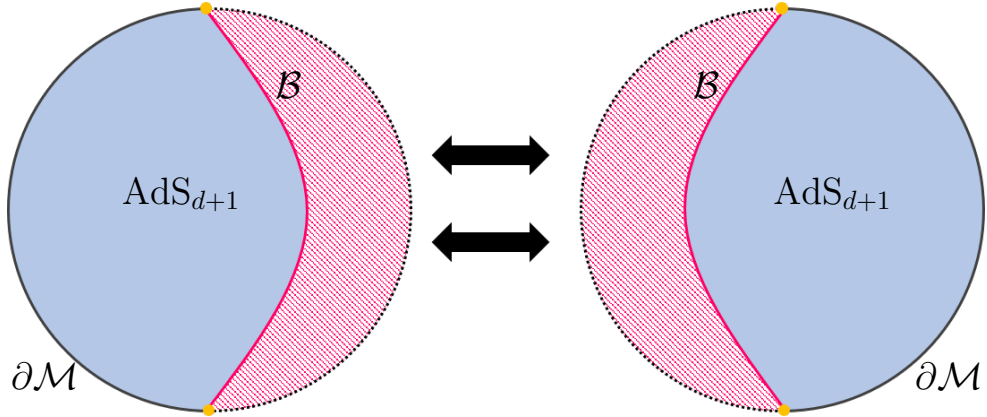
\begin{figure}[t!]
\centering
 \includegraphics[width=13cm]{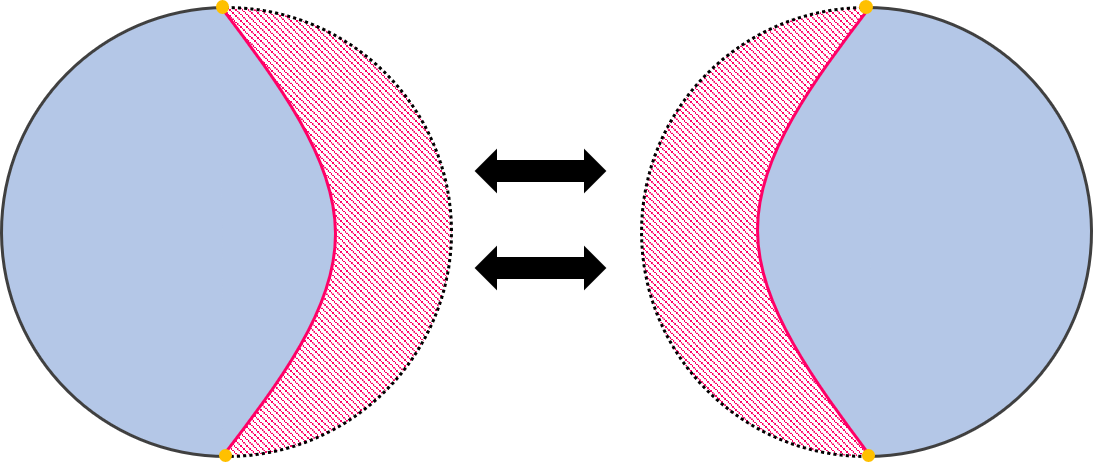}
%\put(-39,80){ $\mathcal{B}$}
\put(-330,80){ \Large{$\text{AdS}_{d+1}$}}
\put(-80,80){\Large{$\text{AdS}_{d+1}$}}
\put(-272,120){ \Large{$\mathcal{B}$}}
\put(-114,120){ \Large{$\mathcal{B}$}}
\put(-30,10){ \Large{$\partial\mathcal{M}$}}
\put(-372,10){ \Large{$\partial\mathcal{M}$}}
\caption{\small \textbf{Braneworld surgery.} Replace the IR cutoff surface with an end-of-the-world (Karch-Randall) brane $\mathcal{B}$ (thick, red line), excising the shaded region from the bulk spacetime. To complete the space, two copies of the spacetime are glued along $\mathcal{B}$ making the brane double-sided. A $\text{BCFT}_{d}$ lives on the $\text{AdS}_{d+1}$ boundary $\partial\mathcal{M}$ and is coupled to a defect $\text{CFT}_{d-1}$ where $\mathcal{B}$ intersects the AdS boundary (yellow dot). The induced brane theory is characterized by a specific higher-derivative gravity coupled to a $\text{CFT}_{d}$ with a UV cutoff.}
\label{fig:braneworldsurgery}\end{figure}

In braneworld holography \cite{deHaro:2000wj}, the bulk IR cutoff surface $\partial\mathcal{M}$ is instead replaced by a $d$-dimensional end-of-the-world (ETW)  Randall-Sundrum \cite{Randall:1999vf,Randall:1999ee} or Karch-Randall \cite{Karch:2000ct,Karch:2000gx} brane $\mathcal{B}$ at a small fixed distance away from the boundary (for a lightning review of braneworlds, see Appendix \ref{app:braneworldbasics}). Hence, the physical space is cutoff at the ETW brane and there are no longer IR divergences to be removed. For simplicity, assume the brane is purely tensional, having an action
\beq 
I_{\tau}=-\tau \int_{\mathcal{B}}d^{d}x\sqrt{-h}\,,
\label{eq:braneaction}\eeq
where $\tau$ is the brane tension. Since a portion of the bulk has been removed, to complete the space, a second copy of $\text{AdS}_{d+1}$ with a brane is sewn to the first cutoff geometry along the cutoff surface (see Figure \ref{fig:braneworldsurgery}). This surgical procedure leads to a discontinuity in the extrinsic curvature $K_{ij}$ across the junction. The Israel junction conditons \cite{Israel:1966rt} relate this discontinuity to the brane stress-tensor $S_{ij}$ via
\beq \Delta K_{ij}-h_{ij}\Delta K=8\pi G_{d+1}S_{ij}=-8\pi G_{d+1}\tau h_{ij}\;,\label{eq:israeljuncconds}\eeq
where $\Delta K_{ij}=K^{+}_{ij}-K_{ij}^{-}$ denoting the difference between the extrinsic curvature across either `$+$' and `$-$' sides of the brane (here we take $K_{ij}^{+}=-K_{ij}^{-}$ such that $\Delta K_{ij}=2K_{ij}$), and the last equality follows from taking the metric $h_{ij}$ variation of the brane action (\ref{eq:braneaction}), i.e., $S_{ij}\equiv-\frac{2}{\sqrt{-h}}\frac{\delta I_{\tau}}{\delta h^{ij}}$.  Thus, the location of the brane in the completed space is determined by the junction conditions (\ref{eq:israeljuncconds}), which in the present case amounts to tuning the brane tension $\tau$.

Moreover, unlike the metric on the AdS boundary, the brane metric is dynamical, governed by a holographically induced higher-curvature theory of gravity coupled to matter. Precisely, the induced brane theory is found by adding the bulk theory (\ref{eq:BulkTheory}) to the brane action (\ref{eq:braneaction}). Integrating out the bulk up to the ETW brane $\mathcal{B}$, as in holographic regularization, leads to an effective induced theory with action $I$
\beq
I\equiv I_{\text{Bgrav}}[\mathcal{B}]+I_{\text{CFT}}[\mathcal{B}]\,.
\label{eq:inductheorygen}\eeq
The brane gravity theory is (cf. \cite{Chen:2020uac,Bueno:2022log})
\beq 
\begin{split} 
I_{\text{Bgrav}}&=2I_{\text{div}}+I_{\tau}\\
&=\frac{1}{16\pi G_{d}}\int_{\mathcal{B}} \hspace{-1mm}d^{d}x\sqrt{-h}\biggr[R-2\Lambda_{d}+\frac{L_{d+1}^{2}}{(d-4)(d-2)}\left(R_{ij}^{2}-\frac{dR^{2}}{4(d-1)}\right)+\cdots\biggr]\;,
\end{split}
\label{eq:Ibgravgen}\eeq
where the factor of two accounts for integrating out the bulk on both sides of the brane, and the ellipsis corresponds to higher curvature densities, entering with higher powers of $L^{2}_{d+1}$. So far the higher-derivative contributions have been computed up to quintic order in curvature for arbitrary $d$, and sextic order for $d=3$ \cite{Bueno:2022log}. In principle, these results could be extended to arbitrary order, even though the calculations might be practically prohibitive. Here $G_{d}$ represents the effective brane Newton's constant induced from the bulk 
\beq G_{d}=\frac{d-2}{2L_{d+1}}G_{d+1} \,,\label{eq:effGd}\eeq
and $\Lambda_{d}=-(d-1)(d-2)/2L_{d}^{2}$ is an effective brane cosmological constant with an induced curvature scale $L_{d}$
\beq
\frac{1}{L_{d}^2}=\frac{2}{L_{d+1}^2}\left(1-\frac{4\pi G_{d+1}L_{d+1}}{d-1}\tau\right)\,.
\label{eq:curvscale}\eeq
As written, it has been assumed the brane has a negative cosmological constant such that the bulk theory is coupled to a Karch-Randall brane \cite{Karch:2000ct}. When coupled to a Randall-Sundrum brane, the brane cosmological constant can be tuned to be positive or zero, as will be considered later. 

Due to the presence of higher-derivative terms in the induced action (\ref{eq:Ibgravgen}), the brane theory of gravity is in general `massive' since a massive graviton bound state will localize on the brane \cite{Karch:2000ct}. This brane graviton mass, however, will become negligible for a brane very near the boundary. Further, general higher-derivative theories of gravity are often sick since they are typically accompanied by ghosts. In the present case, however, provided the series of higher-derivative terms is not truncated, the  brane theory is not expected to inherit these usual pathologies since the starting bulk theory and
the procedure of integrating out bulk degrees of freedom are not pathological.\footnote{See, however, \cite{Aguilar-Gutierrez:2023kfn}, where $I_{\text{Bgrav}}$ alone can generically have infinite towers of massive ghost-like gravitons, and, in particular dimensions, additional tachyonic modes or modes with complex squared-mass. The well-behaved nature of the bulk gravity suggests the induced cutoff CFT resolves these pathologies.}

The action $I_{\text{CFT}}[\mathcal{B}]$, meanwhile, describes the CFT theory, now living on the brane, and corresponds to the finite contribution to the regulated bulk action upon integrating out the bulk. To see this, note that upon integrating out the bulk degrees of freedom on both sides of the brane we have 
\beq I\equiv 2I_{\text{bulk}}^{\text{reg}}+I_{\tau}\;.\eeq
Then add and subtract the $2I_{\text{div}}$, giving
\beq I\equiv (2I_{\text{div}}+I_{\tau})+(2I_{\text{bulk}}^{\text{reg}}-2I_{\text{div}})\;,\eeq
where the first term in parentheses is recognized as $I_{\text{Bgrav}}$ (\ref{eq:Ibgravgen}). The second term is simply $2I_{\text{fin}}\equiv I_{\text{CFT}}$, which, to leading order in the cutoff $\epsilon$, is identified with the quantum effective action of the CFT, $I_{\text{CFT}}=W_{\text{CFT}}+\mathcal{O}(\epsilon)$. In most cases of interest, we work in the limit that $\epsilon$ is small, i.e., when the brane is close to the (now fictitious) $\text{AdS}_{d+1}$ boundary, such that the matter on the brane has an approximate description as a large-$c$ holographic CFT. Roughly speaking, a portion of the conformal $\text{AdS}_{d+1}$ boundary has been pushed into the bulk, such that the dual $\text{CFT}_{d}$ is now residing on the brane -- however, at a cost. Since the brane represents an IR cutoff surface, the CFT has a UV cutoff \cite{Emparan:2006ni,Myers:2013lva}. The cutoff, from the perspective of the boundary $g^{(0)}_{ij}$ metric, is denoted by $\epsilon$, while from the induced brane metric $h_{ij}=(L_{d+1}^{2}/\epsilon) g^{(0)}_{ij}$ the UV cutoff of the CFT is $\delta_{\text{UV}}=L_{d+1}$.

\subsection*{Double holography}

A Karch-Randall braneworld\footnote{A notion of double holography was proposed in a Randall-Sundrum set-up \cite{Emparan:2022ijy}, where the analog of defect $\text{CFT}_{d-1}$ is given by two Euclidean CFTs, disconnected from the boundary viewpoint.} has three equivalent descriptions: (i) \emph{bulk}, (ii) \emph{intermediate}, and (iii) \emph{boundary}. 

\begin{itemize}
    \item \textbf{Bulk:} The bulk perspective is that of classical dynamical gravity in $\text{AdS}_{d+1}$ coupled to an asymptotically $\text{AdS}_{d}$  ETW brane of tension $\tau$. The simplest set-up assumes Einstein gravity plus a purely tensional brane, however, it is in principle possible to include higher-curvature corrections or fields to the bulk or brane actions. Israel-junction conditions determine the location of the brane in the bulk, such that, for a purely tensional brane, tuning the tension constitutes changing the position of the brane. 
    \item \textbf{Intermediate:} The intermediate brane viewpoint describes induced dynamical gravity coupled to a UV cutoff $\text{CFT}_{d}$ UV, which further communicates with a boundary CFT$_{d}$ (BCFT$_{d}$) via transparent boundary conditions. Bulk graviton fluctuations localize on the brane \cite{Randall:1999ee,Randall:1999vf}. A subset of these graviton modes are light states with mass controlled by the tension; hence, the induced brane theory is an example of a massive theory of gravity. The remaining bulk graviton modes appear as a tower of Kaluza-Klein modes with masses set by the effective $\text{AdS}_{d}$ brane length scale $1/L_{d}$.
    \item \textbf{Boundary:} Holographically,  the bulk system has a dual description in terms of a $\text{CFT}_{d}$ with a boundary (where the brane intersects the $\text{AdS}_{d+1}$ boundary), i.e., a boundary $\text{CFT}_{d}$. This set-up constitutes AdS/BCFT \cite{Takayanagi:2011zk,Fujita:2011fp}. This perspective emerges when the brane gravity itself has a dual description in terms of a $(d-1)$-dimensional conformal defect. Upon replacing the brane gravity by a conformal defect (by integrating out the bulk and brane), the boundary perspective is characterized by the $\text{CFT}_{d}$ on a fixed background, coupled to the defect. The boundary perspective is thus a UV/microscopic description of the bulk/brane gravity viewpoints.
\end{itemize}

\noindent Specific models exhibiting this type of `double holography' -- holographic spacetimes dual to a BCFT that have three equivalent descriptions -- have top-down string theoretic realizations \cite{Karch:2022rvr}. We, however,  will work with bottom-up constructions, for which their doubly-holographic nature will play an important role in studying aspects of holographic entanglement and complexity, as we summarize in Section \ref{sec:applications}.\footnote{Concerns that double-holography leads to superluminal signalling \cite{Omiya:2021olc} are ameliorated when the brane description is treated as an effective theory, such that ETW brane models are consistent with causality \cite{Neuenfeld:2023svs}.}

%The bulk perspective is that of dynamical gravity in $\text{AdS}_{d+1}$ coupled to an asymptotically $\text{AdS}_{d}$  ETW brane. Holographically, this bulk system has a dual description in terms of a $\text{CFT}_{d}$ with a boundary (where the brane intersects the $\text{AdS}_{d+1}$ boundary), i.e., a boundary $\text{CFT}_{d}$ ($\text{BCFT}_{d}$). This holographic set-up constitutes AdS/BCFT \cite{Takayanagi:2011zk,Fujita:2011fp}. Second, the intermediate brane viewpoint describes induced dynamical gravity coupled to a $\text{CFT}_{d}$ on the ETW brane, which further communicates with the $\text{BCFT}_{d}$ via transparent boundary conditions. A third viewpoint emerges when the brane gravity itself has a dual description in terms of a $(d-1)$-dimensional conformal defect. Upon replacing the brane gravity by a conformal defect, the boundary perspective is characterized by the $\text{CFT}_{d}$ coupled to the defect. Notably, specific models exhibiting this type of double holography have top-down string theoretic realizations \cite{Karch:2022rvr}. The doubly-holographic nature of some of these braneworld set-ups will play an important role in studying aspects of holographic entanglement and complexity, as we summarize in Section \ref{sec:applications}. 

\subsection{Holographic quantum black holes: a conjecture} \label{ssec:holoqbhsconj}

Two equivalent ways to interpret the theory (\ref{eq:inductheorygen}) are as follows. From the bulk perspective, $I$ characterizes a theory of a $(d+1)$-dimensional system with dynamics ruled by Einstein gravity coupled to an end-of-the-world brane obeying appropriate boundary conditions. Meanwhile, from the (intermediate) brane perspective $I$ represents a specific  higher-curvature gravity in $d$ dimensions coupled to a large-$c$ cutoff CFT that backreacts on the brane metric $h_{ij}$. The tower of higher-order derivative terms to the Einstein-Hilbert contribution represent quantum-corrections induced by the backreaction of the $\text{CFT}_{d}$. We refer to this higher-derivative tower as `corrections' because in most cases of interest one treats the brane action as an effective theory, thereby assuming $L_{d}\gg L_{d+1}$ and guaranteeing the higher-derivative terms are suppressed by at least $\mathcal{O}(L^{2}_{d+1}/L_{d}^{2})$.\footnote{Equally, $L^{2}_{d+1}/L_{d}^{2}\sim \epsilon$, and thus the gravitational brane action is recognized as an expansion in small $\epsilon$. Moreover, from the brane perspective, the short-distance UV cutoff of the CFT$_{d}$ goes like $L_{d+1}$ such that the higher-derivative terms also correspond to an expansion in the UV cutoff.} Consistency between these two viewpoints implies solutions to the classical bulk equations satisfying proper brane boundary conditions exactly correspond to solutions to the semi-classical field equations on the brane.  Therefore, the classical $(d+1)$-dimensional geometry encodes the entire series of quantum-corrections to the $d$-dimensional brane geometry, accounting for all orders in backreaction. Thus, holographic braneworlds provide a distinct computational advantage: rather than directly solving a complicated semi-classical theory of gravity, one may instead solve simpler classical gravitational field equations in one dimension higher.
%\footnote{It is worth emphasizing that exact bulk solutions lead to exact solutions to the entire gravity theory on the brane, including the whole tower of higher-derivative terms. Importantly, while general higher derivative theories of gravity are pathological since they are typically accompanied with ghosts, one does not expect the brane theory to be pathological (assuming one does not truncate the series of terms) since the bulk theory and the procedure of integrating out the bulk are not pathological.}

This philosophy, combined with the observation \cite{Duff:2000mt} that the $\sim 1/r^{3}$ corrections to the four-dimensional Newtonian potential due to massive Kaluza-Klein modes in the Randall-Sundrum model precisely coincide with corrections induced by one-loop quantum effects of the graviton propagator \cite{Duff:1974ud}, suggests braneworld black holes from the brane perspective are quantum-corrected geometries. These insights in part motivated Emparan, Fabbri and Kaloper \cite{Emparan:2002px} to make the following conjecture: 

\vspace{2mm}

\noindent \textbf{Conjecture:} Classical black holes which localize on a brane in $\text{AdS}_{d+1}$ exactly map to $d$-dimensional quantum-corrected black holes including all orders of backreaction.

\vspace{2mm}

\begin{figure}[t!]
\centering
\includegraphics[trim={0.03cm 0 0.02cm 0},clip,width=11cm]{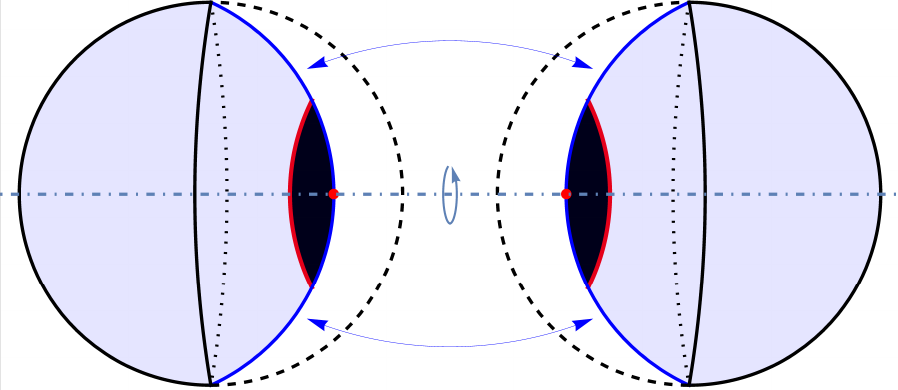}
\put(-36,80){\Large{$\partial\mathcal{M}$}}
%\put(-48,56){$\mathcal{M}$}
\put(-100,100){\Large{{\color{blue}$\mathcal{B}$}}}
\put(-222,100){\Large{{\color{blue}$\mathcal{B}$}}}
%\put(-201,56){$\mathcal{M}$}
\put(-302,80){\Large{$\partial\mathcal{M}$}}
\caption{\small \textbf{Braneworld black hole.} The bulk white region is excised down
to the brane $\mathcal{B}$ (blue line), and glued to a copy of itself. A
bulk black hole with an event horizon (red line) is intersected by (depicted here, Karch-Randall) brane, inducing a horizon on the brane.}
\label{fig:bwbh}\end{figure}
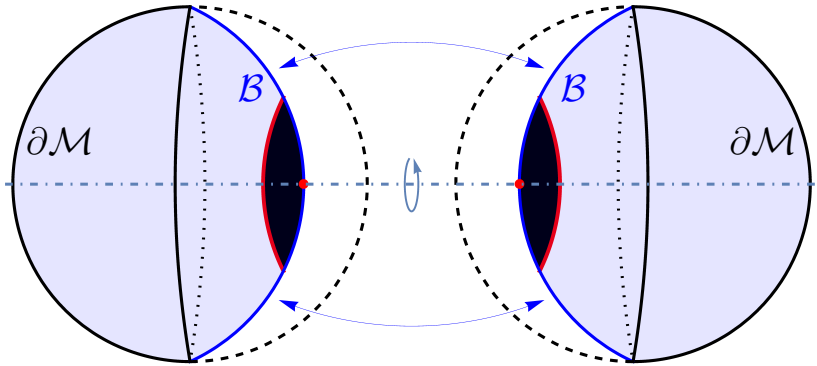

Such quantum-corrected black holes are dubbed `quantum' black holes, though, technically are solutions to the semi-classical theory induced on the brane. An illustration is given in Figure \ref{fig:bwbh}.

Explicit tests of this proposal include the exact localized $\text{AdS}_{4}$ braneworld black holes discovered by Emparan, Horowitz, and Myers \cite{Emparan:1999wa,Emparan:1999fd}, with their projection onto the brane being reinterpreted as three-dimensional quantum black holes.\footnote{Historically, the exact three-dimensional braneworld black holes  \cite{Emparan:1999wa,Emparan:1999fd}, while interpreted as holographic quantum black holes in \cite{Emparan:2002px}, the higher-derivative corrections in the
induced gravity action on the brane were not explicitly accounted for until \cite{Emparan:2002px,Emparan:2020znc,Emparan:2022ijy,Panella:2023lsi}.} As we will see below, at least for the neutral, static geometries, the exact quantum black holes receive the same modifications to their geometry as suggested by the (non-holographic) perturbative analysis summarized in Section \ref{sec:perturbanalysis}. The rotating and charged holographic quantum black holes, meanwhile, do not match to the non-holographic perturbatively-corrected counterparts (cf. \cite{Steif:1993zv,Casals:2019jfo,Panella:2023lsi}). In particular, perturbative backreaction to the rotating BTZ black hole or Kerr-$\text{dS}_{3}$ solution due to a conformally coupled scalar field leads to a wildly more complicated radial dependence than that of a holographic CFT and a quantum stress-tensor with a qualitatively different singularity structure, having an impact on the status of strong cosmic censorship. The remainder of this review will focus on these three-dimensional quantum black holes. 

Before moving on to analyze the three-dimensional black holes, it is worth briefly commenting on the status of the proposal \cite{Emparan:2002px} in higher and lower dimensions. For brane dimensions $d\geq4$, the most physically relevant case being $d=4$, there are still no known exact stationary solutions (see \cite{Tanahashi:2011xx} for a review of analytic and numerical braneworld black holes).\footnote{Alternatively, one can study topological black holes on the brane, e.g., \cite{Chen:2020uac,Chen:2020hmv}. In this context, the bulk is $d+1$-dimensional AdS-Rindler, for which the Rindler horizon induces a Rindler horizon on the brane. Since the bulk geometry is simply vacuum AdS, it evades the no-go theorem of \cite{Bruni:2001fd}.} In fact, there is a no-go theorem \cite{Bruni:2001fd} which alleges the exterior geometry on the brane in $d\geq4$ cannot be static. The lack of exact solutions makes identifying the specific state of the $\text{CFT}_{d}$ more difficult. In \cite{Emparan:2002px,Tanaka:2002rb}, for $d\geq4$, it was qualitatively argued the obstruction to having static quantum black holes can be understood as a consequence of backreaction due to Hawking effects, such that any black hole that localizes on the brane must evaporate. However, static braneworld black holes in higher-dimensions have been found numerically, e.g., \cite{Kudoh:2003xz,Kudoh:2004kf,Fitzpatrick:2006cd,Yoshino:2008rx,Figueras:2011gd,Emparan:2023dxm}, and the qualitative argument was shown to have flaws \cite{Fitzpatrick:2006cd}. We review the status of the conjecture for higher-dimensional quantum black holes in Section \ref{app:evapbhshighd}.

In $d=2$ dimensions, the induced theory on the brane is characterized by a matter quantum effective action, which to leading order in expansion in the UV cutoff is characterized by the Polyakov action \cite{Polyakov:1981rd} coupled to a topological term and a cosmological constant. The precise form of the brane action likewise follows from a modification of holographic renormalization \cite{deHaro:2000vlm,Henningson:1998gx,Skenderis:1999nb}, where the counterterm action exactly truncates (see also \cite{Carlip:2005tz,Nguyen:2021pdz}). Such a theory does not admit black hole solutions by itself. Thus, in order to find two-dimensional braneworld black holes, the theory must be modified by including a Dvali-Gabadadze-Porrati (DGP) term \cite{Dvali:2007hz} to the tensional brane action (\ref{eq:braneaction}). For example, one may replace the brane action (\ref{eq:braneaction}) with a non-minimally coupled dilaton theory, e.g., Jackiw-Teitelboim (JT) gravity \cite{Jackiw:1984je,Teitelboim:1983ux}, as done in \cite{Chen:2020uac}.\footnote{There are other mechanisms to induce two-dimensional dilaton gravity on the brane without explicitly introducing a DGP term. These include, for example, using `wedge holography' to uncover AdS- or dS-JT gravity \cite{Geng:2022tfc,Geng:2022slq,Bhattacharjee:2022pcb,Aguilar-Gutierrez:2023tic}, or via the holography of a deformed braneworld \cite{Neuenfeld:2024gta}, which yields a host of dilaton-gravity models on a two-dimensional brane.} Such dilaton models of gravity admit exactly solvable black hole solutions including semi-classical backreaction, e.g., \cite{Callan:1992rs,Russo:1992ax,Susskind:1993if,Fiola:1994ir}.

Lastly, let us make some general remarks about static black holes localized on the brane. First, a brane with non-vanishing tension is an accelerated trajectory with respect to the bulk, i.e., the brane does not undergo geodesic motion. Thus, a black hole which localizes on the brane is in an accelerating frame; likewise for any observer glued to the brane. Next, a static black hole stuck to the brane will neither eat the brane or slide off it. The reason is as follows \cite{Emparan:2000rs}. To be static, the brane intersects the black hole orthogonally, otherwise the black hole would grow by eating the brane.\footnote{Indeed, a black hole will grow if $T_{ij}k^{i}k^{j}>0$ in the background, for null generator of the horizon $k^{i}$. A black hole thus remains static when $T_{ij}k^{i}k^{j}=0$. Since the brane stress tensor is proportional to the induced metric, the static condition translates to $k^{i}k_{i}=0$, i.e., the $k^{i}$ lies entirely on the brane, which occurs when the radial direction orthogonal to the black hole is tangent to the brane.} Consequently, the brane bends to remain orthogonal to the black hole if the latter is being pulled off the brane (by, say, another black hole in the bulk). Thus, a static black hole localized on the brane experiences a restoring force due to the tension of the brane and does not slide off. Evaporating black holes, on the other hand, eventually slide off the brane.

%%%%%%%%%%%%%%%%%%%%%%%%%%%%%%%%%%%%%%%%%%%%%%%%%%%%%%%%%%%

\section{Quantum black hole taxonomy} \label{sec:qbhtaxonomy}

Having laid the groundwork of braneworld holography in general dimensions, here we focus on the $d=3$ case, i.e., a holographic $\text{AdS}_{4}$ bulk with a three-dimensional ETW brane. In this set-up it is possible to write down a host of analytic braneworld black hole solutions to the bulk field equations plus Israel junction conditions. From the brane perspective, these black holes may be interpreted as `quantum' black holes -- those which exactly solve the induced semi-classical brane gravity including all orders of backreaction due to a large-$c$ holographic $\text{CFT}_{3}$ with a UV cutoff. Below we classify known quantum black holes in three-dimensional (A)dS and Minkowski backgrounds. These include the quantum BTZ family of black holes (static, rotating, and charged), and their de Sitter/Minkowski analogs. Each of these braneworld black holes arise from specific parametrizations of the AdS$_{4}$ C-metric coupled to a Randall-Sundrum or Karch-Randall brane.  

\subsection{Bulk geometry: AdS C-metric} \label{ssec:bulkgeomcmet}

In this review, the bulk $\text{AdS}_{4}$ gravity is taken to be Einstein-Maxwell theory. The most general metric solving the Einstein-Maxwell$-\Lambda$ gravity is the Plebia\'nski-Demia\'nski type-D metric \cite{Plebanski:1976gy}. We are interested in a specific sub-class of solutions, namely the AdS$_{4}$ C-metric, which can be interpreted as a single or pair of accelerating black holes in an $\text{AdS}_{4}$ background, depending on the parameters of the solution.  Here we summarize the essentials for the braneworld construction (see Appendix \ref{app:AdSCmetprops} for a longer review). In particular, the neutral, non-rotating C-metric in Boyer-Lindquist-like coordinates has line element (primarily following the conventions of \cite{Emparan:2020znc})
\beq
ds^{2}=\frac{\ell^{2}}{(\ell+xr)^{2}}\left[-H(r)dt^{2}+\frac{dr^{2}}{H(r)}+r^{2}\left(\frac{dx^{2}}{G(x)}+G(x)d\phi^{2}\right)\right]\;,
\label{eq:AdS4cmetBLstat}\eeq
with metric functions
\beq
H(r)= \frac{r^{2}}{\ell_{3}^{2}}+\kappa-\frac{\mu\ell}{r} \;,\qquad G(x)=1-\kappa x^{2}-\mu x^{3}\;.
\label{eq:metfuncsstatCmet}\eeq
We treat $t$ and $r$ as time and radial coordinates, respectively. However, the range for $r$ is not the usual one for Boyer-Lindquist coordinates, with the $\text{AdS}_{4}$ conformal boundary shifted from its familiar location, $r=\infty$. Rather, the position of the boundary depends on the coordinate $x$: for some values of $x$, the conformal boundary is closer than infinity, while for other values of $x$ the boundary is `further' than infinity. That is, the radial coordinate has range $r\in(-\infty,r_{\text{bdry}})\cup(0,\infty)$, with $x r_{\text{bdry}}=-\ell$ being the location of the asymptotic AdS$_{4}$ boundary, where the conformal factor diverges (this unfamiliar range for the radial coordinate can be seen more readily using the $(t,x,y,\phi)$ coordinates in Appendix \ref{app:AdSCmetprops}). Further, $(x,\phi)$ are angular variables, with $x\in[-1,1]$, analogous to $\cos\theta$, and $\phi$ is a general azimuthal coordinate whose periodicity will be discussed below.

Here $\mu$, $\kappa$, $\ell$ and $\ell_3$, are real parameters characterizing the solution whose physical meaning will become more apparent momentarily. For now, $\kappa=\pm1,0$ describes different possible slicings of the brane geometry, e.g., $\kappa=-1$ will recover the classical BTZ geometry on the brane, however, here we leave $\kappa$ unspecified, thereby describing a family of braneworld black holes. The non-negative parameter $\mu$ is related to the mass of the bulk black hole.
When interpreted as an accelerating black hole, the parameter $\ell\geq0$ equals the inverse acceleration, $A=\ell^{-1}$, and is related to the bulk $\text{AdS}_{4}$ length scale $L_{4}$ via 
\beq
L_4 = \left(\frac{1}{\ell^2}+\frac{1}{\ell_3^2}\right)^{-1/2} \;,
\label{eq:bulkAdS4length}\eeq
Lastly, the positive parameter $\ell_{3}$ will be related to the $\text{AdS}_{3}$ brane curvature scale, however, with $0\leq \ell<\infty$, such that $\ell_{3}>L_{4}$.

%Here we have chosen to parametrise the bulk metric via $\mu$, $\kappa$, $\ell$ and $\ell_3$, following primarily the convention in \cite{Emparan:2020znc}. The C-metric is well known to describe accelerating black holes in AdS$_4$. Even though the parametrisation has been adapted to be easier to interpret from the brane perspective rather than from the bulk one, it is useful to understand what properties of the bulk theory these quantities represent. 

To gain further intuition for the global aspects of the metric (\ref{eq:AdS4cmetBLstat}), first set $\mu=0$ and perform the coordinate transformation \cite{Emparan:2020znc}
\beq \cosh(\sigma)=\frac{\ell_{3}}{L_{4}}\frac{1}{|1+\frac{rx}{\ell}|}\sqrt{1+\frac{r^{2}x^{2}}{\ell_{3}^{2}}}\;,\qquad \hat{r}=r\sqrt{\frac{1-\kappa x^{2}}{1+\frac{r^{2}x^{2}}{\ell_{3}^{2}}}}\;.\label{eq:coordinatetransflat}\eeq
This brings the C-metric line element (\ref{eq:AdS4cmetBLstat}) to the form
\beq ds^{2}=L_{4}^{2}d\sigma^{2}+\frac{L_{4}^{2}}{\ell_{3}^{2}}\cosh^{2}(\sigma)\left[-\left(\kappa+\frac{\hat{r}^{2}}{\ell_{3}^{2}}\right)dt^{2}+\left(\kappa+\frac{\hat{r}^{2}}{\ell_{3}^{2}}\right)^{-1}d\hat{r}^{2}+\hat{r}^{2}d\phi^{2}\right]\;,\label{eq:AdS4empty}\eeq
which is recognized to be empty $\text{AdS}_{4}$, foliated by $\text{AdS}_{3}$ slices with radius $L_{4}\cosh(\sigma)$ at constant $\sigma$. Written this way, the role of $\kappa$ becomes apparent: for constant $\sigma$, $\kappa=+1,-1$ and $0$ respectively correspond to global $\text{AdS}_{3}$, BTZ and Poincar\'e $\text{AdS}_{3}$.

To get a better sense for the parameter $\ell$ and why the geometry (\ref{eq:AdS4cmetBLstat}) describes a black hole, perform the parameter and coordinate rescalings $\mu=2m/\ell$ and $r=\ell_{3}\rho/\ell$. Keeping $m,L_{4}$, and $\rho$ finite, the limit $\ell\to\infty$ results in 
\beq ds^{2}=-f(\rho)dt^{2}+f^{-1}(\rho)d\rho^{2}+\rho^{2}\left(\frac{dx^{2}}{(1-\kappa x^{2})}+(1-\kappa x^{2})d\phi^{2}\right)\;,\quad f(\rho)=\kappa+\frac{\rho^{2}}{L_{4}^{2}}-\frac{2m}{\rho}\;,\label{eq:limitgeom}\eeq
the metric for static, neutral $\text{AdS}_{4}$ black holes (further clarified upon transforming $x=\cos\theta$). Since $\ell^{-1}=A$, the limiting geometry (\ref{eq:limitgeom}) shows the acceleration distorts spherical surfaces parametrized by $(x,\phi)$.

\subsubsection*{Horizons and bulk regularity}

The limiting geometry (\ref{eq:limitgeom}) also shows black holes appear for $\mu\neq0$. More generally, whether the C-metric (\ref{eq:AdS4cmetBLstat}) has a black hole depends on the root structure of the metric functions (\ref{eq:metfuncsstatCmet}). In particular, the roots of $H(r)$ correspond to Killing horizons generated by the time translation Killing vector $\partial_{t}$, where we desire for positive roots if we are to describe
physical black hole horizons. Since the system is accelerating, it will also have non-compact Rindler-like acceleration horizons. To rid ourselves of the acceleration horizon, it is sufficient to work in the case where $\ell>L_{4}$. With this restriction, $H(r)$ has a single positive root $r_{+}$ representing a black hole horizon,\footnote{To see this, rearrange the bulk length scale (\ref{eq:bulkAdS4length}) to find $\ell_{3}^{-2}=\frac{1}{L_{4}^{2}}-\frac{1}{\ell^{2}}$ and $H(r)=\kappa+\frac{r^{2}}{L_{4}^{2}}-\frac{r^{2}}{\ell^{2}}-\frac{\mu\ell}{r}$. It follows the only positive real root to the cubic $H(r)=0$ occurs when $\ell>L_{4}$. Moreover, cast in this way, it is clear the acceleration and negative curvature of $\text{AdS}_{4}$ counteract one another such that the restriction $\ell>L_{4}$ effectively removes the acceleration horizon.} and the C-metric describes a  single `slowly accelerating' black hole suspended away from the center of $\text{AdS}_{4}$ by a cosmic string attached at the horizon \cite{Podolsky:2002nk}. As with other spherical surfaces, the acceleration distorts the horizon into a conical shape.  

Real roots of $G(x)$, meanwhile, correspond to symmetry axes of the Killing vector $\xi^{a}=\partial^{a}_{\phi}$, i.e., $\xi^{2}\sim G(x)$, vanishing at a zero of $G(x)$. These roots characterize the geometry of the horizon in the bulk. 
%For instance, a surface of constant $r$ with $0\leq x\leq x_1$ is a (distorted) half-sphere with disk topology.
To ensure a finite black hole horizon in the bulk, one must be in the parameter space where there exists at least one positive root of $G(x)$, the smallest of which will be denoted $x_{1}$, and then work in the restricted range $0\leq x\leq x_{1}$.
%Then, a surface of constant $r$ with $0\leq x\leq x_1$ is a (distorted) half-sphere with disk topology. 
The general strategy is the following \cite{Emparan:1999wa,Emparan:1999fd}. Treat the root $x_{1}$ as a primary parameter while $\mu$ is derived from $x_{1}$ via $G(x_{1})=0$, i.e., 
\beq \mu=\frac{1-\kappa x_{1}^{2}}{x^{3}_{1}}\;.\label{eq:defmu}\eeq
The desired parameter range follows from taking $x_{1}>0$ and
\beq x_{1}\in(0,1]\quad \text{for}\;\;\kappa=+1\;,\label{eq:paramrangex1v1}\eeq
\beq x_{1}\in(0,\infty)\quad \text{for}\;\;\kappa=-1,0\;.\eeq
 Observe $\mu$ monotonically decreases from $+\infty$ to zero, where $\mu=0$ coincides with the upper limit of $x_{1}$, e.g., when $\kappa=+1$, $\mu\to0$ as $x_{1}\to 1$.
%Later we will see that when $\ell\neq 0$ the allowed value of $\mu$ will be limited above if we want to have a regular black hole horizon.

%Our strategy is to first consider the roots of $G(x)$, where we look for at least one real root. This can be established by specifying a range for the parameter $\mu$ \cite{Emparan:1999wa,Emparan:1999fd}. When $\mu>0$ there will be one positive root, denoted by $x_{1}$, and we restrict the range of $x$ such that $0\leq x\leq x_{1}$. 

\begin{figure}[t!]
\centering
 \includegraphics[width=7cm]{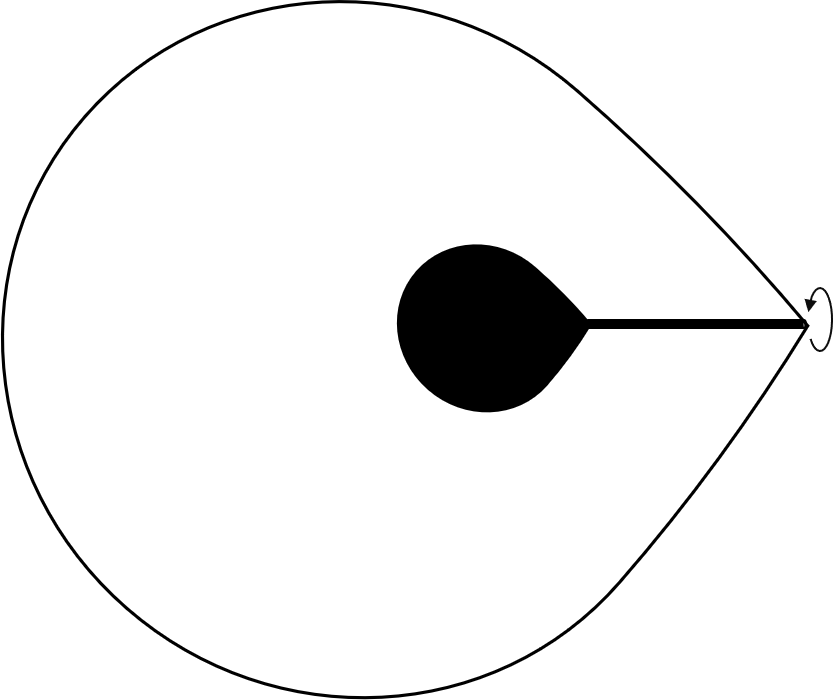}
%\put(-270,-8){\footnotesize $(\mu=0)$}
\put(-1,87){ $\phi$}
\caption{\small \textbf{Accelerating black hole}. A constant $t$ and $\phi$ slice of the $\text{AdS}_{4}$ C-metric in the `slow acceleration' limit. Conical singularities in the C-metric distort the black hole horizon, giving it a conical shape at one pole where a cosmic string is attached. The string pulls the black hole toward the conformal boundary, suspending it away from the center.}
\label{fig:accelbh}\end{figure}

Further, for the range of $\mu$ and $\kappa$ we are interested in, $G(x)$ has three distinct zeros, each of which leads to a distinct conical singularity. The conical singularity at $x=x_{1}$, for example, is removed via the identification
\beq \phi\sim\phi+\Delta\phi\;,\qquad \Delta\phi=\frac{4\pi}{|G'(x_{1})|}=\frac{4\pi x_{1}}{3-\kappa x_{1}^{2}}\;.\label{eq:conicaldef}\eeq
We see for the range of $x_{1}$, the function $G'(x_{1})=-\frac{3-\kappa x_{1}^{2}}{x_{1}}<0$, and that $\Delta\phi$ is independent of $\ell$ and $\ell_{3}$.  Moreover, $\Delta\phi$ grows monotonically from $0$ to $2\pi$. Fixing the period of $\phi$ in this way, the spacetime will have conical singularities at the remaining roots of $G(x)$. The effect of these conical singularities is a distortion to the black hole horizon, which may be viewed as a cosmic string with a tension proportional to the angular deficit pulling the black hole away from the center of AdS$_{4}$ toward the boundary, generating the black hole acceleration (see Figure \ref{fig:accelbh}). Below we will see how the spacetime surgery used to construct a $\mathbb{Z}_{2}$ braneworld will eliminate these conical singularities from the surgically complete bulk geometry.

\subsubsection*{Karch-Randall braneworld construction}

The most advantageous geometric feature of the C-metric (\ref{eq:AdS4cmetBLstat}), for any $\mu$ and $\kappa$, is that the $x=0$ timelike hypersurface is \emph{umbilic}. That is, the extrinsic curvature $K_{ij}$ of the hypersurface is proportional to the induced metric $h_{ij}$ on $x=0$. In particular, the outward unit normal to timelike surfaces of constant $x$ is $n^{i}=-\left(\frac{x}{\ell}+\frac{1}{r}\right)\sqrt{G(x)}\partial^{i}_{x}$, such that\footnote{Our conventions for the sign of the normal vector and junction conditions match with \cite{Emparan:1999wa}. These differ from \cite{Emparan:2020znc}, where the unit normal is $n^{i}=\left(\frac{x}{\ell}+\frac{1}{r}\right)\sqrt{G(x)}\partial^{i}_{x}$ such that $K_{ij}=-\ell^{-1}h_{ij}$, and where the junction conditions are $2(K_{ij}-h_{ij}K)=-8\pi G_{4}S_{ij}=8\pi G_{4}\tau h_{ij}$, resulting in the same tension (\ref{eq:branetension}).} (see Appendix \ref{app:propsofCmet})
\beq K_{ij}=\frac{1}{\ell}h_{ij}\;,\label{eq:umbiliccondx0}\eeq
at $x=0$. Umbilic surfaces automatically satisfy the Israel-junction conditions (\ref{eq:israeljuncconds}). In this case, upon substituting (\ref{eq:umbiliccondx0}), a brane at $x=0$ has tension 
\beq \tau=\frac{1}{2\pi G_{4}\ell}\;.\label{eq:branetension}\eeq
Thus, the tension is proportional to the acceleration of the bulk black hole. The tensionless limit corresponds to $\ell\to\infty$, where $\ell_{3}\to L_{4}$ by virtue of (\ref{eq:bulkAdS4length}).

\begin{figure}[t!]
\centering
 \includegraphics[width=12.3cm]{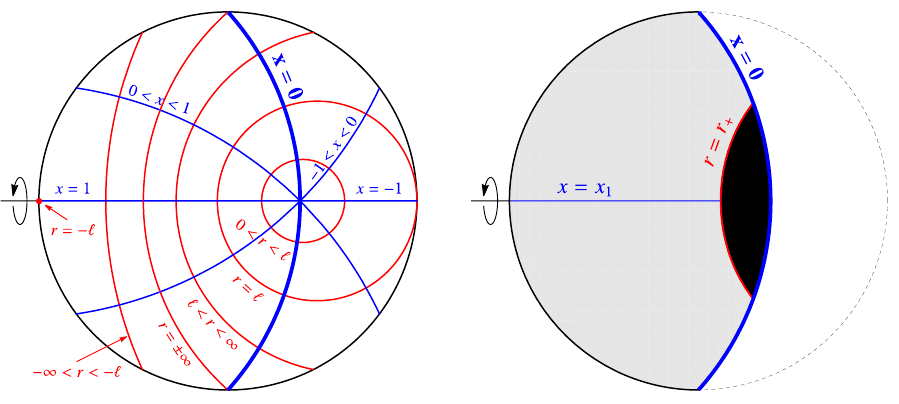}
\put(-270,-8){\footnotesize $(\mu=0)$}
\put(-90,-8){\footnotesize $(\mu\neq0)$}
\caption{\small \textbf{Karch-Randall braneworld}. \emph{Left:} A constant $t$ and $\phi$ slice of the $\text{AdS}_{4}$ C-metric with $\mu=0$ and $\kappa=+1$ (Poincar\'e disc). Lines of constant $x$ are denoted in blue, while lines of constant $r$ are denoted in red. The $\phi$-axis of rotation is at $x=\pm1$. \emph{Right:} Schematic of a Karch-Randall ETW brane at $x=0$ with a static black hole ($\mu\neq0$). Only the (gray) shaded region, $0\leq x\leq x_{1}$ kept where $x=x_{1}$ is the $\phi$-axis of rotation. To complete the space, a second copy of the shaded region is glued along $x=0$, resulting in a $\mathbb{Z}_{2}$-symmetric double-sided braneworld.}
\label{fig:KRbrane}\end{figure}

To see how tuning the tension amounts to changing the position of the brane, recall the empty $\text{AdS}_{4}$ metric (\ref{eq:AdS4empty}). The $x=0$ brane amounts to surfaces of fixed $\sigma=\sigma_{b}$ obeying
\beq
\cosh(\sigma_b) =\frac{\ell_{3}}{L_{4}}= \sqrt{1+\frac{\ell_3^2}{\ell^2}}\;,
\eeq
and the brane geometry is $\text{AdS}_{3}$ with curvature radius $\ell_{3}$, i.e., a Karch-Randall brane. In the tensionless limit, $\ell\to\infty$, then $\sigma_{b}=0$, cutting the bulk in half through the equator. Alternatively, as the tension becomes larger, $\ell \to 0$, it follows $\sigma_b \to \infty$, i.e.,  the brane is pushed to the asymptotic boundary of AdS$_4$. The solution is valid for all $0\leq \ell<\infty$, however, we will be primarily interested in the case where the brane is near the $\text{AdS}_{4}$ boundary (before it has been removed via the ETW brane). See Figure \ref{fig:KRbrane} for an illustration (and refer to Appendix \ref{appsec:poincdisk} for details projecting to the Poincar\'e disk).

More generally, the induced metric at $x=0$ simply follows from setting $x=0$ in the bulk C-metric (\ref{eq:AdS4cmetBLstat}) resulting in
\beq
\label{eq:naiveBTZ}
ds^2|_{x=0} = -\left(\frac{r^{2}}{\ell_{3}^{2}}+\kappa-\frac{\mu\ell}{r} \right)dt^2 +\left(\frac{r^{2}}{\ell_{3}^{2}}+\kappa-\frac{\mu\ell}{r} \right)^{-1}  dr^2 + r^2 d\phi^2 \;,
\eeq
For $\kappa=-1$, the boundary geometry ($\ell\to0$) has a black hole with horizon radius $r_{+}=\ell_{3}$. For $\ell\neq0$, it is clear the geometry (\ref{eq:naiveBTZ}) is capable of describing a static black hole, will ultimately be understood as the quantum BTZ black hole as we detail below.

Recall the bulk geometry will retain conical singularities at the remaining roots of $G(x)$ (not $x=x_{1}$). As it happens, there are no conical singularities in the restricted range  $0\leq x\leq x_{1}$; for $\mu>0$ the remaining conical singularities live in the range $x<0$. Then, treating the $x=0$ hypersurface as a cutoff brane, excising the $x<0$ region leads to a spacetime free of conical singularities. To complete the space, a second copy of the $0\leq x\leq x_{1}$ region is glued along $x=0$, resulting in a $\mathbb{Z}_{2}$-symmetric double-sided Karch-Randall braneworld. Further, with the conical singularities removed, the final bulk solution no longer has a cosmic string. Nonetheless, the static black hole attached to the brane is in an accelerated frame.

\subsubsection*{Karch-Randall braneworld holography}

Applying the formalism of Section \ref{sec:BWholo}, the induced brane action is (setting $d=3$ in (\ref{eq:inductheorygen})) 
\beq I=\frac{1}{16\pi G_{3}}\int_{\mathcal{B}}d^{3}x\sqrt{-h}\left[R-2\Lambda_{3}+L_{4}^{2}\left(\frac{3}{8}R^{2}-R_{ij}^{2}\right)+...\right]+I_{\text{CFT}}\;,\label{eq:inducactqbtzneu}\eeq
with induced Newton's constant
\beq G_{3}=\frac{G_{4}}{2L_{4}}\;\label{eq:indG3}\eeq
and effective (AdS$_{3}$) brane cosmological constant and curvature scale,
\beq \Lambda_{3}=-\frac{1}{L_{3}^{2}}\;,\qquad \frac{1}{L_{3}^{2}}=\frac{2}{L_{4}^{2}}\left(1-\frac{L_{4}}{\ell}\right)\;,\label{eq:indL3}\eeq
where we used tension (\ref{eq:branetension}). Note that while $L_{3}$ appears in the action, the solutions on the brane are characterized by $\ell_{3}$.

In fact, we will primarily be interested in the case when $L_{3}\approx \ell_{3}$. This is because we interested in the case when the three-dimensional graviton becomes effectively massless, i.e., when the brane is near the boundary. Hence, using the bulk length scale (\ref{eq:bulkAdS4length}), it follows
\beq \frac{1}{L_{3}^{2}}=\frac{1}{\ell_{3}^{2}}\left[1+\frac{\ell^{2}}{4\ell_{3}^{2}}+\mathcal{O}\left(\frac{\ell^{4}}{\ell_{3}^{4}}+...\right)\right]\;,\eeq
with $\ell\sim L_{4}\ll \ell_{3}$. In this limit, the induced theory is 
\beq I=\frac{1}{16\pi G_{3}}\int_{\mathcal{B}}d^{3}x\sqrt{-h}\left[R+\frac{2}{\ell_{3}^{2}}+\ell^{2}\left(\frac{3}{8}R^{2}-R_{ij}^{2}\right)+...\right]+I_{\text{CFT}}\;,\label{eq:inducactqbtzneu2}\eeq
such that the higher-curvature expansion can be viewed as an expansion in an effective cutoff scale $\ell$. Notice higher-curvature corrections enter at quadratic order.

The parameter $\ell$ also features into the coupling to the gravity theory of the cutoff $\text{CFT}_{3}$ on the brane. To see this, first note the central charge $c_{3}$ of the $\text{CFT}_{3}$ is normalized such that
\beq c_{3}=\frac{L_{4}^{2}}{G_{4}}=\frac{\ell}{2G_{3}\sqrt{1+\nu^{2}}}\;,\eeq
where we introduced the parameter $\nu\equiv\ell/\ell_{3}$. Expanding for small $(\ell/\ell_{3})$ gives
\beq 2c_{3}G_{3}=L_{4}\approx \ell\left(1-\frac{\ell^{2}}{2\ell_{3}^{2}}+\frac{3}{8}\frac{\ell^{4}}{\ell_{3}^{4}}+...\right)\;,\label{eq:c3G3l}\eeq
and $c_{3}\sim \frac{\ell}{G_{3}}$, which observe enters at linear order in $\ell$. Thus, the matter contributions enter at order $\mathcal{O}(\ell)$ while higher-curvature corrections appear at $\mathcal{O}(\ell^{2})$. Proceeding, the typical value of  $I_{\text{CFT}}$ is $|I_{\text{CFT}}|\sim c_{3}$, while the typical value of the gravitational part of the action (\ref{eq:inducactqbtzneu}) is $|I_{\mathcal{B}\text{grav}}|\sim \frac{L_{3}}{G_{3}}$. Combined, this leads to an effective dimensionless coupling $g_{\text{eff}}$ quantifying the size of the effects the $\text{CFT}_{3}$ has on the brane geometry \cite{Emparan:2021hyr}
\beq g_{\text{eff}}\sim \frac{|I_{\text{CFT}}|}{|I_{\mathcal{B}\text{grav}}|}\sim\frac{G_{3}c_{3}}{L_{3}}\;.\eeq
Thus, for small $\ell<\ell_{3}$, we see $g_{\text{eff}}\approx \nu\ll1$, and backreaction effects are `small', and the induced theory is effectively characterized by the action (\ref{eq:inducactqbtzneu2}). Further, from (\ref{eq:c3G3l}), we see for fixed $c_{3}$ then gravity becomes weak ($G_{3}\to0$) as $\ell\to0$. Alternatively, when $\nu\geq1$, backreaction effects are said to be large.

It is worth comparing the cutoff length scale to the three-dimensional Planck length $L_{\text{P}}=\hbar G_{3}$ (where here we have temporarily restored factors of $\hbar$). From (\ref{eq:c3G3l}) it follows 
\beq \ell\sim c_{3}\hbar G_{3}=c_{3}L_{\text{P}}\gg L_{\text{P}}\;,\label{eq:ellscalLP}\eeq
since the holographic cutoff $\text{CFT}_{3}$ obeys $c_{3}\gg1$.\footnote{Note taking $c_{3}$ large is consistent with keeping $\nu$ small. Indeed, the large central charge limit has $c_{3}\sim \frac{\ell}{\hbar G_{3}}\gg1$ (equivalently, the semi-classical bulk limit $L^{(4)}_{\text{P}}/L_{4}\ll1$ for four-dimensional Planck length $L^{(4)}_{\text{P}}$), which is consistent for solutions with $\nu\ll1$ and when four-dimensional bulk quantum effects are neglected.} Recalling the induced brane metric (\ref{eq:naiveBTZ}), notice the blackening factor includes a $\mathcal{O}(r^{-1})$ term which goes like $\mu\ell$. We can anticipate this term as a semi-classical correction which sets the size of the braneworld black hole to be much larger than the Planck length.

\subsection{Quantum BTZ black holes} \label{ssec:qbtzbhs}

Let us now delve into the geometry of quantum black holes. We begin with the simpler neutral, static quantum BTZ solution to establish its essential features, which we later enrich by adding rotation and charge.

\subsubsection{Static quantum BTZ}

Recall the induced geometry at $x=0$ (\ref{eq:naiveBTZ}). Naively, we might refer to this as the quantum black hole, however, the angular coordinate $\phi$ has period $\Delta\phi$ (\ref{eq:conicaldef}). We thus rescale coordinates $(t,r,\phi)$ to put the naive metric (\ref{eq:naiveBTZ}) in canonically normalized coordinates $(\bar{t},\bar{r},\bar{\phi})$. Specifically,  
\beq
t = \eta \bar{t}\ ,  \qquad r = \frac{\bar{r}}{\eta} \ , \qquad \phi= \eta \bar{\phi} \ ,
\eeq
with
\beq
\eta \equiv \frac{\Delta \phi}{2\pi} = \frac{2 x_1}{3- \kappa x_1^2} \ .
\label{eq:etacanon}\eeq
The line element for the brane metric is now 
\beq
\hspace{-2mm}\boxed{ds^{2}_{\text{qBTZ}} = -\left(\frac{\bar{r}^{2}}{\ell_{3}^{2}}-8 \mathcal{G}_3 M-\frac{\ell F(M)}{\bar{r}} \right)d\bar{t}^2 +\left(\frac{\bar{r}^{2}}{\ell_{3}^{2}}-8 \mathcal{G}_3 M-\frac{\ell F(M)}{\bar{r}} \right)^{-1}  d\bar{r}^2 + \bar{r}^2 d\bar{\phi}^2}
\label{eq:qBTZ}\eeq
Here we have suggestively identified the three-dimensional mass
\beq
M \equiv -\frac{\kappa}{8G_3}\frac{\ell}{L_4}\eta^2 =-\frac{1}{2\mathcal{G}_3}\frac{\kappa x_1^2}{(3-\kappa x_{1}^{2})^2} \ , \qquad \mathcal{G}_3\equiv G_3 \frac{L_4}{\ell} =\frac{G_{3}}{\sqrt{1-\nu^{2}}}\ ,
\label{eq:massqBTZ}\eeq
and 
\beq F(M)\equiv \mu\eta^{3}=8\frac{(1-\kappa x_{1}^{2})}{(3-\kappa x_{1}^{2})^{3}}\;.\label{eq:formfuncFM}\eeq

The mass identification (\ref{eq:massqBTZ}) is primarily motivated by the fact that for in Einstein-AdS gravity, the mass corresponds to the subleading constant term in the $g_{tt}$ metric in appropriate coordinates. However, the induced theory on the brane is a higher-derivative theory, and thus the definition of mass is expected to be modified due to higher-derivative corrections, starting at $\mathcal{O}(\ell^{2})$. When treated as corrections to Einstein gravity, the mass $M$ may still be identified as the constant term in $g_{tt}$, however, now the classical Newton's constant $G_{3}$ is `renormalized' by the higher-derivative terms \cite{Cremonini:2009ih}
\beq
\begin{split} \mathcal{G}_{3}&=\left(1-\frac{\ell^{2}}{2L_{3}^{2}}+\mathcal{O}\left(\frac{\ell}{L_{3}}\right)^{4}\right)G_{3}= \left(1-\frac{\nu^{2}}{2}\right)\frac{G_{4}}{2L_{4}}+\mathcal{O}\left(\frac{\ell}{L_{3}}\right)^{4}\;.
\end{split}
\label{eq:G3renv1}\eeq
Alternatively, here it is assumed that the renormalized Newton's constant be identified as $\mathcal{G}_{3}\equiv G_{3}\frac{L_{4}}{\ell}$ at all orders in $\ell$, though this differs from (\ref{eq:G3renv1}) at order $\mathcal{O}\left(\ell/L_{3}\right)^{4}$.\footnote{Evidence of the $\mathcal{G}_{3}$ identification  in (\ref{eq:massqBTZ}) is that the relation can be derived exactly by integrating the bulk volume in the action with a bulk IR cutoff as $r\to\infty$ \cite{Emparan:1999wa}.}  As such, $\mathcal{G}_{3}$ in mass (\ref{eq:G3renv1}) is interpreted as an all-order resummation of the higher-derivative corrections to the mass \cite{Emparan:2020znc}, whilst $G_{3}$ is the `bare' Newton constant, since it is the physical constant $G_{3}M$ which is being corrected. Further evidence for the identification of the mass will be given when we explore the horizon thermodynamics of the three-dimensional black hole. Lastly, the form function $F(M)$ (\ref{eq:formfuncFM}) is purely a function of mass $M$ because it depends on $M$ only through $x_{1}$ and is otherwise independent of $\nu=\ell/\ell_{3}$.

The metric (\ref{eq:qBTZ}) is highly reminiscent of the perturbative geometry (\ref{eq:pertbackAdS3geom}) incorporating semi-classical backreaction effects of a conformally coupled scalar.\footnote{Perturbative corrections to black hole solutions in semi-classical new massive gravity give rise to logarithmic terms in the blackening factor \cite{Chernicoff:2024dll}. This suggests a resummation of the infinite tower of higher-derivative terms in the induced action (the $\mathcal{O}(\ell^{2})$ term being that of new massive gravity) eliminate the logarithmic dependence.}  We emphasize, however, that the $1/r$ correction in $g_{tt}$ (\ref{eq:pertbackAdS3geom}) is perturbative, while it is \emph{exact} in (\ref{eq:qBTZ}). In particular, the geometry (\ref{eq:qBTZ}) is an exact solution to the gravitational field equations of the semi-classical induced action (\ref{eq:inducactqbtzneu2})
\beq
\begin{split}
8\pi G_{3}\langle T_{ij}\rangle&=R_{ij}-\frac{1}{2}h_{ij}\left(R+\frac{2}{L_{3}^{2}}\right)\\
&+\ell^{2}\biggr[4R_{i}^{\;k}R_{jk}-\frac{9}{4}RR_{ij}-\Box R_{ij}+\frac{1}{4}\nabla_{i}\nabla_{j}R+\frac{1}{2}h_{ij}\left(\frac{13}{8}R^{2}-4R_{kl}^{2}+\frac{1}{2}\Box R\right)\biggr]\;,
\end{split}
\label{eq:semiclasseombrane}\eeq
where $\Box\equiv\nabla^{2}$. Decomposing the holographic stress-tensor in an expansion in $\ell^{2}$, i.e.,  $\langle T^{i}_{\;j}  \rangle = \langle T^{i}_{\;j}  \rangle_0 + \ell^2 \langle T^{i}_{\;j}  \rangle_2 + ...$, it follows
\beq
8\pi  G_3 \langle T^{i}_{\;j}  \rangle_0 = R^{i}_{\;j} - \frac{1}{2}\delta^{i}_{\;j} \left( R+\frac{2}{\ell_3^2} \right)  \ ,
\label{eq:T0stressstat}\eeq
\beq
\begin{split}
8\pi G_{3}\langle T^{i}_{\;j}\rangle_{2}&=4R^{ik}R_{jk}-\Box R^{i}_{\;j}
-\frac{9}{4}RR^{i}_{\;j}+\frac{1}{4}\nabla^{i}\nabla_{j}R+\frac{1}{2}\delta^{i}_{\;j}\left(\frac{13}{8}R^{2}-3R_{kl}^{2}+\frac{1}{2}\Box R-\frac{1}{2\ell_{3}^{4}}\right)\;.
\end{split}
\label{eq:T2stressstat}\eeq
Substituting in the metric (\ref{eq:qBTZ}) we obtain for the renormalized stress-energy tensor,
\beq
\langle T^i_{\;j}  \rangle_0 = \frac{1}{16 \pi G_3}\frac{\ell F(M)}{\bar{r}^3} \text{diag} \{1,1,-2\} \ 
\label{eq:renstressstatqbtz}\eeq
to leading order, and 
\beq
\langle T^i_{\;j}  \rangle_2 = \frac{1}{16\pi G_3}\frac{\ell F(M)}{\bar{r}^3} \left(\frac{1}{2\ell_3^2} \text{diag}\{1,-11,10 \} - \frac{24 \mathcal{G}_3 M}{\bar{r}^2} \text{diag}\{3,1,-4 \}+ \frac{\ell F(M)}{2 \bar{r}^3} \right)
\label{eq:renstressstatqbtz2}\eeq
for the $\mathcal{O}(\ell^{2})$ corrections. Notice $\langle T_{ij}\rangle_{2}$ has a non-zero trace, a consequence of breaking the conformal symmetry due to the cutoff $\ell$. In principle, one could compute $\langle T_{ij}\rangle$ at all orders in backreaction -- unlike the non-holographic perturbative analysis -- however, it proves cumbersome to do so.

Since the backreaction is turned off as $\ell\to0$, the metric (\ref{eq:qBTZ}) in this limit may be interpreted as `classical'; indeed the $1/r$ correction is eliminated and the resulting geometry solves the three-dimensional vacuum Einstein equations. Hence, for $\ell\neq0$ the geometry  (\ref{eq:qBTZ}) is naturally understood as a quantum black hole in AdS$_{3}$, namely, the \emph{quantum BTZ} (qBTZ) black hole. Notice further, in the limit of small backreaction ($\nu<1$), the $1/r$ term in (\ref{eq:qBTZ}) is proportional to $\ell\sim cG_{3}\gg L_{\text{P}}$ (\ref{eq:ellscalLP}), such that the horizon size of the quantum BTZ black hole is large compared to the Planck length. Further, it is worth emphasizing that the solution (\ref{eq:qBTZ}) consistently solves the semi-classical brane equations of motion for any $\ell>0$, including large backreaction effects $(\nu>1)$. 

Unlike the classical BTZ black hole, the quantum BTZ (\ref{eq:qBTZ}) has a curvature singularity at $\bar{r}=0$, as evidenced by the Kretschmann invariant, 
\beq R^{ijkl}R_{ijkl}=\frac{12}{\ell_{3}^{4}}+\frac{6F(M)^{2}\ell^{2}}{\bar{r}^{6}}\;.\eeq
The curvature singularity descends from the curvature singularity of the bulk four-dimensional C-metric at $r=0$, and is hidden behind the (bulk) horizon $r=r_{+}$. Here the singularity at $\bar{r}=0$ sits behind the horizon at $\bar{r}=\bar{r}_{+}$, the largest positive root of the metric function $H(\bar{r})$, where the time-translation Killing vector $\bar{\zeta}^{i}=\eta\partial^{i}_{\bar{t}}$ goes null. Thus, at least for holographic quantum matter, the singularity structure of a classical black hole becomes dramatically altered due to semi-classical backreaction. Note, moreover, relative to $\bar{\zeta}^{i}$, the surface gravity, defined via $\bar{\zeta}^{i}\nabla_{i}\bar{\zeta}_{j}=\kappa_{+}\bar{\zeta}_{j}$, is 
\beq \kappa_{+}=\frac{\eta}{2}|H'(r_{+})|=\frac{1}{2}|H'(\bar{r}_{+})|=\frac{1}{2\ell_{3}^{2}}\biggr|2\bar{r}_{+}+\frac{\mu\ell\eta^{3}}{\bar{r}_{+}^{2}}\biggr|\;,\label{eq:surfgravqbtz}\eeq
In the limit $\ell\to0$, this corresponds to the surface gravity of the classical BTZ black hole.

Aside from the match of the perturbed geometry, the leading order contribution to $\langle T_{ij}\rangle$ (\ref{eq:renstressstatqbtz}) agrees with the renormalized stress-tensor of the free conformal scalar (except $L_{\text{P}}\to \ell$). This match between the holographic and free field results is because transparent boundary conditions have been imposed on the CFT or the free scalar. In the former, these boundary conditions are naturally selcted because bulk fluctuations, which are dual to CFT excitations, move freely throughout the bulk. For the free field computation, however, transparent boundary conditions were an explicit choice. In either case, notice the leading order contribution to the stress-tensor is not of the standard thermal type, i.e., $\text{diag}\{-2,1,1\}$. Nonetheless, the CFT is in a thermal state. Indeed, the Green's function from which $\langle T_{ij}\rangle$ is derived (in the perturbative treatment) is periodic in imaginary time with a period given by the (inverse) Tolman temperature \cite{Lifschytz:1993eb}. Further, the Green function obeys analyticity properties shared by the Hartle-Hawking state,  a state describing a black hole in thermal equilibrium with its own radiation.

\vspace{3mm}

\noindent \textbf{A family of quantum black holes.} From the mass identification (\ref{eq:massqBTZ}), $M$ takes values in the finite range \cite{Emparan:1999wa,Emparan:1999fd}, 
\beq -\frac{1}{8\mathcal{G}_{3}}\leq M\leq \frac{1}{24\mathcal{G}_{3}}\;,\label{eq:massbhrange}\eeq
where $\ell$ is held fixed. The upper bound follows from studying for the end behavior of $M(x_{1})$, with $x_{1}=\sqrt{3}$ (excluding $x_{1}=-\sqrt{3}$ on account of the parameter range (\ref{eq:paramrangex1v1})) at $\kappa=-1$ being a maximum. The lower bound, meanwhile, occurs for $\kappa=+1$ and $x_{1}=1$ (where $\mu=0$ according to (\ref{eq:defmu})), and $M=0$ for either $x_{1}=0$ or $\kappa=0$. Thus, the qBTZ solution has negative and positive masses. 

In fact, the solution may be categorized into physically distinct branches:
\beq
\begin{split}
    &\text{Branch 1:} \qquad -1<-\kappa x_1^2<3 \ , \\
    &\text{Branch 2:} \qquad 3<-\kappa x_1^2<\infty \ .
\end{split}
\label{eq:2branchesqbtz}\eeq
To understand how these branches arise, it is first natural to distinguish solutions with negative mass versus those with positive mass. All negative mass solutions occur for $\kappa=+1$ and for $x_{1}\in(0,1]$, while all positive mass solutions occur for $\kappa=-1$ and $x_{1}\in(0,\infty)$. Note further $M=0$ for either $x_{1}=0$ (for both $\kappa=\pm1$) or $x_{1}\to\infty$ (and $\kappa=+1$). This suggests the positive mass solutions subdivide along the range $x_{1}\in(0,\sqrt{3})$ and $x_{1}\in(\sqrt{3},\infty)$, smoothly joining at $x_{1}=\sqrt{3}$.  Further, since it is the combination $\kappa x_{1}^{2}$ that appears in all physical quantities of interest, the negative mass branch and the positive mass branch with $x_{1}\in(0,\sqrt{3})$ smoothly connect at $x_{1}=0$. Succinctly, the qBTZ solution is characterized by the two branches (\ref{eq:2branchesqbtz}), however, a finer subdivision of the two branches is \cite{Emparan:2020znc}
\beq
\begin{split}
    &\text{Branch 1a:} \qquad \kappa=+1\,,\quad  0<x_{1}<1 \ ,  \\
    &\text{Branch 1b:} \qquad  \kappa=-1\,,\quad 0<x_{1}<\sqrt{3} \ , \\
    &\text{Branch 2:} \qquad \kappa=-1\,,\quad \sqrt{3}<x_{1}<\infty \ .
    \end{split}
\label{eq:2branchesqbtzv2}\eeq

Although each of the mass branches (\ref{eq:2branchesqbtzv2}) smoothly connect, each branch has a different physical interpretation. This can be better understood by considering the types of solutions belonging to each branch in the limit of vanishing backreaction. Firstly,  branch 1a contains the ground state $M=-\frac{1}{8\mathcal{G}_{3}}$, where $F(M)=0$ (substitute $x_{1}=1$ into (\ref{eq:formfuncFM})). Geometrically, the qBTZ line element (\ref{eq:qBTZ}) takes the form of global $\text{AdS}_{3}$. Notably, however, this ground state is valid for any $\ell\geq0$, and thus for $\ell\neq0$ can be thought of as three-dimensional \emph{quantum} anti-de Sitter spacetime, $\text{qAdS}_{3}$, accounting for a large-$c$ cutoff $\text{CFT}_{3}$ living in AdS$_{3}$ with vanishing renormalized stress-tensor. Above this ground state, negative mass solutions (branch 1a) correspond to AdS$_{3}$ conical defects in the limit $\ell\to0$, such that for $\ell\neq0$ the negative mass family of solutions are understood to be quantum-corrected conical singularities, though are still referred to as quantum `black holes'. Classically, the conical defects are horizonless, while when $\ell>0$ the quantum Casimir stress tensor shrouds the conical singularity in a horizon.

Branches 1b and 2, having $M\geq0$, both describe quantum corrections to the classical BTZ black hole geometry. The source of the corrections of these two branches, however, differ. Since branch 1a and 1b connect at $x_{1}=0$ (where $M=0$ and $F(0)=8/27$) and 1a black holes form due to backreaction of Casimir stress-energy, so too are the corrections resulting in the 1b black holes dominated by Casimir energy. Alternatively, the $M=0$ state of the branch 2 black holes (where $x_{1}\to\infty$) have zero $F(M)$ and hence stress-energy. This suggests the non-zero stress-energy of the $M>0$ states among the branch 2 black holes is due to Hawking radiation in thermal equilibrium with the finite temperature black hole (where the dominant Casimir energy has been subtracted from the quantum state appearing in $\langle T_{ij}\rangle$ \cite{Emparan:2020znc}). 

\begin{figure}[t!]
\centering
 \includegraphics[width=12.3cm]{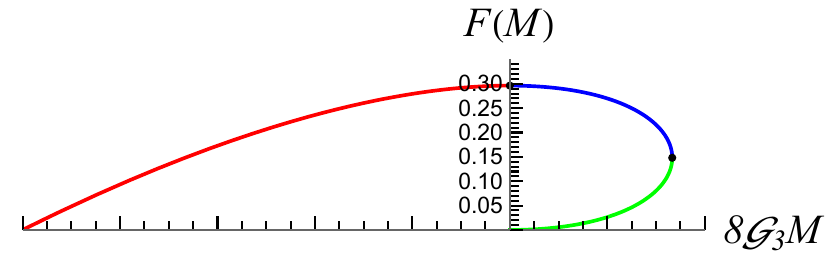}
\put(-230,65){1a}
\put(-95,23){2}
\put(-95,72){1b}
\caption{\small \textbf{The  qBTZ family of solutions}. The qBTZ black hole consists of quantum-dressed conical singularities, branch 1a (red) and quantum-corrected BTZ black holes, branch 1b (blue) and branch 2 (green). Branches 1a and 1b join at $M=0$ and $F=8/27$ ($x_{1}=0$), while branches 1b and 2 coincide at $8G_{3}M=1/3$ and $F=4/27$ ($x_{1}=\sqrt{3}$). Another branch describes (uncorrected) BTZ black holes with mass $M\geq0$ induced from the bulk BTZ black string (the positive horizontal axis).}
\label{fig:qbtzfam}\end{figure}

Classically, the negative mass conical singularities and positive mass BTZ black holes are disconnected sets of solutions to Einstein-AdS$_{3}$ gravity -- there is a `mass gap' \cite{Banados:1992gq}. Evidently, semi-classical backreaction smoothly connects these solutions, such that the quantum BTZ geometry represents a family of quantum black holes. We illustrate this family in Figure \ref{fig:qbtzfam}.

It is natural to wonder if there exist any other bulk solutions which give rise to BTZ black hole on the brane with masses $M\geq0$, including those that exceed the upper bound in the range (\ref{eq:massbhrange}). Indeed, recall the AdS$_{4}$ geometry (\ref{eq:AdS4empty}) for which $\mu=0$. When $\kappa=-1$ this four-dimensional geometry is dubbed the `BTZ black string' since sections of constant $\sigma$ contain a BTZ black hole. Despite $\mu=0$, a BTZ black hole lives on the brane at $x=0$ with mass $M\geq0$ in the canonically normalized coordinates $(\bar{t},\bar{r},\bar{\phi})$, however, the geometry is uncorrected from its classical counterpart since the quantum stress tensor vanishes (on account of $F(M)=0$ for all $M\geq0$). Unlike the three branches (\ref{eq:2branchesqbtzv2}), this braneworld black hole has no upper bound on $M$ and hence solutions with $M>1/24\mathcal{G}_{3}$ are induced by a BTZ black string localized on the AdS$_{3}$ brane.

\subsubsection{Rotating quantum BTZ} \label{sssec:rotqbtz}

\subsubsection*{Bulk and brane geometry}

 It is reasonably straightforward to find rotating quantum BTZ black holes. The starting point is the rotating AdS C-metric. describing accelerating Kerr-AdS$_{4}$ black holes,
%We can move on from the static case, by allowing the bulk black hole to have non-zero angular momentum. The stationary AdS C-metric describing rotating black holes is given by:
\begin{equation}
\begin{split}
 ds^2 = \frac{\ell^2}{(\ell + xr)^2} \biggr[ & -\frac{H(r)}{\Sigma(x,r)} \left(dt+ax^2 d\phi \right)^2 + \frac{\Sigma(x,r)}{H(r)} dr^2   \\
 &  + r^2 \left(\frac{\Sigma(x,r)}{G(x)}dx^2 + \frac{G(x)}{\Sigma(x,r)}\left( d\phi- \frac{a}{r^2} dt \right)^2  \right) \biggr] \ ,
 \end{split}
\label{eq:rotatingCmetqbtz}\end{equation}
where
\beq
\begin{split}
    &H(r)= \frac{r^2}{\ell_3^2}+ \kappa -\frac{\mu \ell}{r}+ \frac{a^2}{r^2}  \ , \quad G(x)= 1-\kappa x^2-\mu x^3+ \frac{a^2}{\ell_3^2}x^4 \ , \\
   &\Sigma(x,r)= 1 + \frac{a^2 x^2}{r^2} \ . 
\end{split}
\label{eq:metfuncsrotqbtz}\eeq
Here $a$ controls the rotation (the angular momentum per unit mass) and for $a=0$ we recover the static C-metric (\ref{eq:AdS4cmetBLstat}). 
%In the zero acceleration limit $\ell \to \infty$, instead, one recovers the Kerr-AdS$_4$ solution.
By evaluating the Kretschmann scalar invariant $R^{abcd}R_{abcd}$, there is a curvature singularity when $r^2 \Sigma = r^2 + a^2 x^2=0$, i.e., where $r=x=0$. This is the familiar ring singularity of Kerr black holes. 

Despite complicating the geometry by including rotation, the $x=0$ hypersurface remains umbilic, such that $K_{ij}=\ell^{-1}h_{ij}$ (\ref{eq:umbiliccondx0}) and a Karch-Randall brane at $x=0$ has tension (\ref{eq:branetension}). The geometry at $x=0$ is 
%Whilst more complicated in the details, most of the analysis goes through without major changes for the rotating case. First, the $x=0$ surface remains totally umbilic, meaning that we can place the brane on that hypersurface whose tension still obeys Equation \ref{eq:tension_l}. The geometry induced on the brane is given by:
\beq
\label{eq:rotating_qbtz_naive}
ds^2|_{x=0} = -H(r) dr^2 + H^{-1}(r) dr^2 + r^2 \left(d\phi - \frac{a}{r^2} dt\right)^2 \ .
\eeq
At first glance, this line element looks like a  rotating black hole in Boyer-Lindquist-like coordinates. As in the static case, however, this geometry does not reflect the whole story. In particular, the `naive metric' (\ref{eq:rotating_qbtz_naive}) unexpectedly no longer has a ring singularity, but instead a curvature singularity at $r=0$. Of course, we should not yet expect the geometry (\ref{eq:rotating_qbtz_naive}) to describe a black hole since bulk regularity conditions to deal with the conical nature of the bulk solution have not yet been imposed. Unlike the static system, however, bulk regularity conditions for the rotating solution are more subtle because they affect more than just the periodicity of angular coordinate $\phi$. Our treatment below follows the analysis in \cite{Panella:2023lsi}.

%one must be careful in taking Equation \ref{eq:rotating_qbtz_naive} too seriously. Indeed, non-trivial features of the spacetime are encoded in the periodicity of the angular coordinate due to the conical singularities in the bulk. To uncover them, we need to enforce bulk regularity of the rotating C-metric.

Firstly, notice that the Killing vector $\partial_{\phi}$ of the bulk C-metric (\ref{eq:rotatingCmetqbtz}) no longer has vanishing norm at a zero $x_{i}$ of $G(x)$. Rather, it is the Killing vector 
\beq \xi^{b}=\partial^{b}_{\phi}-ax_{i}^{2}\partial^{b}_{t}\;,\label{eq:phitvec}\eeq
which obeys $\xi^{2}|_{x_{i}}=0$. Thus, avoiding conical defects at $x=x_{i}$ requires one to identify points along the integral curves of (\ref{eq:phitvec}) with an appropriate period. To determine the correct periodicity, consider the rotating C-metric (\ref{eq:rotatingCmetqbtz}) near a zero $x=x_{i}$ such that $G(x)\sim G'(x_{i})(x-x_{i})$. Removal of a conical singularity at, say $x=x_{1}$ (denoting the smallest positive root of $G(x)$) has us simultaneously perform a coordinate transformation $\tilde{t}=t+ax_{i}^{2}\phi$ together with the periodicity condition for $\phi$ 
\beq \phi\sim\phi+\Delta\phi\;,\quad \Delta\phi=\frac{4\pi}{|G'(x_{1})|}=\frac{4\pi x_{1}}{3-\kappa x_{1}^{2}-\tilde{a}^{2}}\;,\label{eq:periodx1KdS}\eeq
where to arrive to the second equality we recast the parameter $\mu$ in terms of $x_{1}$ and define
\beq \mu=\frac{1-\kappa x_{1}^{2}+\tilde{a}^{2}}{x_{1}^{3}}\;,\quad \tilde{a}\equiv\frac{ax_{1}^{2}}{\ell_{3}}\;.\label{eq:muatdef}\eeq
In other words, identifying points along the orbits of (\ref{eq:phitvec}) are made on surfaces of constant 
\beq \tilde{t}\equiv t+ax_{1}^{2}\phi\;.\label{eq:tildet}\eeq
As in the static braneworld construction, the remaining zeros $x_{i}\neq x_{1}$ are effectively removed by gluing a second copy of the spacetime with the end-of-the-world brane at $x=0$ such that the complete bulk spacetime has the restricted range $0\leq x\leq x_{1}$. 

%Similarly, in the static case roots of $H(r)$ correspond to the Killing horizons of the Killing vector $\partial_{t}$, whereas with rotation,  the Killing vector 
%\beq \zeta^{b}=\partial_{t}-\frac{a}{r_{i}^{2}}\partial_{\phi}\;\label{eq:zetavecv1}\eeq
%becomes null at roots $r_{i}$ of $H(r)$. Thus, the roots $r_{i}$ of $H(r)$ correspond to rotating horizons with angular velocity $\Omega=-ar_{i}^{2}$. Working in coordinate frame $(\tilde{t},\phi)$, as required for bulk regularity, the angular velocity is $\tilde{\Omega}=-a/(r_{i}^{2}+a^{2}x_{1}^{2})$ \cite{Emparan:1999fd}, such that non-zero $a$ leads to an additional root corresponding to an inner horizon. 
%We will analyze these horizons in detail momentarily.

Now return to the naive metric (\ref{eq:rotating_qbtz_naive}) and consider the limit $r\to\infty$. The metric is asymptotic to `rotating $\text{AdS}_{3}$', where the $dtd\phi$ metric component is a constant. The coordinates $(t,r,\phi)$, however, are not canonically normalized due to the periodicity in $\phi$ (\ref{eq:periodx1KdS}). Further, since points along orbits of (\ref{eq:phitvec}) are identified, the $\phi$-periodicity returns one to a different point in time $t$: indeed, from (\ref{eq:tildet}) we see that assuming $\tilde{t}\sim\tilde{t}$ then $t\sim t-2\pi\eta ax_{1}^{2}$, for $\eta\equiv \Delta\phi/2\pi$. Unfortunately, this means we cannot merely rescale coordinates $(t,r,\phi)\to(\bar{t},\bar{r},\bar{\phi})$  as in the static case. Moreover, the periodicity alters the asymptotic form of the metric such that the $dtd\phi$ would instead grow as $r^{2}$, implying a diverging angular momentum.\footnote{This can be easily seen by performing the coordinate transformation $t\to \tilde{t}-ax_{1}^{2}\tilde{\phi}$ and $\phi\to\tilde{\phi}$ in the brane geometry (\ref{eq:rotating_qbtz_naive}). For large $r$ the $h_{\tilde{t}\tilde{\phi}}$ component of the geometry diverges as $r^{2}$.}

The $r^{2}$-divergence can be ameliorated by changing coordinates $(t,\phi)$ to $(\tilde{t},\tilde{\phi})$ where,
\beq t=\tilde{t}-ax_{1}^{2}\tilde{\phi}\;,\quad \phi=\tilde{\phi}-\frac{\tilde{a}}{\ell_{3}}\tilde{t}\;.\eeq
The coefficient $-\frac{\tilde{a}}{\ell_{3}}$ in $\phi$ is carefully chosen such that the $r^{2}$ divergence in the $\tilde{t}-\tilde{\phi}$ component of the naive brane metric is eliminated. Even still, the angular coordinate $\tilde{\phi}$ is not periodic in $2\pi$. Luckily, this is now easily resolved by the simple rescaling, $\tilde{t}=\eta\bar{t}$ and $\tilde{\phi}=\eta\bar{\phi}$, such that the transformation \cite{Emparan:2020znc}
\beq t=\eta(\bar{t}-\tilde{a}\ell_{3}\bar{\phi})\;,\quad \phi=\eta\left(\bar{\phi}-\frac{\tilde{a}}{\ell_{3}}\bar{t}\right)\;,\label{eq:coordtbphib}\eeq
places the brane geometry (\ref{eq:rotating_qbtz_naive}) in a more canonical form. It proves useful to also have the inverted coordinate transformation (\ref{eq:coordtbphib}), 
\beq
\label{eq:ads_coordtrans}
\bar{t}=\frac{1}{\eta \left(1-\tilde{a}^2\right)}\left(t+\tilde{a}\ell_3 \phi \right) \ , \qquad \bar{\phi}= \frac{1}{\eta \left(1-\tilde{a}^2\right)} \left(\phi+\frac{\tilde{a}}{\ell_3}t \right) \ .
\eeq 
From here, the Killing vectors transform as
\beq
\partial_t = \frac{1}{\eta \left(1-\tilde{a}^2\right)} \left(\partial_{\bar{t}} + \frac{\tilde{a}}{\ell_3}\partial_{\bar{\phi}}\right) \ , \qquad \partial_\phi = \frac{1}{\eta \left(1-\tilde{a}^2\right)} \left(\partial_{\bar{\phi}} + \tilde{a}\ell_3\partial_{\bar{t}}\right) \ .
\label{eq:Killvectrans}\eeq
Notice in the coordinates $(\bar{t},\bar{\phi})$, the Killing vector (\ref{eq:phitvec}) is now $\xi^b = \eta^{-1} \partial_{\bar{\phi}}$.

With the coordinate change (\ref{eq:coordtbphib}), the brane metric does not quite have the canonical asymptotic form of a rotating AdS black hole. To do so, one introduces a normalized radial coordinate $\bar{r}$ such that \cite{Emparan:2020znc}. 
\beq
r^2 = \frac{\bar{r}^2-r_s^2}{(1-\tilde{a}^2)\eta^2} \ , \qquad r_s\equiv \ell_3 \frac{\tilde{a}\eta}{x_1}\sqrt{2-\kappa x_1^2} = \ell_3 \frac{2 \tilde{a}\sqrt{2-\kappa x_1^2}}{3-\kappa x_1^2 - \tilde{a}^2} \ ,
\label{eq:radialqbtz}\eeq
Altogether, having imposed bulk regularity conditions, the brane geometry (\ref{eq:rotating_qbtz_naive}) in the canonically normalized coordinates $(\bar{t},\bar{r},\bar{\phi})$ is
\beq
\begin{split}
ds^{2}|_{x=0}&=-\left(\kappa\eta^{2}\left(1+\tilde{a}^{2}-\frac{4\tilde{a}^{2}}{\kappa x_{1}^{2}}\right)+\frac{\bar{r}^{2}}{\ell_{3}^{2}}-\frac{\mu\ell\eta^{2}}{r}\right)d\bar{t}^{2}\\
&+\left(\kappa\eta^{2}\left(1+\tilde{a}^{2}-\frac{4\tilde{a}^{2}}{x_{1}^{2}}\right)+\frac{\bar{r}^{2}}{\ell_{3}^{2}}-\frac{\mu\ell(1-\tilde{a}^{2})^{2}\eta^{4}r}{\bar{r}^{2}}+\frac{\ell_{3}^{2}\tilde{a}^{2}\mu^{2}x_{1}^{2}\eta^{4}}{\bar{r}^{2}}\right)^{-1}d\bar{r}^{2}\\
&+\left(\bar{r}^{2}+\frac{\mu\ell\tilde{a}^{2}\ell_{3}^{2}\eta^{2}}{r}\right)d\bar{\phi}^{2}-\ell_{3}\tilde{a}\mu x_{1}\eta^{2}\left(1+\frac{\ell}{x_{1}r}\right)(d\bar{\phi}d\bar{t}+d\bar{t}d\bar{\phi})\;,
\end{split}
\label{eq:branegeomv1}\eeq
where we have kept both $r$ and $\bar{r}$ when convenient (treating $r=r(\bar{r})$).

\subsubsection*{Quantum black hole}

As in the static case, we  reexpress the metric (\ref{eq:branegeomv1}) as
\beq
\hspace{-6mm} \boxed{
\begin{split}
ds^2 = &-\left(\frac{\bar{r}^2}{\ell_3^2}-8\mathcal{G}_3 M -\frac{\ell \mu \eta^2}{r} \right) d\bar{t}^2 + \left(\frac{\bar{r}^2}{\ell_3^2}-8\mathcal{G}_3 M + \frac{(4 \mathcal{G}_3 J)^2}{\bar{r}^2}- \ell \mu (1-\tilde{a}^2)^2 \eta^4 \frac{r}{\bar{r}^2} \right)^{\hspace{-2mm}-1} \hspace{-1mm}d\bar{r}^2  \\
& + \left(\bar{r}^2 + \frac{\mu \ell\tilde{a}^2 \ell_3^2 \eta^2}{r} \right) d\bar{\phi}^2 - 8 \mathcal{G}_3 J \left( 1+ \frac{\ell}{x_1 r} \right) d\bar{\phi}d \bar{t}
\end{split}
}
\label{eq:qbtzrotatmet}\eeq
where we have suggestively identified the mass and angular momentum $J$ of the black hole,
\beq
\begin{split}
8 \mathcal{G}_3 M &= -\kappa\eta^2 \left(1+ \tilde{a}^2-\frac{4 \tilde{a}^2}{\kappa x_1^2} \right) = 4 \frac{- \kappa x_1^2 + \tilde{a}^2(4-\kappa x_1^2)}{(3-\kappa x_1^2 - \tilde{a}^2)^2}
\end{split}
\label{eq:massrotqbtz}\eeq 
\beq
\begin{split}
4 \mathcal{G}_3 J &= \ell_3\eta^2 \tilde{a}\mu x_1=\frac{4 \ell_{3}\tilde{a}(1-\kappa x_1^2 + \tilde{a}^2)}{(3 - \kappa x_1 ^2 -\tilde{a}^2)^2}\;.
\end{split}
\label{eq:rotJqbtz}\eeq
The above identifications are made on geometric grounds: in the asymptotic limit $\bar{r}\to\infty$, the terms proportional to $\ell$ decay faster than than the constant $-8\mathcal{G}_{3}M$ or the $J$ in the $d\bar{t}d\bar{\phi}$ component.  Further, $\mathcal{G}_{3}\equiv L_{4}G_{3}/\ell$ is again the renormalized Newton's constant, which now also plays the role of accounting for higher-derivative corrections to the angular momentum.

\vspace{3mm}

\noindent \textbf{Horizon and singularity structure.} In the static case, roots of the bulk metric function $H(r)$ correspond to the bulk Killing horizon of the time-translation Killing vector $\partial_{t}$. Including rotation in the bulk, the Killing vector 
\beq \zeta^{b}=\partial_{t}+\frac{a}{r_{i}^{2}}\partial_{\phi}\;,\label{eq:genzetabtz}\eeq
for $r_{i}$ finite, has modulus $\zeta^{2}=g_{ab}\zeta^{a}\zeta^{b}=-\frac{\ell^{2}}{(\ell+x r)^{2}}H(r_{i})\Sigma(x,r_{i})$ at $r=r_{i}$. 
Taking $r_{i}$ to be real, $\zeta^{2}=0$ when $H(r_{i})=0$, where $r_{i}$ are positive real roots of $H(r)$ (\ref{eq:metfuncsrotqbtz})  (restricting to a coordinate range where there exists at least one real root). Let $r_{+}$ denote the largest positive, real root of $H(r)$, corresponding to the radius of the outer black hole. As with Kerr-AdS$_{4}$, there is a second positive real root $r_{-}$, satisfying $r_{-}< r_{+}$, which corresponds to the inner black hole horizon. Using $H(r_{\pm})=0$ and assuming $r_{+}\neq r_{-}$, it is straightforward to express
\beq
\begin{split} 
&\mu\ell=\frac{(r_{+}+r_{-})}{\ell_{3}^{2}}\left[(r_{+}^{2}+r_{-}^{2})+\ell_{3}^{2}\kappa\right]\;,\\
&a^{2}=\frac{r_{+}r_{-}}{\ell_{3}^{2}}\left[r_{+}^{2}+r_{-}^{2}+r_{+}r_{-}+\ell_{3}^{2}\kappa\right]\;.
\end{split}
\label{eq:muellarotqbtz}\eeq
The limit of vanishing angular momentum, $a\to0$, coincides with $r_{-}\to0$. 

In canonically normalized coordinates $(\bar{t},\bar{r},\bar{\phi})$, the inner and outer horizons of the quantum black hole are generated by orbits of the canonically normalized generator (\ref{eq:genzetabtz}), i.e., 
\beq \bar{\zeta}^{b}_{\pm}\equiv \frac{\eta(1-\tilde{a}^{2})}{1+\frac{a^{2}x_{1}^{2}}{r_{\pm}}}\zeta^{b}=\frac{\partial}{\partial \bar{t}}+\Omega_{\pm}\frac{\partial}{\partial\bar{\phi}}\;,\eeq
where we used the Killing vector transformation (\ref{eq:Killvectrans}).
Here $\Omega_{\pm}$ is the angular velocity of the horizons $r_{\pm}$ relative to a \emph{non-rotating} frame at spatial infinity
\beq \Omega_{\pm}\equiv \frac{a}{r_{\pm}^{2}+a^{2}x_{1}^{2}}\left(1+\frac{r_{\pm}^{2}x_{1}^{2}}{\ell_{3}^{2}}\right)\;.\label{eq:Omqbtznonrotinf}\eeq
For $x_{1}=1$ this coincides with the familiar angular velocity of Kerr-AdS$_{4}$ (e.g., \cite{Caldarelli:1999xj,Gibbons:2004ai}). Meanwhile, the angular velocity $\Omega'$ relative to a \emph{rotating} frame at spatial infinity is 
\beq \Omega'_{\pm}\equiv\frac{a}{r_{\pm}^{2}+a^{2}x_{1}^{2}}\left(1-\frac{a^{2}x_{1}^{4}}{\ell_{3}^{2}}\right)\;,\eeq
obeying $\Omega_{\pm}-\Omega'_{\pm}=ax_{1}^{2}/\ell_{3}^{2}$. Further, relative to $\bar{\zeta}^{b}$, the surface gravity of the inner and outer horizons is 
\beq \kappa_{\pm}=\frac{\eta(1-\tilde{a}^{2})}{(r_{\pm}^{2}+a^{2}x_{1}^{2})}\frac{r_{\pm}^{2}}{2}|H'(r_{\pm})|=\frac{\eta(1-\tilde{a}^{2})}{(r_{\pm}^{2}+a^{2}x_{1}^{2})}\frac{1}{2\ell_{3}^{2}r_{\pm}}|\ell_{3}^{2}\mu\ell r_{\pm}+2r_{\pm}^{4}-2a^{2}\ell_{3}^{2}|\;,\eeq
where we used the definition of the surface gravity $\kappa_{\pm}$ on a Killing horizon generated by $\bar{\zeta}$, $\bar{\zeta}^{b}\nabla_{b}\bar{\zeta}^{c}\equiv\kappa_{\pm} \bar{\zeta}^{c}$. In particular, using parameters (\ref{eq:muellarotqbtz})
\beq 
\begin{split} 
&\kappa_{+}=\frac{\eta(1-\tilde{a}^{2})}{2\ell_{3}^{2}(r_{+}^{2}+a^{2}x_{1}^{2})}(r_{+}-r_{-})|(r_{+}^{2}+r_{-}^{2}+\ell_{3}^{2}\kappa)+2r_{+}(r_{+}+r_{-})|\;,\\
&\kappa_{-}=\frac{\eta(1-\tilde{a}^{2})}{2\ell_{3}^{2}(r_{-}^{2}+a^{2}x_{1}^{2})}(r_{-}-r_{+})|(r_{+}^{2}+r_{-}^{2}+\ell_{3}^{2}\kappa)+2r_{-}(r_{+}+r_{-})|\;.
\end{split}
\label{eq:kapparotqbtz}\eeq
In the limit of vanishing rotation, $\kappa_{+}$ becomes the surface gravity of the static qBTZ, (\ref{eq:surfgravqbtz}). Further, the surface gravities vanish in the extremal limit, i.e., where the inner and outer black hole horizons coincide, $r_{+}=r_{-}$.

The horizons shroud a ring curvature singularity at $r=0$, i.e., $\bar{r}=r_{s}$. Further, near the ring singularity, there exists the possibility of closed timelike curves. Indeed, consider 
%As often is the case with rotating solutions, there is the possibility of closed timelike curves. To see this, it is first important to note that there exist two real roots for the blackening factor $H(\bar{r})$, giving rise to the typical horizon structure of an AdS rotating black hole with an inner horizon at $\bar{r}_-$ and an outer horizon at $\bar{r}_+\geq \bar{r}_-$. The region of spacetime where closed timelike curves become possible is the one close to the ring singularity. Indeed, the 
the norm of the axial Killing vector $\partial_{\bar{\phi}}$ of (\ref{eq:qbtzrotatmet}) is
\beq
\partial_{\bar{\phi}}^2 = h_{\bar{\phi}\bar{\phi}} = \bar{r}^2 + \frac{\mu \ell \tilde{a}^2 \ell_3^2 \eta^2}{r}= \bar{r}^2 + \frac{\mu \ell \tilde{a}^2 \ell_3^2 \eta^{3}}{\sqrt{\bar{r}^{2}-r_{s}^{2}}}\sqrt{1-\tilde{a}^{2}}\ ,
\eeq
meaning that if one gets to small and negative enough radii $r$, the axial vector becomes timelike and its orbits are closed curves around the rotation axis. To ensure $\partial^{2}_{\phi}\geq0$, requires an additional restriction of the solution parameters: (i) $1-\tilde{a^2}\geq 0$, and (ii)
%which can be seen as forbidding time inversion in the coordinate transformation given by Equation \ref{eq:ads_coordtrans};
 $\eta>0$, which implies $-\kappa x_1^2 > \tilde{a}^2-3$. Notice the second condition is always implied by (i) if $\kappa=-1$.

\vspace{2mm}

\noindent \textbf{Quantum stress tensor.} As an exact solution to the full semi-classical theory (\ref{eq:inducactqbtzneu2}), the metric (\ref{eq:qbtzrotatmet}) is known as the rotating quantum BTZ black hole. 
%We can indeed identify those combinations as the mass $M$ and angular momentum $J$ of the black hole to all orders in $\ell$ as the quantum correction terms (i.e. those proportional to $\ell$) die faster than the ones that are used to read off the properties of the black hole at $\bar{r} \to \infty$ by matching to the classical Kerr solutions in $2+1$-dimensions. 
%There are two limits of interests. On the one hand, it is straightforward to check that $a\to 0$ (and hence $J=0$) does indeed give back the qBTZ solution derived from the static C-metric. Moreover, much like in the static case, taking the no-backreaction limit of $\ell \to 0$, one sees that the quantum corrections vanish as the CFT decouples from the gravitational degrees of freedom. Then, for $\kappa=-1$ one obtains the classical BTZ black hole with non-zero angular momentum, whilst $\kappa=1$ gives a rotating conical singularity. $\mu=0$ corresponds to $x_1=\sqrt{(1-a^2)/\kappa}$, meaning that it can only be reached in the parameter space of interest (i.e. $x_1$ real and positive) for $\kappa=1$, we obtain vanishing angular momentum and $8\mathcal{G}_3 M=-1$, corresponding to quantum-corrected global AdS$_3$.
 In the limit of vanishing backreaction, $\ell\to0$, the classical rotating BTZ black hole is recovered for $\kappa=-1$, while for $\kappa=+1$ the geometry is that of a rotating AdS$_{3}$ conical defect. For $\ell\neq0$, the non-vanishing components of the renormalized quantum-stress tensor (\ref{eq:T0stressstat}) are found to be \cite{Emparan:2020znc}
\beq
\begin{split}
\langle T^{\bar{t}}_{\;\bar{t}}  \rangle_0 &= \frac{1}{16 \pi G_3}\frac{\ell \mu}{(1-\tilde{a}^2) r^3} \left(1+2\tilde{a}^2+\frac{3 \tilde{a}^2 \ell_3^2}{x_1^2 r^2} \right) \ , \\
\langle T^{\bar{r}}_{\;\bar{r}}  \rangle_0 &= \frac{1}{16 \pi G_3} \frac{\ell \mu}{r^3} \ , \\
\langle T^{\bar{\phi}}_{\;\bar{\phi}}  \rangle_0 &=- \frac{1}{16 \pi G_3}\frac{\ell \mu}{(1-\tilde{a}^2) r^3}\left(2+\tilde{a}^2+\frac{3\tilde{a}^2 \ell_3^2}{x_1^2 r^2} \right) \ , \\
\langle T^{\bar{t}}_{\;\bar{\phi}}  \rangle_0 &=- \frac{1}{16 \pi G_3}\frac{3 \ell_3 \ell \mu \tilde{a}}{(1-\tilde{a}^2) r^3}\left(1+\frac{\tilde{a}^2 \ell_3^2}{x_1^2 r^2} \right) \ , \\
\langle T^{\bar{\phi}}_{\;\bar{t}}  \rangle_0 &= \frac{1}{16 \pi G_3}\frac{3 \ell \mu \tilde{a}}{\ell_3 (1-\tilde{a}^2) r^3}\left(1+\frac{\ell_3^2}{x_1^2 r^2} \right) \ .
\end{split}
\label{eq:stresstenscomponentsqbtzrot}\eeq
These are most efficiently computed by first computing the components of the stress-tensor in the naive coordinates (\ref{eq:rotating_qbtz_naive}), and then performing the appropriate coordinate transformations imposing bulk regularity. The $\mathcal{O}(\ell^{2})$ components $\langle T_{ij}\rangle_{2}$ (\ref{eq:T2stressstat}) are cumbersome to express in canonically normalized coordinates, however, in naive coordinates $\langle T_{ij}\rangle_{2}$ is not traceless, indicating the presence of the a UV cutoff breaking the conformal symmetry of the CFT$_{3}$. 

Notice the mass (\ref{eq:massrotqbtz}) and angular momentum (\ref{eq:rotJqbtz}) only depend on $\tilde{a}$ and $\kappa x_{1}^{2}$ and not on $\ell$. Meanwhile, $\ell$ only enters as an overall prefactor in the stress-tensor components (\ref{eq:stresstenscomponentsqbtzrot}). Hence, the leading order stress-tensor only depends on backreaction effects via $\mathcal{G}_{3}$ and not $\mathcal{G}_{3}M$ or $\mathcal{G}_{3}J$. Further, unlike the static solution (\ref{eq:renstressstatqbtz}), the stress-tensor for the rotating qBTZ black hole is not characterized by a single function $F(M)$. Consider, however, the large-$\bar{r}$ asymptotics of the components (\ref{eq:stresstenscomponentsqbtzrot}). For example, substituting normalized coordinate (\ref{eq:radialqbtz}) into $\langle T^{\bar{t}}_{\;\bar{t}}\rangle_{0}$, yields 
\beq \lim_{\bar{r}\to\infty}\langle T^{\bar{t}}_{\;\bar{t}}\rangle_{0}\to\frac{\eta^{3}\mu\ell\sqrt{1-\tilde{a}^{2}}}{16\pi G_{3}\bar{r}^{3}}(1+2\tilde{a}^{2})\;.
\eeq
Peeling off $\ell/16\pi G_{3}\bar{r}^{3}$, identify the form function
\beq F(M,J)\equiv \eta^{3}\mu\sqrt{1-\tilde{a}^{2}}(1+2\tilde{a}^{2})=\frac{8\sqrt{1-\tilde{a}^{2}}(1+2\tilde{a}^{2})(1-\kappa x_{1}^{2}+\tilde{a}^{2})}{(3-\kappa x_{1}^{2}-\tilde{a}^{2})^{3}}\;,\label{eq:FMJqbtz}\eeq
such that for large $\bar{r}$
\beq \langle T^{\bar{t}}_{\;\bar{t}}\rangle_{0}\to\frac{1}{16\pi G_{3}}\frac{\ell F(M,J)}{\bar{r}^{3}}\;.
\eeq
Due to the $\sim\mu/r^{3}$ behavior in each of the remaining components, up to unimportant factors, the large-$\bar{r}$ structure of (\ref{eq:stresstenscomponentsqbtzrot}) essentially depends on $F(M,J)$ as defined in (\ref{eq:FMJqbtz}). We will use $F(M,J)$ to better understand the family of rotating quantum black holes momentarily. 

It is worth comparing the structure of the holographic stress-tensor (\ref{eq:stresstenscomponentsqbtzrot}) to the stress-tensor of a conformally coupled scalar field  found by solving the semi-classical Einstein equations perturbatively (in particular, see \cite{Casals:2019jfo,Emparan:2020znc}). The non-holographic stress-tensor may be expressed as an infinite sum over images, e.g., 
\beq 8\pi G_{3}\langle T^{\bar{t}}_{\;\bar{t}}\rangle=\sum_{n=1}^{\infty}\frac{1}{r_{n}^{3}}\left(A_{n}+\frac{\bar{A}_{n}}{r_{n}^{2}}\right)\;,\label{eq:infsumTrot}\eeq
with $r_{n}=\sqrt{D_{n}\bar{r}^{2}+\bar{D}_{n}}$, where coefficients $A_{n},\bar{A}_{n}$, $D_{n}$ and $\bar{D}_{n}$ are all complicated functions of $M$ and $J$. An analogous structure holds for the remaining components. Coefficients aside, each term of the infinite sum has a similar radial dependence as the holographic stress-tensor. When radial dependence of the entire infinite sum (\ref{eq:infsumTrot}), however, is far more complicated than the radial dependence of the holographic stress-tensor (\ref{eq:stresstenscomponentsqbtzrot}). Notably, while the holographic stress-tensor is manifestly non-singular away from the ring singularity $\bar{r}=r_{s}$, it is not clear whether the same is true for the perturbative stress-tensor.

\vspace{2mm}

\noindent \textbf{Black hole branches revisited.} In the non-rotating case, there are three branches of quantum black holes, branches 1a, 1b, and 2 (\ref{eq:2branchesqbtzv2}). There is an analogous set of branches for non-vanishing $J$, where, in particular, branches 1b and 2 meet at a maximum value of $M$ for fixed $J$. This occurs when 
\beq x_{1}^{2}+\tilde{a}^{2}=3\;,\quad M=\frac{1}{8\mathcal{G}_{3}}\left(\frac{12}{x_{1}^{4}}-1\right)\;,\quad J=\frac{\ell_{3}}{\mathcal{G}_{3}}\frac{\sqrt{3-x_{1}^{2}}}{x_{1}^{4}}\;.\label{eq:MJextbranch1}\eeq
At $x_{1}=\sqrt{2}$ one attains an extremal bound, where $M=J/\ell_{3}=1/4\mathcal{G}_{3}$. There is another extremal bound among the branch 2 black holes, founding by minimizing the mass $M$ for fixed $J$,
\beq \tilde{a}=1\;,\quad M=\frac{J}{\ell_{3}}=\frac{1}{\mathcal{G}_{3}(2+x_{1}^{2})}\;,\label{eq:classextlim}\eeq
which coincides with the bound (\ref{eq:MJextbranch1}) at $x_{1}=\sqrt{2}$.  Classically, the rotating BTZ black hole obeys the extremality bound $M\geq J/\ell_{3}$. For the quantum black hole, however, for any value of $J$, this classical extremality bound is violated, $M\leq J/\ell_{3}$, when $-\kappa x_{1}^{2}<2\tilde{a}^{2}$, giving rise to `super-extremal' black holes among the branch 1 solutions. Pictorially, the branches of quantum black holes have a similar representation as Figure \ref{fig:qbtzfam} (see Figure 6 of \cite{Emparan:2020znc}).

\subsubsection{Charged quantum BTZ} 

\subsection*{Bulk and brane geometry}

It is relatively straightforward to generalize the neutral quantum BTZ solutions to a charged system. Now the bulk is characterized by the charged AdS$_{4}$ C-metric. For simplicity, setting rotation to zero,  the line element is of the same form as (\ref{eq:AdS4cmetBLstat}), except the metric functions (\ref{eq:metfuncsstatCmet}) receive an additional term:
\beq ds^{2}=\frac{\ell^{2}}{(\ell+xr)^{2}}\left[-H(r)dt^{2}+\frac{dr^{2}}{H(r)}+r^{2}\left(\frac{dx^{2}}{G(x)}+G(x)d\phi^{2}\right)\right]\;,\label{eq:chargedcmet}\eeq
\beq H(r)=\frac{r^{2}}{\ell_{3}^{2}}+\kappa-\frac{\mu\ell}{r}+\frac{q^{2}\ell^{2}}{r^{2}}\;,\quad G(x)=1-\kappa x^{2}-\mu x^{3}-q^{2}x^{4}\;.\eeq
This is a solution to four-dimensional Einstein-Maxwell gravity with negative cosmological constant $-2\Lambda_{4}=6/L_{4}^{2}$, 
\beq I=\frac{1}{16\pi G_{4}}\int d^{4}x\sqrt{-\hat{g}}\left[\hat{R}+\frac{6}{L_{4}^{2}}-\frac{\ell_{\ast}^{2}}{4}F^{2}\right]\;, \quad \ell^{2}_{\ast}=\frac{16\pi G_{4}}{g_{\ast}^{2}}\label{eq:EinMaxact}\eeq
where $\ell_{\ast}$ is a coupling constant with dimensions of length, $g_{\ast}$ is the dimensionless gauge coupling constant, and  $F_{ab}=\partial_{a}A_{b}-\partial_{b}A_{a}$ is the Maxwell field tensor. The $U(1)$ gauge field $A_{a}$ for the charged C-metric (\ref{eq:chargedcmet})
\beq A=A_{a}dx^{a}=\frac{2\ell}{\ell_{\ast}}\left[e\left(\frac{1}{r_{+}}-\frac{1}{r}\right)dt+g(x-x_{1})d\phi\right]\;,\label{eq:gaugpot}\eeq
where $e$ and $g$ respectively denote the electric and magnetic charge parameters of the accelerating black hole such that $q^{2}=e^{2}+g^{2}$. A gauge has been chosen for the gauge potential (\ref{eq:gaugpot}) such that it remains regular at the largest root of $H(r)=0$, denoted $r_{+}$, and $x=x_{1}$.

As for the neutral black hole, real roots of $G(x)$ correspond to symmetry axes of the Killing vector $\partial^{a}_{\phi}$, and we work in the restricted regime $0\leq x\leq x_{1}$. The conical singularity at $x=x_{1}$ is removed via the identification 
\beq \phi\sim\phi+\Delta\phi\;,\quad \Delta\phi=\frac{4\pi}{|G'(x_{1})|}=\frac{4\pi}{|-3+\kappa x_{1}^{2}-q^{2}x_{1}^{4}|}\;,\eeq
where we treat $\mu$ as a `derived' parameter following $G(x_{1})=0$, 
\beq \mu=\frac{1-\kappa x_{1}^{2}-q^{2}x_{1}^{4}}{x_{1}^{3}}\;,\label{eq:muderivcbtz}\eeq
while $x_{1}$ and $q$ are primary parameters.

Lastly, as usual, the $x=0$ hypersurface is totally umbilic, such that the Israel junction conditions fix the brane tension to be $\tau=(2\pi G_{4}\ell)^{-1}$. The Maxwell field strength also has junction conditions to obey. In particular, let $n^{a}$ denote the normal to the brane at $x=0$ (pointing towards increasing values of $x$) and let $e^{a}_{i}$ be a basis for tangent vectors to the brane. Then, projecting $F_{ab}$ onto the brane such that $F_{ij}\equiv F_{ab}e^{a}_{i}e^{b}_{j}$ and $f_{i}\equiv F_{ab}e^{a}_{i}n^{b}$, one has the following junction conditions \cite{Lemos:2021jtm} for a purely tensional brane
\beq 
\begin{split}
&\Delta F_{ij}=F^{+}_{ij}-F^{-}_{ij}=0\;,\\
&\Delta f_{i}=f^{+}_{i}-f^{-}_{i}=4\pi j_{i}\;,
\end{split}
\label{eq:junccondEM}\eeq
where $j_{i}$ is the electromagnetic surface current.

\subsection*{Induced brane theory}

As in the neutral examples, the induced theory on the brane essentially follows from replacing the IR bulk cutoff in holographic renormalization with a brane. When the bulk theory of gravity is Einstein-Maxwell, the induced brane action (\ref{eq:inducactqbtzneu}) receives corrections due to the bulk Maxwell contribution, such that the induced brane action is now 
\beq I=\frac{1}{16\pi G_{3}}\int_{\mathcal{B}}d^{3}x\sqrt{-h}\left[R-2\Lambda_{3}+L_{4}^{2}\left(\frac{3}{8}R^{2}-R_{ij}^{2}\right)+...\right]+I_{\text{EM}}+I_{\text{CFT}}\;,\label{eq:inducactqbtzchar}\eeq
where the electromagnetic term $I_{\text{EM}}$ is 
\beq I_{\text{EM}}=2\int d^{3}x\sqrt{-h}A_{i}j^{i}+I^{\text{ct}}_{\text{EM}}\;.\eeq
The first contribution is a boundary term with respect to the bulk Maxwell term necessary to keep the Dirichlet variational problem well-posed. The second contribution describes local counterterms associated with the four-dimensional bulk Maxwell action that are included in holographic renormalization\footnote{In addition to the local counterterms in pure gravity, $p$-form fields $F_{p}$ (where $p=2$ corresponds to Maxwell) may require local counterterm subtraction. As reported in \cite{Taylor:2000xw}, for a $d+1$-dimensional bulk, when $d<2p$ there are no divergences while a logarithmic divergence appears for $d=2p$ and there will be divergences for $d>2p$. Further, for $d=2p+2n$ with $n\in\mathbb{Z}^{+}$, derivatives of $F_{p}$ and its coupling to curvature appear in the conformal anomaly such that counterterms are needed for $d>2p+2n$. Thus, the four-dimensional Maxwell action has no divergences as the IR cutoff $\epsilon\to0$. Such terms, however, contribute on the brane because the brane effective action keeps the cutoff finite and non-zero.} \cite{Taylor:2000xw}
\beq
\begin{split}
 I^{\text{ct}}_{\text{EM}}=\frac{L_{4}\ell^{2}_{\ast}}{8\pi G_{4}}&\int d^{3}x\sqrt{-h}\biggr[-\frac{5}{16}F^{2}+L_{4}^{2}\biggr(\frac{1}{288}RF^{2}-\frac{5}{8}R^{i}_{\;j}F_{ik}F^{jk}\\
&+\frac{3}{98}F^{ij}(\nabla_{j}\nabla^{k}F_{ki}-\nabla_{i}\nabla^{k}F_{kj})+\frac{5}{24}\nabla_{i}F^{ij}\nabla_{k}F^{k}_{\;j}\biggr)+\mathcal{O}(L_{4}^{3})\biggr]\;.
\end{split}
\eeq
Then, treating $\ell<\ell_{3}$ such that $L_{4}\sim \ell$, the effective theory on the brane (\ref{eq:inducactqbtzneu2}) is now
\beq
\begin{split}
 I&=\frac{1}{16\pi G_{3}}\int d^{3}x\sqrt{-h}\biggr[R+\frac{2}{\ell_{3}^{2}}-\frac{\tilde{\ell}_{\ast}^{2}}{4}F^{2}+16\pi G_{3}A_{i}j^{i}\\
&+\ell^{2}\left(\frac{3}{8}R^{2}-R_{ij}^{2}\right)+\frac{4}{5}\ell^{2}\tilde{\ell}^{2}_{\ast}\biggr(\frac{1}{288}RF^{2}-\frac{5}{8}R^{i}_{\;j}F_{ik}F^{jk}\\
&+\frac{3}{98}F^{ij}(\nabla_{j}\nabla^{k}F_{ki}-\nabla_{i}\nabla^{k}F_{kj})+\frac{5}{24}\nabla_{i}F^{ij}\nabla_{k}F^{k}_{\;j}\biggr)+\mathcal{O}(\ell^{3})\biggr]+I_{\text{CFT}}\;.
\end{split}
\label{eq:indactbranecharge}\eeq
where in addition to the induced scales (\ref{eq:indG3}) $G_{3}=G_{4}/2L_{4}$, and $L_{3}$ (\ref{eq:indL3}), there is an effective three-dimensional gauge coupling
\beq \tilde{\ell}_{\ast}^{2}=\frac{16\pi G_{3}}{g_{3}^{2}}\;,\quad g_{3}^{2}=\frac{2}{5}\frac{g_{\ast}^{2}}{L_{4}}\approx \frac{2}{5}\frac{g_{\ast}^{2}}{\ell}\;,\eeq
such that $\ell_{\ast}^{2}=4\tilde{\ell}^{2}_{\ast}/5$. It is clear that in the limit $\ell\to0$, the coupling $g_{3}\to\infty$ becomes non-dynamical. Indeed, we will see how this `charge' contribution to the metric disappears. 

The metric equations of motion of the induced theory to order $\mathcal{O}(\ell^{2})$ are as the neutral case, except now with an additional contribution coming from the $F^{2}$ contribution in the action. That is, 
\beq 
\begin{split}
8\pi G_{3}\langle T_{ij}\rangle&=G_{ij}-\frac{1}{L_{3}^{2}}h_{ij}-\frac{\tilde{\ell}^{2}_{\ast}}{2}\left(F^{k}_{i}F_{jk}-\frac{1}{4}h_{ij}F^{2}\right)+16\pi G_{3} A_{k}j^{k}h_{ij}+...\;,
\end{split}
\label{eq:stresstenscqbtzlead}\eeq
where the ellipsis constitutes terms at higher order in $\ell$; the $\mathcal{O}(\ell^{2})$ contribution is precisely the same as in the neutral case (\ref{eq:T2stressstat}), while $\mathcal{O}(\ell^{3})$ contributions arise from the $\ell^{2}\tilde{\ell}_{\ast}^{2}$ term in the action (\ref{eq:indactbranecharge}). Further, varying with respect to the gauge field $A_{i}$, we find the analog of the semi-classical Maxwell equations, 
\beq
\begin{split}
\langle J^{j}\rangle&=j^{j}+\frac{\tilde{\ell}^{2}_{\ast}}{16\pi G_{3}}\biggr\{\nabla_{i}F^{ji}+\frac{16}{5}\ell^{2}\biggr(-\frac{1}{72}R\nabla_{i}F^{ji}+\frac{11}{18}F^{j}_{\;\,i}\nabla^{i}R+\frac{209}{294}R^{ij}\nabla_{k}F_{i}^{\;k}\\
&+\frac{5}{4}R^{ik}\nabla_{k}F^{j}_{\;\,i}+\frac{5}{4}F^{ik}\nabla_{k}R^{j}_{\;i}+\frac{317}{588}\nabla_{i}\nabla^{i}\nabla_{k}F^{jk}+\frac{317}{588}\nabla^{j}\nabla^{k}\nabla_{i}F^{ik}\biggr)+\mathcal{O}(\ell^{3})\biggr\}\;.
\end{split}
\label{eq:sccurrentdens}\eeq

\subsection*{Quantum black hole}

Upon imposing bulk regularity conditions (equivalent to those for the neutral, static geometry), the induced geometry on the brane at $x=0$, in terms of canonically normalized coordinates $(t,r,\phi)=(\eta\bar{t},\eta^{-1}\bar{r},\eta\bar{\phi})$, is \cite{Climent:2024nuj} (see also \cite{Feng:2024uia})
\beq
\boxed{
\begin{split}
ds^2_{\text{cqBTZ}} = & -H(\bar{r})d\bar{t}^{2}+H^{-1}(\bar{r})d\bar{r}^{2}+\bar{r}^{2}d\bar{\phi}^{2}\;,\\
&H(\bar{r})=-8M\mathcal{G}_{3}+\frac{\bar{r}^{2}}{\ell_{3}^{2}}-\frac{\ell F(M,q)}{\bar{r}}+\frac{\ell^{2}Z(M,q)}{\bar{r}^{2}}\;.
\end{split}
}
\label{eq:qbtzchargeatmet}\eeq
Here the mass is identified to be
\beq M\equiv-\frac{\kappa}{8\mathcal{G}_{3}}\eta^{2}=-\frac{\kappa}{8\mathcal{G}_{3}}\frac{4x_{1}^{2}}{(3-\kappa x_{1}^{2}+q^{2}x_{1}^{4})^{2}}\;,\label{eq:masscqbtz}\eeq
and form functions
\beq F(M,q)\equiv \mu\eta^{3}=8\frac{1-\kappa x_{1}^{2}-q^{2}x_{1}^{4}}{(3-\kappa x_{1}^{2}+q^{2}x_{1}^{4})^{3}}\;,\eeq
\beq Z(M,q)\equiv q^{2}\eta^{4}=\frac{16 q^{2}x_{1}^{4}}{(3-\kappa x_{1}^{2}-q^{2}x_{1}^{4})^{4}}\;. \eeq
As an exact solution to the semi-classical theory (\ref{eq:indactbranecharge}), the geometry (\ref{eq:qbtzchargeatmet}) is recognized as the `charged' version of the quantum BTZ black hole. Notice, however, in the limit $\ell\to0$ the $q$-dependent correction vanishes, indicating the charge of the braneworld black is a consequence of backreaction. Notably, the classical geometry ($\ell=0$) is not that of the charged BTZ metric \cite{Martinez:1999qi}, we will return to this point momentarily. 
%which has a logarithmic correction to the blackening factor \cite{Martinez:1999qi}, and classical gauge field is logarithmic in the radial coordinate. One way of understanding the lack of logarithmic behavior in the backreacted solution is that four-dimensional gauge field does not localize on the brane. As such, the correction proportional to $Z(M,q)$ does not go like $\sim Q^{2}/\bar{r}^{2}$, where $Q$ is the charge. 

Substituting the metric (\ref{eq:qbtzchargeatmet}) into the stress-tensor at leading order in $\ell$ is equivalent to the neutral static qBTZ stress-tensor (\ref{eq:renstressstatqbtz}), while the effects of charge arise at $\mathcal{O}(\ell^{2})$. Further, in coordinates $(\bar{t},\bar{r},\bar{\phi})$, the projected components of the electromagnetic tensor are
\beq F_{\bar{r}\bar{t}}=\frac{2e\ell}{\ell_{\ast}\bar{r}^{2}}\eta^{2}\;,\quad f_{\bar{\phi}}=-\frac{2g\ell}{\ell_{\ast}\bar{r}}\eta^{2}\;.\eeq
Using the junction conditions (\ref{eq:junccondEM}), the induced current density is 
\beq j^{\bar{\phi}}=\frac{g\ell\eta^{2}}{\pi\ell_{\ast}\bar{r}^{3}}\;.\eeq
With these, the leading order $\mathcal{O}(\ell)$ contribution to the semi-classical current density (\ref{eq:sccurrentdens}) has components
\beq
\begin{split}
&\langle J^{\bar{t}}\rangle=-\frac{\ell\tilde{\ell}_{\ast}^{2}}{8\pi G_{3}\ell_{\ast}}\frac{e\eta^{2}}{\bar{r}^{3}}\propto \frac{e\ell\sqrt{c}}{g_{\ast}\bar{r}^{3}}\;,\\
& \langle J^{\bar{\phi}}\rangle=\frac{\ell}{\ell_{\ast}}\frac{g\eta^{2}}{\pi\bar{r}^{3}}\propto \frac{gg_{\ast}\sqrt{c}}{\bar{r}^{3}}\;.
\end{split}
\eeq
Interestingly, the temporal component vanishes in the limit $\ell\to0$ while the azimuthal component is independent of backreaction. This is consistent with three-dimensional dyonic defect in conical AdS$_{3}$ or a BTZ black hole \cite{Climent:2024nuj}.

Unlike the rotating case, the limit of vanishing backreaction does not return a classically AdS$_{3}$ geometry charged under three-dimensional Maxwell theory. Indeed, in three dimensions, the (electric) gauge field $A_{t}$ has a logarithmic dependence, $A_{t}\sim q\log(r)$, producing a logarithmic correction to the three-dimensional blackening factor of classical charged BTZ \cite{Martinez:1999qi}. This lack of logarithmic behavior arises from the fact the bulk four-dimensional gauge field does not localize on the brane in the same way as a gravity \cite{Climent:2024nuj}. Still, the quantum black hole (\ref{eq:qbtzchargeatmet}) is charged. As with mass or angular momentum, computing charge $Q$ from the brane perspective would require a resummation of the infinite tower of higher-derivative terms appearing in the induced theory. Alternatively, the bulk theory performs this resummation, and the charge of the brane black hole is identified with the electric charge of the bulk black hole \cite{Climent:2024nuj} (see also \cite{Emparan:2000fn})
\beq Q=\frac{2}{g_{\ast}^{2}}\int\star F\;,\eeq
where the factor of two in the first equality is due to the $\mathbb{Z}_{2}$ symmetry and $\star F$ refers to the Hodge dual of the bulk Maxwell tensor, and the integration is taken to be at the boundary. In particular, the electric charge is 
\beq Q_{e}=\frac{2e\ell}{g_{\ast}^{2}\ell_{\ast}}\int_{0}^{2\pi\eta}\hspace{-2mm}d\phi\int_{0}^{x_{1}}\hspace{-1mm}dx=\frac{8\pi\ell e \eta x_{1}}{g_{\ast}^{2}\ell_{\ast}}\;,
\label{eq:echargeqbtz}\eeq
where $\star F=r^{2}F_{rt}d\phi dx$. Similarly, the magnetic charge $Q_{g}$ is 
\beq Q_{g}=\frac{2}{g_{\ast}^{2}}\int F=\frac{8\pi\ell g \eta x_{1}}{g_{\ast}^{2}\ell_{\ast}}\;,\eeq
and $Q^{2}=Q_{e}^{2}+Q_{g}^{2}$. 
%We will see further evidence this is the correct form of the brane black hole charge when we study the horizon thermodynamics. 
Notice $Q^{2}$ does not explicitly appear in the blackening factor (\ref{eq:qbtzchargeatmet}).

Associated with the electric and magnetic charges are their respective potentials $\mu_{e}$ and $\mu_{g}$. Since, holographically, the bulk Einstein-Maxwell theory (\ref{eq:EinMaxact}) has a dual interpretation in terms of a CFT$_{3}$ with a chemical potential, it is natural to refer to $\mu_{e}$ and $\mu_{g}$ as chemical potentials. In particular, the electric chemical potential equals the electric component of the bulk gauge field (\ref{eq:gaugpot}) at the boundary intersecting the brane at $x=0$ (where $r_{\text{bdry}}\to-\infty$), 
\beq \lim_{r_{\text{bdry}}\to-\infty}A_{\bar{t}}\equiv \mu_{e}=\frac{2\ell e\eta}{r_{+}\ell_{\ast}}\;.\label{eq:elecpot}\eeq
Similarly, under electromagnetic duality of the bulk solution, the magnetic chemical potential $\mu_{g}$ has the same form as (\ref{eq:elecpot}), except with $e\leftrightarrow g$. Notice $\mu_{g}Q_{e}-\mu_{e}Q_{g}=0$.

\subsubsection*{A family of charged quantum black holes}

Having included $q$, the parameter $x_{1}$ depends on $q$ belongs to a different range than the static (\ref{eq:paramrangex1v1}) or rotating case. To determine this range, two conditions are imposed \cite{Climent:2024nuj}: (i) the bulk has a horizon $r_{0}>0$, and (ii) $x_{1}$ is finite, having a maximum $x_{1}^{\text{max}}$, such that constant $(t,r)$-surfaces are compact. The finite maximum value of $x_{1}$ is linked to the fact the bulk (and brane) geometry has an extremal limit. Demanding there are no naked singularities requires $\mu$ to be bounded below by $\mu^{(\kappa)}_\text{ext}$, i.e., the value of $\mu$ when the bulk black hole becomes extremal. The parameter $x_{1}=x_{1}^{\text{max}}$ when $\mu=\mu_{\text{ext}}^{(\kappa)}$.

Explicitly,  for arbitrary $\kappa$, the extremal mass is \cite{Climent:2024nuj} 
\beq
\label{eq:ext_funccharge}
\mu^{(\kappa)}_{\text{ext}} = \sqrt{\frac{2}{3}}\frac{\sqrt{\sqrt{\kappa^2 + 12 \nu^2 q^2}-\kappa}\left( 2\kappa+\sqrt{\kappa^2 + 12 \nu^2 q^2}\right)}{3 \nu} \ .
\eeq
This can be derived as follows. First, the condition of extremality two positive real roots of $H(r)$ be coincident, i.e., $r_{+}=r_{-}=r_{0}$ implies $H(r)=(r-r_{0})^{2}f(r)$ for some differentiable function $f(r)$ such that $H(r_{0})=H'(r_{0})=0$. A little algebra yields  
\beq
\label{eq:extremal_funcrad}
    \mu_{\text{ext}}^{(\kappa)}(r_0) = \frac{2 r_0}{\ell}\left( \kappa +\frac{2 r_0^2}{\ell_3^2}\right) \ , \qquad q_{\text{ext}}^{(\kappa)}(r_0) = \frac{r_0}{\ell}\sqrt{\kappa+\frac{3 r_0^2}{\ell_3^2}} \ .
\eeq
Inverting $q^{(\kappa)}_{\text{ext}}(r_{0})$ for $r_{0}$ gives 
\beq r_{0}^{(\kappa)}=\frac{\ell_{3}\sqrt{\sqrt{\kappa^{2}+12\nu^{2}q^{2}}-\kappa}}{\sqrt{6}}\;.\eeq
Substituting this into $\mu_{\text{ext}}^{(\kappa)}$ (\ref{eq:extremal_funcrad}) yields (\ref{eq:ext_funccharge}). Evidently, extremality occurs when backreaction is non-vanishing ($\nu\neq0$) or charge is non-zero for $\kappa=+1$. Alternatively, for $\kappa=-1$, there exists an extremal radius, $r_{0}^{(-1)}=\ell_{3}/\sqrt{3}$, even when $q=0$. and where $\mu_{\text{ext}}^{(-1)}<0$. In fact, generally $\mu_{\text{ext}}^{(-1)}$ can become negative (whenever $q<\frac{1}{\sqrt{12}\nu}$), as does $F(M,q)$, for $\kappa=-1$; meanwhile $\kappa=+1,0$ gives $\mu_{\text{ext}}^{(-1)}\geq0$.
%Geometrically, the near-horizon geometry of the extremal black hole takes the form AdS$_{2}\times S^{1}$. 

\begin{figure}[t]
\centerline{\includegraphics[width=0.7\textwidth]{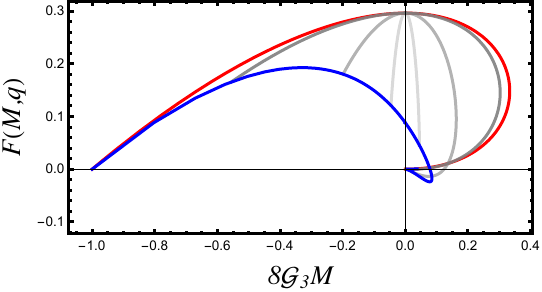}}
\caption{\small \textbf{Branches of charged qBTZ}. Metric function $F(M,q)$ as a function of mass $M$ for fixed $q$ and $\nu=1/5$. The blue (bottom) and red (top) lines correspond to, respectively, extremal and neutral qBTZ solutions. The gray lines are non-extremal charged solutions with, from darkest to lightest, $q=1/5,1,5$.}
\label{fig:branches}\end{figure}

Further, $x_{1}^{\text{max}}$ is determined by substituting the extremal parameter (\ref{eq:ext_funccharge}) into (\ref{eq:muderivcbtz}) and solving for the resulting $x_{1}$ in terms of $q$ and $\nu$.
%Physically, it is easier to interpret the parameter space if one has an expression for the extremal mass in terms of the physical charge of the black hole, rather than its horizon radius $r_+$. To derive \eqref{eq:ext_funccharge} it suffices to invert \eqref{eq:extremal_funcrad} to find the radius of the extremal black hole as a function of charge, and substitute that back in into the expression for the extremal mass.
%For AdS branes there is however no upper bound. 
%To see this, notice that $G(x)$ is a monotonic function of $\mu$ and $|q|$, decreasing at fixed $x$ with increasing mass and charge. It is therefore easy to extract the physical range for the value of $x_1$. For fixed charge, the maximum value $x^{\text{max}}_1$ is attained when the bulk black hole is extremal, since it corresponds to the smallest allowed mass that is cloaked by a horizon. Moreover, as $q\to 0$ the lower bound for the mass approaches zero, making larger and larger values of $x_1$ accessible.
%The minimum value $x^{\text{min}}_1=0$ is obtained in the large $\mu$ limit instead. \EP{As we'll see later, this is no longer true in the de Sitter case.} \AS{Is mentioning $x_{\text{min}}$ important here?}
%Substituting this into $G(x)$ and solving for the smallest positive root gives the maximum allowed value of $x_1$ for fixed $q$ and $\nu$, which can be done analytically. 
An analytic expression for $x_{1}^{\text{max}}$ is possible but cumbersome and not particularly illuminating. Some intuition can be gained, however, by looking at the limit when both $q$ and $\nu$ are small. Three cases worth highlighting are:
\begin{itemize}
    \item $\kappa = 1$: for $q=0$, $x_1$ is bounded above by 1, as $\mu \to 0$. For $q\ll1$, this upper bound is lowered to
    \beq
    x_1^{\text{max,1}} = 1-q+2 q^2 .
    \eeq
    \item $\kappa = -1$: for $q=0$, $x_1$ covers the whole real line, approaching $x_1 = 0$ from above as $\mu\to \infty$ and vice versa. Turning on $q\ll1$, the maximum $x_1^{\text{max}}$ is
    \beq
    x_1^{\text{max,0}} = \frac{2}{3\sqrt{3}\nu q^2}+ \frac{\sqrt{3}\nu}{2}-\frac{21 \sqrt{3}\nu^3 q^2}{8}+ ... \ .
    \eeq
\item $\kappa = 0$: in the neutral case, the parameter range is equivalent to the $\kappa=-1$ case, however, for $q\ll1$ then
    \beq
    x_1^{\text{max,0}} = \frac{h(\nu)}{\sqrt{q}} \ ,
    \eeq
with $h(\nu) = 1-\frac{\sqrt{\nu}}{3^{3/4}} + \frac{\nu}{2\sqrt{3}}+\mathcal{O}(\nu^{3/2})+...$
\end{itemize}
A visual representation of the branches for the charged qBTZ black hole can be found in Figure \ref{fig:branches}. Notably, the branches have qualitatively similar features as the neutral qBTZ. This is because the finite mass range of the charged black holes is a subset of the mass range of the neutral qBTZ (\ref{eq:massbhrange}). The branch with $M<0$ correspond to the classically horizonless charged defects, now with a horizon induced by quantum backreaction.  Note, however, unlike the neutral or rotating quantum BTZ solutions, here $F(M,q)$ can go negative for any $\nu$ when charge parameter $q$ is large enough.

\begin{figure}[t]
\centerline{\includegraphics[width=0.5\textwidth]{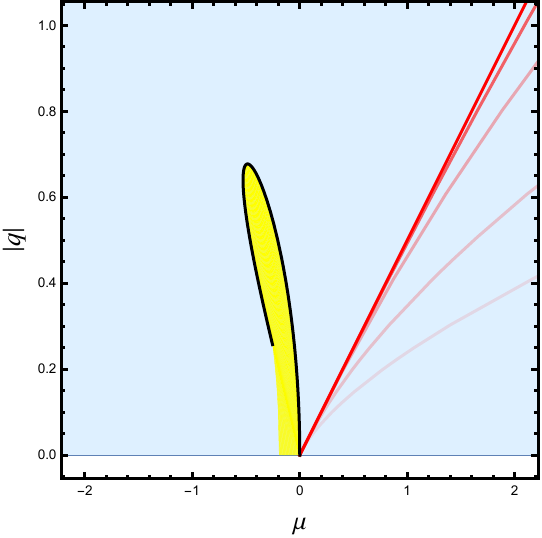}}			
\caption{\small \textbf{Parameter space for charged quantum BTZ}. The blue square and yellow `banana' regions correspond to solutions with a conical defect and excess, respectively, delimited by a thick black contour. The rightward directed (red) lines denote the extremal family of solutions at fixed $\nu$ (from darkest to lightest, $\nu = 0,1/3,1,5,20$). Here $\kappa=1$. For details, refer to \cite{Climent:2024nuj}.}
%which bounds from the left the region of parameter space where singularities are cloaked by a horizon. }
\label{fig:banana}\end{figure}

The limit of vanishing backreaction, $\nu\to0$, is more subtle than the neutral quantum BTZ solutions. As is usual, in this limit the brane is sent closer to the asymptotic AdS$_{4}$ boundary, where both gravity and the gauge theory become frozen. The geometry becomes that of a charged defect in conical AdS$_{3}$ or Mink$_{3}$ ($\kappa=+1$) or a black hole ($\kappa=-1$). Since the latter is a black hole geometry, backreaction ($\nu\neq0$) produces a quantum-corrected charged black hole, i.e., the horizon is not induced solely due to backreaction. Alternatively, for the conical charged defects, horizons can only arise via quantum effects. Unlike the neutral set-up, however, whether a horizon appears depends on a balance between $\mu$ and $q$ -- backreaction does not always dress the conical defects with a horizon. Specifically, when the non-backreacted solution has $q\geq\mu/2$, then backreaction will result in a spacetime possessing a  naked (timelike) singularity \cite{Climent:2024nuj}. Meanwhile, for $q<\mu/2$, backreaction will produce a quantum black hole provided backreaction is not too large. As $\nu$ increases for a fixed $q$, an extremal black hole forms, and even stronger backreaction results in naked singularities. 
%Note that the region where $q>\mu/2$ is akin to when $Q>M$ for standard RN black holes. Such geometries characterize the gravitational field of a charged particle down to distances where quantum electrodynamic effects modifies the electromagnetic stress-tensor such that it is less singular. Backreaction, in fact, can even produce naked singularities when there were none in the non-backreaction conical defect geometries.
For an illustration see Figure \ref{fig:banana}.

%%%%%%%%%%%%%%%%%%%%%%%%%%%%%%%%%%%%%%%%%%%%%%%%%%%%%%%%%%%%%%%%%%

\subsection{Quantum dS black holes} \label{ssec:qdsbhs}

In Section \ref{sec:perturbanalysis} we saw, perturbatively, semiclassical backreaction modifies conical dS$_{3}$ geometries to induce a black hole horizon (in addition to its classical cosmological horizon). This observation is bolstered using braneworld holography. The construction and analysis are similar to the quantum BTZ black hole, however, with some essential differences. 

\subsubsection{Bulk set-up}

The starting point in the bulk is again the AdS$_{4}$ C-metric (\ref{eq:AdS4cmetBLstat}) with metric functions (\ref{eq:metfuncsstatCmet}), except with fixed $\kappa=+1$ and $\ell_{3}^{2}=-R_{3}^{2}$ (or Wick rotate $\ell_{3}\to iR_{3}$). The AdS$_{4}$ length scale is then 
\beq \frac{1}{L_{4}^{2}}=\frac{1}{R_{3}^{2}}-\frac{1}{\ell^{2}}\;.\label{eq:L4bulkRS}\eeq
Unlike the qBTZ set-up, keeping $L_{4}^{2}>0$ requires $R_{3}^{2}>\ell^{2}$, placing an upper bound on $\ell$ for fixed $R_{3}$. Further, rearranging (\ref{eq:L4bulkRS}) yields $\ell<L_{4}$ -- the opposite of the quantum BTZ construction. This condition is the first crucial difference between the quantum BTZ and dS set-ups, as maintaining $\ell<L_{4}$ implies the acceleration horizon is not effectively eliminated. To see this explicitly, consider the analog of the coordinate transformation (\ref{eq:coordinatetransflat}) \cite{Emparan:2022ijy}
\beq \sinh(\sigma)=\frac{R_{3}}{L_{4}}\frac{1}{|1+\frac{rx}{\ell}|}\sqrt{1-\frac{x^{2}r^{2}}{R_{3}^{2}}}\;,\quad \hat{r}=r\sqrt{\frac{1-x^{2}}{1-\frac{x^{2}r^{2}}{R_{3}^{2}}}}\;,\label{eq:coordtransdsslice}\eeq
such that the line element of the C-metric (\ref{eq:AdS4cmetBLstat}) (with $\mu=0$) becomes
\beq ds^{2}=L_{4}^{2}d\sigma^{2}+\frac{L_{4}^{2}}{R_{3}^{2}}\sinh^{2}(\sigma)\left[-\left(1-\frac{\hat{r}^{2}}{R_{3}^{2}}\right)dt^{2}+\left(1-\frac{\hat{r}^{2}}{R_{3}^{2}}\right)^{-1}d\hat{r}^{2}+\hat{r}^{2}d\phi^{2}\right]\;.\label{eq:dS3fols}\eeq
Clearly, constant-$\sigma$ slices give dS$_{3}$ in static patch coordinates with radius $R_{3}=L_{4}\sinh(\sigma)$, and cosmological horizon at $\hat{r}=R_{3}$.

The cosmological horizon at $\hat{r}=R_{3}$ (for constant $\sigma$) is in fact identified with the bulk acceleration horizon. This can be seen via the coordinate transformation 
\beq \frac{t}{R_{3}}=\frac{t_{\text{R}}}{L_{4}}\;,\quad \cosh(\sigma)=\frac{\rho}{L_{4}}\cosh(\vartheta)\;,\quad \frac{\hat{r}^{2}}{R_{3}^{2}}=\frac{\rho^{2}\sinh^{2}(\vartheta)}{\rho^{2}\cosh^{2}(\vartheta)-L_{4}^{2}}\;,\eeq
which brings the line element (\ref{eq:dS3fols}) to AdS$_{4}$-Rindler form, 
\beq ds^{2}=-\left(\frac{\rho^{2}}{L_{4}^{2}}-1\right)dt_{\text{R}}^{2}+\left(\frac{\rho^{2}}{L_{4}^{2}}-1\right)^{-1}d\rho^{2}+\rho^{2}(d\vartheta^{2}+\sinh^{2}(\vartheta)d\phi^{2})\;.\eeq
Orbits of the Rindler-time translation Killing vector $\partial_{t_{\text{R}}}$ correspond to uniformly accelerating trajectories. The cosmological horizon in (\ref{eq:dS3fols}) corresponds to the (non-compact) acceleration horizon $\rho=L_{4}$, with horizon temperature $T_{\text{R}}=(2\pi L_{4})^{-1}$. 

As in the qBTZ black hole, when $\mu\neq0$ the root structure of the metric function $H(r)$ results in a black hole horizon. Meanwhile, real roots of $G(x)$ correspond to orbits of $\partial_{\phi}$, and to ensure a finite black hole horizon we again restrict to the range $0\leq x\leq x_{1}$, where $x_{1}$ is the smallest root of $G(x)$. Specifically, since $\kappa=+1$, it follows $x_{1}\in(0,1]$ (\ref{eq:paramrangex1v1}), where $\mu$, via (\ref{eq:defmu}), is monotonically decreasing from $+\infty$ to zero, with $\mu=0$ for $x_{1}=1$. Further, as before, the conical singularity at $x=x_{1}$ is removed via the identification (\ref{eq:conicaldef}) with $\kappa=+1$. 

\begin{figure}[t!]
\begin{center}
\includegraphics[width=.2\textwidth, height=.32\textwidth]{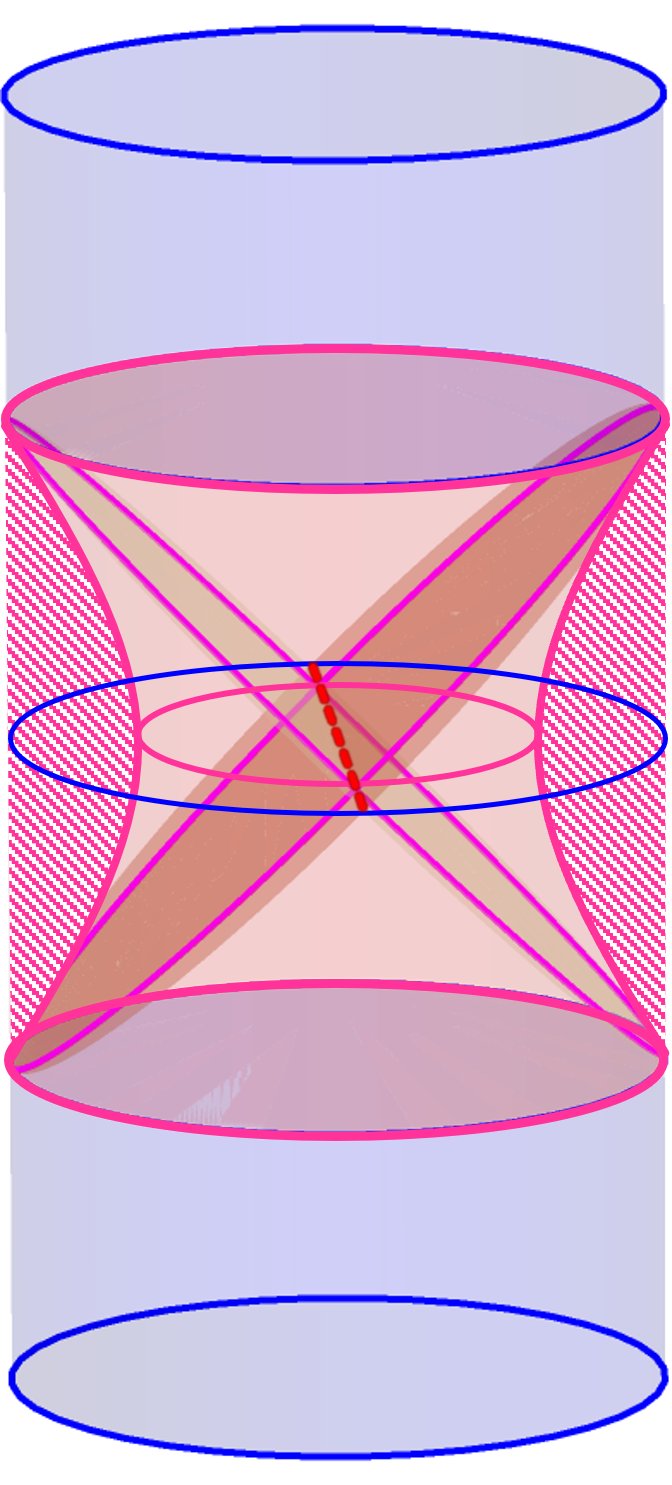} $\quad\quad\quad$ \includegraphics[width=.32\textwidth]{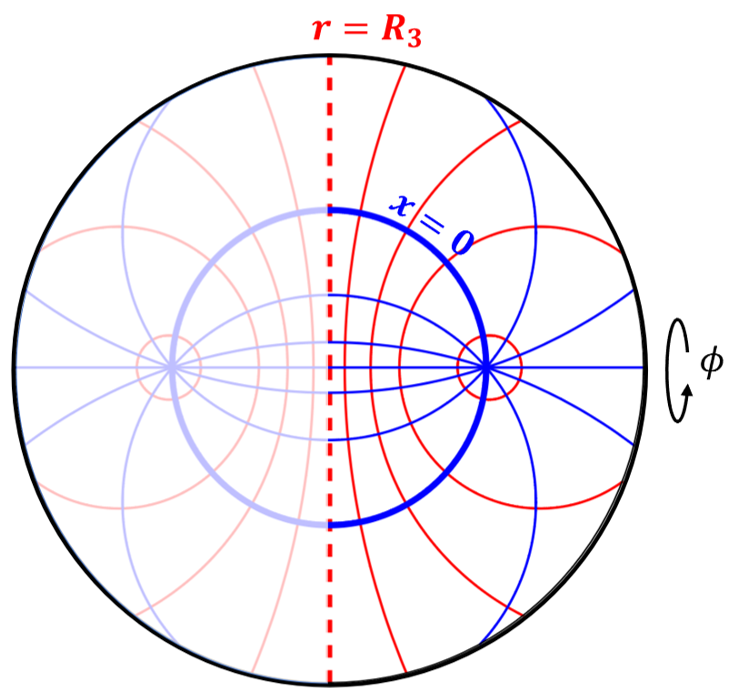} $\quad\quad\quad$ \includegraphics[width=.3\textwidth]{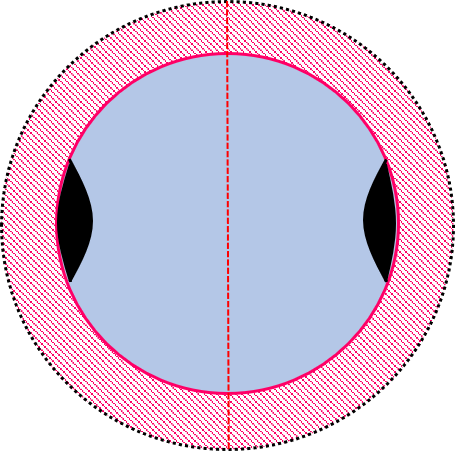}
\end{center}
\caption{\small \textbf{Randall-Sundrum braneworld.} \emph{Left:} Bulk $\text{AdS}_{4}$ with a dS$_3$ brane. The brane is represented as a (magenta) hyperboloid. The bulk region up to the brane ($x<0$, dashed magenta region) is excluded. To complete the construction, glue a second copy along the two-sided brane. Cosmological horizons on the brane correspond to bulk acceleration horizons intersecting the brane (red dashed line). \emph{Center:} AdS$_4$ C-metric with $\mu=0$, in $(r,x)$ coordinates in a slice at $t=0$ and constant $\phi$. Lines of constant $x$ are blue arcs; lines of
constant $r$ are red arcs (full circles for $0 < r <\ell$). The thick blue circle $x=0$ is where we place the dS$_3$ brane; its interior is $0<x\leq1$, with $x=1$ the $\phi$ axis of rotation. The exterior region $x<0$ is excluded in the braneworld construction. The vertical red dashed line is the horizon at $r=R_3$. Its intersection with the brane yields a dS$_3$ cosmological horizon.  The coordinates only cover half of the disk, with the other half being obtained through analytic continuation. \emph{Right:} Constant $t$-time slice of a single $\text{AdS}_{4}$ cylinder with a de Sitter brane (thick red circle) containing black holes. The coordinates cover half of the disk, containing a single black hole and cosmological horizon, where the other half is obtained via analytic continuation.}
\label{fig:dSbranebhs} 
\end{figure}

The next essential difference between the quantum AdS and dS black hole constructions is that an asymptotically $\text{dS}_{3}$ Randall-Sundrum brane is embedded AdS$_{4}$ at the umbilic $x=0$ hypersurface. Israel junction conditions again fix the brane tension to be $\tau=(2\pi G_{4}\ell)^{-1}$. In terms of the empty dS$_{3}$ geometry (\ref{eq:dS3fols}), the brane sits at 
\beq \sinh(\sigma_{b})=\frac{R_{3}}{L_{4}}\;,\eeq
excluding the region $\sigma>\sigma_{b}$. Notice, moreover, the area of the bulk horizon at $\hat{r}=R_{3}$ is finite, 
\beq 2\pi L^{2}_{4}(\cosh(\sigma_{b})-1)=2\pi R^{2}_{3}\frac{\ell}{\ell+R_{3}}\;.\eeq
That is, while the bulk acceleration horizon is generally non-compact (since it extends to the asymptotic boundary at $\sigma\to\infty$), its intersection with the brane is compact. Thus, the bulk acceleration horizon induces a (compact) cosmological horizon on the dS$_{3}$ brane. When $\mu\neq0$, the bulk black hole horizon is projected onto the brane resulting in an induced geometry
\beq ds^{2}|_{x=0}=-\left(1-\frac{r^{2}}{R_{3}^{2}}-\frac{\mu\ell}{r}\right)dt^{2}+\left(1-\frac{r^{2}}{R_{3}^{2}}-\frac{\mu\ell}{r}\right)^{-1}dr^{2}+r^{2}d\phi^{2}\;,\label{eq:x0ds3naive}\eeq
with a single black hole and cosmological horizon. See Figure \ref{fig:dSbranebhs} for an illustration. 

Another caveat about the Randall-Sundrum brane construction is that the brane geometry will contain Big Bang and Big Crunch singularities in the asymptotic past and future. The reason follows because of the amount the brane radiates as it accelerates in the bulk.
From the brane perspective, the time to reach the asymptotic past or future dS boundaries is infinite. Thus the brane emits an infinite amount of radiation, thereby causing a piling of rays at future/past Cauchy horizons.  The formation of the cosmological singularities is essentially the same phenomenon that enforces (strong) cosmic censorship at the inner Cauchy horizon in charged or rotating black holes. Alternatively, the Randall-Sundrum braneworld can be understood in terms of false vacuum decay \cite{Maldacena:2010un,Barbon:2010gn,Barbon:2011ta} mediated by Coleman and de Luccia bubbles \cite{Coleman:1980aw}  of `nothing'. This vacuum decay process inevitably leads to Big Bang/Big Crunch singularities.
%\footnote{It is nature to wonder why such an event does not happen in usual AdS spacetimes, given that we can have the same dS slicing of empty AdS as well. The key difference lies in the amount of radiation one produces: in order to model the brane radiation, one would need to employ infinite shocks from the boundary in order to produce the same singular effect. Of course, such a state would be pathological to begin with, and so we have no (naturally formed) Big Bang and Big Crunch singularities.} 

\subsubsection{Quantum Schwarzschild-de Sitter}

The naive metric (\ref{eq:x0ds3naive}) is not cast in canonically normalized coordinates, as $\phi$ has periodicity $\phi\sim\phi+\Delta\phi$. Rescaling coordinates $(t,r,\phi)\to(\eta\bar{t},\eta^{-1}\bar{r},\eta\bar{\phi})$ with $\eta$ given in (\ref{eq:etacanon}) (for $\kappa=+1$), the geometry on the dS$_{3}$ brane is 
\beq
\hspace{-2mm}\boxed{ds^{2}_{\text{qSdS}} = -H(\bar{r})d\bar{t}^2 + H(\bar{r})^{-1}  d\bar{r}^2 + \bar{r}^2 d\bar{\phi}^2\;,\quad H(\bar{r})=\left(1-8\mathcal{G}_{3}M-\frac{\bar{r}^{2}}{R_{3}^{2}}-\frac{\ell F(M)}{\bar{r}} \right)}
\label{eq:qSdS}\eeq
where we identify
\beq
8\mathcal{G}_{3}M \equiv 1-\eta^{2}=1-\frac{4x_{1}^{2}}{(3-x_{1}^{2})^{2}}\;,\qquad F(M)\equiv 8\frac{(1- x_{1}^{2})}{(3- x_{1}^{2})^{3}}
\label{eq:massqSdSFM}\eeq
with renormalized Newton's constant $\mathcal{G}_3\equiv G_3 \frac{L_4}{\ell}$. The mass identification is motivated by how the mass of de Sitter black holes in Einstein-dS gravity is typically given by the subleading constant term in the $g_{tt}$ component of the metric. 

Analogous to the quantum BTZ set-up, the dS metric (\ref{eq:qSdS}) has the same form as the perturbative solution (\ref{eq:pertbackdS3geom}) to the semi-classical Einstein equations. Again, here, the $1/r$ correction is exact, and the metric is an exact solution to the whole tower of higher-derivative terms of the induced action. Substituting the metric into the gravitational equations of motion (see (\ref{eq:T0stressstat}) -- (\ref{eq:T2stressstat}) with $\ell_{3}^{2}=-R_{3}^{2}$), yields the same structure of the stress-energy tensor $\langle T^{i}_{\;j}\rangle_{0}$ (\ref{eq:renstressstatqbtz}) and $\langle T^{i}_{\;j}\rangle_{2}$ (\ref{eq:renstressstatqbtz2}). In the limit of vanishing backreaction, the geometry (\ref{eq:qSdS}) takes the form of classical Schwarzschild-dS$_{3}$, a conical defect with a single cosmological horizon. For $\ell\neq0$, the metric is interpreted as a static quantum black hole in dS$_{3}$, i.e., a three-dimensional \emph{quantum Schwarzschild-de Sitter} black hole (qSdS) \cite{Emparan:2022ijy}. As in the qBTZ system, the qSdS black hole for $\ell>0$ has a curvature singularity at $\bar{r}=0$ that is hidden behind a black hole horizon induced by semi-classical backreaction. 

In contrast with the qBTZ system,  the qSdS solution does not subdivide into several branches of quantum black holes. Indeed, since $\kappa=+1$, here $0<x_{1}<1$. In this range, given the mass identification (\ref{eq:massqSdSFM}), the qSdS always has $M\geq0$, with $M=1/8\mathcal{G}_{3}$ and $F=8/27$ at $x_{1}=0$, and $M=F=0$ at $x_{1}=1$. The $M=0$ solution with $\ell\neq0$ is dubbed \emph{quantum} dS$_{3}$, accounting for a large-$c$ cutoff CFT living in dS$_{3}$. Meanwhile, the $M=1/8\mathcal{G}_{3}$ solution is the analog of the upper bound on the classical Schwarzschild-dS$_{3}$ conical defect. 

The quantum SdS black hole, in fact, has a more stringent upper bound on mass than the conical defect mass. This is a consequence of the fact that the geometry (\ref{eq:qSdS}) has both a black hole and cosmological horizon, and is in complete analogy with classical higher-dimensional Schwarzschild-de Sitter black holes. It is well known that classical SdS black holes (in four or higher spacetime dimensions) have an upper bound on their mass in order to avoid the presence of a naked singularity. This largest mass black hole is known as the Nariai black hole \cite{Nariai99}, with mass $M_{\text{N}}$. Geometrically, the Nariai black hole is one in which the (typically smaller) black hole horizon $r_{h}$ coincides with the (typically larger) cosmological horizon $r_{c}$, i.e., $r_{h}=r_{c}\equiv r_{\text{N}}$, the Nariai horizon radius. In the context of the AdS$_{4}$ C-metric, the four-dimensional geometry has a Nariai limit that induces a Nariai geometry on the brane at $x=0$ \cite{Emparan:2022ijy}
\beq ds^{2}_{\text{N}}=-\left(1-\frac{\rho^{2}}{\bar{r}_{\text{N}}^{2}}\right)d\tau^{2}+\left(1-\frac{\rho^{2}}{\bar{r}_{\text{N}}^{2}}\right)^{-1}d\rho^{2}+\bar{r}_{\text{N}}^{2}d\bar{\phi}^{2}\;.\label{eq:qNariaigeom}\eeq
Here $\tau$ and $\rho$ are time and radial coordinates, respectively, and $\bar{r}_{\text{N}}=\eta r_{\text{N}}$ with $r_{\text{N}}=R_{3}/\sqrt{3}$. The (bulk) Nariai black hole places an upper bound on $(\mu\ell)$ in the C-metric; specifically $(\mu\ell)\leq(\mu\ell)_{\text{N}}=2r_{\text{N}}/3$. This bulk upper bound places an upper bound on the mass $M$ of the qSdS black hole, denoted $M_{\text{N}}$. The precise form of $M_{\text{N}}$ is complicated, however, the Nariai mass was found to live in the finite range \cite{Emparan:2022ijy}
\beq \frac{11}{27}<8\mathcal{G}_{3}M<1\;,\eeq
where the upper limit corresponds to when $\ell/R_{3}\to0$, while the lower limit occurs for $\nu\approx 1$.
Therefore, the Nariai mass bound $M\leq M_{\text{N}}$ is generally more restrictive than the conical defect bound $M\leq 1/8\mathcal{G}_{3}$. Notably, the classical SdS$_{3}$ conical defect does not have a Nariai limit. Thus, semi-classical backreaction induces an upper limit on the amount of mass allowed in dS$_{3}$ which does not
saturate the maximum conical deficit angle. 

Finally, the Nariai solution puts a bound on the quantum backreaction due to the CFT for which a quantum dS$_{3}$ black hole exists. In particular, for non-vanishing backreaction $(\ell\neq0)$, the form function $F(M)$ has a maximum value. Correspondingly, the angular deficit $\Delta\phi$ has a maximum value. For too large of angular deficits,  backreaction creates a black hole with a mass $M>M_{\text{N}}$ such that it is too large to fit inside the dS$_{3}$ static patch. For such deficits, the quantum SdS solution no longer exists, and the resulting geometry is described by a naked conical defect spacetime, %(whose holographic description is given by an `unexcited' CFT), 
analogous to the braneworld geometry induced by the bulk BTZ black string.

\subsubsection{Quantum Kerr-de Sitter}

As in the static case, rotating quantum dS$_{3}$ black holes can be constructed in essentially the same way as the rotating qBTZ (\ref{eq:qbtzrotatmet}). Starting from the rotating C-metric (\ref{eq:rotatingCmetqbtz}) with $\kappa=+1$ and replacing $a\to-a$, and imposing bulk regularity, the metric on the brane in canonically normalized coordinates is \cite{Panella:2023lsi}\footnote{The rotating qBTZ metric (\ref{eq:qbtzrotatmet})  follows from the double replacement $\ell_{3}\to iR_{3}$ and $a_{\text{AdS}_{3}}\to-a_{\text{dS}_{3}}$, such that $\tilde{a}_{\text{AdS}_{3}}\to i\tilde{a}_{\text{dS}_{3}}$.}
\beq
\boxed{
\begin{split}
ds^{2}_{\text{qKdS}}&=-\left(1-8\mathcal{G}_{3}M-\frac{\bar{r}^{2}}{R_{3}^{2}}-\frac{\mu\ell\eta^{2}}{r}\right)d\bar{t}^{2}\\
&+\left(1-8\mathcal{G}_{3}M-\frac{\bar{r}^{2}}{R_{3}^{2}}+\frac{(4\mathcal{G}_{3}J)^{2}}{\bar{r}^{2}}-\frac{\mu\ell(1+\tilde{a}^{2})^{2}\eta^{4}r}{\bar{r}^{2}}\right)^{-1}d\bar{r}^{2}\\
&+\left(\bar{r}^{2}+\frac{\mu\ell\tilde{a}^{2}R_{3}^{2}\eta^{2}}{r}\right)d\bar{\phi}^{2}-4\mathcal{G}_{3}J\left(1+\frac{\ell}{x_{1}r}\right)(d\bar{\phi}d\bar{t}+d\bar{t}d\bar{\phi})\;.
\end{split}
}
\label{eq:branegeomv2}\eeq
with mass and angular momentum identified as
\beq 8\mathcal{G}_{3}M\equiv 1-\eta^{2}\left(1-\tilde{a}^{2}+\frac{4\tilde{a}^{2}}{x_{1}^{2}}\right)=1-\frac{4[x_{1}^{2}-\tilde{a}^{2}(x_{1}^{2}-4)]}{(3-x_{1}^{2}+\tilde{a}^{2})^{2}}\;,\label{eq:Massid}\eeq
\beq 4\mathcal{G}_{3}J\equiv-R_{3}\tilde{a}\mu x_{1}\eta^{2}=\frac{4R_{3}\tilde{a}(x_{1}^{2}+\tilde{a}^{2}-1)}{(3-x_{1}^{2}+\tilde{a}^{2})^{2}}\;.\label{eq:angJ}\eeq
The metric (\ref{eq:branegeomv2}) is dubbed  the quantum Kerr-dS$_{3}$ black hole since when $\ell\to0$ the geometry of the classical Kerr-ds$_{3}$ conical defect is recovered.  
%Before we analyze the brane geometry (\ref{eq:branegeomv2}) in more detail, there are a few special limits to consider. First, clearly, when the rotation $a\to0$, then $J=0$ and the geometry reduces to the static metric (\ref{eq:metqsds}), the quantum Schwarzschild-de Sitter black hole \cite{Emparan:2022ijy}. Next, in the limit of vanishing backreaction $\ell\to0$, in which the gravitational effects of the cutoff CFT are suppressed (where $\mathcal{G}_{3}\to G_{3}$), the metric (\ref{eq:branegeomv2}) takes the form of the classical Kerr-$\text{dS}_{3}$ conical defect spacetime (see Appendix \ref{app:KerrdS3}). Thirdly,
A couple of limits worth noting, (i) when the parameter $\mu$ (\ref{eq:muatdef}) vanishes, i.e., $x_{1}=\sqrt{1-\tilde{a}^{2}}$, both $M=J=0$, resulting in the empty $\text{dS}_{3}$ geometry, and (ii) mass $M$ also vanishes when $x_{1}=\sqrt{9-\tilde{a}^{2}}$, however, $J\neq0$ and $\mu\neq0$, resulting in 
%\beq 4\mathcal{G}_{3}J=\frac{32\tilde{a}R_{3}}{(6-2\tilde{a}^{2})^{2}}\;,\quad \mu=\frac{512 \tilde{a}R_{3}}{(\tilde{a}^{2}-9)^{3}(\tilde{a}^{2}-3)^{2}}\;,\eeq
 quantum rotating $\text{dS}_{3}$.

 Substituting the geometry (\ref{eq:branegeomv2}) into the gravity field equations results in a holographic stress-tensor with components of essentially the same form as (\ref{eq:stresstenscomponentsqbtzrot}) -- see \cite{Panella:2023lsi} for explicit details. Now the function $F(M,J)$ from evaluating the large-$\bar{r}$ behavior of $\langle T^{\bar{t}}_{\;\bar{t}}\rangle_{0}$ is
 \beq F(M,J)\equiv\mu\eta^{3}\sqrt{1+\tilde{a}^{2}}(1-2\tilde{a}^{2})=\frac{8\sqrt{1+\tilde{a}^{2}}(1-2\tilde{a}^{2})(1-x_{1}^{2}-\tilde{a}^{2})}{(3-x_{1}^{2}+\tilde{a}^{2})^{3}}\;.\eeq
 Notice $F(M,J)$ will vanish either when $\mu=0$, i.e., $x_{1}^{2}=1-\tilde{a}^{2}$. The zero $\tilde{a}^{2}=1/2$, meanwhile, is unique only to the $\langle T^{\bar{t}}_{\;\bar{t}}\rangle_{0}$ component, with the remaining components of the holographic stress-tensor being non-zero for this value of $\tilde{a}$. This stands in contrast with the rotating qBTZ black hole, for which every component vanishes when $\tilde{a}^{2}=1$. Further, as in the qBTZ case, the radial dependence of the renormalized stress-tensor due to a conformally coupled scalar field (found perturbatively in \cite{Panella:2023lsi}) is significantly more complicated than the holographic stress-tensor.

\subsubsection*{Horizon structure}

Horizons in the bulk correspond to positive real roots $r_{i}$ of $H(r)$, where the Killing vector
\beq \zeta^{b}=\partial_{t}-\frac{a}{r_{i}^{2}}\partial_{\phi}\;\label{eq:zetavecv1}\eeq
becomes null. To classify the types of horizons, define the function $Q(r)\equiv r^{2}H(r)$. Since $Q(r)$ is a quartic polynomial in $r$, it will generally have either four, two, or zero real roots. Consider the case when there are four real roots: three positive roots correspond to the cosmological horizon $r_{c}$, the outer black hole horizon $r_{+}$, and inner black hole horizon $r_{-}$, obeying $r_{-}\leq r_{+}\leq r_{c}$, while the fourth root, $r_{n}=-(r_{c}+r_{+}+r_{-})<0$ and resides behind the singularity at $r=0$. Using $H(r_{c})=0$, and $H(r_{\pm})=0$, we can express
\beq
\begin{split}
&R_{3}^{2}=r_{c}^{2}+r_{+}^{2}+r_{c}r_{+}+r_{-}(r_{c}+r_{+}+r_{-})\;,\\
&\mu\ell=\frac{(r_{c}+r_{+})(r_{c}+r_{-})(r_{+}+r_{-})}{r_{c}^{2}+r_{+}^{2}+r_{c}r_{+}+r_{-}(r_{c}+r_{+}+r_{-})}\;,\\
&a^{2}=\frac{r_{c}r_{+}r_{-}(r_{c}+r_{+}+r_{-})}{r_{c}^{2}+r_{+}^{2}+r_{c}r_{+}+r_{-}(r_{c}+r_{+}+r_{-})}\;.
\end{split}
\label{eq:paramsintermsofri}\eeq
%The blackening factor $H(r)$ factorizes as 
%\beq 
%\begin{split}
%H(r)&=
%=\frac{1}{R_{3}^{2}r^{2}}(R_{3}^{2}r^{2}-r^{4}+a^{2}R_{3}^{2}-r\mu\ell R_{3}^{2})\\
%\frac{1}{R_{3}^{2}r^{2}}(r_{c}-r)(r-r_{+})(r-r_{-})(r+r_{c}+r_{+}+r_{-})\;.
%\end{split}
%\eeq
The limit $r_{-}\to0$ coincides with $a=0$, while $r_{+}=r_{-}=0$ corresponds to $\mu\to0$, resulting in the Kerr-$\text{dS}_{3}$ geometry with a single cosmological horizon. 

The positive roots $r_{i}$ to $H(r)$ correspond to rotating horizons with rotation $\Omega_{i}$, 
\beq \Omega_{i}\equiv \frac{a}{R_{3}^{2}}\frac{(x_{1}^{2}r_{i}^{2}-R_{3}^{2})}{(r_{i}^{2}+a^{2}x_{1}^{2})}\;,\label{eq:Omi}\eeq
generated by the (canonically normalized) Killing vector
%\beq \zeta^{b}=\frac{\left(1+\frac{a^{2}x_{1}^{2}}{r_{i}^{2}}\right)}{\eta(1+\tilde{a}^{2})}\left(\partial^{b}_{\bar{t}}+\frac{a}{R_{3}^{2}}\frac{(x_{1}^{2}r_{i}^{2}-R_{3}^{2})}{(r_{i}^{2}+a^{2}x_{1}^{2})}\partial^{b}_{\bar{\phi}}\right)\;.\eeq
\beq \bar{\zeta}^{b}\equiv\frac{\eta(1+\tilde{a}^{2})}{\left(1+\frac{a^{2}x_{1}^{2}}{r_{i}^{2}}\right)}\zeta^{b}=\partial_{\bar{t}}^{b}+\Omega_{i}\partial_{\bar{\phi}}^{b}\;.\label{eq:zetakillvec}\eeq
Moreover, the surface gravity $\kappa_{i}$ associated with each horizon $r_{i}$ is given by 
\beq \kappa_{i}=\frac{\eta(1+\tilde{a}^{2})}{\left(r_{i}^{2}+a^{2}x_{1}^{2}\right)}\frac{r_{i}^{2}}{2}|H'(r_{i})|=\frac{\eta(1+\tilde{a}^{2})}{\left(r_{i}^{2}+a^{2}x_{1}^{2}\right)}\frac{1}{2R_{3}^{2}r_{i}}|R_{3}^{2}\mu\ell r_{i}-2r_{i}^{4}-2a^{2}R_{3}^{2}|
%=\frac{\eta(1+\tilde{a}^{2})}{2\left(r_{i}^{2}+a^{2}x_{1}^{2}\right)}\biggr|\mu\ell-\frac{2r_{i}^{3}}{R_{3}^{2}}-\frac{2a^{2}}{r_{i}}\biggr|
\;,\label{eq:surfacegravs}\eeq
where we used the definition $\bar{\zeta}^{b}\nabla_{b}\bar{\zeta}^{c}=\kappa\bar{\zeta}^{c}$. Explicitly, 
\beq 
\begin{split}
&\kappa_{c}=-\frac{\eta(1+\tilde{a}^{2})}{2R_{3}^{2}\left(r_{c}^{2}+a^{2}x_{1}^{2}\right)}(r_{c}-r_{+})(r_{c}-r_{-})(r_{+}+r_{-}+2r_{c})\;,\\
&\kappa_{+}=\frac{\eta(1+\tilde{a}^{2})}{2R_{3}^{2}\left(r_{+}^{2}+a^{2}x_{1}^{2}\right)}(r_{c}-r_{+})(r_{+}-r_{-})(r_{c}+r_{-}+2r_{+})\;,\\
&\kappa_{-}=-\frac{\eta(1+\tilde{a}^{2})}{2R_{3}^{2}\left(r_{-}^{2}+a^{2}x_{1}^{2}\right)}(r_{c}-r_{-})(r_{+}-r_{-})(r_{c}+r_{+}+2r_{-})\;.
\end{split}
\label{eq:surfacegravsv2}\eeq
Notice the cosmological horizon surface gravity $\kappa_{c}$ vanishes when $r_{c}=r_{+}$ or $r_{c}=r_{-}$, and similarly for the other surface gravities. When $r_{-}\to0$, i.e., $\kappa_{c}$ and $\kappa_{+}$ simplify to the surface gravities of the cosmological horizon and black hole horizon of the qSdS black hole \cite{Emparan:2022ijy}. 
%Additionally, in the limit of vanishing backreaction, where $r_{\pm}\to0$, only the 

In the naive coordinates, a computation of the Kretschmann scalar reveals a curvature singularity at $r=0$, corresponding to a ring singularity at $\bar{r}=r_{s}$ in canonically normalized coordinates. As in the rotating qBTZ black hole, closed timelike curves can appear near the ring singularity. These closed timelike curves can be eliminated via an appropriate periodic identification \cite{Booth:1998gf}, such that constant $\bar{t}$ hypersurfaces are closed and span two black hole regions with opposite spin, cutting through intersections of $r=r_{c}$ and $r=r_{+}$.

\vspace{3mm}

\noindent \textbf{Ergoregions.} As with classical Kerr-dS spacetimes, the qKdS black hole has a stationary limit surface and two ergoregions associated with the outer black hole and cosmological horizons. Explicitly, the Killing vector $\partial_{t}$ in the naive metric has norm $\mathcal{N}$
\beq \mathcal{N}=-H(r)+\frac{a^{2}}{r^{2}}\;.\eeq
Thus, at the outer and cosmological horizons, the time-translation Killing vector $\partial_{t}$ becomes spacelike. The locus of points where $\mathcal{N}=0$ yields a stationary limit surface, satisfying $r(R_{3}^{2}-r^{2})=R_{3}^{2}\mu\ell$. Since there exist regions in between the outer and cosmological horizons where $\partial_{t}$ is timelike, there are two ergoregions, where an observer is forced to move in the direction of rotation of the outer black hole horizon or cosmological horizon. In principle, the Penrose process of energy extraction in the qKdS solution in morally the same way as a classical four-dimensional Kerr-de Sitter black hole (see, \emph{e.g.}, \cite{Bhattacharya:2017scw}). 
%At least for small backreaction, it is expected the Penrose process in the cosmological ergoregion is not possible.  

\vspace{3mm}

\noindent \textbf{Extremal limits.} As with the four-dimensional Kerr-de Sitter black hole, the quantum Kerr-$\text{dS}_{3}$ has a number of limiting geometries. Specifically, (i) extremal or `cold' limit, where $r_{+}=r_{-}$; (ii) rotating Nariai limit, where $r_{c}=r_{+}$, and (iii) the `ultracold' limit where $r_{c}=r_{+}=r_{-}$. In the near-horizon regime, the geometries appear as fibered products of a circle and two-dimensional anti-de Sitter, de Sitter, and Minkowski space, respectively. For details, see \cite{Panella:2023lsi}. These limiting geometries, moreover, have the same qualitative features as (warped) $\text{dS}_{3}$ black hole solutions to topologically massive gravity, cf. \cite{Nutku:1993eb,Anninos:2009jt,Anninos:2011vd}. There is also a `lukewarm' limit, where the surface gravities $\kappa_{c}=\kappa_{+}$ at a value different from the surface gravity of the Nariai black hole.

% \subsubsection{Charged quantum dS}

% blah

\subsection{Quantum black holes in flat space} \label{ssec:qbhsmink}

A point mass in three-dimensional Minkowski space, Mink$_{3}$, is described by a conical singularity with no horizon; the Schwarzschild solution in three dimensions is not a black hole. Again, quantum backreaction effects alter the geometry such that the black hole horizon appears to hide the conical singularity \cite{Souradeep:1992ia}. The only known way to consistently construct an exact quantum black hole in asymptotic Mink$_{3}$ space is via braneworld holography. 

The bulk set-up is the same as for quantum dS$_{3}$ black holes: the AdS$_{4}$ C-metric with a two-sided Randall-Sundrum brane. In some respects, the asymptotically flat quantum black holes may be viewed as a special limit, $R_{3}\to\infty$, of the dS$_{3}$ black holes. Though they were the first exact three-dimensional braneworld black hole solution to be discovered \cite{Emparan:1999wa},  quantum black holes in Mink$_{3}$ have received less attention than their (A)dS cousins \cite{Emparan:2002px}. For this reason, we present a fresh take on quantum Mink$_{3}$ black holes.

\subsubsection{Bulk set-up}

Consider the static C-metric (\ref{eq:AdS4cmetBLstat}) with metric functions (\ref{eq:metfuncsstatCmet}), except $\kappa=+1$ and without the (A)dS factor, such that $H(r)$ has the form
\beq H(r)=1-\frac{\mu\ell}{r}\;.\eeq
It is straightforward to verify the bulk geometry is an Einstein metric with a negative cosmological constant where the bulk length scale is 
\beq L_{4}=\ell\;.\label{eq:L4bulkMink}\eeq
This is the first notable difference compared to the (A)dS examples, and will have ramifications as we proceed. This means the acceleration horizon is present in the spacetime. However, it has a different effect than in the dS case where $\ell<L_{4}$. 

For $\mu\neq0$, the bulk geometry has a black hole horizon at $H(r_{+})=0$, 
\beq r_{+}=\mu\ell\;,\eeq
hiding the curvature singularity at $r=0$. There is a second horizon at $r\to\infty$ corresponding to the bulk AdS$_{4}$ horizon.\footnote{By AdS `horizon' we mean the null hypersurface infinitely far from the brane in spacelike directions, but can nonetheless be reached by an observer in finite proper
time.} To see this, consider the bulk geometry with $\mu=0$ (such that $H(r)=1$ and $G(x)=1-x^{2}$). The coordinate transformation 
\beq w\equiv \ell+xr\;,\quad \hat{r}\equiv r\sqrt{1-x^{2}}\;,\label{eq:Poinccoord}\eeq
brings the bulk metric to empty AdS$_{4}$ in Poincar\'e form
\beq ds^{2}=\frac{\ell^{2}}{w^{2}}\left(-dt^{2}+d\hat{r}^{2}+\hat{r}^{2}d\phi^{2}+dw^{2}\right)\;.\eeq
Evidently,  $r=0$ no longer represents a curvature singularity but instead a non-singular worldline $w=\ell$, $\hat{r}=0$. Meanwhile, $xr=-\ell$ maps to the asymptotic boundary $w=0$ and $r\to\infty$ to the AdS$_{4}$ horizon $w\to\infty$. 

As is now standard, real roots of $G(x)$ correspond to orbits of $\partial_{\phi}$. Let $x_{1}$ denote the smallest root of $G(x)$ and  again restrict to the range $0\leq x\leq x_{1}$. With $\kappa=+1$, $x_{1}\in(0,1]$ (\ref{eq:paramrangex1v1}), where $\mu$ is defined in (\ref{eq:defmu}). The conical singularity at $x=x_{1}$ is removed via the identification (\ref{eq:conicaldef}) with $\kappa=+1$.

\begin{figure}[t!]
\centering
 \includegraphics[width=13.3cm]{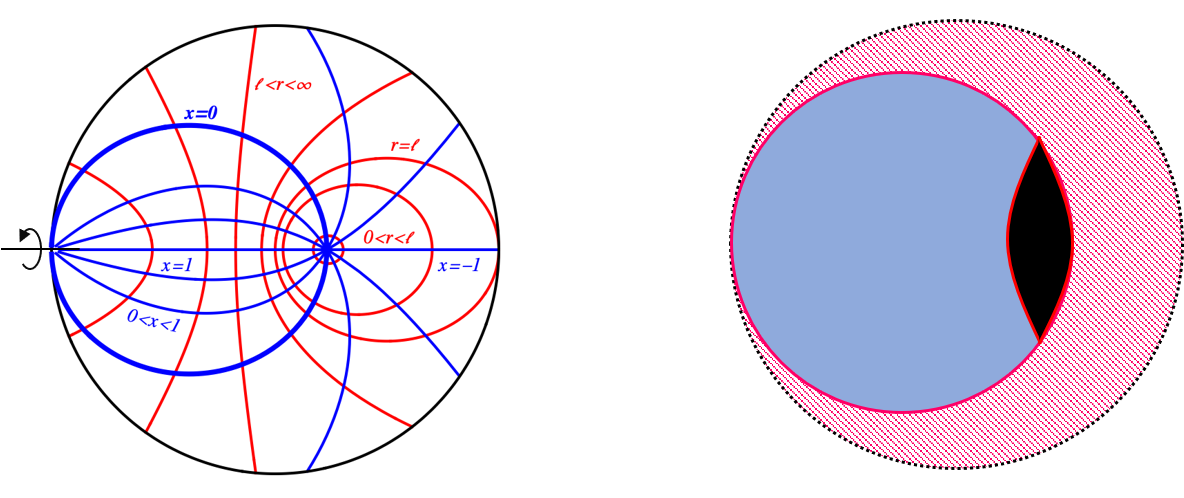}
%\put(-270,-8){\footnotesize $(\mu=0)$}
%\put(-90,-8){\footnotesize $(\mu\neq0)$}
\caption{\small \textbf{Randall-Sundrum braneworld}. \emph{Left:} A constant $t$ and $\phi$ slice of the $\text{AdS}_{4}$ C-metric with $\mu=0$ and $\kappa=+1$ (Poincar\'e disc). Lines of constant $x$ are in blue while lines of constant $r$ are in red. The $\phi$-axis of rotation is at $x=\pm1$. The thick blue circle is $x=0$. \emph{Right:} Schematic of a asymptotically flat Randall-Sundrum ETW brane at $x=0$ with a static black hole ($\mu\neq0$). The shaded magenta region is excluded. The acceleration horizon for the dS$_{3}$ braneworld has been effectively pushed to the AdS$_{4}$ boundary.}
\label{fig:RSflatbrane}\end{figure}

The $x=0$ hypersurface is umbilic. In Poincar\'e coordinates with $\mu=0$ (\ref{eq:Poinccoord}), this surface is located at $w=\ell$. We place a two-sided ETW Randall-Sundrum brane at $x=0$, retaining only the region $0\leq x\leq x_{1}$ (Figure \ref{fig:RSflatbrane}). To complete space, we glue another copy of the region $0\leq x\leq x_{1}$ along the brane at $x=0$. Israel junction conditions fix the brane tension to be $\tau=(2\pi G_{4}\ell)^{-1}$. Notably, since $\ell=L_{4}$, the tension is said to be at its critical value. As we will see momentarily, this condition leads to a vanishing induced cosmological constant. The induced geometry at $x=0$ is 
\beq ds^{2}_{x=0}=-\left(1-\frac{\mu\ell}{r}\right)dt^{2}+\left(1-\frac{\mu\ell}{r}\right)^{-1}dr^{2}+r^{2}d\phi^{2}\;.\label{eq:naiveflat}\eeq
For $\ell\neq0$, this geometry describes a black hole with horizon radius $r_{+}=\mu\ell$. In terms of coordinate $w$, the black hole horizon is located at $w_{+}=\ell+xr_{+}$. Therefore, the brane at $w=\ell$ cuts through the bulk black hole horizon such that the black hole extends only slightly off of the brane to a distance $w_{\text{max}}=\ell+r_{+}x_{1}\leq \ell+r_{+}$ (since $x_{1}\leq 1$). 

The induced brane theory is \ref{eq:inducactqbtzneu}), however, since $L_{4}=\ell$, the induced brane cosmological constant (\ref{eq:indL3}) is zero while the induced Newton's constant (\ref{eq:indG3}) is $G_{3}=G_{4}/2\ell$. The central charge of the cutoff CFT$_{3}$ is $c_{3}=\frac{L^{2}_{4}}{G_{4}}=\frac{\ell}{2G_{3}}$. Higher-derivative terms enter at order $\mathcal{O}(\ell^{2})$, and treating the cutoff as small, $\ell\leq 1$, the higher-derivative contributions serve as corrections to three-dimensional Einstein gravity.

\subsubsection{Quantum Schwarzschild black hole}

The angular coordinate $\phi$ in the naive metric (\ref{eq:naiveflat}) has periodicity $\phi\sim\phi+\Delta\phi$. Working with rescaled coordinates $(t,r,\phi)\to(\eta\bar{t},\eta^{-1}\bar{r},\eta\bar{\phi})$ for $\eta$  (\ref{eq:etacanon}) (with $\kappa=+1$), the geometry on the Mink$_{3}$ brane is\footnote{In \cite{Emparan:1999wa,Emparan:2002px} only the naive brane metric is analyzed. While our analysis is qualitatively similar, some of the precise expressions differ.} 
\beq
\hspace{-2mm}\boxed{ds^{2}_{\text{qSchw}} = -H(\bar{r})d\bar{t}^2 + H(\bar{r})^{-1}  d\bar{r}^2 + \bar{r}^2 d\bar{\phi}^2\;,\quad H(\bar{r})=1-8G_{3}M-\frac{\ell F(M)}{\bar{r}}}
\label{eq:qSchwarz}\eeq
where we identified the mass as in the de Sitter context (\ref{eq:massqSdSFM})
\beq
M \equiv \frac{1}{8G_{3}}\frac{(x_{1}^{2}-1)(x_{1}^{2}-9)}{(3-x_{1}^{2})^{2}}\;,\qquad F(M)\equiv 8\frac{(1- x_{1}^{2})}{(3- x_{1}^{2})^{3}}\;.
\label{eq:massqSFM}\eeq
Since $L_{4}=\ell$, the renormalized Newton's constant $\mathcal{G}_{3}=G_{3}$.

The geometry (\ref{eq:qSchwarz}) is an exact solution to the induced semi-classical equations of motion (\ref{eq:semiclasseombrane}) (with $\ell_{3}\to\infty$), with a quantum stress tensor as in (\ref{eq:renstressstatqbtz}) and (\ref{eq:renstressstatqbtz2}) (with $\ell_{3}\to\infty$). Thus, we interpret the solution as the three-dimensional \emph{quantum Schwarzschild} black hole. A curvature singularity at $\bar{r}=0$ is hidden behind a black hole event horizon at 
\beq \bar{r}_{+}=\frac{\ell F(M)}{(1-8G_{3}M)}\;.\eeq
At the horizon the time-translation Killing vector $\bar{\zeta}^{i}=\eta\partial_{\bar{t}}^{i}$ goes null, having surface gravity 
\beq \kappa_{+}=\frac{1}{2}|H'(\bar{r}_{+})|=\frac{\ell F(M)}{\bar{r}_{+}^{2}}=\frac{(1-8G_{3}M)^{2}}{\ell F(M)}\;.\label{eq:surfgravmink}\eeq

\subsubsection{Quantum Kerr black hole}

Rotating black holes in Mink$_{3}$ are constructed by embedding a Randall-Sundrum brane with critical tension inside  the rotating C-metric (\ref{eq:rotatingCmetqbtz}), such that the metric functions are
\beq
\begin{split} 
&H(r)= 1-\frac{\mu\ell}{r}+\frac{a^{2}}{r^{2}}\;,\quad G(x)=1-x^{2}-\mu x^{3}\;,\\
&\Sigma(x,r)=1+\frac{a^{2}x^{2}}{r^{3}}\;.
\end{split}
\eeq
Formally, the bulk regularity analysis is the same, however, some of the expressions are simpler because in this limit $\tilde{a}=0$, a consequence of the $G(x)$ function losing its quartic term compared to its (A)dS siblings. Thus, the metric on the brane in canonically normalized coordinates is 
\beq
\boxed{
\begin{split}
ds^{2}_{\text{qKerr}}&=-\left(1-8G_{3}M-\frac{\mu\ell\eta^{2}}{r}\right)d\bar{t}^{2}+\left(1-8G_{3}M+\frac{(4G_{3}J)^{2}}{\bar{r}^{2}}-\frac{\mu\ell\eta^{4}r}{\bar{r}^{2}}\right)^{-1}d\bar{r}^{2}\\
&+\left(\bar{r}^{2}+\frac{\mu\ell a^{2}x_{1}^{4}\eta^{2}}{r}\right)d\bar{\phi}^{2}-4G_{3}J\left(1+\frac{\ell}{x_{1}r}\right)(d\bar{\phi}d\bar{t}+d\bar{t}d\bar{\phi})\;,
\end{split}
}
\label{eq:branegeomMinkq}\eeq
with canonical coordinates
\beq 
\begin{split}
 &\bar{t}=\eta^{-1}(t-ax_{1}^{2}\phi)\;,\qquad \bar{\phi}=\eta^{-1}\phi\;,\\
 &\bar{r}^{2}=\eta^{2}r^{2}+r_{s}^{2}\;,\qquad r_{s}\equiv-\frac{2ax_{1}^{2}\sqrt{2-x_{1}^{2}}}{(3-x_{1}^{2})}\;,
\end{split}
\eeq
where $\eta=2x_{1}/(3-x_{1}^{2})$. Further, the angular momentum $J$ is identified as
\beq 4G_{3}J\equiv\frac{4a x_{1}^{2}(x_{1}^{2}-1)}{(3-x_{1}^{2})^{2}}\;.\label{eq:angJmink}\eeq
while the mass is identified as in the static case (\ref{eq:massqSFM}), and is thus independent of the rotation parameter $a$. A ring singularity appears at $\bar{r}=r_{s}$.

The metric (\ref{eq:branegeomMinkq}) is dubbed  the quantum Kerr$_{3}$ black hole. In the limit of vanishing backreaction one recovers the classical conical Kerr geometry. At leading order in a small-$\ell$ expansion, the stress-tensor of the holographic CFT$_{3}$ is 
\beq 
\begin{split}
 &\langle T^{\bar{t}}_{\;\bar{t}}\rangle_{0}=\frac{\mu\ell}{16\pi G_{3}r^{3}}\left(1+\frac{3a^{2}x_{1}^{2}}{r^{2}}\right)\;,\\
 &\langle T^{\bar{r}}_{\;\bar{r}}\rangle_{0}=\frac{\mu\ell}{16\pi G_{3}r^{3}}\;,\\
 &\langle T^{\bar{\phi}}_{\;\bar{\phi}}\rangle_{0}=-\frac{\mu\ell}{16\pi G_{3}r^{3}}\left(2+\frac{3a^{2}x_{1}^{2}}{r^{2}}\right)\;,\\
 &\langle T^{\bar{t}}_{\;\bar{\phi}}\rangle_{0}=\frac{3\mu\ell ax_{1}^{2}}{16\pi G_{3}r^{3}}\left(1+\frac{a^{2}x_{1}^{2}}{r^{2}}\right)\;,\\
 &\langle T^{\bar{\phi}}_{\;\bar{t}}\rangle_{0}=-\frac{3\mu\ell a}{16\pi G_{3}r^{5}}\;.
\end{split}
\eeq
Notice now that function $F(M,J)$ from evaluating the large-$\bar{r}$ behavior of $\langle T^{\bar{t}}_{\;\bar{t}}\rangle_{0}$ is exactly equal to $F(M)=\mu\eta^{3}$, as in the static case.

\subsection*{Horizon structure}

Bulk black hole horizons correspond to real roots $r_{i}$ of $H(r)$, i.e., 
\beq r_{\pm}=\frac{1}{2}(\mu\ell\pm\sqrt{(\mu\ell)^{2}-4a^{2}})\;.\eeq
These correspond to rotating horizons with rotation 
\beq \Omega_{\pm}=-\frac{a}{(r_{\pm}^{2}+a^{2}x_{1}^{2})}\;,\eeq
generated by Killing vector $\bar{\zeta}^{j}=\partial^{j}_{\bar{t}}+\Omega_{\pm}\partial^{j}_{\bar{\phi}}$. The surface gravities are
\beq \kappa_{\pm}=\frac{\eta}{(r_{\pm}^{2}+a^{2}x_{1}^{2})}\frac{r_{\pm}^{2}}{2}|H'(r_{\pm})|=\frac{\eta}{2(r_{\pm}^{2}+a^{2}x_{1}^{2})}\biggr|\mu\ell-\frac{2a^{2}}{r_{\pm}}\biggr|\;.\label{eq:flatsurfgrav}\eeq
There is an extremal black hole when the inner and outer horizons coincide, $r_{+}=r_{-}\equiv r_{\text{ex}}$, i.e., when $\mu\ell=2a$. In this limit the surface gravities (\ref{eq:flatsurfgrav}) vanish.

%\subsection*{Quantum Reissner–Nordström black hole}

%%%%%%%%%%%%%%%%%%%%%%%%%%%%%%%%%%%%%%%%%%%%%%%%%%%%%%%%%%%

\section{Quantum black hole thermodynamics} \label{sec:qbhthermo}

Black hole thermodynamics \cite{Bekenstein:1972tm,Bekenstein:1973ur,Hawking:1974sw,Hawking:1976de} reveals an interplay between geometry, quantum mechanics, and thermodynamics. This is encapsulated by the Bekenstein-Hawking entropy-area relation, 
\beq S=\frac{k_{\text{B}}c^{3}}{\hbar G}\frac{A[H]}{4}\;,\label{eq:Bekhawkarealawgen}\eeq
for Boltzmann constant $k_{\text{B}}$, and we have temporarily restored factors of $\hbar$ and speed of light $c$. The area law states black holes carry a thermodynamic entropy proportional to the area of a codimension-2 cross-section of their event horizon $H$. Hence, black holes may be treated as genuine thermal systems with energy, entropy, and temperature, and other thermodynamic variables depending on the type of black hole. In the case of stationary black holes, the laws of black hole mechanics \cite{Bardeen:1973gs} may be reinterpreted as laws of thermodynamics. For example, the first law relates a variation in mass $M$ to variations of the other thermodynamic variables
\beq dM=TdS+...\;,\label{eq:firstlawgeneralbhs}\eeq
with Hawking temperature $T$, and where the ellipsis implies variations of other possible thermodynamic variables, e.g., electric/magnetic charge, rotation, and so forth.

As we review below, just as the bulk geometry imprints itself on the brane, so too does its thermal description. Thus, classical thermodynamics of the bulk black hole geometry is interpreted as semi-classical thermodynamics of the quantum black hole system. This allows for an exact study of quantum black hole thermodynamics at any order in backreaction.

\subsection{Bulk thermodynamics}

When we think of mapping out the properties of a thermodynamic system, we typically think of one near equilibrium. Likewise, in black hole thermodynamics often the first task is to check whether the system has a well-defined notion of thermal equilibrium. Such is the case, for example, of Kerr-Newman black holes in AdS, where thermal equilibrium is unambiguously defined \cite{Hawking:1982dh}. Alternatively, the thermodynamics of black holes in de Sitter space is conceptually more subtle because a static patch observer encounters a system with two horizons at different temperatures (except in special limits, e.g., Nariai), such that the system is not generally in equilibrium. Similarly, due to the presence of an acceleration and black hole horizon, the thermodynamics of the C-metric is generally more subtle than non-accelerating black holes -- even when the accelerating black holes are embedded in AdS. Further, although the C-metric line element does not display time dependence, a uniformly accelerating black holes will deliver non-vanishing radiation at asymptotic infinity \cite{Podolsky:2003gm}. Thus, it is not obvious how an accelerating black hole could possibly be in equilibrium. 

The AdS$_{4}$ C-metric, does, however, have a distinct advantage over the flat or dS C-metrics. Working in a particular regime of parameters, the negative cosmological constant has the effect of essentially removing the acceleration horizon such that it can be consistently neglected. This regime is precisely the `slowly accelerating' black hole \cite{Podolsky:2002nk} where the (inverse) acceleration and AdS$_{4}$ length scale satisfy 
\beq \textbf{Slowly accelerating:} \quad \ell>L_{4}\;.\label{eq:slowaccelbh}\eeq
In such situations, the AdS$_{4}$ C-metric is described by a single black hole suspended away from the origin by a cosmic string attached to the AdS boundary (recall Figure \ref{fig:accelbh}). Consequently, being able to ignore the acceleration allows for the temperature of the black hole system to be defined in a straightforward way. Even still, due to the presence of cosmic string, the black hole is not isolated. This makes consistently carrying out the thermodynamic analysis of the slowly accelerating black hole non-trivial. Indeed, one must account for both the black hole and cosmic string, leading to a modified first law including variations of the tension of the cosmic string \cite{Appels:2016uha,Appels:2017xoe,Anabalon:2018ydc,Appels:2018jcs}.\footnote{A first law for the charged C-metric with vanishing cosmological constant was derived in \cite{Ball:2020vzo} using covariant phase space methods, where, moreover, `boost time'  was treated as canonical time.}

While the starting point is the same, the thermodynamics of the bulk black hole system used in the braneworld black hole constructions in Section \ref{sec:qbhtaxonomy}, is ultimately different from the usual AdS$_{4}$ C-metric. Firstly, in the braneworld construction, there is no cosmic string; instead, there is a brane. For a purely tensional brane of constant tension, the brane does not play an obvious role in the bulk black hole thermodynamics (we will revisit this when we allow for variable tension).  Second, the bulk geometry is regular in that the zeros to the metric function $G(x)$ are removed, along with the conical defects they induce. These two facts make the thermodynamic analysis less subtle than the standard C-metric. 

In general, however, the question of thermal equilibrium remains due to the presence of the bulk acceleration horizon. In fact, this question distinguishes the treatment of the Karch-Randall and Randall-Sundrum braneworld constructions.  For the former, the bulk system started with a slowly accelerating black hole (\ref{eq:slowaccelbh}), such that the acceleration horizon does not imprint itself on the brane. Alternatively, for a dS$_{3}$ Randall-Sundrum brane, the bulk acceleration horizon cannot ignored because one is not in the slow acceleration regime thus complicating the thermal analysis. Instead, $\ell\leq L_{4}$.
%\beq \textbf{Not slowly accelerating} \quad \ell\leq L_{4}\;.\eeq
The edge case $\ell=L_{4}$ corresponds to the flat Randall-Sundrum brane. Recall in this scenario the acceleration horizon does not localize on the brane (as in the dS$_{3}$ Randall-Sundrum brane) such that the acceleration horizon may be ignored. We will return to these differences momentarily.

Moving forward, let us first consider the thermodynamics of the bulk system with an AdS$_{3}$ Karch-Randall braneworld. In the literature on the exact three-dimensional braneworld black holes, typically one assumes the Bekenstein-Hawking entropy formula (\ref{eq:Bekhawkarealawgen}) and Hawking temperature. The energy is then identified by demanding the first law hold, which is subsequently found to coincide with the mass $M$ identified geometric construction. For example, consider the static, neutral black hole construction (\ref{eq:AdS4cmetBLstat}). It proves useful to introduce the real and non-negative parameter \cite{Emparan:1999fd}
\beq z\equiv \frac{\ell_{3}}{r_{+}x_{1}}\;,\label{eq:zdef}\eeq
for black hole horizon radius $r_{+}$. Given the range of $x_{1}$, generally, $z\in[0,\infty)$. It is possible to express the parameters $x_{1},\mu$ and $r_{+}$ solely in terms of $z$ and $\nu\equiv\ell/\ell_{3}$. In particular, solving $H(r_{+})=0$ for $x_{1}^{2}$ yields
\beq x_{1}^{2}=-\frac{1}{\kappa}\frac{(1-\nu z^{3})}{z^{2}(1+\nu z)}\;.\label{eq:x1sqz}\eeq
Rearranging $z$ (\ref{eq:zdef}) and substituting in $x_{1}^{2}$ above gives
\beq r_{+}^{2}=-\ell_{3}^{2}\kappa\frac{(1+\nu z)}{(1-\nu z^{3})}\;.\label{eq:rpsqz}\eeq
Further, with (\ref{eq:x1sqz}) the parameter $\mu$ (\ref{eq:defmu}) obeys
\beq \mu x_{1}=-\kappa\frac{(1+z^{2})}{(1-\nu z^{3})}\;.\label{eq:mux1z}\eeq

The Hawking temperature $T$ of the bulk black hole horizon is proportional to the surface gravity relative to the canonical timelike Killing vector $\partial_{\bar{t}}$ (\ref{eq:surfgravqbtz}). Using  (\ref{eq:x1sqz}) --- (\ref{eq:mux1z}), the temperature can be recast in terms of $z$ and $\nu$, 
\beq T=\frac{\kappa_{+}}{2\pi}=\frac{1}{2\pi\ell_{3}}\frac{z(2+3\nu z+\nu z^{3})}{1+3z^{2}+2\nu z^{3}}\;.\label{eq:bulktempqbtz}\eeq
The Bekenstein-Hawking entropy, meanwhile, is (setting $c=\hbar=k_{\text{B}}=1$)
\beq 
\begin{split}
S=\frac{\text{Area}(r_{+})}{4G_{4}}&=\frac{2}{4G_{4}}\int_{0}^{2\pi\eta}d\phi\int_{0}^{x_{1}}dxr_{+}^{2}\frac{\ell^{2}}{(\ell+x r_{+})^{2}}\\
&=\frac{8\pi\ell_{3}^{2}}{4G_{4}}\frac{\nu z}{1+3z^{2}+2\nu z^{3}}\;,
\end{split}
\label{eq:bulkentqbtz}\eeq
where the factor of two appearing in the second equality is because the brane is two-sided. 

Keeping parameters $\ell_{3}$ and $\nu$ fixed, we can identify the energy $E$ via
\beq \partial_{z}E=T\partial_{z}S\;,\eeq
such that the first law 
\beq dE=TdS\;\label{eq:firstlawbulkgen}\eeq
is obeyed. In particular, the energy $E$ is explicitly found to be 
\beq E=\frac{\sqrt{1+\nu^{2}}}{2G_{3}}\frac{z^{2}(1-\nu z^{3})(1+\nu z)}{(1+3z^{2}+2\nu z^{3})^{2}}=M\;,\label{eq:bulkenergqbtz}\eeq
where in the second equality substituted (\ref{eq:x1sqz}) --- (\ref{eq:mux1z}) into the identified mass (\ref{eq:massqBTZ}). This confirms $M$ should indeed be identified as the mass of the bulk black hole.

\subsection*{Gravitational path integral approach}

The above method is sufficient for studying the thermodynamics, assuming the identification of the thermodynamic variables is correct. A more fundamental approach would be to evaluate the quantum gravitational canonical partition function in the semi-classical limit via the on-shell Euclidean action \cite{Gibbons:1976ue}. Such an approach was taken in \cite{Kudoh:2004ub} (see Appendix \ref{app:onshellaction} for details).

Formally, the partition function $Z(\beta)$ is given by a Euclidean path integral whose fixed boundary data on field configurations corresponds to thermodynamic data defining the thermal ensemble. At leading order in a stationary phase approximation, this becomes
\beq Z(\beta)\approx e^{-I_{\text{on-shell}}}\;,\label{eq:partfuncgen}\eeq
where $\beta$ is the (inverse) temperature of the system and $I_{\text{on-shell}}$ is the on-shell Euclidean action. In say, the Schwarzschild black hole, $\beta$ is introduced as the periodicity of the Euclidean time circle to make the Euclidean solution regular at the horizon. In the case of the AdS$_{4}$ warped geometry, additional care is needed. Firstly, one Wick rotates the Lorentzian geometry (\ref{eq:AdS4cmetBLstat}) $t_{E}=it$ for Euclidean time $t_{E}$. To avoid a conical singularity at $r=r_{+}$, the Euclidean time direction is compactified into a circle, $t_{E}\sim t_{E}+\Delta t_{E}$, with period 
\beq \Delta t_{E}=\frac{4\pi}{|H'(r_{+})|}\;.\eeq
Further, the bulk regularity conditions eliminating the conical singularity at $x_{1}$ must be respected. Combined, one works with Euclideanized canonically normalized time coordinate $\bar{t}_{E}$ with periodicity
\beq \bar{t}_{E}\sim \bar{t}_{E}+\beta\;,\quad \beta=\frac{\Delta t_{E}}{\eta}\;.\label{eq:betagen}\eeq
Working at fixed $\beta$ thus defines working in a canonical ensemble of fixed temperature $T=\beta^{-1}$. Indeed, the periodicity (\ref{eq:betagen}) is the inverse of the temperature (\ref{eq:bulktempqbtz}). 

With respect to the partition function (\ref{eq:partfuncgen}), the thermodynamic energy $E$ and entropy $S$ are defined as 
\beq 
\begin{split}
&E\equiv -\partial_{\beta}\log Z\approx \partial_{\beta}I_{\text{on-shell}}\;,\\
&S\equiv \beta E+\log Z\approx (\beta\partial_{\beta}-1)I_{\text{on-shell}}\;.
\end{split}
\eeq
With some work the on-shell action for the static, neutral black hole is 
\beq I_{\text{on-shell}}=-\frac{2\pi \ell^{2}z}{G_{4}\nu}\frac{[1+2\nu z+\nu z^{3}(2+\nu z)]}{(2+3\nu z+\nu z^{3})(1+3z^{2}+2\nu z^{3})}\;,\eeq
from which one recovers the mass (\ref{eq:bulkenergqbtz}) and entropy (\ref{eq:bulkentqbtz}). In summary, the thermodynamics of the bulk black hole directly follow from a Euclidean gravitational path integral.

Critical to this approach is being able to identify a system in thermal equilibrium. The black hole system in question has multiple non-degenerate horizons, as in the case with the dS$_{3}$ Randall-Sundrum brane. Thus, upon Wick rotating to Euclidean signature, the geometry will have multiple conical singularities, one associated with each horizon. In such cases, one removes a single conical singularity by fixing the periodicity of the Euclidean time coordinate for the associated horizon. Consequently, one is only able to treat a part of the entire system (neglecting the other horizons). Strictly speaking, however, in such situations, the complete system is not in thermal equilibrium and it is not clear how to define a thermal partition function (without further modification).

\subsection*{Identifying bulk and brane thermodynamics}

The thermodynamics of the classical four-dimensional black hole is reinterpreted as the semiclassical thermodynamics of the three-dimensional quantum black hole. This is a by-product of the fact that the bulk black hole horizon localizes on the brane, such that the temperature of the bulk black hole coincides with the temperature of the horizon induced on the brane. To wit, consider a $(d+1)$-dimensional static bulk geometry with line element
\beq ds^{2}=-A(r)dt^{2}+A^{-1}(r)dr^{2}+r^{2}d\Omega_{d-1}^{2}\;,\eeq
where $r=r_{h}$ denotes the event horizon of the black hole, equal to the largest root of $A(r)=0$, and is generated by the time-translation Killing vector $\partial^{a}_{t}$. Let $\Phi=\Phi(r,\phi^{i})$ for $(d-1)$ Gaussian normal coordinates $\{\phi^{i}\}$ denote the hypersurface equation of the brane $\mathcal{B}$. The induced metric on $\mathcal{B}$ is \cite{Myers:2024zhb}
\beq 
\begin{split} 
ds^{2}_{\mathcal{B}}&=-A(r)dt^{2}+\left(A^{-1}(r)+r^{2}(\partial_{r}\Phi)^{2}\right)dr^{2}+r^{2}\gamma_{ij}d\phi^{i}d\phi^{j}\\
&\equiv -f(r)dt^{2}+g^{-1}(r)dr^{2}+r^{2}\gamma_{ij}d\phi^{i}d\phi^{j}\;,
\end{split}
\eeq
having identified
\beq f(r)\equiv A(r)\;,\quad g(r)\equiv \frac{A(r)}{1+A(r)r^{2}(\partial_{r}\Phi)^{2}}\;.\eeq
The bulk event horizon at $r=r_{h}$ corresponds to a horizon on $\mathcal{B}$, i.e., the bulk and brane blackening factors have the same root structure. Further, assuming $\partial_{r}\Phi$ is regular such that $A(r)(\partial_{r}\Phi)^{2}$ vanishes at $r=r_{h}$, the temperature of the bulk black hole coincides with the horizon induced on the brane:
\beq T^{h}_{\text{bulk}}=\frac{|A'(r)|}{4\pi}\biggr|_{r=r_{h}}=\frac{\sqrt{f'(r)g'(r)}}{4\pi}\biggr|_{r=r_{h}}=T^{h}_{\mathcal{B}}\;.\eeq
Geometrically, the identification of the bulk and brane horizon temperatures is because the bulk time-translation Killing vector remains a time-translation Killing vector on the brane.\footnote{More carefully, the projection of the bulk Killing vector $\zeta^{\mu}=\partial^{\mu}_{t}$ is $k^{a}=h^{a}_{\mu}\zeta^{\mu}$, for projector $h^{a}_{\mu}$. Consequently, the surface gravity $\kappa$ on the brane coincides with the bulk surface gravity: $k^{a}D_{a}k^{b}=\kappa k^{b}$, for projected covariant derivative $D_{a}$, implies $h^{a}_{\nu}(\zeta^{\mu}\nabla_{\mu} \zeta^{\nu})=h^{a}_{\nu}(\kappa \zeta^{\nu})$.}

\vspace{3mm}

Below we review the thermodynamics for each type of quantum black hole explored in Section \ref{sec:qbhtaxonomy}, starting with the static neutral quantum BTZ family of black holes. 

\subsection{Quantum BTZ black holes}

\subsubsection{Static quantum BTZ}

The thermodynamic variables of the static bulk black hole  worked out above are
\beq 
\begin{split}
 &M=\frac{\sqrt{1+\nu^{2}}}{2G_{3}}\frac{z^{2}(1-\nu z^{3})(1+\nu z)}{(1+3z^{2}+2\nu z^{3})^{2}}\;,\\
 &T=\frac{1}{2\pi\ell_{3}}\frac{z(2+3\nu z+\nu z^{3})}{1+3z^{2}+2\nu z^{3}}\;,\\
 &S=\frac{8\pi\ell_{3}^{2}}{4G_{4}}\frac{\nu z}{1+3z^{2}+2\nu z^{3}}\;.
\end{split}
\label{eq:qbtzstathermo}\eeq
The claim is that these thermodynamic variables are to be interpreted as the thermodynamic quantities of the quantum black hole. Above we already saw how the temperatures coincide. From the brane perspective, determining the mass is a highly non-trivial task due to the higher-derivative nature of the gravity action. Let us therefore first focus on the entropy $S$. 

\subsubsection*{Entropy}

From the bulk perspective, $S$ is the classical four-dimensional Bekenstein-Hawking entropy, $S=S_{\text{BH}}^{(4)}$ (\ref{eq:bulkentqbtz}). Alternatively, from the brane point of view, $S$ must be a sum of gravitational entropy and the von Neumann entropy of the holographic CFT$_{3}$. That is, $S$ is identified as the \emph{generalized} entropy \cite{Bekenstein:1974ax}:
\beq S_{\text{BH}}^{(4)}\equiv S_{\text{gen}}^{(3)}=\frac{4\pi\ell_{3}}{4G_{3}}\frac{z\sqrt{1+\nu^{2}}}{1+3z^{2}+2\nu z^{3}}\;,\label{eq:S4Sgenqbtz}\eeq
where we replaced $G_{4}=2G_{3}L_{4}=2G_{3}\ell/\sqrt{1+\nu^{2}}$. Because we have the full bulk solution, $S_{\text{gen}}^{(3)}$ is exact, valid for all $\nu$. To parse the gravitational and matter contributions to the entropy, however, it is useful to expand (\ref{eq:S4Sgenqbtz}) in a small $\nu$ expansion, 
\beq S^{(3)}_{\text{gen}}=\frac{4\pi\ell_{3}z}{4G_{3}(1+3z^{2})}-\frac{8\pi\ell_{3}z^{4}}{4G_{3}(1+3z^{2})^{2}}\nu+\frac{2\pi\ell_{3}z(1+6z^{2}+9z^{4}+8z^{6})}{4G_{3}(1+3z^{2})^{3}}\nu^{2}+\mathcal{O}(\nu^{3})\;.\label{eq:expentqbtz}\eeq
The first term can be understood as the three-dimensional Bekenstein-Hawking entropy of the classical BTZ black hole\footnote{It is not obvious Eq. (\ref{eq:SBTZz}) is geometrically equal to the BTZ black hole entropy. Indeed, substituting $z=\ell_{3}/x_{1}r_{+}$ into (\ref{eq:SBTZz}) does not yield the usual $S_{\text{BTZ}}=2\pi r_{+}/4G_{3}$. This is because the qBTZ represents a family of black hole solutions and a specific $x_{1}$ must be chosen to match to the classical black hole. Nonetheless, for any $z$ the relation (\ref{eq:SBTZz}) holds.} 
\beq S_{\text{BTZ}}=\frac{4\pi\ell_{3}z}{4G_{3}(1+3z^{2})}=\frac{\pi^{2}\ell_{3}^{2}}{G_{3}}T_{\text{BTZ}}=\pi\ell_{3}\sqrt{\frac{2M_{\text{BTZ}}}{G_{3}}}\;.\label{eq:SBTZz}\eeq
where $T_{\text{BTZ}}\equiv\lim_{\nu\to0}T$ and similarly for $M_{\text{BTZ}}$. The second term in (\ref{eq:expentqbtz}) is proportional to  $\ell\sim c_{3}$. Since higher-derivative contributions to the entropy enter at order $\nu^{2}$, the $\mathcal{O}(\nu)$ term can only correspond to the von Neumann entropy of the CFT$_{3}$ in the limit of weak backreaction. The $\mathcal{O}(\nu^{2})$ and higher order terms are in principle a combination of the matter and higher-derivative effects that are difficult to distinguish. 

Since the brane theory is in general a higher-derivative theory, in principle the entire gravitational entropy can be computed using the Iyer-Wald entropy functional \cite{Wald:1993nt,Iyer:1994ys},
\beq S_{\text{IW}}\equiv-2\pi\int_{H}\hspace{-1mm} d^{D-2}x\sqrt{q}\;\frac{\partial\mathcal{L}}{\partial R^{ijkl}}\epsilon_{ij}\epsilon_{kl}\;.\eeq
Here $q_{ij}$ is the induced metric of the codimension-2 cross-section $H$ of the horizon with binormal $\epsilon_{ij}$, and $\mathcal{L}$ is the scalar Lagrangian scalar characterizing the gravity theory, e.g., for the Einstein-Hilbert Lagrangian, $S_{\text{IW}}=\frac{A[H]}{4G_{D}}=S_{\text{BH}}$. In the case of the semi-classical induced theory on the brane (\ref{eq:inducactqbtzneu2}), the Iyer-Wald entropy is \cite{Emparan:2020znc}\footnote{Strictly speaking, the entropy (\ref{eq:IWentqbtz}) follows from an application of the field redefinition method for computing entropy of higher-curvature theories \cite{Jacobson:1993vj}; see their Eq. (20) with $a_{1}=-1$, $a_{2}=\frac{3}{8}$ and $16\pi G_{3}\lambda=\ell^{2}$.}
\beq S^{(3)}_{\text{IW}}=\frac{1}{4G_{3}}\int dx\sqrt{q}\left[1+\ell^{2}\left(\frac{3}{4}R-g^{ij}_{\perp}R_{ij}\right)+\mathcal{O}(\ell^{4}/\ell_{3}^{6})\right]\;,\label{eq:IWentqbtz}\eeq
with $g_{ij}^{\perp}=g_{ij}-q_{ij}$ being the metric in the directions orthogonal to the horizon.\footnote{Note $q^{ij}=h^{ij}+ n^{i}n^{j}- u^{i}u^{j}$ for spacelike and timelike unit normals $n^{i}$ and  $u^{i}$, respectively. The binormal $\epsilon_{ij}=(n_{i}u_{j}-n_{j}u_{i})$ satisfies $\epsilon^{2}=-2$, and $\epsilon^{k}_{\;j}\epsilon_{kl}=(u_{j}u_{l}-n_{l}n_{j})\equiv g^{\perp}_{jl}=g_{jl}-q_{jl}$ for bulk metric $g_{ij}$.} The leading contribution is the three-dimensional Bekenstein-Hawking entropy, 
\beq S^{(3)}_{\text{BH}}=\frac{1}{4G_{3}}\int_{H}dx\sqrt{q}=\frac{1}{4G_{3}}\int_{0}^{2\pi}d\bar{\phi}\bar{r}_{+}=\frac{2\pi\bar{r}_{+}}{4G_{3}}\;.\label{eq:3DBHqbtz}\eeq
Substituting in parameters (\ref{eq:x1sqz}) and (\ref{eq:rpsqz}), notice 
\beq S_{\text{BH}}^{(3)}=\frac{(1+\nu z)}{\sqrt{1+\nu^{2}}}S^{(3)}_{\text{gen}}=S_{\text{BTZ}}+\frac{\pi\ell_{3}z^{2}(1+z^{2})}{G_{3}(1+3z^{2})^{2}}\nu+\mathcal{O}(\nu^{2})\;.\eeq
Due to its dependence on $\nu$, $S_{\text{BH}}^{(3)}$ contains semi-classical backreaction effects; only when $\nu=0$ does (\ref{eq:3DBHqbtz}) coincide with the classical entropy in (\ref{eq:expentqbtz}). Notice the difference
\beq S^{(3)}_{\text{BH}}-S_{\text{BTZ}}=\frac{\nu z(1+z^{2})}{1+3z^{2}}S_{\text{BTZ}}\;,\eeq
It is natural to interpret this difference as the leading contribution to the CFT entropy.

Evaluating the Iyer-Wald entropy (\ref{eq:IWentqbtz}) on the quantum BTZ background (\ref{eq:qBTZ}) yields
\beq S_{\text{IW}}^{(3)}=\left[1-\frac{\nu^{2}}{2}-\frac{z(1+z^{2})}{1+\nu z}\nu^{3}+\mathcal{O}(\nu^{4})\right]S_{\text{BH}}^{(3)}\;.\label{eq:IWexpqbtz}\eeq
Thus, the higher-derivative contributions to the gravitational entropy enter at order $\mathcal{O}(\nu^{2})$.

With the Iyer-Wald and generalized entropies, the CFT$_{3}$ entropy can be determined. This is because the generalized entropy associated with a black hole horizon is, generally, 
\beq S_{\text{gen}}=S_{\text{IW}}+S_{\text{vN}}^{\text{mat}}\;,\label{eq:sgengen}\eeq
where $S_{\text{vN}}^{\text{mat}}\equiv-\text{tr}\rho\log\rho$ is the von Neumann entropy of state $\rho$ of quantum fields living on the classical background confined to one side of the horizon. Typically, the matter entropy is UV divergent due to vacuum entanglement just across the horizon. The leading order divergence is of the form $A[H]/\epsilon^{D-2}$ for UV regulator $\epsilon$, while there will also be subleading divergences in $\epsilon$. The Bekenstein-Hawking contribution to $S_{\text{IW}}$ (with renormalized Newton's constant) regularizes the area divergence of the matter entropy, while the subleading divergences are regulated via the higher-derivative contributions to $S_{\text{IW}}$. Thence, the generalized entropy is UV finite and independent of the UV cutoff \cite{Susskind:1994sm,Solodukhin:2011gn,Cooperman:2013iqr}. Therefore, the matter entanglement entropy can be formally computed by taking the difference of the generalized and gravitational entropies. Explicitly to leading order in $\nu$\footnote{If we replace $S^{(3)}_{\text{BH}}$ by $S_{\text{BTZ}}$ in the Iyer-Wald entropy (\ref{eq:IWexpqbtz}), the leading contribution in $S^{(3)}_{\text{CFT}}$ is precisely the second term in (\ref{eq:expentqbtz}).}
\beq S_{\text{CFT}}^{(3)}\equiv S_{\text{gen}}^{(3)}-S_{\text{IW}}^{(3)}\approx -\frac{4\pi\ell_{3} z^{2}\nu}{4G_{3}(1+3z^{2})}=-\nu z S_{\text{BTZ}}\;.\label{eq:matent}\eeq
Note that the minus here does not imply the entanglement entropy is negative. Rather, here $S_{\text{CFT}}^{(3)}$ corresponds to the finite contribution to the entanglement entropy after the leading
piece has been absorbed in a renormalization of $G_{3}$ (which differs from the renormalization of $G_{3}$ due to higher-derivatives effects on the mass). 

Observe when $z\ll1$ the matter entropy (\ref{eq:matent}) takes the form
\beq S^{(3)}_{\text{CFT}}\big|_{z\ll1}\approx -\frac{\pi\ell_{3}\nu z^{2}}{G_{3}}=-2\pi c_{3}(\pi\ell_{3}T)^{2}\;,\label{eq:thermalcftent}\eeq
where we used $T_{\text{BTZ}}|_{z\ll1}\approx z/\pi\ell_{3}$ and $\ell\approx 2c_{3}G_{3}$. The proportionality to $T^{2}$ is consistent with the behavior of a $2+1$-dimensional conformal gas, implying the entropy is thermal. Meanwhile, for large-$z$ but $\nu z\ll1$,
 \beq S^{(3)}_{\text{CFT}}\big|_{z\gg1}\approx -\frac{\pi\ell_{3} \nu}{3G_{3}}=-\frac{2\pi c_{3}}{3}\;,\label{eq:nonthermalent}\eeq
a non-thermal entropy. Comparing the two limits of the matter entropy, we can infer $z\ll1$ characterizes states where thermal effects dominate, while $z\gg1$ describes states dominated by non-thermal effects. In fact, as we will see momentarily, the large and small-$z$ limits respectively coincide with quantum states having large and small Casimir effects.

\begin{figure}[t!]
\centering
 \includegraphics[scale=.45]{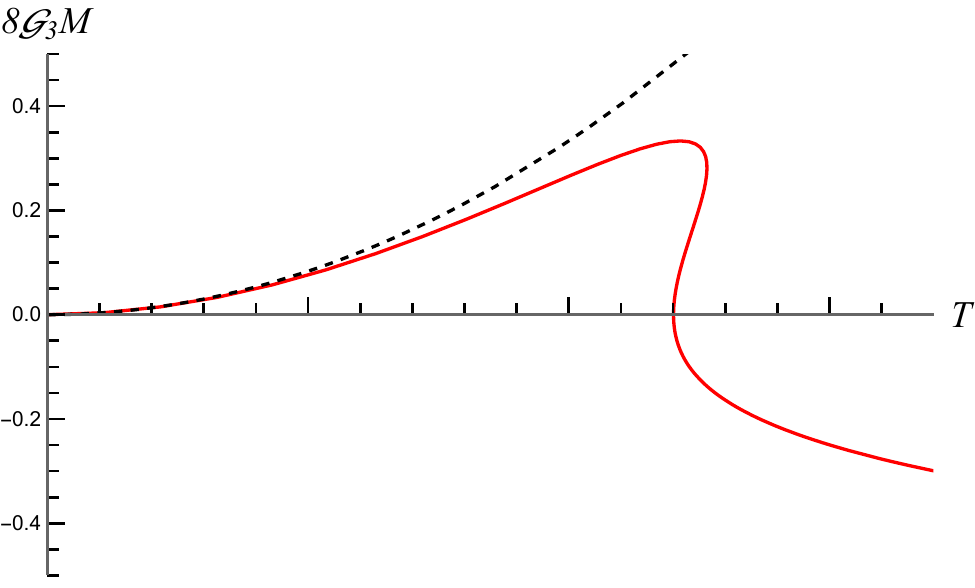}$\qquad$  \includegraphics[scale=.5]{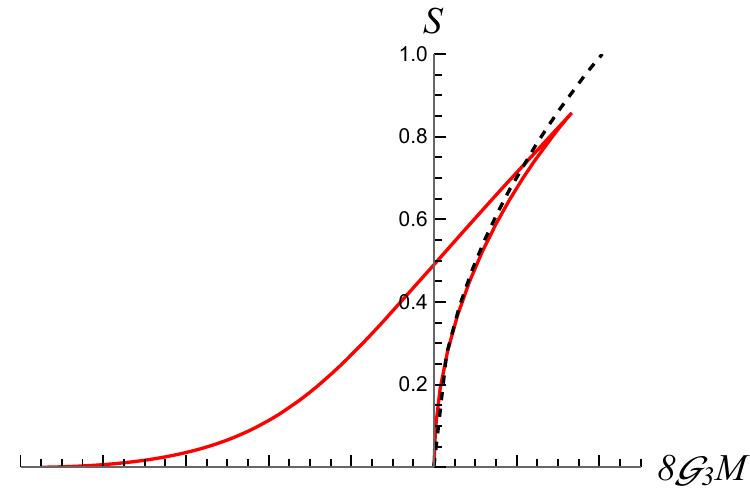}
%\put(-205,20){$S_{\text{gen}}$}
%\put(-185,40){$S_{\text{BH}}$}
\caption{\small \textbf{Quantum BTZ thermodynamics}. \emph{Left:} Mass (solid, red curve) versus temperature for $\nu=1/3$. \emph{Right:} Generalized entropy (solid, red curve) versus mass for $\nu=1/3$. The black dashed lines correspond to the classical BTZ behavior.}
\label{fig:STvsMstatqbtz}\end{figure}

\subsubsection*{The first law of thermodynamics} 

Having identified the bulk horizon entropy with the generalized entropy on the brane (\ref{eq:S4Sgenqbtz}), the bulk first law (\ref{eq:firstlawbulkgen}) from the brane perspective reads
\beq dM=TdS^{(3)}_{\text{gen}}\;.\label{eq:firstlawqbtz}\eeq
This observation is consistent with two-dimensional quantum black holes (another context where the backreaction problem can be exactly solved) \cite{Pedraza:2021cvx,Svesko:2022txo}. Thus, accounting for semi-classical backreaction, the standard first law of horizon thermodynamics is modified by replacing `classical' entropy with $S_{\text{gen}}$. It is worth emphasizing this first law is exact and valid for all $\nu$, i.e., for small or large backreaction. 

From the brane perspective, since the theory includes an infinite tower of higher-derivative contributions, an unambiguous definition of the mass is lacking. Therefore, the first law (\ref{eq:firstlawqbtz}) provides another route to determining the mass. That is, we demand that the quantum BTZ black hole satisfy the semi-classical first via which the right-hand side defines the mass $M$ (\ref{eq:qbtzstathermo}). Notice that in the large-$z$ limit, the mass hits its minimum value
\beq \lim_{z\to\infty}M=-\frac{1}{8\mathcal{G}_{3}}\;.\eeq
This mass can be thought of as the negative Casimir energy of the cutoff CFT$_{3}$, thus explaining the non-thermality of the matter entropy (\ref{eq:nonthermalent}). Alternatively,  for small $\nu$ and $z$ the Casimir effects are suppressed, leading to a thermal matter entropy (\ref{eq:thermalcftent}). 

Recall that the qBTZ solution parametrizes a family of quantum black holes with three branches (\ref{eq:2branchesqbtz}).\footnote{The parametrization depends on $\kappa=\text{sign}(\nu z^{3}-1)$, such that the range of the three branches is covered by imposing $0\leq \nu,z\leq\infty$.} As depicted in Figure \ref{fig:STvsMstatqbtz}, all quantum black holes have higher temperatures than classical BTZ with the same mass. Incidentally, the branch 1a quantum dressed conical singularities ($-1/(8\mathcal{G}_{3})<M<0$) have negative heat capacity, $\partial M/\partial T<0$, except at $M=0$ where the heat capacity diverges. In this range, where Casimir effects are dominant, the entropy $S$ is most naturally understood to be entanglement entropy due to vacuum fluctuations across the horizon. Meanwhile, the largest black holes in branch 1b also have negative heat capacity, $M_{\text{c}}<M<1/24\mathcal{G}_{3}$, where $M_{\text{c}}\neq0$ is some critical value of the mass where the heat capacity diverges. Further, the branch 1b black holes have a larger entropy than branch 2, likely due to the fact that the branch 1b black holes are formed due to backreaction of the Casimir energy.

\subsubsection{Rotating quantum BTZ}

Using the analysis of the static quantum BTZ as a guide, it is in principle straightforward to analyze the thermal properties of the rotating quantum BTZ black hole (\ref{eq:qbtzrotatmet}). The essential new feature here is that there is an outer and inner black hole horizon, $r_{+}$ and $r_{-}$, with $r_{-}<r_{+}$. Thus, there is a parameter $z$ (\ref{eq:zdef}) for each horizon. Below we report only the thermodynamics for the outer horizon as the analysis of the inner horizon follows \emph{mutatis mutandis}. In addition to $z=\ell_{3}/r_{+}x_{1}$, also  introduce the rotation parameter
%\footnote{For $\kappa=+1$, i.e., negative mass quantum dressed cones belonging to branch 1a, it follows $\alpha^{2}<0$. In this case rotation can lead to naked closed timelike curves. Such solutions deserve further study, but will be ignored in what follows.} 
\beq \alpha\equiv \tilde{a}/\sqrt{-\kappa x_{1}}\;,\label{eq:rotpara}\eeq
where recall $\tilde{a}\equiv ax_{1}^{2}/\ell_{3}$. 

Following the logic leading to the parameters (\ref{eq:x1sqz}) -- (\ref{eq:mux1z}), now
\beq 
\begin{split}
&x_{1}^{2}=-\frac{1}{\kappa}\frac{1-\nu z^{3}}{z^{2}[1+\nu z-\alpha^{2}(z-\nu)]}\;,\\
& r_{+}^{2}=-\ell_{3}^{2}\kappa\frac{1+\nu z-\alpha^{2}z(z-\nu)}{1-\nu z^{3}}\;,\\
&\mu x_{1}=-\kappa\frac{(1+z^{2})(1+\alpha^{2}(1-z^{2}))}{1-\nu z^{3}}\;.
\end{split}
\eeq
Substituting these into the expressions for mass (\ref{eq:massrotqbtz}), rotation (\ref{eq:rotJqbtz}), surface gravity (\ref{eq:kapparotqbtz}), and angular velocity (\ref{eq:Omqbtznonrotinf}), the thermodynamic variables are \cite{Emparan:1999fd,Emparan:2020znc}.
\beq 
\begin{split}
&M=\frac{\sqrt{1+\nu^{2}}}{2G_{3}}\frac{(1-\nu z^{3})[z^{2}(1+\nu z)+\alpha^{2}(1+4\nu z^{3}(1+\alpha^{2})-(1+4\alpha^{2})z^{4})]}{[1+3z^{2}+2\nu z^{3}-\alpha^{2}(1+4\nu z^{3}+3z^{4})]^{2}}\;,\\
&T=\frac{1}{2\pi \ell_{3}}\frac{[z^{2}(1+\nu z)-\alpha^{2}(1-2\nu z^{3}+z^{4})][2+3\nu z(1+\alpha^{2})-4\alpha^{2} z^{2}+\nu z^{3}+\alpha^{2}\nu z^{5}]}{z(1+\nu z)[1+\alpha^{2}(1-z^{2})][1+3z^{2}+2\nu z^{3}-\alpha^{2}(1-4\nu z^{3}+3z^{4})]}\;,\\
&S=\frac{\pi\ell_{3}\sqrt{1+\nu^{2}}}{G_{3}}\frac{z(1+\alpha^{2}(1-z^{2}))}{[1+3z^{2}+2\nu z^{3}-\alpha^{2}(1+4\nu z^{3}+3z^{4})]}\;,\\
&J=\frac{\ell_{3}\sqrt{1+\nu^{2}}}{G_{3}}\frac{\alpha z(1+z^{2})[1+\alpha^{2})(1-z^{2})]\sqrt{(1-\nu z^{3})[1+\nu z-\alpha^{2}z(z-\nu)]}}{[1+3z^{2}+2\nu z^{3}-\alpha^{2}(1+4\nu z^{3}+3z^{4})]^{2}}\;,\\
&\Omega=\frac{\alpha(1+z^{2})}{\ell_{3}}\frac{\sqrt{(1-\nu z^{3})[1+\nu z-\alpha^{2}z(z-\nu)]}}{z(1+\nu z)[1+\alpha^{2}(1-z^{2})]}\;.
\end{split}
\label{eq:thermorotqbtz}\eeq
As with the static case, these variables serve as the thermal quantities of the bulk AdS$_{4}$ black hole and are identified to be the thermodynamic variables of the quantum black hole. In particular, the Bekenstein-Hawking entropy of the bulk black hole is 
\beq 
\begin{split}
S^{(4)}_{\text{BH}}&=\frac{2}{4G_{4}}\int_{0}^{2\pi\eta}d\phi\int_{0}^{x_{1}}dx\frac{r_{+}^{2}\ell^{2}}{(\ell+r_{+}x)^{2}}\eta\left(1+\frac{a^{2}x_{1}^{2}}{r_{+}^{2}}\right)\\
&=\frac{\pi}{G_{4}}\eta\frac{\ell x_{1}(r_{+}^{2}+a^{2}x_{1}^{2})}{\ell+r_{+}x_{1}}\;,
\end{split}
 \eeq
and is identified to be the generalized entropy of the black hole on the brane $S_{\text{gen}}^{(3)}$. Notice that in the limit of vanishing backreaction the entropy reduces to the classical BTZ entropy
\beq \lim_{\nu\to0}S_{\text{gen}}^{(3)}=\frac{\pi\ell_{3}}{\sqrt{2G_{3}}}\left(\sqrt{M+\frac{J}{\ell_{3}}}+\sqrt{M-\frac{J}{\ell_{3}}}\right)\;.\eeq
The Wald entropy is formally no different than before, and the matter entropy (\ref{eq:matent}) obeys the same relation at leading order in $\nu$.

It can be explicitly verified that the thermodynamic variables (\ref{eq:thermorotqbtz}) satisfy (for fixed $\nu$)
\beq 
\begin{split}
&\partial_{z}M-T\partial_{z}S-\Omega\partial_{z}J=0\;,\\
&\partial_{\alpha}M-T\partial_{\alpha}S-\Omega\partial_{\alpha}J=0\;.\\
\end{split}
\eeq
Consequently, from the brane perspective, the semi-classical first law is
\beq dM=TdS^{(3)}_{\text{gen}}+\Omega dJ\;.\eeq
Again, n.b., mass $M$ and angular momentum $J$, from the point of view of the brane, include the infinite tower of higher-derivative contributions in the gravity action (encoded in $\mathcal{G}_{3}$), exactly resummed due to our knowledge of the bulk theory. 

While the first law holds for any range of parameters, not all ranges are physically sensible. Thus,  restrictions are made when exploring the thermodynamics of the rotating solution \cite{Emparan:2020znc}. In particular, for non-extremal solutions, one takes
\beq 0\leq \alpha^{2}\leq \frac{1+\nu z}{z(z-\nu)}\;.\eeq
The lower bound is chosen to avoid naked closed timelike curves ($\kappa=+1$ negative mass quantum-dressed cones belonging to branch 1a have  $\alpha^{2}<0$ (\ref{eq:rotpara})). The upper bound follows from demanding the outer black hole event horizon $r_{+}$ be real and positive.
%, and can be understood as the analog of the classical Kerr bound on how fast a black hole can spin. 
For $\kappa=-1$ and $\nu<z<\nu^{-1/3}$, the upper bound implies $1+\alpha^{2}(1-z^{2})>0$. Even with this bound, however, the temperature $T$ of the black hole can go negative without further restricting the range of parameters. The classical BTZ extremal limit (\ref{eq:classextlim}) occurs when
\beq \alpha^{2}=\alpha^{2}_{\text{ext}}=\frac{z^{2}(1+\nu z)}{1-2\nu z^{3}+z^{4}}\;,\label{eq:extboundrot}\eeq
where temperature $T=0$. 

\begin{figure}[t]
\centerline{\includegraphics[width=0.4\textwidth]{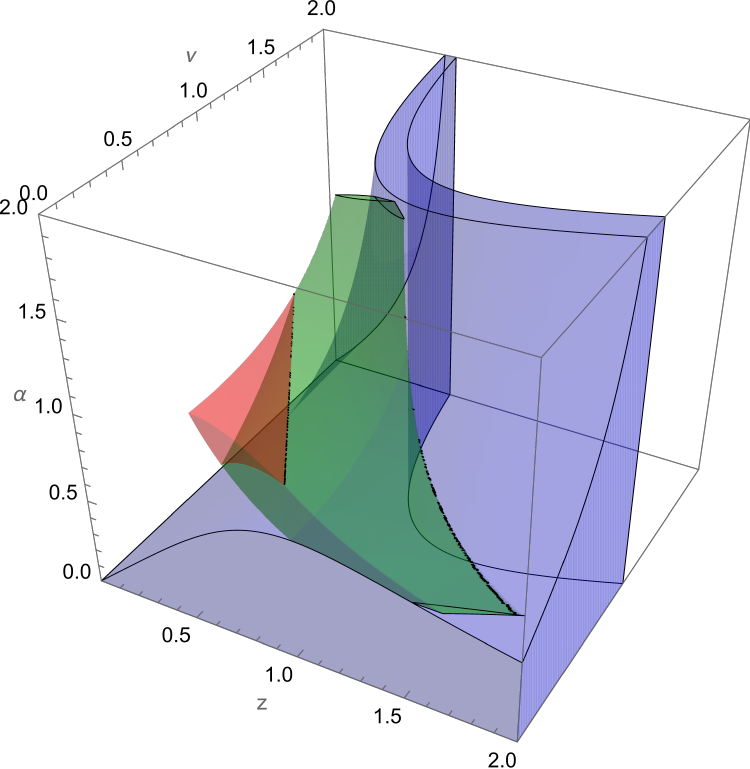}}
			\caption{\small \textbf{Parameter space for rotating quantum BTZ}. The blue region covers solutions that obey the extremality bound of classical rotating BTZ black holes, $0\leq\alpha\leq\alpha_{\text{ext}}$. Green and red regions correspond to superextremal black holes, $\alpha>\alpha_{\text{ext}}$. The red region contains black holes that violate the quantum reverse isoperimetric inequality (\ref{eq:qRII}) and are thermodynamically unstable.}
\label{fig:supp}\end{figure}

As noted in section \ref{sssec:rotqbtz}, the rotating quantum black holes can exist beyond the classical extremality bound (\ref{eq:extboundrot}). This is because, the temperature $T$ and angular momentum $J$ are in general non-monotonic with respect to $\alpha$. There are thus two distinct types of rotating black holes: (i) those which respect the classical extremality bound, $0\leq\alpha\leq \alpha_{\text{ext}}$, where $T$ and $J$ are monotonic in $\alpha$, and (ii) the `superextremal', i.e., $\alpha> \alpha_{\text{ext}}$, where $T$ and $J$ are non-monotonic. In Figure \ref{fig:supp} we illustrate these distinct families, where the blue region corresponds to black holes respecting the extremality bound, while the green and red regions correspond to `superextremal' solutions. The superextremal black hole solutions are possible due to the combined non-linear effects of the rotation and quantum backreaction, and are consequently dubbed nonperturbative rotating black holes \cite{Frassino:2024bjg}. Among these solutions (those belonging to the red region) are black holes which are both thermodynamically unstable (having a negative heat capacity) and violate the so-called quantum reverse isoperimetric inequality (see Eq. (\ref{eq:qRII})).

\subsubsection{Charged quantum BTZ}

As with the rotating case, the charged quantum BTZ black hole (\ref{eq:qbtzchargeatmet}) has an outer and inner horizon. In addition to the parameter $z=\ell_{3}/r_{+}x_{1}$, it is useful to introduce 
\beq \gamma\equiv qx_{1}^{2}\;,\eeq
along with $\gamma_{e}\equiv ex_{1}^{2}$ and $\gamma_{g}\equiv gx_{1}^{2}$, obeying $\gamma^{2}=\gamma_{e}^{2}+\gamma_{g}^{2}$. Consequently, following the logic yielding parameters (\ref{eq:x1sqz}) -- (\ref{eq:mux1z}) gives \cite{Feng:2024uia}\footnote{Start with $H(r_{+})=0$, replace $r_{+}=\ell_{3}/z x_{1}$, $qx_{1}^{2}=\gamma$, and $\mu=(1-\kappa x_{1}^{2}-\gamma^{2})/x_{1}^{3}$, and then rearrange.}
\beq 
\begin{split}
 &x_{1}^{2}=-\frac{1}{\kappa}\frac{(1-\nu z^{3}+\gamma^{2}\nu z^{3}+\gamma^{2}\nu^{2}z^{4})}{z^{2}(1+\nu z)}\;,\\
 & r_{+}^{2}=-\frac{\kappa\ell_{3}^{2}(1+\nu z)}{(1-\nu z^{3}+\gamma^{2}\nu z^{3}+\gamma^{2}\nu^{2}z^{4})}\;,\\
 &\mu x_{1}=-\frac{\kappa(1+z^{2}-\gamma^{2}z^{2}+\gamma^{2}\nu^{2}z^{4})}{(1-\nu z^{3}+\gamma^{2}\nu z^{3}+\gamma^{2}\nu^{2}z^{4})}\;.
\end{split}
\label{eq:chargedparams}\eeq
In terms of these parameters, the thermodynamic quantities are \cite{Climent:2024nuj}
\beq
\begin{split}
&M=\frac{\sqrt{1+\nu^{2}}}{2G_{3}}\frac{z^{2}(1+\nu z)[1-\nu z^{3}+\gamma^{2}\nu z^{3}(1+\nu z)]}{[1+3z^{2}+2\nu z^{3}+\gamma^{2}z^{2}(1+\nu z)^{2}]^{2}}\;,\\
&T=\frac{z}{2\pi\ell_{3}}\frac{2+3\nu z-\nu z^{3}(\gamma^{2}(1+\nu z)^{2}-1)}{[1+3z^{2}+2\nu z^{3}+\gamma^{2}z^{2}(1+\nu z)^{2}]}\;,\\
&S=\frac{\pi\ell_{3}\sqrt{1+\nu^{2}}}{G_{3}}\frac{z}{[1+3z^{2}+2\nu z^{3}+\gamma^{2}z^{2}(1+\nu z)^{2}]}\;,\\
&Q_{e}=\sqrt{\frac{16\pi}{5g_{3}^{2}G_{3}}}\frac{\gamma_{e} z^{2}(1+\nu z)\sqrt{1+\nu^{2}}}{[1+3z^{2}+2\nu z^{3}+\gamma^{2}z^{2}(1+\nu z)^{2}]}\;,\\
&\mu_{e}=\sqrt{\frac{5 g_{3}^{2}}{4\pi G_{3}}}\frac{\gamma_{e}\nu z^{3}(1+\nu z)}{[1+3z^{2}+2\nu z^{3}+\gamma^{2}z^{2}(1+\nu z)^{2}]}\;.
\end{split}
\label{eq:chargeqbtzthermo}\eeq
More specifically, parameters (\ref{eq:chargedparams}) are substituted into mass (\ref{eq:masscqbtz}), electric charge (\ref{eq:echargeqbtz}) and potential (\ref{eq:elecpot}) (the form of the magnetic charge $Q_{g}$ and potential $\mu_{g}$ are of the same form but with $\gamma_{e}\to\gamma_{g}$).
% with 
% $$\frac{8\pi\ell}{g_{\ast}^{2}\ell_{\ast}}=\sqrt{\frac{4\pi}{5g_{3}^{2}G_{3}}}\sqrt{1+\nu^{2}}$$
% where we used $\ell_{\ast}=\frac{\sqrt{16\pi G_{4}}}{g_{\ast}}$, $G_{4}=2L_{4}G_{3}$, $g_{\ast}=g_{3}\sqrt{\frac{5L_{4}}{2}}$, and $L_{4}=\ell/\sqrt{1+\nu^{2}}$.
Further, the temperature follows from the surface gravity, $T=\frac{\kappa_{+}}{2\pi}=\frac{|H'(\bar{r}_{+})|}{4\pi}$, while the entropy is
\beq 
\begin{split}
S&=S_{\text{BH}}^{(4)}=\frac{2}{4G_{4}}\int_{0}^{2\pi\eta}d\phi\int_{0}^{x_{1}}dx\frac{\ell^{2}r_{+}^{2}}{(\ell+x r_{+})^{2}}=\frac{4\pi\ell r_{+}^{2}x_{1}\eta}{G_{4}(\ell+r_{+}x_{1})}\;, 
\end{split}
 \eeq
 and is identified as the three-dimensional generalized entropy $S_{\text{gen}}^{(3)}$. It can be easily verified that the thermodynamic quantities (\ref{eq:chargeqbtzthermo}) satisfy
 \beq 
 \begin{split}
& \partial_{z}M-T\partial_{z}S-\mu_{e}\partial_{z}Q_{e}-\mu_{g}\partial_{z}Q_{g}=0\;,\\
 & \partial_{\nu}M-T\partial_{\nu}S-\mu_{e}\partial_{\nu}Q_{e}-\mu_{g}\partial_{\nu}Q_{g}=0\;,\\
 &\partial_{\gamma_{i}}M-T\partial_{\gamma_{i}}S-\mu_{e}\partial_{\gamma_{i}}Q_{e}-\mu_{g}\partial_{\gamma_{i}}Q_{g}=0\;,
 \end{split}
\eeq
where $\gamma_{i}=\gamma_{e},\gamma_{g}$. Thus, from the brane perspective first law of thermodynamics is
\beq dM=TdS_{\text{gen}}^{(3)}+\mu_{e}dQ_{e}+\mu_{g}dQ_{g}\;.\eeq

Notice that in the limit $\nu\to0$ the chemical potentials $\mu_{e}$ and $\mu_{g}$ vanish (as does the charge, which is readily apparent from (\ref{eq:echargeqbtz})). This is indicative of the fact the charge of the brane black hole is a quantum effect, and the charged quantum BTZ black hole does not reduce to the classical charged quantum BTZ black hole. Further, in the limit of vanishing backreaction $\nu\to0$, the classical relation (\ref{eq:SBTZz}) between $S_{\text{BTZ}}, T_{\text{BTZ}}$ and $M_{\text{BTZ}}$ continues to hold. In fact, the generalized entropy has a qualitatively similar behavior as the entropy of the neutral, static qBTZ (Figure \ref{fig:STvsMstatqbtz}). The only essential difference between entropies of the charged and neutral black holes is that, rather than reaching zero, the charged black hole entropy ends at a finite, non-zero value -- the entropy of the extremal black hole \cite{Climent:2024nuj}.  Meanwhile, in stark contrast with the neutral qBTZ, with $q\neq0$, as the mass decreases monotonically below zero, the temperature reaches a finite maximum value and then quickly tends to zero at the extremal limit. For a more complete treatment of the thermodynamics of the extremal black hole, see \cite{Climent:2024nuj}. 

\begin{figure}[t!]
\centering
 \includegraphics[scale=.6]{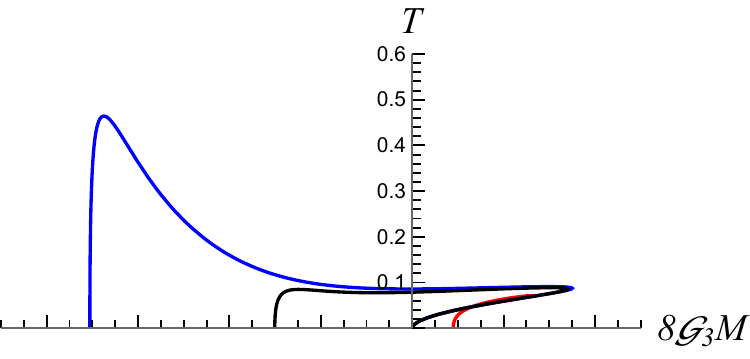}$\qquad$  \includegraphics[scale=.5]{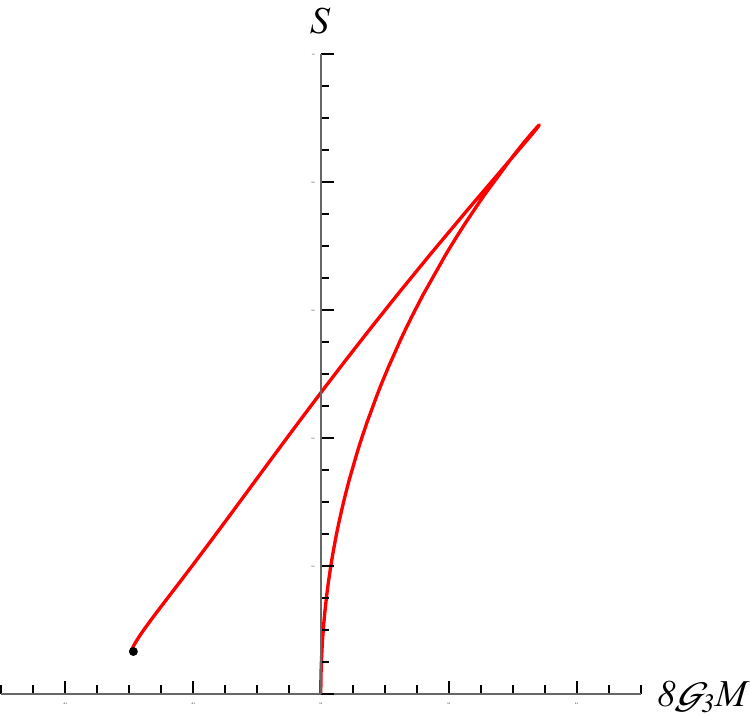}
%\put(-205,20){$S_{\text{gen}}$}
%\put(-185,40){$S_{\text{BH}}$}
\caption{\small \textbf{Charged quantum BTZ thermodynamics}. \emph{Left:} Temperature versus mass for $\nu=1/3$ and $q=.1,.3,1$ (left to right; blue, black, red). \emph{Right:} Generalized entropy versus mass for $\nu=1/3$ and $q=.3$. The entropy ends at a non-zero finite value (black dot). }
\label{fig:STvsMcqbtz}\end{figure}

\subsection{Quantum dS black holes}

Geometrically, the chief difference between the quantum BTZ and dS$_{3}$ black holes is that the latter are equipped with a cosmological horizon. As with classical de Sitter black holes, the black hole horizon is generally hotter than the cosmological horizon. According to a static patch observer, then, the system is characterized by two generally unequal temperatures and thus not in thermal equilibrium. This complicates and enriches the thermodynamic analysis of quantum de Sitter black holes. 

\subsubsection{Quantum Schwarzschild-de Sitter}

Along with $\nu\equiv \ell/R_{3}$, it is useful to introduce the dimensionless parameter \cite{Emparan:2022ijy}
\beq z\equiv\frac{R_{3}}{r_{i}x_{1}}\;,\label{eq:paramsforSdS}\eeq
where $r_{i}$ refers to either the black hole or cosmological horizon, $r_{h}$ and $r_{c}$ respectively. 
%Two limits worth noting are: the limit of empty quantum de Sitter, which corresponds to $\mu=0$, i.e., $x_1=1$ and $r_+=R_3$, so we recover it for $z=1$; and the zero-backreaction limit, $\ell\to0$, in which there is no black hole, \emph{i.e.,} $r_{h}=0$, and $z_{h}\to+\infty$. 
We can now express $x_{1}$, $\mu$, and $r_{+}$ solely in terms of $\nu$ and $z$
\beq
\begin{split}
&x_{1}^{2}=\frac{1}{z^{2}}\frac{1+\nu z^{3}}{1+\nu z}\;,\\
&r_{+}^{2}=R_{3}^{2}\frac{1+\nu z}{1+\nu z^{3}}\;,\\
&\mu x_{1}=\frac{z^{2}-1}{1+\nu z^{3}}\;.
\end{split}
\label{eq:paramsqsds}\eeq
and are the Wick rotated counterparts of the qBTZ solution parameters (\ref{eq:x1sqz}) --- (\ref{eq:mux1z}).
%\footnote{With $\ell_{3}^{2}\to -R_{3}^{2}$, then $z^{2}\to -z^{2}$, $\nu^{2}\to-\nu^{2}$ and $\nu z\to \nu z$.} Lastly, in terms of $\nu$ and $z$, the relation between the bare Newton's constants $G_{4}$ and $G_{3}$ is
Further, $G_{4}=2\ell_{4}G_{3}=2G_{3}\ell/\sqrt{1-\nu^{2}}$. In terms of parameters (\ref{eq:paramsqsds}), the mass (\ref{eq:massqSdSFM}) is
\beq 
M %=\frac{1}{8\mathcal{G}_{3}}\left(1-\frac{4x_{1}^{2}}{(3-x_{1}^{2})^{2}}\right)
=\frac{1}{8G_{3}}\sqrt{1-\nu^{2}}\frac{(z^{2}-1)(9z^{2}-1+8\nu z^{3})}{(3z^{2}-1+2\nu z^{3})^{2}}\;.
\label{eq:mass3}\eeq
In the quantum de Sitter limit, where $z=1$,  it follows $M=0$.
Also note the mass $M$ vanishes at large $z$, $\lim_{z\to\infty} M\approx \frac{1}{4\mathcal{G}_{3}\nu z}+\mathcal{O}(1/z^{2})$.
For fixed $x_{1}\neq 1$ and for $z=z_{h}$, it is natural to think of the large $z_{h}$ limit as a \emph{small} quantum Schwarzschild black hole, where $R_{3}\gg r_{h}  x_{1}$.

\subsection*{Temperature}

On the brane the black hole and cosmological horizons will appear to emit radiation at the Hawking and Gibbons-Hawking temperatures $T_{h},T_{c}$, respectively.
Explicitly,
\beq 
T_{h } = -\frac{z_h}{2\pi R_{3}}\frac{ 2+3\nu z_{h}-\nu z_{h}^{3} }{3z_{h}^{2}-1+2\nu z_{h}^{3}}\;, \qquad T_{c } =\frac{z_c}{2\pi R_{3}}\frac{ 2+3\nu z_{c}-\nu z_{c}^{3} }{3z_{c}^{2}-1+2\nu z_{c}^{3}}\,.\label{eq:hortempsSdS}\eeq
 In the limit $\nu\to0$,   the black hole   temperature vanishes, since $z_h$ diverges, while the cosmological horizon temperature reduces to 
\beq \lim_{\nu\to0}T_{ c}= \frac{1}{\pi R_{3}}\frac{z_{c}}{3z_{c}^{2}-1}\equiv T_{\text{SdS}_{3}}\;.\label{eq:Tnu0lim}\eeq
 Since in general $r_{h}<r_{c}$, and hence $z_{c}<z_{h}$, the black hole horizon has a higher temperature than the cosmological horizon, $T_{h}>T_{c}$. Thus, as usual for Schwarzschild-de Sitter spacetimes, the black hole and cosmological horizons are not in thermal equilibrium. Only in the Nariai limit do the temperatures of the two horizons coincide. More on this below.

\subsection*{Entropy and the first law}

The four-dimensional Bekenstein-Hawking entropy of the bulk black hole is 
\beq
\begin{split}
S_{\text{BH}}^{(4)}&=\frac{\text{Area}(r_{+})}{4G_{4}}=\frac{2}{4G_{4}}\int_{0}^{2\pi\eta}d\phi\int_{0}^{x_{1}}dx r_{+}^{2}\frac{\ell^{2}}{(\ell+xr_{+})^{2}}\\
%&=\frac{8\pi}{4G_{4}}\frac{2x_{1}^{2}r_{+}^{2}\ell}{R_{3}(3-x_{1}^{2})}\frac{z}{(1+\nu z)}\\
%&=\frac{2\pi R_{3}^{2}}{G_{4}}\frac{z\nu}{(3z^{2}-1+2\nu z^{3})}\\
&=\frac{\pi R_{3}}{G_{3}}\frac{z\sqrt{1-\nu^{2}}}{3z^{2}-1+2\nu z^{3}}\;.
\end{split}
\label{eq:Sgen}\eeq
 As with the AdS$_{3}$ quantum black holes, from the brane perspective the bulk entropy is interpreted as the three-dimensional generalized entropy $S_{\text{gen}}^{(3)}$. In the limit $z=1$, the generalized entropy of the quantum de Sitter solution 
%\beq S_{\text{gen}}^{(3)}\big|_{z=1}=\frac{2\pi R_{3}}{4G_{3}}\frac{\sqrt{1-\nu^{2}}}{1+\nu}\;,\label{eq:SgenqdS}\eeq
is proportional to the Gibbons-Hawking entropy of the $\text{dS}_{3}$ cosmological horizon, i.e., the sum of gravitational entropy and entanglement entropy due to the CFT living outside of the cosmological horizon. Further, the generalized entropy is related to the three-dimensional Bekenstein-Hawking entropy $S_{\text{BH}}^{(3)}$ of the horizon(s) on the brane via
\beq
S_{\text{gen}}^{(3)} = \frac{\sqrt{1- \nu^2}}{1 + \nu z}  S_{\text{BH}}^{(3)}\;,
\label{eq:SgenasScl}\eeq
where $S_{\text{BH}}^{(3)}=\frac{2\pi r_{+}\eta}{4G_{3}}$. In the limit of vanishing backreaction, the three-dimensional entropies coincide and are equal to the cosmological horizon entropy of classical Schwarzschild-dS$_{3}$ \cite{Spradlin:2001pw}
\beq \lim_{\nu\to0}S_{\text{gen}}^{(3)}=\frac{\pi R_{3}}{G_{3}}\frac{z_c}{3z_c^{2}-1}=\frac{\pi^{2}R_{3}^{2}}{G_{3}}T_{\text{SdS}_{3}}=\frac{\pi R_{3}}{2G_{3}}\sqrt{1-8G_{3}M}=S_{\text{SdS}_{3}}\;.\eeq

%  Finally, in the quantum de Sitter limit ($z=1$), the area entropy is simply equal to the Gibbons-Hawking entropy
% \beq S_{\text{BH},c}^{(3)}|_{z=1}=\frac{2\pi R_{3}}{4G_{3}}=\frac{2\pi R_{3}}{4\mathcal{G}_{3}}\frac{1}{\sqrt{1-\nu^{2}}}\;,\eeq
% where $\mathcal{G}_{3}$ is the renormalized Newton's constant. Therefore, the Gibbons-Hawking entropy of $\text{qdS}_{3}$ scales like the classical entropy of $\text{dS}_{3}$. 

% A plot of each of these entropies $S_{\text{gen}}^{(3)},S_{\text{BH}}^{(3)}$, and $S_{\text{SdS}_{3}}$ is given in Fig.~\ref{fig:SvsM}. We observe that the sum of the black hole and cosmological horizon entropies $S_{\text{gen},h}^{(3)}$ and $S_{\text{gen},c}^{(3)}$ produces an approximately linear curve always equal to or less than the entropy of the quantum de Sitter solution (\ref{eq:SgenqdS}), see Fig.~\ref{fig:Sgenwithtot}. This is reminiscent of the observation  in    \cite{Visser:2019muv,Morvan:2022ybp} for the classical SdS solution that the   sum of the horizon entropies is approximately a linear function of the mass.   We will return to this point in Sec.~\ref{sec:entdeficit}, as it will prove useful when computing the nucleation rate of quantum dS black holes.

In a perturbative series expansion of small $\nu$, the linear order  $\mathcal{O}(\nu)$ contribution to the generalized entropy (\ref{eq:Sgen}) captures the CFT matter entropy while quadratic and higher order contributions include the effects of the higher-derivative corrections. Formally, the matter entropy is given by the difference of the generalized entropy and Iyer-Wald entropy, which, to leading order and for large-$z$ has the same non-thermal result as static quantum BTZ (\ref{eq:nonthermalent}).

Putting together the mass (\ref{eq:mass3}), temperatures (\ref{eq:hortempsSdS}) and entropy (\ref{eq:Sgen}), there is a separate first law for the black hole and cosmological horizons,
\beq dM=T_hdS^{(3)}_{\text{gen},h}\;,\label{eq:bhfirstlaw}\eeq
and
\beq
dM = - T_c dS^{(3)}_{\text{gen},c}\;.
\label{eq:cfirstlaw}\eeq
In combination, notice
\beq
0=T_h dS^{(3)}_{\text{gen},h} + T_c d S^{(3)}_{\text{gen},c}\;.
\label{eq:firstlawsum}\eeq
Thus, as the generalized entropy attributed to the black hole increases, the generalized entropy of the cosmological horizon decreases. This is a consequence of the minus sign appearing in the first law for the cosmological horizon (\ref{eq:cfirstlaw}), i.e.,  the entropy of the cosmological horizon decreases as the mass increases. Akin to classical de Sitter space, this suggests quantum $\text{dS}_{3}$ represents a maximum entropy state with a finite number of degrees of freedom. Hence, quantum de Sitter black holes behave as instantons constraining the states
of the original de Sitter degrees of freedom (see, e.g., \cite{Morvan:2022ybp,Morvan:2022aon,Draper:2022xzl}).

\subsection*{Nariai limit}

\begin{figure}[t]
\begin{center}
\includegraphics[scale=.5]{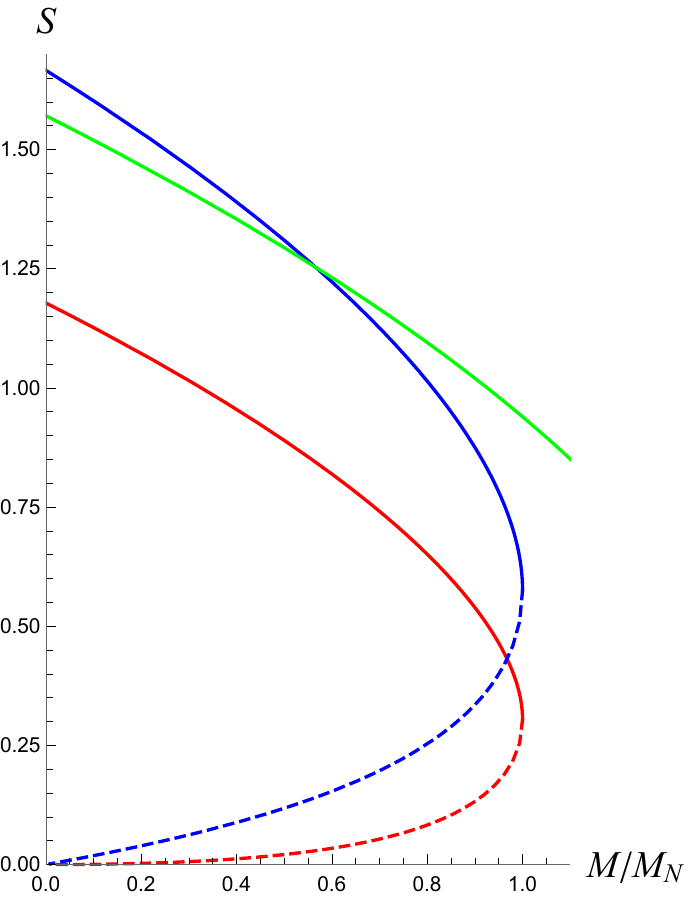}$\qquad$\includegraphics[scale=.7]{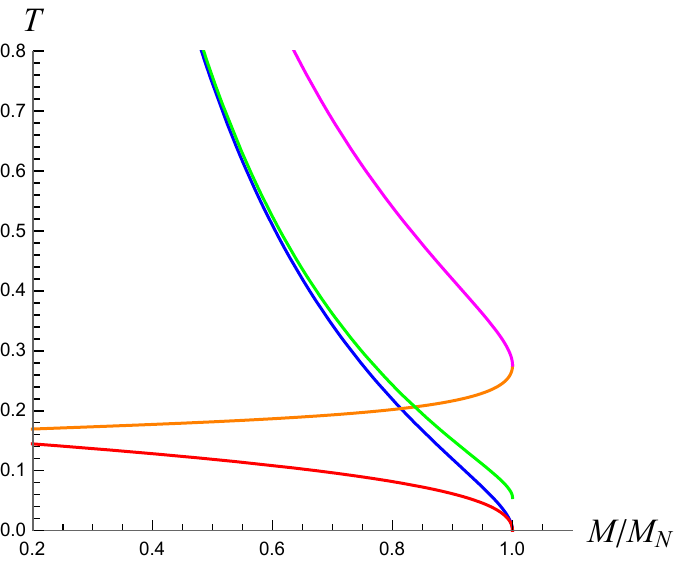}
\end{center}
\caption{\small \textbf{Quantum SdS$_{3}$ thermodynamics.} \emph{Left:} Plot of $S_{\text{gen}}^{(3)}$ (red), $S_{\text{BH}}^{(3)}$ (blue) and $S_{\text{SdS}_{3}}$ (green) as a function of mass $M$ and $\nu=1/3$. Dashed curves refer to black hole entropies $S_{h}$, while the solid curves denote the entropies associated with the cosmological horizon $S_{c}$. \emph{Right:}Temperature as a function of mass $M$ for $\nu=1/3$. Blue and red curves respectively correspond to temperatures $T_{h}$ and $T_{c}$,  while the magenta and orange curves denote $\bar{T}_{h}$ and $\bar{T}_{c}$, respectively. The green curve represents the temperature of the Schwarzschild limit ($R_{3}\gg r_{h}$).
}
\label{fig:qsdsthermo} 
\end{figure}

As noted above, in general a static patch observer sees the quantum SdS$_{3}$ black hole as one with two unequal temperatures. A special limit where the temperatures coincide is the (quantum) Nariai solution (\ref{eq:qNariaigeom}), i.e., the largest mass black hole able to fit inside the cosmological horizon, where $r_{h}=r_{c}=r_{\text{N}}$. Recall that in this limit $(\mu\ell)$ attains a maximum 
\beq \mu_{\text{N}}=\frac{2}{3\sqrt{3}}\frac{1}{\nu}\;.\label{eq:muNnu}\eeq
such that  
\beq z_{\text{N}}=\frac{R_{3}}{r_{\text{N}}x_{1}^{\text{N}}}=\frac{\sqrt{3}}{x_{1}^{\text{N}}}\;.\eeq
In this case $x_{1}^{\text{N}}$ is the particular value of $x_{1}$ in the Nariai limit found by solving $\mu_{\text{N}}=(1-x_{1}^{2})/x_{1}^{3}$ for $x_{1}$. For arbitrary $\nu$, there will generally be one real solution for of $x_{1}$ and one real solution, which for small $\nu$ takes the form
\beq x_{1}^{\text{N}}=\sqrt{3}\left(\frac{\nu}{2}\right)^{1/3}-\frac{\sqrt{3}}{2}\nu+\mathcal{O}(\nu^{5/3})\;,\eeq
vanishing in the limit $\nu\to0$ (as one would expect). In terms of $x_{1}^{\text{N}}$, the mass of the Nariai solution $M_{\text{N}}$ is therefore defined by $M_{\text{N}}\equiv M|_{z=z_{\text{N}}}$.  Similarly, the entropy of the Nariai solution is defined as $S_{\text{N}}^{(3)}\equiv S_{\text{gen}}^{(3)}|_{z=z_{\text{N}}}$. See Figure \ref{fig:qsdsthermo} for a plot of the horizon entropies as a function of mass $M$ normalized with respect to the Nariai mass. 

Regarding the temperature, the Nariai limit is subtle. Naively, setting $r_{c}=r_{h}$ leads to the horizon temperatures (\ref{eq:hortempsSdS}) vanishing. This is a consequence of working with surface gravities  $\kappa_{h}$ and $\kappa_{c}$ defined with respect to the time-translation Killing vector $\xi=\partial_{\bar{t}}$:
\beq 
\begin{split}
&\kappa_{h}=\frac{1}{2}H'(\bar{r}_{h})=\frac{1}{2\bar{r}_{h}r_{\text{N}}^{2}}(\bar{r}_{\text{N}}^{2}-\bar{r}_{h}^{2})\;,\quad \kappa_{c}=-\frac{1}{2}H'(\bar{r}_{c})=-\frac{1}{2\bar{r}_{c}r_{\text{N}}^{2}}(\bar{r}_{\text{N}}^{2}-\bar{r}_{c}^{2})\;,
\end{split}
\eeq
which clearly vanish when $\bar{r}_{h,c}\to \bar{r}_{\text{N}}=\frac{R_{3}}{\sqrt{3}}\eta$. However, for Schwarzschild-de Sitter there is a more natural choice of normalization of the time-translation Killing vector, owed to Bousso and Hawking  \cite{Bousso:1996au}, resulting in a non-vanishing Nariai temperature. Technically, the Bousso-Hawking normalization is chosen such that $\xi^2 =-1$ at the radius $\bar r_0$ where the blackening factor $H(\bar r)$ obtains a maximum,
\beq H'(\bar{r}_{0})=0 \quad \Longrightarrow \quad  \bar r^3_0=\frac{\bar{r}_{i}}{2}(3\bar{r}_{\text{N}}^{2}-\bar{r}_{i}^{2}) \;.\eeq
for positive real root $\bar{r}_{i}$ of $H(\bar{r})$. Physically, the radius $\bar{r}_{0}$ corresponds to the location where an observer can stay in place without accelerating, i.e., the position where the acceleration due to the gravitational attraction from the black hole balances out the cosmological acceleration. This is consistent with the asymptotically flat Schwarzschild black hole where $\bar{r}_{0}\to\infty$ (since $R_{3}\to\infty$), and empty de Sitter where $\bar{r}_{0}\to0$ (via $\bar{r}_{i}\to \eta R_{3}$).

In terms of radius $\bar{r}_{0}$, where  
\beq
\begin{split}
H(\bar r_0) &=\frac{1}{r_{\text{N}}^{2}}(\bar{r}_{\text{N}}^{2}-\bar{r}_{0}^{2})\;. 
\end{split}
\eeq
the analogous Bousso-Hawking temperature is
\beq
\bar{T}= \frac{T}{\sqrt{H(\bar r_0)}}\;.\eeq
In terms of the horizon radii $\bar{r}_{h,c}$,
\beq 
\begin{split}
&\bar{T}_{h,c}%=\frac{T_{h}}{\sqrt{H(\bar{r}_{0})}}
=\mp\frac{1}{4\pi r_{\text{N}}\bar{r}_{h,c}}\frac{\bar{r}_{\text{N}}^{2}-\bar{r}_{h}^{2}}{\sqrt{\bar{r}_{\text{N}}^{2}-\left(\frac{\bar{r}_{h,c}}{2}(3\bar{r}^{2}_{\text{N}}-\bar{r}_{h,c}^{2})\right)^{2/3}}}\;, %\\
%&\bar{T}_{c}%=\frac{T_{c}}{\sqrt{H(\bar{r}_{0})}}
%=\frac{1}{4\pi r_{\text{N}}\bar{r}_{c}}\frac{\bar{r}_{\text{N}}^{2}-\bar{r}_{c}^{2}}{\sqrt{\bar{r}_{\text{N}}^{2}-\left(\frac{\bar{r}_{c}}{2}(3\bar{r}^{2}_{\text{N}}-\bar{r}_{c}^{2})\right)^{2/3}}}\;.
\end{split}
\eeq
where the minus sign corresponds to the black hole temperature, and the plus sign to the cosmological horizon temperature.
Carefully taking the limit $\bar{r}_{\text{N}}\approx \bar{r}_{h,c}$,  the temperature $\bar{T}_{h,c}$ of both horizons approaches the Nariai temperature $T_{\text{N}}=1/(2\pi r_{\text{N}})$ (see, e.g., Appendix B of \cite{Svesko:2022txo}). The expression for $\bar{T}$ in terms of $z$ and $\nu$ is cumbersome to leading order for small $\nu$ is equal to the dS$_{3}$ Gibbons-Hawking temperature \cite{Emparan:2022ijy}. 
Figure \ref{fig:qsdsthermo} displays the temperatures $T_{c},T_{h}$, and $\bar{T}$  as a function of mass $M$ normalized by the Nariai mass. The behavior is essentially identical to classical four-dimensional Schwarzschild-de Sitter black holes (see, e.g. Figure 2 of \cite{Morvan:2022ybp}), reflecting the holographic character of the induced geometry. 
%In particular, notice for small mass $M$ black holes, the black hole temperature diverges such that it may be approximated by the temperature of a (quantum) Schwarzschild black hole.  
%Collectively, the three-dimensional quantum Schwarzschild-de Sitter system, at least thermodynamically, behaves like a four-dimensional classical SdS spacetime, reflecting the holographic character of our setup.

\subsubsection{Quantum Kerr-de Sitter}

The new feature of the quantum Kerr-de Sitter black hole (\ref{eq:branegeomv2}) compared to the quantum SdS is that now there is an outer and inner black hole horizon obeying $r_{-}<r_{+}<r_{c}$. In addition to $z$ (\ref{eq:paramsforSdS}) introduce the rotation parameter
\beq \alpha\equiv\frac{ax_{1}}{R_{3}}=\frac{\tilde{a}}{x_{1}}\;,\label{eq:paramsforKdS}\eeq
where $\{r_{i}\}=\{r_{\pm},r_{c}\}$.  We can express $x_{1}$, $\mu$ and $r_{i}$ solely in terms $z,\nu,\alpha$:
\beq 
\begin{split}
&x_{1}^{2}=\frac{1+\nu z^{3}}{z^{2}[1+\nu z+\alpha^{2}z(z+\nu)]}\;,\\
&r_{i}^{2}=R_{3}^{2}\frac{1+\nu z+\alpha^{2}z(z+\nu)}{1+\nu z^{3}}\;,\\
&\mu x_{1}=\frac{(z^{2}-1)(1+\alpha^{2}(1+z^{2}))}{1+\nu z^{3}}\;.
\end{split}
\label{eq:paramssub}\eeq
Using the parameters (\ref{eq:paramssub}), the thermodynamic variables are \cite{Panella:2023lsi}
%we can recast the mass $M$ (\ref{eq:Massid}) and angular momentum $J$ (\ref{eq:angJ}) 
\beq 
\begin{split} 
&M=\frac{1}{8G_{3}}\sqrt{1-\nu^{2}}\frac{(z^{2}-1)[1+\alpha^{2}(1+z^{2})][9z^{2}-1+8\nu z^{3}+\alpha^{2}(9z^{4}-1+8\nu z^{3})]}{(3z^{2}-1+2\nu z^{3}+\alpha^{2}(1+4\nu z^{3}+3z^{4}))^{2}}\;,\\
&J=\frac{\alpha R_{3}}{G_{3}}\sqrt{1-\nu^{2}}\frac{z(z^{2}-1)[1+\alpha^{2}(1+z^{2})]\sqrt{(1+\nu z^{3})(1+\nu z+\alpha^{2}z(z+\nu))}}{(3z^{2}-1+2\nu z^{3}+\alpha^{2}(1+4\nu z^{3}+3z^{4}))^{2}}\;,\\
&\Omega_{i}=\frac{\alpha}{R_{3}}\frac{(z^{2}-1)\sqrt{(1+\nu z^{3})(1+\nu z+\alpha^{2}z(z+\nu))}}{z(1+\nu z)(1+\alpha^{2}(1+z^{2}))}\;,\\
& T_{i}=\frac{1}{2\pi R_{3}}\frac{(z^{2}(1+\nu z)+\alpha^{2}(1+2\nu z^{3}+z^{4}))|(2+3\nu z-\nu z^{3}+\alpha^{2}(4z^{2}+\nu z(z^{4}+3)))|}{z(1+\nu z)(1+\alpha^{2}(1+z^{2}))(3z^{2}-1+2\nu z^{3}+\alpha^{2}(1+3z^{4}+4\nu z^{3}))}\;,\\
&S=\frac{\pi R_{3}}{G_{3}}\frac{\sqrt{1-\nu^{2}}z(1+\alpha^{2}(1+z^{2}))}{(3z^{2}-1+2\nu z^{3}+\alpha^{2}(1+3z^{4}+4\nu z^{3}))}\;,
\end{split}
\eeq
 which follow from substituting the parameters (\ref{eq:paramssub}) into mass $M$ (\ref{eq:Massid}), angular momentum $J$ (\ref{eq:angJ}), rotation $\Omega_{i}$ (\ref{eq:Omi}), surface gravity (\ref{eq:surfacegravs}), and the Bekenstein-Hawking area formula.\footnote{The thermodynamics of the rotating qBTZ solution follows from the reassignments $\ell_{3}^{2}\to-R_{3}^{2}$ and $a\to-a$, such that $z^{2}\to -z^{2}$, $\nu^{2}\to-\nu^{2}$ and $\nu z\to \nu z$.} Collectively, the variables obey the first law
% \beq \partial_{z}M-T_{i}\partial_{z}S^{(4)}_{\text{BH}}-\Omega_{i}\partial_{z}J=0\;,\quad \partial_{\alpha}M-T_{i}\partial_{\alpha}S^{(4)}_{\text{BH}}-\Omega_{i}\partial_{\alpha}J=0\;,\eeq
%and hence the first law
\beq dM=T_{i}dS_{i}+\Omega_{i}dJ\;,\label{eq:classfirstlawkerrds}\eeq
for all values of the parameters. On the brane $S$ has the usual interpretation as the generalized entropy $S_{\text{gen}}^{(3)}$.

The quantum Kerr-dS$_{3}$ black hole has limits where two or more horizons become degenerate. These include the extremal ($r_{+}=r_{-}$) black hole, where $T_{\text{ext}}=0$, the lukewarm black hole ($T_{c}=T_{+}$), and the Nariai geometry ($r_{c}=r_{+}$). As with quantum SdS$_{3}$, the Nariai black hole will have a vanishing temperature using the standard normalization of the Killing vector. However, in its near horizon geometry, the temperature of the Nariai black hole and cosmological horizons are equal to the same non-zero temperature $T_{\text{N}}$.

Lastly, two limiting cases include the quantum de Sitter, at $z=1$ or $\mu=0$, leading to 
\beq M=J=\Omega_{i}=0\;,\eeq
\beq S=\frac{2\pi R_{3}}{4G_{3}}\frac{\sqrt{1-\nu^{2}}}{1+\nu}\;,\quad T_{c}=\frac{1}{2\pi R_{3}}\;,\eeq
and the limit of vanishing backreaction, where
\beq 
\begin{split}
&M=\frac{1}{8G_{3}}\frac{(z^{2}-1)(1+\alpha^{2}(1+z^{2}))(9z^{2}-1+\alpha^{2}(9z^{4}-1))}{(3z^{2}-1+\alpha^{2}(1+3z^{4}))^{2}}\;,\\
&J=\frac{\alpha R_{3}}{G_{3}}\frac{z(z^{2}-1)(1+\alpha^{2}(1+z^{2}))\sqrt{1+\alpha^{2}z^{2}}}{(3z^{2}-1+\alpha^{2}(1+3z^{4}))^{2}}\;,\\
&\Omega_{c}=\frac{\alpha(z^{2}-1)\sqrt{1+\alpha^{2}z^{2}}}{R_{3}z(1+\alpha^{2}(1+z^{2}))}\;,\\
&T_{c}=\frac{1}{2\pi R_{3}}\frac{2(1+2\alpha^{2}z^{2})(z^{2}+\alpha^{2}(1+z^{4}))}{z(3z^{2}-1+\alpha^{2}(1+3z^{4}))(1+\alpha^{2}(1+z^{2}))}\;,\\
&S_{c}=\frac{\pi R_{3}}{G_{3}}\frac{z(1+\alpha^{2}(1+z^{2}))}{(3z^{2}-1+\alpha^{2}(1+3z^{4}))}\;,
\end{split}
\label{eq:thermonu0}\eeq
with $z=z_{c}$ since there are no black holes. It is straightforward to verify the resulting thermodynamic variables (\ref{eq:thermonu0}) agree with classical Kerr-$\text{dS}_{3}$ \cite{Panella:2023lsi}.

\subsection{Quantum black holes in flat space}

As with the quantum de Sitter black holes, the thermodynamics of the asymptotically flat quantum black holes follow from the bulk AdS$_{4}$ black hole geometry with a Randall-Sundrum brane. Notably, however, in this case, the tension is tuned to its critical value (since $L_{4}=\ell$) such that no cosmological horizon appears. Therefore, only the thermodynamics of the black hole horizons in the bulk are imprinted on the brane.

\subsubsection{Quantum Schwarzschild black hole}

In a sense, the thermodynamics of the quantum Schwarzschild black hole (\ref{eq:qSchwarz}) is given by the $R_{3}\to\infty$ limit of the thermodynamics of the quantum Schwarzschild-de Sitter. Some care must be taken, however, since in this limit $z=R_{3}/r_{+}x_{1}$ diverges while $\nu=\ell/R_{3}$ goes to zero. To be more illustrative, recall the mass (\ref{eq:massqSFM}). Rearranging $\mu x_{1}=(1-x_{1}^{2})/x_{1}^{2}$ such that $x_{1}^{2}=1/(1+\mu x_{1})$, yields
\beq M=\frac{\ell\hat{x}(8+9\hat{x})}{16 G_{4}\left(1+\frac{3}{2}\hat{x}\right)^{2}}\;,\label{eq:massschrv2}\eeq
where 
\beq \hat{x}\equiv \mu x_{1}\;.\label{eq:xhatdefschwar}\eeq
It is easy to verify $\mu^{2}=\hat{x}^{2}(1+\hat{x})$. Meanwhile, from the surface gravity (\ref{eq:surfgravmink}) the temperature of the bulk horizon is
\beq T=\frac{1}{4\pi\ell\hat{x}}\frac{1}{\left(1+\frac{3}{2}\hat{x}\right)}\;.\label{eq:tempscharv2}\eeq
Further, the four-dimensional Bekenstein-Hawking entropy is 
\beq 
\begin{split}
S_{\text{BH}}^{(4)}&=\frac{4\pi\eta}{4G_{4}}\int_{0}^{x_{1}}dx r_{+}^{2}\frac{\ell^{2}}{(\ell+x r_{+})^{2}}=\frac{4\pi}{4G_{4}}\frac{r_{+}^{2}\ell x_{1}}{\ell+r_{+}x_{1}}\\
&=\frac{4\pi\ell^{2}}{4G_{4}}\frac{\hat{x}^{2}}{\left(1+\frac{3}{2}\hat{x}\right)}\;,
\end{split}
\label{eq:entschwarv2}\eeq
where to get to the final line we used $r_{+}=\mu\ell$.
%It is straightforward to check the thermodynamic quantities $M$, $T$, and $S$ follow from those of the quantum SdS$_{3}$ in the limit $R_{3}\to\infty$ and making the replacement (\ref{eq:xhatdefschwar}). 

In summary, the thermodynamic variables of the bulk AdS$_{4}$ black hole are mass (\ref{eq:massschrv2}), temperature (\ref{eq:tempscharv2}), and entropy (\ref{eq:entschwarv2}), 
and obey the first law\footnote{Note the mass (\ref{eq:massschrv2}) and temperature (\ref{eq:tempscharv2}) differ from those reported in \cite{Emparan:1999wa,Emparan:2002px}, a consequence of us working in canonically normalized coordinates. Moreover, the mass in \cite{Emparan:1999wa} is identified as the result of explicitly integrating $TdS_{\text{BH}}^{(4)}=T\partial_{\hat{x}}S_{\text{BH}}^{(4)}d\hat{x}$.}
\beq dM=TdS_{\text{BH}}^{(4)}\;.\eeq
From the brane perspective, these thermodynamic quantities are identified as the thermodynamic variables of the quantum Schwarzschild black hole, upon substituting $G_{4}=2\ell G_{3}$, and where $S_{\text{BH}}^{(4)}=S_{\text{gen}}^{(3)}$. 
%It is easy to verify these thermodynamic quantities follow from those of the quantum SdS$_{3}$ in the limit $R_{3}$. 
% \beq 
% \begin{split}
% &M=\frac{\hat{x}(8+9\hat{x})}{32 G_{3}\left(1+\frac{3}{2}\hat{x}\right)^{2}}\;,\\
% &T=\frac{1}{4\pi\ell\hat{x}}\frac{1}{\left(1+\frac{3}{2}\hat{x}\right)}\;,\\
% &S_{\text{gen}}^{(3)}=\frac{4\pi\ell}{8G_{3}}\frac{\hat{x}^{2}}{\left(1+\frac{3}{2}\hat{x}\right)}
% \end{split}
% \label{eq:thermoqschr}\eeq
% using $G_{4}=2\ell G_{3}$. 

It is easy to verify the mass is a monotonically increasing function of $\hat{x}$, with a minimum $M=0$ (at $\hat{x}=r_{+}=0$) and a maximum at $M=\frac{1}{8G_{3}}$, at $\hat{x}\to\infty$. Further, for small $\hat{x}$ the temperature goes like $T\sim 1/\hat{x}$ and $T\sim1/\hat{x}^{2}$ for large $\hat{x}$, while the entropy behaves as 
\beq 
\begin{split}
&S_{\text{gen}}^{(3)}\approx \frac{\pi\ell\hat{x}^{2}}{2G_{3}}\qquad (\text{small} \;\hat{x})\;,\\
&S_{\text{gen}}^{(3)}\approx \frac{\pi\ell\hat{x}}{3G_{3}}\qquad\;(\text{large} \;\hat{x})\;.
\end{split}
\label{eq:qschwarzlimitent}\eeq
It is useful to think about these limits in terms of the bulk parameter $\mu$, where $\hat{x}\approx \mu$ for small $\hat{x}$ and $\hat{x}\approx \mu^{2/3}$ for large $\hat{x}$.\footnote{For $\mu\ll1$, the real positive root $x_{1}$ to $G(x)=0$ is $x_{1}\approx 1-\mu$, such that $\hat{x}=\mu x_{1}\approx \mu$. Meanwhile, the real positive root $x_{1}$ for $\mu\gg1$ is $x_{1}\approx \mu^{-1/3}$.} So, for small $\mu$, the entropy $S_{\text{BH}}^{(4)}\approx \frac{\pi(\mu\ell)^{2}}{G_{4}}$, having the behavior of a classical four-dimensional Schwarzschild black hole of mass $2MG_{4}=\mu\ell$ and temperature $T=(4\pi\mu\ell)^{-1}$. Alternatively, for $\mu\gg1$ the black hole extends a distance $\ell$ off of the brane, looking like a `flattened pancake' \cite{Emparan:1999wa}. 

Lastly, notice that in the limit of large $\ell$, the temperature (\ref{eq:tempscharv2}) vanishes while the entropy (\ref{eq:entschwarv2}) diverges, whereas for small $\ell$, the temperature diverges and the entropy decreases. The latter is consistent with the fact that for vanishing backreaction there is no black hole.

\subsubsection{Quantum Kerr black hole}

The (outer) horizon thermodynamics of the quantum Kerr black hole (\ref{eq:branegeomMinkq}) are readily worked out to be
\beq 
\begin{split}
&M=\frac{(r_{+}^{2}+a^{2})x_{1}[8r_{+}\ell+9x_{1}(r_{+}^{2}+a^{2})]}{8G_{3}[3x_{1}(r_{+}^{2}+a^{2})+2r_{+}\ell]^{2}}\;,\\
&T=\frac{\ell}{2\pi r_{+}}\frac{(r_{+}^{2}-a^{2})(x_{1}(r_{+}^{2}+a^{2})+r_{+}\ell)}{(r_{+}^{2}+a^{2})x_{1}(r_{+}x_{1}+\ell)[3(r_{+}^{2}+a^{2})x_{1}+2r_{+}\ell]}\;,\\
&S_{\text{gen}}^{(3)}=\frac{\pi r_{+}(r_{+}^{2}+a^{2})x_{1}^{2}}{G_{3}[3x_{1}(r_{+}^{2}+a^{2})+2r_{+}\ell]}\,\\
&J=\frac{ar_{+}(r_{+}^{2}+a^{2})x_{1}^{2}}{G_{3}[3x_{1}(r_{+}^{2}+a^{2})+2r_{+}\ell]^{2}}\sqrt{\frac{\ell[x_{1}(r_{+}^{2}+a^{2})+r_{+}\ell]}{r_{+}}}\;,\\
&\Omega=\frac{a}{x_{1}(r_{+}^{2}+a^{2})(r_{+}x_{1}+\ell)}\sqrt{\frac{\ell[x_{1}(r_{+}^{2}+a^{2})+r_{+}\ell]}{r_{+}}}\;.
\end{split}
\eeq
Alternatively, using $(r_{+}^{2}+a^{2})=r_{+}\mu\ell$, $\hat{x}\equiv \mu x_{1}$, and introducing $\hat{a}\equiv a/(\mu\ell)$, the thermodynamic variables become
\beq 
\begin{split}
&M=\frac{\hat{x}(8+9\hat{x})}{32G_{3}\left(1+\frac{3}{2}\hat{x}\right)^{2}}\;,\\
&T=\frac{(1+\hat{x})}{4\pi\ell\hat{x}\left(1+\frac{3}{2}\hat{x}\right)}\frac{(1-4\hat{a}^{2}+\sqrt{1-4\hat{a}^{2}})}{\left[1+\frac{\hat{x}}{2}(1+\sqrt{1-4\hat{a}^{2}})\right]\left[1-2\hat{a}^{2}+\sqrt{1-4\hat{a}^{2}}\right]}\;,\\
&S_{\text{gen}}^{(3)}=\frac{\pi\ell}{4G_{3}}\frac{\hat{x}^{2}}{\left(1+\frac{3}{2}\hat{x}\right)}(1+\sqrt{1-4\hat{a}^{2}})\;,\\
&J=\frac{\hat{a}\ell\hat{x}^{2}}{4G_{3}\left(1+\frac{3}{2}\hat{x}\right)^{2}}\sqrt{1+\hat{x}}\;,\\
&\Omega=\frac{4\hat{a}\sqrt{1+\hat{x}}}{\hat{x}\ell[\hat{x}(1+\sqrt{1-4\hat{a}^{2}})+2][1+\sqrt{1-4\hat{a}^{2}}]}\;,
\end{split}
\eeq
and obey the first law
\beq dM=TdS_{\text{gen}}^{(3)}+\Omega dJ\;.\eeq
The quantum Schwarzschild black hole thermodynamics is recovered in the $\hat{a}\to0$ limit. As before, $T\sim 1/\hat{x}$ for small $\hat{x}$ and $T\sim 1/\hat{x}^{2}$ for large $\hat{x}$, while $S_{\text{gen}}^{(3)}$ has the same $\hat{x}$-dependence as in the non-rotating case (\ref{eq:qschwarzlimitent}).

\subsection{Extended black hole thermodynamics} \label{ssec:extbhthermo}

While black holes have a thermodynamic description, they are peculiar in that the first law (\ref{eq:firstlawgeneralbhs})  lacks a pressure-volume work term. This is because for general black holes there is no clear notion of pressure or volume. For black holes in spacetimes with a cosmological constant, the situation changes dramatically. A first glimpse of this comes from Euler's theorem of homogeneous functions implies black holes with a non-zero cosmological constant $\Lambda_{d+1}$ obey 
obey \cite{Kastor:2009wy}
\beq (d-2)G_{d+1}M=(d-1)TS-2P_{d+1}V+...\;,\label{eq:extSmarr}\eeq
for temperature $T$, Bekenstein-Hawking entropy $S$, and the ellipsis refers to other possible thermodynamic variables, e.g., $\Omega dJ$. The essential new feature is to identify the cosmological constant as thermodynamic pressure
\beq P_{d+1}\equiv-\frac{\Lambda_{d+1}}{8\pi G_{d+1}}\;.\eeq
Then $V$ is the conjugate variable to the pressure, dubbed the `thermodynamic volume'
\cite{Kastor:2009wy,Dolan:2010ha,Cvetic:2010jb} formally equal to\footnote{The interpretation of the thermodynamic volume remains fairly mysterious. In simple cases, e.g., static black holes in $d+1\geq4$, $v$ coincides with the geometric volume occupied by the black hole, i.e., the amount of spacetime volume excluded by the black hole horizon. Generally, however, the thermodynamic volume (\ref{eq:thermovolgen}) differs from the geometric volume \cite{Dolan:2010ha,Cvetic:2010jb,Johnson:2014xza}.} 
\beq V\equiv\left(\frac{\partial M}{\partial P}\right)_{S,...}\;.\label{eq:thermovolgen}\eeq
 For spacetimes with vanishing cosmological constant, the relation  (\ref{eq:extSmarr}) reduces to the Smarr formula, however, for $\Lambda_{d+1}\neq0$ the $P-V$ term is required for consistency.

Going one step further, treating the cosmological constant as a dynamical variable leads to an extended framework of black hole thermodynamics, resulting in the first law of extended black hole thermodynamics 
\beq dM=TdS+VdP_{d+1}+...\;.\label{eq:extfirstlawgenbhs}\eeq
Glossing over the details, allowing for the cosmological constant to be a dynamical pressure has the thermodynamics of anti-de Sitter black holes, in particular, acquire a richer structure than their asymptotically flat counterparts; behaving as Van der Waals fluids \cite{Chamblin:1999tk,Kubiznak:2012wp}, polymers \cite{Kubiznak:2014zwa}, and allowing for the construction of black hole heat engines \cite{Johnson:2014yja}. Thus, the extended thermodynamics of AdS black holes offer a rich gravitational perspective on everyday phenomena (for a review, see \cite{Kubiznak:2016qmn}). 

Extended black hole thermodynamics is not without its criticisms. A common critique is treating the cosmological constant as a \emph{variable} pressure. Indeed, while the $P-V$ term in the Smarr formula (\ref{eq:extSmarr}) is required for consistency, its appearance does not imply the pressure should be made variable, at least from a gravitational perspective. Assuming AdS/CFT duality, it is more natural to allow for varying the cosmological constant since variations in $\Lambda_{d+1}$ are dual to variations in the number of degrees of freedom of the dual theory \cite{Kastor:2009wy,Johnson:2014yja,Dolan:2014cja,Kastor:2014dra,Caceres:2015vsa,Karch:2015rpa,Caceres:2016xjz,Rosso:2020zkk}.
%even beyond the infinite central charge limit \cite{Karch:2015rpa,Visser:2021eqk}.

Holographic braneworlds suggest a higher-dimensional origin for extended black hole thermodynamics \cite{Frassino:2022zaz}. In particular, a dynamical cosmological constant on the brane naturally follows from tuning the brane tension. In fact, keeping  other bulk parameters $L_{d+1}$ and $G_{d+1}$ fixed, varying the tension alone  corresponds to varying the induced brane cosmological constant on the brane $\Lambda_{d}$:
\beq
\delta \tau 
=  \frac{\delta \Lambda_d}{8 \pi G_{d}}\,.
\eeq
Hence, classical black hole thermodynamics in the bulk including work done by the brane, induces extended thermodynamics of quantum black holes on the brane.

This observation can be made explicit in the context of the braneworld constructions considered here. Specifically, when treating the brane tension $\tau$ as a thermodynamic variable akin to  the surface tension of liquids, the 
 first law of the bulk black hole  is
\be
dM = TdS + A_{\tau} d\tau \,,
\label{eq:firstlawbulk}\ee
where $A_{\tau}\equiv\left(\frac{\partial M}{\partial \tau}\right)_{\hspace{-1mm}S}$ is the variable conjugate to $\tau$.
%We note that the black hole mass plays the role of    enthalpy $H=E+ \tau A_\tau$, since   enthalpy in standard thermodynamics satisfies a first law of the form \eqref{eq:firstlawbulk}, whereas the   internal energy $E$ satisfies the usual first law   $dE = T dS - \tau dA_\tau$. 
%Our approach contrasts previous work on (extended) thermodynamics of accelerating black holes \cite{Appels:2016uha,Anabalon:2018ydc,Anabalon:2018qfv,Gregory:2019dtq, Ball:2020vzo} as we include a brane and work in an ensemble where $P_{4}$ is fixed.
Consequently, tension variation $d\tau$ induces extended thermodynamics on the brane. Particularly, the bulk first law (\ref{eq:firstlawbulk}) maps to  the extended first law on the brane
\beq \label{eq:branefirstlaw1ext} dM=TdS_{\text{gen}}+VdP_{d}+...\,,\eeq
thus extending the quantum first law, and where the pressure $P_{d}$ is the pressure of the quantum black hole and $V$ is its conjugate thermodynamic volume. 

In summary, since the scales of the brane theory are induced, holographic braneworlds provide a gravitational motivation for treating the cosmological constant as a variable.  For example,  the pressure and volume of the static quantum BTZ black hole are \cite{Frassino:2022zaz}
\beq 
\begin{split}
&P_{3}\equiv-\frac{\Lambda_{3}}{8\pi G_{3}}=\frac{\sqrt{1+\nu^{2}}}{4\pi \nu^{2}\ell_{3}^{2}G_{3}}(\sqrt{1+\nu^{2}}-1)\;,\\
&V\equiv \left(\frac{\partial M}{\partial P_{3}}\right)_{\hspace{-1mm}S_{\text{gen}},c_{3}}=-\frac{2\pi\ell_{3}^{2}z^{2}[-2+\nu^{2}+3\nu^{3}z^{3}+\nu^{4} z^{4} +\nu z(\nu^{2}-4)]}{(1+3z^{2}+2\nu z^{3})^{2}}\,.
\end{split}
\eeq
Together with the standard thermodynamic variables (\ref{eq:qbtzstathermo}), it is straightforward to verify the extended first law (\ref{eq:branefirstlaw1ext}) is obeyed. Extended thermodynamics for charged and rotating qBTZ black holes were computed in \cite{Frassino:2024bjg} (see also \cite{Feng:2024uia} for charged BTZ). 

\subsection*{Quantum reverse isoperimetric inequality}

One of the more puzzling aspects of extended black hole thermodynamics is the thermodynamic volume $V$. In simple cases, e.g., $D\geq4$ AdS-RN, volume $V$ coincides with the geometric volume of the black hole, i.e., the amount of spacetime volume excluded by the black hole horizon. In general, however, the thermodynamic volume is not the geometric volume, cf \cite{Dolan:2010ha,Cvetic:2010jb,Johnson:2014xza}.\footnote{This is not to say the thermodynamic volume does not have a geometric character. Indeed, in classical gravity, the thermodynamic volume has a geometric definition in terms of Komar integrals. In this sense, moreover, the definition of the thermodynamic volume is independent of treating the cosmological constant as a thermodynamic variable.} Nonetheless, the thermodynamic volume plays a crucial role in understanding black hole thermodynamics. A sharp example of this is that there is strong evidence that AdS black holes obey the reverse isoperimetric inequality \cite{Cvetic:2010jb}
\be 
\mathcal{R} \equiv \left(\frac{(D-1) V}{\Omega_{D-2}} \right)^{\frac{1}{D-1}} \left(\frac{\Omega_{D-2}}{A_{\text{BH}}}\right)^{\frac{1}{D-2}}\geq1\;.
\label{eq:classRII}\ee
Here $\Omega_{D-2}$ is the volume of a unit $(D-2)$ sphere, of $D$-dimensional AdS and $A_{\text{BH}}$ is the area of the black hole horizon. Notably, it is the thermodynamic volume for which this inequality holds, not the geometric volume. Refined generalizations of the inequality (\ref{eq:classRII}), inspired by the classical Penrose inequality \cite{Penrose:1968ar,Penrose:1969pc}, have been conjectured and tested for a plethora of examples \cite{Amo:2023bbo}.

Physically, the reverse isoperimetric inequality states an asymptotically AdS black hole with fixed thermodynamic volume has an entropy no larger than Schwarzschild-AdS of the same volume, i.e., Schwarzschild-AdS is a maximal entropy state at fixed (thermodynamic) volume. While there is no general proof of the inequality (\ref{eq:classRII}), there are very few known counterexamples\footnote{In $D\geq4$, `ultraspinning' black holes \cite{Klemm:2014rda,Hennigar:2014cfa,Hennigar:2015cja} were initially thought to obey $\mathcal{R}<1$, thus violating (\ref{eq:classRII}), however, this has been called into question \cite{Appels:2019vow}. In $D=3$, only the electrically charged BTZ black hole violates the reverse isoperimetric inequality \cite{Frassino:2015oca}.} and all such `superentropic' black holes have a negative heat capacity at constant volume, $C_{\text{V}}<0$, and are thus thermodynamically unstable \cite{Johnson:2019mdp}. Via AdS$_{3}$/CFT$_{2}$ duality, black hole superentropicity can be microscopically understood as an overcounting of the (naive) Cardy entropy of the CFT$_{2}$ \cite{Johnson:2019wcq}. 

When semi-classical quantum effects are accounted for, the classical inequality (\ref{eq:classRII}) is known to be violated \cite{Frassino:2022zaz}. A natural quantum generalization of (\ref{eq:classRII}) is proposed to be \cite{Frassino:2024bjg}
\be 
\mathcal{R}_{\text{Q}} \equiv \left(\frac{(D-1) V_{\rm th}}{\Omega_{D-2}} \right)^{\frac{1}{D-1}} \left(\frac{\Omega_{D-2}}{4\mathcal{G}_{D}S_{\text{gen}}}\right)^{\frac{1}{D-2}}\geq1\;.
\label{eq:qRII}\ee
Here, the classical area has been replaced for the generalized entropy and $V_{\text{th}}$ is the Casimir-subtracted thermodynamic volume,
\beq V_{\text{th}}=V-V_{\text{cas}}\;,\label{eq:cassubvol}\eeq
where, in analogy with Casimir mass, $V_{\text{cas}}$ is the thermodynamic volume assigned to empty AdS space. In the case of $D=3$, the quantum reverse isoperimetric inequality (\ref{eq:qRII}) has been shown to hold for all AdS$_{3}$ quantum black holes at all orders of backreaction -- except for a subspace of rotating black holes (those belonging to the red region in Figure \ref{fig:supp}), which are all found to be thermodynamically unstable, in accordance with Johnson's conjecture \cite{Johnson:2019mdp} regarding the classical inequality. This implies there exists a maximum entropy state among thermodynamically stable quantum black holes at fixed volume.

.

\subsection{Phase transitions of quantum black holes} \label{ssec:ptsqbhs}

As with ordinary thermodynamic systems, black holes in AdS can undergo phase transitions. A paradigmatic example is the Hawking-Page (HP) phase transition  \cite{Hawking:1982dh}: below a certain temperature $T_{\text{HP}}$, large AdS black holes in equilibrium with their radiation transition to thermal AdS. At the level of the quantum gravitational partition function, this transition signals an exchange between dominant contributions in the Euclidean path integral. The classical BTZ black hole also undergoes a HP phase transition \cite{Maldacena:1998bw,Birmingham:2002ph}.
To wit, the canonical free energy $F_{\text{BTZ}}=-T\log Z$ of a static BTZ black hole is
\begin{equation}
    F_{\text{BTZ}} =M - TS= - \frac{\pi^2 \ell_3^2}{2 G_{3}} T^2\;,
\end{equation}
for temperature $T\!=\!r_{+}/2\pi\ell^{2}_{3}$. Comparing to the free energy of thermal AdS, $ F_{\text{AdS}}=M_{\text{AdS}}= -1/8G_{3}$, a first-order phase transition  occurs at a temperature
\beq T_{\text{HP}} = \frac{1}{2 \pi \ell_3}\;.\eeq 
When $T<T_{\text{HP}}$, thermal AdS has a lower free energy than the black hole and is, therefore, the dominant contribution to the partition function; for $T>T_{\text{HP}}$ the black hole has lower free energy and becomes the dominant contribution to the partition function.

\begin{figure}[t]
\begin{center}
\hspace{-1.5cm}\includegraphics[scale=.55]{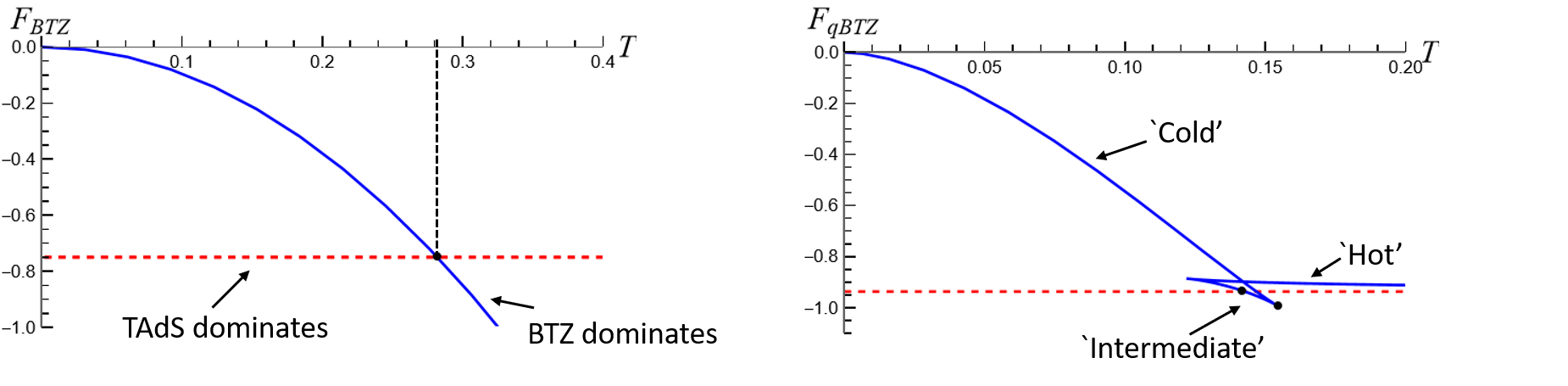}
\end{center}
\caption{\small \textbf{Phases of BTZ black holes.} \emph{Left:} Hawking-Page transition of classical BTZ. For temperatures $T<T_{\text{HP}}$, thermal AdS has a lower free energy (red, dashed line) than the black hole (blue, solid curve). At $T=T_{\text{HP}}$ there is a first-order phase transition, and for $T>T_{\text{HP}}$ the black hole has a lower free energy, leading to an exchange in dominance in the quantum gravitational partition function. \emph{Right:} Reentrant phase transition of quantum BTZ black hole with large backreaction. Below a critical temperature (the analog of the HP temperature), quantum TAdS has the lower free energy. At this critical temperature there is a first-order transition between qTAdS and intermediate black holes. Between this HP-like temperature and a second critical temperature, the qBTZ has a lower free energy. At the second critical temperature there is a zeroth-order transition back to TAdS. 
}
\label{fig:qbhpts} 
\end{figure}

Quantum AdS black holes also undergo thermal phase transitions \cite{Kudoh:2004ub,Frassino:2023wpc}. In fact, large backreaction effects can trigger new transitions unseen by their classical counterparts. Consider, for example, static, neutral quantum BTZ. Working in the standard canonical ensemble, where $c$ and $P$ are held fixed,\footnote{Other ensembles are equally intriguing. For a discussion on the constant $c$ and $V$ ensemble, characterized by the Gibbs free energy $G=M+PV-TS$, see for example \cite{Johnson:2023dtf,HosseiniMansoori:2024bfi}.} the free energy is
 \beq
 \begin{split}
 \label{freeenergy}
F_{\text{qBTZ}}&\equiv M-TS_{\text{gen}}=-\frac{z^{2}\sqrt{1+\nu^{2}}}{2G_{3}}\frac{[1+2\nu z+\nu z^{3}(2+\nu z)]}{(1+3z^{2}+2\nu z^{3})^{2}}\;.
\end{split}
\eeq
As with the classical scenario, compare to  `quantum' thermal AdS$_3$ (qTAdS), i.e., 
 pure $\text{AdS}_{3}$ including backreaction due to the cut-off $\text{CFT}_{3}$, with free energy
 \beq F_{\text{TqAdS}}= M_{\text{qTAdS}}=-\frac{1}{8\mathcal{G}_{3}}\;.\label{eq:TqAdSfree}\eeq
 In Figure \ref{fig:qbhpts} we present a side-by-side comparison of the canonical free energy of the classical BTZ black hole and quantum BTZ for large backreaction, $\nu>1$. Focusing on the quantum BTZ black hole, as the temperature monotonically increases, there are \emph{reentrant phase transitions} from thermal AdS to qBTZ and back to thermal AdS: (i) for temperatures up to a critical temperature (where $\Delta F\equiv F_{\text{qBTZ}}-F_{\text{qTAdS}}=0$), thermal AdS has lower free energy, until at the critical temperature there is a discontinuity in the slope of the free energy, i.e., a first-order phase transition and a quantum analog of the Hawking-Page phase transition; (ii) After this temperature, the qBTZ black hole has a lower free energy until there is a jump discontinuity in the free energy, a zeroth-order phase transition, beyond which TAdS always has a lower free energy. The reentrant phase transition is the combination of the first- and zeroth-order phase transitions as the temperature monotonically varies, and is exhibited by an (inverse) swallow tail. 
 
 As displayed in Figure \ref{fig:qbhpts}, reentrant phase transitions do not occur for the classical BTZ black hole. Reentrant phase transitions, however, are known to appear in (classical) higher-derivative theories of gravity in higher dimensions and Born-Infeld gravity \cite{Gunasekaran:2012dq,Altamirano:2013ane,Frassino:2014pha,Ahmed:2023dnh}. In all such cases, the transitions are between different phases of the black hole, e.g., large to small and back to large black holes. For the quantum black hole, the reentrant phase is a reentrant Hawking-Page phase transition, moving between thermal AdS and the same black hole.  Again,  the zeroth-order phase transitions only occur for large enough backreaction. In this regime, the brane has decreasing tension and the gravitational theory on the brane becomes more massive and effectively four-dimensional. Meanwhile, for small backreaction, thermal AdS dominates at all temperatures, until the $\nu=0$ limit where one recovers phase behavior of classical BTZ.

Even though the phase structure admits a region where the black hole is thermodynamically favored, it may not contribute to the Euclidean path integral if it is thermally unstable. The heat capacity serves as a diagnostic to determine whether the black hole is stable against thermal fluctuations. For example, classically, the Hawking-Page transition is between thermal AdS and stable black holes (small AdS black holes are unstable). Likewise, in the case of quantum BTZ, a computation of the heat capacity \cite{Frassino:2024fin} reveals reentrant phase transitions occur in the canonical ensemble between thermal AdS and a branch of thermally stable quantum black holes. More precisely, a study of the temperature of the qBTZ solution reveals three branches\footnote{Notably, the three branches described here do not coincide with branches $1a$, $1b$, and $2$ characterizing the mass of the qBTZ \cite{Emparan:2020znc}. A more detailed study of these phases, including a stability analysis and their critical behavior was given in \cite{Kudoh:2004ub,Johnson:2023dtf,HosseiniMansoori:2024bfi}. Interestingly, in a fixed $c$ and $V$ ensemble, phase transitions between cold and hot black holes occur, demonstrating continuous critical phenomena along the coexistence curve, with critical exponents that deviate from those observed in mean-field Van der Waals fluids \cite{HosseiniMansoori:2024bfi}.} of continuously connected black holes: (A) `cold' black hole, (B) `intermediate' black hole, and (C) `hot' black hole. Black holes belonging to branches A and C have a negative heat capacity and are thus unstable, while branch B black holes have a positive heat capacity and are thermally stable. The reentrant phase transitions occur only between thermal AdS and these intermediate black holes.

%%%%%%%%%%%%%%%%%%%%%%%%%%%%%%%%%%%%%%%%%%%%%%%%%%%%%%%%%%%%%%%

\section{Braneworld black holes in higher dimensions} \label{app:evapbhshighd}

The majority of this review focused on exact constructions of three-dimensional quantum black holes. From the perspective of the non-holographic perturbative treatment, working in three spacetime dimensions rather $d+1\geq4$ was solely to simplify the problem. Indeed, the expectation value of the renormalized quantum stress tensor $\langle T_{ij}\rangle$ is not known in general. Exceptions do exist.\footnote{Even in a fixed background, explicit computations of $\langle T_{ij}\rangle$ are complicated and limited, cf. \cite{Birrell:1982ix,Anderson:1994hg}.} For example, for even-dimensional homogeneous and isotropic FRW cosmologies, the trace anomaly for conformal quantum fields is known \cite{Hawking:2000bb,Herzog:2013ed}.
%four-dimensional homogeneous and isotropic FRW spacetimes, the trace anomaly for conformal quantum fields, 
%\beq g^{ij}\langle T_{ij}\rangle= a\mathcal{G}+bW^{2}\;,\eeq
%with Gauss-Bonnet tensor $\mathcal{G}=R^{2}-4R_{ij}^{2}+R_{ijkl}^{2}$ and $W^{2}$ is the square of the Wely curvature, is known, e.g., \cite{Hawking:2000bb}. In fact, for arbitrary even higher-dimensional Weyl-flat ($W=0$) backgrounds the vacuum stress-tensors of quantum CFTs are known via the trace anomaly \cite{Herzog:2013ed}.  
In the case of static, spherically symmetric backgrounds, however, the renormalized stress-tensor is only known up to an arbitrary function of an appropriate radial coordinate \cite{Christensen:1977jc}.

One possible way to circumvent this issue is to instead work in a truncated s-wave sector of the matter theory and use the two-dimensional conformal anomaly to fix the form of the quantum-stress tensor and find quantum corrections to, say, the Schwarzschild black hole \cite{Fabbri:2005zn,Fabbri:2005nt,Ho:2017joh,Arrechea:2019jgx}. Another approach is to solve the (semi-classical) Tolman-Oppenheimer-Volkoff equations for a massless conformally coupled scalar field \cite{Beltran-Palau:2022nec}. Working perturbatively in $\hbar$, one can construct analytic asymptotically flat, static, spherically symmetric solutions, while numerics yields non-perturbative corrections in $\hbar$.\footnote{In \cite{Cai:2009ua} it was argued that one can construct exact analytic static, spherically symmetric solutions (see \cite{Fernandes:2023vux} for a stationary generalization) if one imposes an equation of state on the quantum stress-tensor such that the type-B trace anomaly vanishes.} Thus, the situation for solving the backreaction problem in $d+1\geq4$ is more complicated and subject to quantum gravitational corrections. This again motivates the use of holographic techniques as employed in the three-dimensional case. However, as we now review, the status of finding higher-dimensional quantum black holes via braneworlds is more complicated.

\subsection{Higher-dimensional quantum black holes?}

Given the success of studying three-dimensional quantum black holes starting from the AdS$_{4}$ C-metric, it is natural to wonder if a similar analysis follows from a higher-dimensional C-metric. Notably, however, no exact solutions to Einstein's equations describing accelerating black holes in dimension $D=d+1>4$ are known.\footnote{There are, however,  accelerating black holes in AdS$_{3}$, cf. \cite{Astorino:2011mw,Xu:2011vp,Arenas-Henriquez:2022www,Arenas-Henriquez:2023hur,Tian:2024mew}.} As reasoned in \cite{Camps:2010sn}, this is because in $D\geq5$, the string generating the acceleration has a singularity along its symmetry axis passing through the black hole which is distinctly less mild than in $D=4$. Put another way, in four dimensions there exist more general accelerating black holes, accelerated by Levi-Civita strings or rods. The C-metric is a special limit of such a geometry, singled out as a metric where the string has a milder singularity along its axis and better-behaved asymptotics than its brethren. In $D\geq5$, it seems there is no such special limit for accelerating black holes. Alternately, one can perturbatively construct a C-metric in $D>4$ by perturbing a higher-dimensional Schwarzschild black hole to give it uniform acceleration \cite{Kodama:2008wf}. However, it was found such a solution with constant string tension does not allow for localized braneworld black holes (they do not satisfy the Israel junction conditions). Moreover, allowing for non-uniform string tension results in infinitely many localized braneworld black hole solutions. 

Historically, the first attempt at finding a four-dimensional braneworld black hole embedded in a five-dimensional bulk was carried out by Chamblin, Hawking and Reall \cite{Chamblin:1999by}. Their starting point takes the Randall-Sundrum model and replaces the Minkowski line element on the brane with a four-dimensional Ricci flat metric, e.g., the Schwarzschild black hole, 
\beq ds^{2}_{5}=dy^{2}+a^{2}(y)g_{ij}^{\text{Schw}}dx^{i}dx^{j}\;.\label{eq:CHRbs}\eeq
Here $y$ denotes the bulk extra direction, and $a^{2}(y)$ denotes the warp factor (for the RS-II scenario, $a^{2}(y)=e^{-2|y|/L_{5}}$ for bulk AdS$_{5}$ length scale $L_{5}$). The brane is located at $y=0$, such that, from the brane perspective the geometry is exactly the static, four-dimensional Schwarzschild black hole. Note, however, the five-dimensional Chamblin, Hawking, Reall `black string' (\ref{eq:CHRbs}) suffers from a classical \emph{dynamical} instability \cite{Gregory:2000gf} analogous to the Gregory-Laflamme instability of Kaluza-Klein black strings \cite{Gregory:1993vy,Gregory:1994bj}. We will return to this instability momentarily. Further, the black string extends to the AdS$_{5}$ horizon at $y=\infty$ and where the black hole horizon becomes singular with diverging scalar curvature invariants. 

The Chamblin, Hawking, Reall braneworld black hole thus cannot describe the end state of gravitational collapse. In fact, whilst searching for examples of Oppenheimer-Synder gravitational collapse
%\footnote{Recall in the Oppenheimer-Synder model of collapsing dust a FRW cosmology interior is matched to an exterior Schwarzschild solution.}
of braneworld black holes, a no-go theorem was posed \cite{Bruni:2001fd}: the exterior geometry of the dust cloud cannot be static. This theorem, and the dearth of evidence of exact static black holes localizing in the RS-II construction (circa 2002), in part motivated Tanaka \cite{Tanaka:2002rb} and Emparan, Fabbri, and Kaloper \cite{Emparan:2002px} to conjecture higher-dimensional braneworld black holes must be time-dependent (see also \cite{Emparan:2002jp}). Their reasoning utilized AdS/CFT holography and was argued to be consistent with the proposal that black holes that localize on the brane may be interpreted as quantum black holes.

%-- where the time-dependent nature of higher-dimensional braneworld black holes is a result of Hawking radiation. 

\subsection*{Predictions from holography}

Let us review the arguments \cite{Tanaka:2002rb,Emparan:2002px} predicting higher-dimensional braneworld black holes must be time-dependent. 
%Our presentation follows the analysis presented in \cite{Fitzpatrick:2006cd}. 
Take the bulk to be classical AdS$_{5}$ general relativity with a four-dimensional RS brane. According to braneworld holography, the induced theory on the brane describes a higher-derivative theory of gravity coupled to a large-$N$ gauge theory in the 't Hooft planar limit with large 't Hooft coupling $\lambda=N g_{\text{YM}}^{2}$ (we refer to this matter theory as a large-$c$ CFT$_{4}$ with an ultraviolet cutoff owed to the presence of the brane). In particular, the full AdS/CFT duality in this case is between type IIB string theory on $\text{AdS}_{5}\times S^{5}$ and $\mathcal{N}=4$ super Yang-Mills $SU(N)$ gauge theory, with AdS length $L_{5}= L_{\text{P}}^{(10)}(g_{s}N)^{1/4}$ (for string coupling $g_{s}$ and ten-dimensional Planck length $L_{\text{P}}^{(10)}$) and 't Hooft coupling $\lambda=g_{s}N$. The effective number of CFT degrees of freedom is\footnote{To arrive at the second equality, use that in $d$-spacetime dimensions $\hbar G_{d}=(L_{\text{P}}^{(d)})^{d-2}$ and the relation between the brane and bulk Newton's constants (\ref{eq:effGd}). Then, for $d=4$, $(L_{\text{P}}^{(5)})^{3}=L_{5}(L_{\text{P}}^{(4)})^{2}$.}
\beq c\sim N^{2}=\left(\frac{L_{5}}{L_{\text{P}}^{(5)}}\right)^{3}=\left(\frac{L_{5}}{L_{\text{P}}^{(4)}}\right)^{2}\;,\eeq
such that large $N=L_{5}/L_{\text{P}}^{(4)}$ coincides with the four-dimensional Planck length going to zero. Further, notice the combination 
\beq N^{2}\hbar=\left(\frac{L_{5}}{L_{\text{P}}^{(4)}}\right)^{2}\frac{(L_{\text{P}}^{(4)})^{2}}{G_{4}}=\frac{L_{5}^{2}}{G_{4}}\;,\label{eq:fixedcombo}\eeq
remains fixed as $\hbar\to0$ and keeping $L_{5},G_{4}$ fixed. 

Now, according to the conjecture \cite{Emparan:2002px}, the braneworld black hole must be a solution to the semi-classical theory sourced by the CFT stress-tensor $\langle T_{ij}\rangle$ in some quantum state. For a black hole background, in principle, the quantum state could be in any of the common choices of the vacuum state, i.e., the Hartle-Hawking, Boulware, or Unruh state. Recall that the Hartle-Hawking state describes a black hole in a thermal bath in equilibrium with its own radiation, producing a non-dynamical configuration. On the other hand, the Unruh vacuum describes a time-dependent evaporating black hole (there is a thermal flux of radiation at future null infinity). To infer whether a black hole in the Hartle-Hawking state is possible -- without finding an explicit solution -- one can instead estimate the evaporation time of a radiating black hole; if the time for evaporation is finite, then a static black hole solution is not possible. To this end, consider a weakly coupled CFT in four dimensions.
%To leading order in $\hbar$, the trace anomaly is 
%\beq \langle T^{i}_{\;i}\rangle=\frac{\hbar}{16\pi^{2}}(aW^{2}+bE+c\Box R)\;,\eeq
%where $W^{2}=R_{ijkl}^{2}-2R_{ij}^{2}+R^{2}/3$ is the square of the Weyl tensor and  $E=R^{2}-4R_{ij}^{2}+R_{ijkl}^{2}$ is the Gauss-Bonnet term.  The real coefficients $a,b,c$ depend on the specific matter content of the theory.
In particular, the trace anomaly for $\mathcal{N}=4$ $SU(N)$ super-Yang-Mills theory to leading order in $\hbar$ is
%the coefficients conspire such that $R_{ijkl}^{2}$ cancels leaving
\beq \langle T^{i}_{\;i}\rangle=\frac{\hbar(N^{2}-1)}{32\pi^{2}}\left(R_{ij}^{2}-\frac{1}{3}R^{2}\right)\approx \frac{\hbar N^{2}}{32\pi^{2}}\left(R_{ij}^{2}-\frac{1}{3}R^{2}\right)\;,\label{eq:weakcoupTijSYM}\eeq
taking $N\gg1$. Thus, the anomaly for such a CFT is simply the free field result enhanced by a $\mathcal{O}(N^{2})$ factor. With this in mind, the power emitted by Hawking quanta modeled by the large-$N$ CFT will be 
\beq \frac{dM}{dt}\sim \frac{N^{2}\hbar}{R_{0}^{2}}\;,\eeq
for initial horizon radius $R_{0}=2\sim G_{4}M$.
%\footnote{Assuming the far-field outgoing metric encoding a flux of Hawking radiation is approximated by the outgoing Vaidya metric.}
Then, the time for evaporation $t_{\text{evap}}$ heuristically goes as
\beq t_{\text{evap}}^{-1}\equiv\frac{1}{R_{0}}\frac{dR_{0}}{dt}=\frac{2G_{4}}{R_{0}}\frac{dM}{dt}\sim \frac{2}{R_{0}^{3}}\hbar G_{4}N^{2}= \frac{2L_{5}^{2}}{R_{0}^{3}}\;,\label{eq:evaptime}\eeq
where we substituted in (\ref{eq:fixedcombo}). Thus, the evaporation time is finite (even as $\hbar\to0$). Moreover, such a black hole would evaporate rapidly due to the $\mathcal{O}(N^{2})$ enhancement of the free theory result. Altogether, this suggests black holes cannot remain static on the brane: they shrink and evaporate in a finite time.

Assuming the bulk/brane correspondence holds, the semi-classical evaporation of the black hole localized on the brane should have a classical bulk signature. One possibility, originally put forth in  \cite{Tanaka:2002rb} (see also \cite{Chamblin:2004vr}) is that the semi-classical evaporation is linked to a classical instability of the bulk five-dimensional solution. The intuition is as follows. Consider the five-dimensional black Chamblin, Hawking, Reall black string (\ref{eq:CHRbs}), and have a Randall-Sundrum brane intersect it. The radius of the black hole will exponentially shrink with the AdS$_{5}$ length $L_{5}$ as one moves away from the brane. Further, the black string suffers from a Gregory-Laflamme instability when the horizon radius becomes smaller than $L_{5}$. As a result of the instability, a portion of the horizon localized on the brane will pinch off and fall into a region of the bulk black hole away from the brane. Thus, while the horizon area of the bulk black hole does not shrink, the area of the horizon localized on the brane shrinks. This five-dimensional deformation due to dynamical instability implies a type of classical evaporation of braneworld black holes.

Another view is that semi-classical radiation corresponds to gravitational radiation of the bulk black hole. Indeed, from the brane perspective, the Hawking quanta are modeled by the large-$c$ CFT, corresponding to bulk Kaluza-Klein gravitons. Further, a black hole stuck to a brane is accelerating away from the center of AdS, thus producing gravitational waves. Thus, Hawking radiation from the brane perspective corresponds to bulk gravitational bremsstrahlung \cite{Emparan:2002px}. Qualitatively, moreover, the classical gravitational waves emitted into the bulk have a characteristic frequency $\omega$ which from the brane perspective is estimated to be $\omega\sim (G_{4}M)^{-1}$, the four-dimensional Hawking temperature. This reasoning suggests why the classical bulk gravitational radiation appears as thermal radiation on the brane.

\subsection*{Counterexamples}

The essential problem of the holographic argument leading to the finite evaporation time (\ref{eq:evaptime}) is that the result assumes the large-$N$ CFT is weakly coupled. Of course, for the holographic construction of the bulk/brane set-up to be valid, the cutoff CFT on the brane is strongly coupled, i.e., large 't Hooft coupling. Thus, as first recognized by Fitzpatrick, Randall, and Wiseman \cite{Fitzpatrick:2006cd}, it is not clear the radiation has access to all of its $\mathcal{O}(N^{2})$ degrees of freedom. Importantly, being at strong coupling can lead to a reduction to the accessible degrees of freedom, from $\mathcal{O}(N^{2})$ to $\mathcal{O}(1)$ via confinement (see also \cite{Hubeny:2009rc}).\footnote{Roughly, the argument of \cite{Fitzpatrick:2006cd} is as follows. Consider $\mathcal{N}=4$ $SU(N)$ super Yang-Mills theory on a sphere of radius $R$. There are weakly interacting states (`glueballs') with energies $ER\ll N^{2}$, and strongly coupled states with energies $ER\gg N^{2}$. At large ’t Hooft coupling $\lambda$, AdS/CFT says the field theory is dual to closed string theory, where the glueball states correspond to perturbative string excitations in the ambient spacetime. Further, in this limit, the energy separation for weakly interacting states goes like $\Delta E\sim \lambda^{1/4}/R$. Hence, the glueball spectrum is lifted to infinite energy apart from the $\mathcal{O}(1)$ massless states dual
to the supergravity modes of the string, and gravitational perturbations dual to $\mathcal{O}(1)$ of the $\mathcal{O}(N^{2})$ states. A caveat to this reasoning, however, is that in the flat space limit $R\to\infty$ the mass gap might disappear.} Such a reduction thus makes it plausible that static braneworld black holes in higher dimensions do exist. 

In fact, there are several examples of higher-dimensional static braneworld black holes, both analytically and numerically constructed \cite{Kudoh:2003xz,Kudoh:2004kf,Fitzpatrick:2006cd,Tanahashi:2007wt,Yoshino:2008rx,Gregory:2008br,Figueras:2011va,Figueras:2011gd,Abdolrahimi:2012qi,Abdolrahimi:2012pb,Figueras:2013jja,Banerjee:2021qei,Biggs:2021iqw,Emparan:2023dxm}. The question is whether these static solutions are to be viewed as quantum black holes on the brane. To explore this point, reconsider the 5D/4D Randall-Sundrum construction with the five-dimensional Schwarzschild black string (\ref{eq:CHRbs}). Despite its dynamical instability, this is an example of a static braneworld black hole, where the black hole on the brane is simply the four-dimensional Schwarzschild geometry. Seemingly then, the cutoff CFT on the brane does not modify the geometry: from the looks of it, the brane geometry consists of a non-trivial zero mode and no excited Kaluza-Klein modes \cite{Fitzpatrick:2006cd}. Nonetheless, Fitzpatrick, Randall, and Wiseman argued the dual quantum black hole description might be consistent with the existence of such static localized black hole solutions. It is simply that the quantum corrections to the geometry are suppressed.\footnote{Fitzpatrick, Randall, and Wiseman also argue that the dynamical instability dual to semi-classical evaporation is unlikely to occur \cite{Fitzpatrick:2006cd}. Even if it did, the timescales of the bulk Gregory-Laflamme instability and thermodynamic instability of the Schwarzschild black hole via Hawking radiation are different \cite{Fabbri:2007kr}.}

\begin{figure}[t!]
\begin{center}
\includegraphics[width=.35\textwidth]{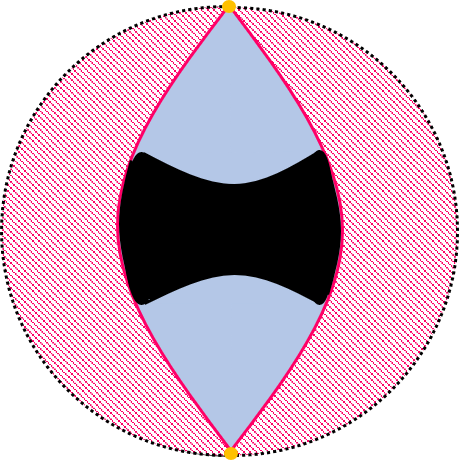}
\end{center}
\caption{\small \textbf{Black string with  Karch-Randall branes.} An AdS$_{5}$ black string stretches between two two end-of-the-world AdS$_{4}$ branes, such that all magenta regions (including the AdS$_{5}$ boundary) is removed. The black string can be replaced by a uniform black funnel, and the ETW branes may be replaced with Randall-Sundrum branes.}
\label{fig:bsbf} 
\end{figure}

To appreciate this last point, consider the set-up of \cite{Gregory:2008br}, consisting of an asymptotically AdS$_{5}$ bulk spacetime, with two positive tension Karch-Randall branes with an AdS$_{5}$ Schwarzschild black string stretching between the two branes (see Figure \ref{fig:bsbf}). 
%For a certain range of mass, this black string is stable, however, even when the bulk solution is unstable it remains regular such that it should have a dual CFT description. 
In particular, the bulk geometry has line element
\beq ds^{2}_{5}=\frac{L_{5}^{2}}{\cos^{2}(u)}\left[du^{2}+\frac{1}{\ell_{4}^{2}}g_{ij}dx^{i}dx^{j}\right]\;,\label{eq:5Dbstring}\eeq
with bulk AdS$_{5}$ radius $L_{5}$, and induced four-dimensional radius $\ell_{4}=L_{5}\text{sec}(u_{0})$. The AdS$_{5}$ boundary is located at $u=\pm\pi/2$, while the two AdS$_{4}$ branes are at $u=\pm u_{0}$, each with the same positive tension $\tau=\frac{6\sin(u_{0})}{8\pi G_{5}L_{5}}$, and separated a finite distance apart. Further, take the four-dimensional metric $g_{ij}$ to be the AdS$_{4}$-Schwarzschild black hole, 
\beq g_{ij}dx^{i}dx^{j}=-f(r)dt^{2}+f^{-1}(r)dr^{2}+r^{2}(d\theta^{2}+\sin^{2}\theta d\phi^{2})\;,\quad f=1+\frac{r^{2}}{\ell_{4}^{2}}-\frac{2G_{4}M}{r}\;.\label{eq:AdS4schw}\eeq
Without branes, this system  has an instability when $r_{+}/\ell_{4}<1$ (small black holes of horizon radius $r_{+}$), while is stable for large black holes, $r_{+}/\ell_{4}>1$ \cite{Hirayama:2001bi}. The stability structure remains the same with branes included \cite{Gregory:2008br}. 

The classical bulk is stable for large mass black holes that localize on the branes (focus on only a single brane as a being living on one brane would not directly experience the other brane). From the brane perspective, the induced geometry is simply the AdS$_{4}$-Schwarzschild black hole. Seemingly, the geometry has no quantum corrections due to the backreacting holographic field theory living on the brane. This runs counter to what would be expected based on a weakly-coupled analysis of the quantum-stress tensor.\footnote{A detailed comparison of holographic versus non-holographic stress-tensors was done in \cite{Gregory:2008br} using an approach developed by Page \cite{Page:1982fm} to compute the renormalized stress-tensor of a (non-holographiic) conformally coupled scalar field in AdS$_{4}$-Schwarzschild.} Indeed, recall the trace anomaly of weakly coupled $\mathcal{N}=4$ SU(N) super Yang-Mills theory (\ref{eq:weakcoupTijSYM}). Substituting in the AdS$_{4}$-Schwarzschild geometry (\ref{eq:AdS4schw}), the anomaly reads
\beq \langle T^{i}_{\;i}\rangle=-\frac{3\hbar(N^{2}-1)}{8\pi^{2}\ell_{4}^{4}}\;.\label{eq:anomaly4D}\eeq
Since $\langle T^{i}_{\;i}\rangle$ is non-vanishing, one expects the AdS$_{4}$-Schwarzschild black hole to become quantum-corrected due to backreaction. In view of the conjecture of holographic braneworld black holes, the question thus becomes in what sense can the AdS$_{4}$-Schwarzschild black hole on the brane be viewed as a `quantum' black hole. 

The resolution of this apparent tension with the conjecture is the following.  The quantum stress tensor of the strongly coupled CFT (as required for a consistent holographic description of the bulk gravity) simply does not correct the geometry, at least at leading order. To see this, consider the boundary CFT stress-tensor, which can be computed in full using holographic renormalization methods, leading to \cite{Gregory:2008br}
\beq \langle T_{ij}\rangle =-\frac{3\hbar N^{2}}{32\pi^{2}\ell_{4}^{4}}g_{ij}\;.\eeq
Taking the trace returns the anomaly (\ref{eq:anomaly4D}). The main takeaway is that the stress tensor is proportional to the metric. In such an instance, the CFT only renormalizes the effective brane cosmological constant, and thus does not lead to any quantum-corrected geometry. Notice that in the limit the brane cosmological constant vanishes, $\ell_{4}\to\infty$, so too does the quantum stress-tensor, $\langle T_{ij}\rangle\to0$. This is consistent with the Schwarzschild black string with a Randall-Sundrum brane construction in \cite{Fitzpatrick:2006cd}.

A conclusion, then, is that, formally, the AdS$_{4}$-Schwarzschild braneworld black hole can be interpreted as a quantum black hole. The cutoff CFT, however, is special in that it does not explicitly backreact on the brane geometry, and only serves to renormalize the induced cosmological constant. Geometrically,  this is a manifestation of the bulk spacetime (\ref{eq:5Dbstring}) being foliated by the AdS$_{4}$-Schwarzschild black hole. Consequently, the classical Kaluza-Klein graviton modes are not excited in the background solution.  Observe, moreover, the correction to the cosmological constant is independent of the black hole mass, and hence its temperature. The brane black hole thus does not radiate, consistent with no excited Kaluza-Klein modes. This is at first puzzling from the perspective of AdS/CFT, as one would have expected a component of the holographic stress tensor to correspond to a
thermal plasma of CFT degrees of freedom outside the black hole. As noted in \cite{Hubeny:2009rc}, this is a consequence of the large-$N$ super-Yang-Mills theory confining to $\mathcal{O}(1)$ degrees of freedom. 

It is worth emphasizing the implications of the static solution found from the AdS$_{5}$ black string \cite{Gregory:2008br}. There is no Hawking emission, and, up to the term generating the Weyl anomaly, the quantum stress tensor is everywhere vanishing. Despite the lack of thermal radiation near infinity, the static state of the CFT is in thermal equilibrium with the black hole since they nonetheless obey the Kubo-Martin-Schwinger (KMS) condition. This suggests, from the bulk perspective, such braneworld black holes remain stuck to the brane, while from the brane perspective, black hole evaporation is disallowed due to the large-$N$ and large 't Hooft coupling of the holographic CFT.

%There are two types of bulk constructions which lead to localized black holes on the branes: (i) a single Karch-Randall brane intersects a bulk black hole such that it localizes on the brane, or (ii) a black string which stretches 

%allowing for a direct comparison between backreaction due to a strongly coupled holographic CFT and Hawking radiation at weak-coupling. 

\subsection{Evaporating braneworld black holes}

Putting the status of the higher-dimensional static quantum black holes aside, braneworld holography has proven useful in studying black hole evaporation. Indeed, while some bulk black holes get stuck on the brane and never shrink (`black droplets') \cite{Figueras:2011gd} and hence dynamically stable, there are other bulk black solutions that evaporate due to a Gregory-Laflamme instability \cite{Emparan:2023dxm}, as originally suggested in \cite{Tanaka:2002rb,Emparan:2002px}. The essential physical insight is that evaporation takes place when the brane black hole is coupled to an appropriate thermal bath modeled by thermalized $\mathcal{O}(N^{2})$ degrees of freedom of the CFT. From the bulk perspective, this means the horizon of the black hole on the brane is to be connected to another horizon in the bulk. The bulk system will have a Gregory-Laflamme-like instability such that the black hole attached to the brane slides off the brane into the bulk, forming a (typically) larger black hole horizon. Thus, while the horizon on the brane reduces its size (evaporation), the horizon in the bulk does not shrink, consistent with the classical evolution of black hole horizons.

There are in fact several examples of evaporating braneworld black holes (see \cite{Emparan:2023dxm} for details). In all cases, the bulk system considers $D$-dimensional black holes in the large-$D$ limit \cite{Emparan:2013moa,Bhattacharyya:2015dva,Emparan:2020inr} that localize on either one or a pair of Karch-Randall branes.\footnote{The appeal of the large-$D$ limit is that the effective dynamics of the (bulk) black hole horizon is encoded in a set of two partial differential equations which can be solved numerically, quickly. A large-$D$ expansion of the Israel-junction conditions is then used to determine the location of the brane intersecting the bulk black hole, and amounts to imposing a simple pair of Neumann boundary conditions. Note, however, a limitation of the large-$D$ effective theory is that the induced brane theory is not well characterized by a $(D-1)$-dimensional gravity. This is because with these limits, the induced theory has large corrections on the brane from the higher-derivative terms such that the brane gravity behaves more like a $D$-dimensional theory. A possible interpretation is that the induced theory describes a semi-classical theory where backreaction effects are large.} In particular: (i) a small AdS black string stretched between two branes with black holes localizing on both branes. The string instability triggers either the evaporation of both black holes into a single bulk black hole, or the evaporation of one brane black hole into the other brane black hole; (ii) A small AdS black hole on the brane connected to a larger bulk black hole (bath) such that the brane black hole evaporates entirely into the bulk, and (iii) A large unstable AdS black droplet connected to a non-gravitating boundary (bath) via a thin black funnel, such that the droplet evaporates into the boundary. 

In short, the exhaustive large-$D$ analysis reveals holographic CFTs coupled to black holes have distinctive traits over their weakly interacting counterparts. According to \cite{Emparan:2023dxm,Emparan:2021ewh}, their semi-classical dynamics are always dual to classical Gregory-Laflamme-like dynamics of bulk black holes.

%%%%%%%%%%%%%%%%%%%%%%%%%%%%%%%%%%%%%%%%%%%%%%%%%%%%%%%%%%%

\section{Outlook and applications} \label{sec:applications}

We have seen how holographic braneworlds provide a means to exactly study the problem of semi-classical backreaction. In particular, the precise construction of quantum-corrected black holes and an exploration of their horizon thermodynamics. But this is only a small cross-section of the utility of braneworld holography. Phenomenological considerations aside -- a historical motivation to study braneworld physics -- holographic braneworlds have a number of applications, particularly in the interdisciplinary field of holographic information theory. Below we very briefly give a non-exhaustive and biased list of applications of quantum black holes, and, more broadly, holographic braneworlds.

\section*{Applications}

\subsection*{Holographic entanglement and gravitational entropy}

For a quantum mechanical system in state $\rho$, the von Neumann entropy is defined as
\be
S_{\text{vN}}\equiv-\text{tr}\,[\rho \log \rho]\,.
\label{eq:vNent}\ee
If the system is in a pure quantum state, the entropy (\ref{eq:vNent}) is identically zero. When the system is comprised of smaller subsystems,  $S_{\text{vN}}$ quantifies how entangled the subsystems are. 
In quantum field theory, subsystems are often chosen to be spatial regions $\Sigma$ of the entire quantum system. Intriguingly, for ground states of local Hamiltonians, the entanglement entropy, while divergent in the ultraviolet, generally adheres to an area law (in $d+1$-spacetime dimensions)
\beq S_{\text{vN}}=c_{0}\frac{\text{Area}[\partial\Sigma]}{\epsilon^{d-1}}+...\;,\label{eq:entarea}\eeq
for some constant $c_{0}$, UV regulator $\epsilon$, and the ellipsis refers to subleading UV divergences whose precise terms are state-dependent.
Thus, $S_{\text{vN}}$ scales with the boundary area of the subsystem rather than its volume, in contrast to the volume law typically observed in thermal states.  
The entropy-area relation (\ref{eq:entarea}) is clearly reminiscent of the Bekenstein-Hawking entropy formula for black holes, suggesting black hole entropy arises from vacuum entanglement due to quantum fields across the horizon, tracing out the interior degrees of freedom \cite{Sorkin:1984kjy,Bombelli:1986rw,Srednicki:1993im,Frolov:1993ym,Callan:1994py,Solodukhin:2011gn}. In particular, in scenarios of gravity induced due to matter loops \cite{Sakharov:1967pk,Visser:2002ew}, the Bekenstein-Hawking entropy is solely due to entanglement due to vacuum fluctuations \cite{Jacobson:1994iw,Frolov:1996aj,Frolov:1997up}, where the UV-cutoff dependence can be absorbed in a renormalization of Newton's constant \cite{Susskind:1994sm}.

Gravitational entropy, therefore, is imbued with an information-theoretic character. This is crystallized in AdS/CFT, where
%Holographically, 
the entanglement entropy of a  holographic CFT can be computed using the Ryu-Takayanagi (RT) formula \cite{Ryu:2006bv,Ryu:2006ef}
\be\label{RTformula}
S^{\text{vN}}_A=\frac{\text{Area}(\gamma)}{4G}\,.
\ee
Here, $\gamma$ denotes a codimension-2 minimal surface in the bulk asymptotically AdS space, anchored to region $A$ on the AdS boundary, such that $\partial\gamma=\partial A$ and $\gamma$ is homologous to $A$. The RT prescription generalizes the Bekenstein-Hawking entropy formula for black holes: a black hole horizon is an example of a minimal surface, and the bulk AdS spacetime need not contain a black hole.\footnote{The RT formula can be directly derived from the gravitational R\'enyi entropy \cite{Lewkowycz:2013nqa} (see also \cite{Takayanagi:2019tvn,Botta-Cantcheff:2020ywu,Kastikainen:2023yyk,Kastikainen:2023omj}).} As such, it has led to numerous insights into gravitational physics, notably in understanding the emergence of gravity from quantum entanglement \cite{Lashkari:2013koa,Faulkner:2013ica,Haehl:2017sot,Swingle:2014uza,Agon:2020mvu,Agon:2021tia}\footnote{See also \cite{Cooper:2019rwk} for a similar treatment in the context of braneworlds.} following analogous insights from coarse-grained thermodynamics \cite{Jacobson:1995ab,Parikh:2009qs,Guedens:2011dy,Parikh:2017aas,Parikh:2018anm,Svesko:2018qim}.

When the bulk theory includes higher curvature and semi-classical quantum corrections, the RT formula (\ref{RTformula}) generalizes to the following extremization prescription
\cite{Engelhardt:2014gca}
\be\label{QES}
S^{\text{vN}}_A=\underset{\gamma}{\text{min}}\,\underset{\gamma}{\text{ext}}\left[\frac{\text{Area}(\gamma)}{4G}+\mathcal{S}_{\text{DC}}(\gamma)+S^{\text{bulk}}_A(\Sigma_\gamma)\right]\,.
\ee
Here $\mathcal{S}_{\text{DC}}$ denote the Dong-Camps anomaly terms \cite{Dong:2013qoa,Camps:2013zua}, analogous to the Wald corrections to the Bekenstein-Hawking entropy, and $S^{\text{bulk}}_A$ is the entanglement entropy of bulk fields across the entanglement wedge of $A$, $\Sigma_\gamma$, such that the quantity in brackets is recognized to be the generalized entropy $S_{\text{gen}}(\gamma)$.  Further, in the event there are multiple extremal surfaces, the prescription says the von Neumann entropy $S_{A}^{\text{vN}}$ is given by choosing the extremal surface which minimizes the generalized entropy $S_{\text{gen}}(\gamma)$. Notably, the von Neumann entropy can undergo phase transitions as extremal surface configurations exchange dominance.

Holographic braneworlds provide additional evidence that gravitational entropy can be interpreted as entanglement entropy. Indeed, the gravity on the brane is induced from the cut-off CFT.\footnote{Explicit realizations of the equivalence between gravitational and entanglement entropies in two-dimensional braneworld models pre-date the Ryu-Takayanagi prescription \cite{Hawking:2000da,Fursaev:2000ym}.} In combination with the RT prescription, it readily follows that the area-entropy of black holes localized on an ETW brane is exactly equal to entanglement entropy \cite{Emparan:2006ni} -- the minimal surface coincides with the bulk black hole horizon intersecting the brane. Additionally, the braneworld set-up naturally resolves various subtleties in identifying Bekenstein-Hawking and entanglement entropies  \cite{Jacobson:1999mi}. In particular, the UV cutoff $\epsilon$ for the CFT -- equal to the bulk AdS length scale $L_{d+1}$ --  fixes the number of species of the dual CFT such that Newton's constant is correctly reproduced (cf. Eq. (\ref{eq:S4Sgenqbtz})), thus resolving the `species problem'. Heuristically, consider bulk AdS$_{4}$ with a black hole. The contribution of a single field to the CFT$_{3}$ entanglement entropy is of the order $S_{A}^{\text{vN}}\sim\frac{c_{0}}{L_{4}}$. Then, for a large number of fields $c_{3}=L_{4}^{2}/G_{4}\sim L_{4}/G_{3}$, the total entanglement entropy is $S_{A}^{\text{vN}}=c_{3}c_{0}/L_{4}\sim c_{0}/G_{3}$.

\subsection*{The entropy of Hawking radiation}

 Hawking's discovery that black holes emit thermal radiation leads to the information puzzle~\cite{Hawking:1976ra}: do black holes evolve unitarily or not? According to Hawking's original calculation, the fine-grained von Neumann entropy of radiation $S_{\text{vN}}^{\text{rad}}$ grows in time indefinitely, paradoxically surpassing the coarse-grained entropy of the black hole. Alternatively, if the black hole plus radiation obeys standard quantum principles, the radiation entropy instead follows a unitary Page curve \cite{Page:1993wv}, never exceeding the black hole entropy. Previously thought to be a problem only quantum gravity would solve, the paradox can be addressed in semi-classical gravity, for which the extremization prescription (\ref{QES}) plays a prominent role \cite{Penington:2019npb,Almheiri:2019psf,Almheiri:2019qdq}. In particular, a variant of (\ref{QES})  known as the `island formula',
 \beq S_{\text{vN}}(\Sigma_X)=\underset{X}{\text{min}}\,\underset{X}{\text{ext}}\left[\frac{\text{Area}(X)}{4G}+\mathcal{S}_{\text{DC}}(X)+S^{\text{sc}}_{\text{vN}}(\Sigma_{X})\right] \label{eq:QESformula}\eeq
 can be used to explicitly compute unitary Page curves in evaporating or eternal black hole backgrounds. Here, $S_{\text{vN}}(\Sigma_X)$ is the fine-grained entropy in the full quantum theory, $S^{\text{sc}}_{\text{vN}}$ is the von Neumann entropy of bulk quantum fields in the semi-classical approximation, and $\Sigma_{X}$ is a codimension-1 slice bounded by a codimension-2 quantum extremal surface (QES) $X$ and a cutoff surface. Generally, $\Sigma_{X}$ is disconnected, $\Sigma_{X}=\Sigma_{R}\cup I$,
where $\Sigma_{R}$ is the region outside the black hole collecting radiation and $I$ is an `island' (with $X=\partial I$) lying primarily inside the black hole.

Originally achieved for models of two-dimensional dilaton gravity,\footnote{The island rule has been derived in two-dimensional Jackiw-Teitelboim gravity  using the Euclidean gravitational path integral to compute the R\'enyi entropy \cite{Almheiri:2019qdq,Penington:2019kki,Goto:2020wnk}, or the microcanonical action \cite{Pedraza:2021ssc,Svesko:2022txo}.}
holographic braneworlds confirm unitary Page curves arise in higher-dimensional gravity \cite{Almheiri:2019hni,Chen:2020hmv}. This is accomplished using `double holography' (recall the description before section \ref{ssec:holoqbhsconj}). For simplicity, consider the boundary perspective, i.e., the vacuum state of the $d$-dimensional boundary CFT on $\mathbb{R}\times S^{d-1}$ coupled to a $(d-1)$-dimensional conformal defect along the equator of the sphere $S^{d-1}$. The entanglement entropy of the holographic boundary CFT can be computed using the RT prescription. Taking the bulk to be  AdS$_{d+1}$ with an AdS$_{d}$ Dvali-Gabadadze-Porrati (DGP) brane \cite{Dvali:2000hr} (where the brane action includes, e.g., its own Einstein-Hilbert term), the entropy of the boundary CFT vacuum reduced to a boundary region $A$ is given by \cite{Chen:2020uac}
 \beq S_{\text{vN}}(A)=\underset{\gamma}{\text{min}}\,\underset{\gamma}{\text{ext}}\left[\frac{\text{Area}(\gamma)}{4G_{d+1}}+\frac{\text{Area}(\gamma\cap\text{brane})}{4G_{d}}\right] \;,\label{eq:branert}\eeq
 for bulk extremal surface $\gamma$ homologous to $A$ and $\partial A=\partial \gamma$. The first term on the right-hand side is the familiar RT formula while the second term is the gravitational contribution that arises when the bulk RT surface intersects the DGP brane. There are topologically distinct configurations for the bulk extremal surface $\gamma$: surfaces $\gamma$ which do not intersect the brane and (ii) surfaces that intersect the brane; only with (ii) does the second term in the brackets contribute. As both are candidate extremal surfaces, the entropy $S_{\text{vN}}(A)$ is computed using the surface which gives the smallest value on the right-hand side (\ref{eq:branert}). Transitions between `disconnected' and `connected' phases will occur depending on the size of $\gamma$ and the brane tension and gravitational coupling. 

 The holographic entropy formula (\ref{eq:branert}) can be understood as a relation between the boundary and bulk perspectives. It is through the brane perspective that the right-hand side can be reinterpreted as the QES or island formula.\footnote{The homology constraint for the RT surfaces depends on which perspective in double holography is being employed \cite{Neuenfeld:2021bsb}.} Qualitatively, the disconnected phase corresponds to the situation where, according to beings confined to the brane, no quantum extremal islands are formed, while in the connected phase extremal islands appear.\footnote{The argument of \cite{Almheiri:2019hni,Chen:2020uac} requires DGP couplings be turned on, couplings which can affect the growth rate of entanglement entropy of subregions of the dual CFT \cite{Lee:2022efh}. Alternatively, the intersection term naturally arises using deformed braneworlds \cite{Neuenfeld:2024gta}.} By applying this doubly-holographic reasoning to higher-dimensional topological black holes \cite{Chen:2020hmv}, black strings \cite{Geng:2021mic,Karch:2023ekf}, and de Sitter space \cite{Geng:2021wcq,Aguilar-Gutierrez:2023tic}, the prescription (\ref{eq:branert}) yields unitary Page curves, except in cases where the horizon is extremal with a vanishing temperature.
 
 Notably, thus far, all $3+1$-dimensional braneworld models that have been explored and exhibit island formation are described by \emph{massive} gravity theories; islands disappear in the limit the graviton on the brane becomes massless \cite{Geng:2020qvw} (see, e.g., \cite{Geng:2023qwm} for a review). Further, long-range gravity theories have that the energy of an excitation localized to the island can be detected outside
the island, inconsistent with the principle that operators in an entanglement wedge commute with operators in its complement \cite{Geng:2021hlu}. Combined, this suggests the phenomenon of island formation is a feature of massive gravity, and is even inconsistent in massless theories.

\subsection*{Holographic complexity}

Quantum complexity is another useful diagnostic in information theory. Loosely speaking, complexity measures the difficulty of constructing an arbitrary quantum state from a reference state using a specific set of resources. This concept has proven highly valuable in computer science and quantum computation, where the resources are given by some elementary operations (`gates') and the mapping between reference and target states defines a quantum circuit.
Complexity has also been fruitful in gravitational physics, particularly as a tool to quantify the information processing of black holes.  This application touches upon the foundational aspects of gravity and its interplay with quantum mechanics, with some evidence suggesting that gravity itself may emerge from the minimization of complexity ---that is, from efficient quantum computation \cite{Czech:2017ryf,Caputa:2018kdj,Susskind:2019ddc,Pedraza:2021mkh,Pedraza:2021fgp,Pedraza:2022dqi,Carrasco:2023fcj}. 

Historically, computational complexity was proposed as a new entry in the holographic dictionary to address the puzzle of why black hole interiors continue to grow after scrambling \cite{Susskind:2014moa}. Various conjectures for its precise dual have been proposed, with the initial contenders being `complexity=volume' (CV) \cite{Stanford:2014jda,Susskind:2014rva} and `complexity=action' (CA) \cite{Brown:2015lvg,Brown:2015bva}. CV posits that the complexity of a CFT state $|\Psi\rangle$ defined on a Cauchy slice $\sigma$ is given by the maximal volume of a codimension-one bulk surface $\Sigma$ anchored on the boundary slice $\sigma$,
	\be
	\mathcal{C}_{V}\left(|\Psi\rangle\right)= \dfrac{\text{Vol}(\Sigma)}{G L}\,.\hfill\label{eq:CV}
	\ee 
On the other hand, CA asserts that complexity is defined by the on-shell gravitational action $I$ evaluated within a codimension-zero bulk region known as the Wheeler-deWitt (WdW) patch $\mathcal{W}$, anchored at the given boundary slice $\sigma$,
	\be
	\mathcal{C}_A\left(|\Psi\rangle\right)= \dfrac{I(\mathcal{W})}{\pi} \,.\hfill \label{eq:CA}
	\ee
Conceptually, CV is arguably more intuitive than CA, resembling the computational complexity of a tensor network circuit that captures the entanglement structure of the CFT state \cite{Swingle:2009bg,Swingle:2012wq,Bao:2018pvs,Jahn:2021uqr}.   For high-temperature thermofield double states, however, CV and CA complexities have the same late time behavior
\be\label{latetimeComp}
\left.\dfrac{d\,\mathcal{C}_{V,A}}{dt}\,\right |_{t \gg \beta} \;\sim \, TS\,,
\ee
consistent with the expectation of complexity in such states.

More generally, an infinite number of gravitational duals to holographic complexity are technically possible \cite{Belin:2021bga,Belin:2022xmt}.
%proposed an infinite number of generalizations, 
Dubbed `complexity=anything', these generalizations are designed to encapsulate key aspects of complexity, including the anticipated linear growth following scrambling and the observed `switchback effect' ---a delay in the onset of linear growth caused by perturbations. While rigorous proof connecting any of these proposals to a concrete field theory definition of complexity is currently lacking, the notion is that they may be associated with ambiguities inherent in its definition. These ambiguities include the arbitrary choice of elementary gates, cost factors, and the reference state, among others.

Aspects of holographic complexity proposals have been explored using braneworlds \cite{Hernandez:2020nem,Emparan:2021hyr,Chen:2023tpi,Aguilar-Gutierrez:2023tic,Aguilar-Gutierrez:2023ccv}, including the influence of backreaction effects using the static quantum BTZ black hole as a guide \cite{Emparan:2021hyr}.
%Specifically, let us focus here on the key findings of \cite{Emparan:2021hyr}, which investigated complexity for the the static qBTZ black hole. 
In summary,
\begin{itemize}
		\item{In the braneworld effective theory, CV admits a semiclassical expansion of the form
  \be\label{genCV}
	\mathcal{C}_V\left(|\Psi\rangle\right)= \dfrac{\text{Vol}(\Sigma)}{GL} +\dfrac{\delta \text{Vol}(\Sigma)}{GL}+ \mathcal{V}(\Sigma)+\mathcal{C}_V^{\text{bulk}}\left(|\phi\rangle\right)+\dots\,,
	\ee
 with the leading term being the complexity of the classical black hole, $\delta \text{Vol}(\Sigma)$ is the change in the volume due to the quantum backreaction, $\mathcal{V}(\Sigma)$ are higher curvature terms akin to the Wald corrections to the Bekentein-Hawking entropy, and $\mathcal{C}_V^{\text{bulk}}\left(|\phi\rangle\right)$ is the complexity of the bulk state $|\phi\rangle$ defined on $\Sigma$. Note that the structure of (\ref{genCV}) resembles an expansion of the QES prescription (\ref{QES}). Furthermore, (\ref{genCV}) does satisfy the late time growth (\ref{latetimeComp}), upon replacing $S$ with the generalized entropy $S_{\text{gen}}$.}
		
	\item{Conversely, CA does \textit{not} simplify to the classical three-dimensional CA proposal plus corrections. The action involves cancellations among the bulk, boundary, and joint terms, making the late-time growth highly sensitive to quantum effects. As a result, the late time behavior \eqref{latetimeComp} is not reproduced. This discrepancy arises because the WdW patch extends to the singularity, whose structure is significantly altered by quantum backreaction, leading to substantial quantum contributions to CA.}
\end{itemize}
Analogous features have been observed for the rotating quantum BTZ \cite{Chen:2023tpi}. Based on these preliminary studies, quantum black holes advocate the need for a semi-classical generalization of holographic complexity.\footnote{Semi-classical extensions of CV complexity have been proposed and explored with \cite{Hernandez:2020nem,RafaInPrep} and without \cite{Carrasco:2023fcj} holographic braneworlds.}

\subsection*{Singularity probes and quantum cosmic censorship}

Black hole singularities reflect the fact that classical gravity is incomplete. Further, the Hawking-Penrose singularity theorems imply their inevitability. Due to their unphysical consequences, Penrose conjectured there must exist a type of cosmic censorship \cite{Penrose:1969pc}, known as weak cosmic censorship, where a horizon must shroud the singularity from null infinity. Singularities moreover, mark a lack of predictive power in an otherwise classically deterministic theory. The strong cosmic censorship conjecture \cite{1974IAUS...64...82P,Penrose:1980ge}, which is independent of weak cosmic censorship, asserts classical general relativity should remain deterministic. In technical terms, for generic smooth initial data, the Cauchy evolution of a Cauchy hypersurface is inextendible beyond the Cauchy horizon. For example,  charged and rotating black holes, whose inner horizons are Cauchy horizons, the past timelike singularity heralds a lack of predictivity beyond the inner horizon. Conceivably, however, the infinitely blue-shifted energy of a particle entering from the exterior of the black hole should result in a large backreaction so as to turn the smooth (regular) inner Cauchy horizon into a non-smooth barrier \cite{Simpson:1973ua,Poisson:1990eh}. 

It is natural to wonder whether either form of cosmic censorship holds when semi-classical backreaction effects are accounted for. Thus far, there is agreement that the expectation value of the quantum stress tensor of a free scalar field diverges at the inner horizon of a RN or Kerr black hole (in $D=d+1=4$) on approach from the exterior \cite{Lanir:2018vgb,Zilberman:2019buh,Dias:2019ery,Hollands:2019whz}, suggesting a type of strong cosmic censorship. A notable exception, however, is the classical rotating BTZ black hole, where the backreaction is mild enough such that the inner horizon does not become singular, indicating a violation in strong cosmic censorship \cite{Dias:2019ery,Hollands:2019whz}.\footnote{Alternatively, leading-order perturbative backreaction was found to yield singular inner horizons in \cite{Casals:2016ioo,Casals:2019jfo}.} A limitation of these works is that the analysis was only accomplished at leading order in backreaction. The rotating quantum BTZ black hole, on the other hand, which exactly accounts for all orders of backreaction, has essentially the structure of an AdS$_{4}$-Kerr black hole. Consequently, as argued in \cite{Emparan:2020rnp}, and verified in \cite{Kolanowski:2023hvh} using the techniques of \cite{Hollands:2019whz}, rotating quantum BTZ has a singular inner horizon. Thus, for the BTZ black hole, strong cosmic censorship still holds once backreaction effects beyond leading order are accounted for.  

There is an intuitive sense that quantum effects induce a weak form of cosmic censorship. Indeed, backreaction of the Casimir stress tensor modifies the conical (A)dS$_{3}$ and Mink$_{3}$ defect geometries such that the naked singularities become hidden behind a horizon. A standard test of (classical) weak cosmic censorship is to try to overspin or overcharge near-extremal black holes such that they shed their horizons. So far, Kerr-Newman black holes do not \cite{Wald:1974hkz,Sorce:2017dst}. Likewise, the rotating quantum BTZ black hole cannot be overspun \cite{Frassino:2024fin}. 

A necessary condition of weak cosmic censorship is the conjectured classical Penrose inequality \cite{Penrose:1973um}. Loosely speaking, assuming there are no naked singularities and collapsing matter settles to a Kerr black hole, then the total mass $M$ for a $D\geq4$-dimensional asymptotically flat \cite{Huisken01,Bray01,Bray:2007opu,Mars:2009cj} or AdS spacetime \cite{Itkin:2011ph,Folkestad:2022dse} with a marginally trapped surface $\sigma$ is bounded below by the area $A[\sigma]$ 
\beq \frac{16\pi G_{D}M}{(D-2)\Omega_{D-2}}\geq \left(\frac{A[\sigma]}{\Omega_{D-2}}\right)^{\hspace{-1mm}\frac{D-3}{D-2}}+\ell^{-2}_{D}\left(\frac{A[\sigma]}{\Omega_{D-2}}\right)^{\hspace{-1mm}\frac{D-1}{D-2}}.
\label{eq:classAdSPI}\eeq
Here $G_{D}$ and $\ell_{D}$ denote the $D$-dimensional Newton's constant and curvature scale, respectively, and
%$\Omega_{k}$ is the volume of a $(D-2)$-dimensional unit sphere ($k=1$), plane ($k=0$), or hyperbolic space ($k=-1$); we will be interested in the spherical case where, 
$\Omega_{n}\equiv 2\pi^{(n+1)/2}/\Gamma[(n+1)/2]$ is the volume of a unit $n$-sphere. In the limit $\ell_{D}\to\infty$ the Penrose inequality for asymptotically flat space is recovered. While the Penrose inequality has not been proven in general, any spacetime violating (\ref{eq:classAdSPI}) is believed to be a counterexample to weak cosmic censorship. Notably, the classical Penrose inequality can be violated at leading order in perturbative backreaction \cite{Bousso:2019var,Bousso:2019bkg}. This motivates  the need for a semi-classical generalization of (\ref{eq:classAdSPI}), i.e., a quantum Penrose inequality \cite{Bousso:2019var,Bousso:2019bkg}
\beq
\begin{split} 
\hspace{-4mm}\frac{16\pi \mathcal{G}_{D}M}{(D-2)\Omega_{D-2}}&\geq \left(\frac{4\mathcal{G}_{D}S_{\text{gen}}}{\Omega_{D-2}}\right)^{\hspace{-1mm}\frac{D-3}{D-2}}\hspace{-2mm}+\ell^{-2}_{D}\left(\frac{4\mathcal{G}_{D}S_{\text{gen}}}{\Omega_{D-2}}\right)^{\hspace{-1mm}\frac{D-1}{D-2}}
\end{split}
\label{eq:qAdSPI}\eeq
where the area $A[\sigma]$ has been replaced for the generalized entropy $S_{\text{gen}}$ associated with a (quantum) marginally trapped surface. To test whether the quantum inequality (\ref{eq:qAdSPI}) holds even for large backreaction effects, a three-dimensional inequality was proposed in \cite{Frassino:2024bjg}
\beq 8\pi\mathcal{G}_{3}(M-M_{\text{cas}})\geq\ell_{3}^{-2}\left(\frac{4\mathcal{G}_{3}S_{\text{gen}}}{2\pi}\right)^{2}\;,\label{eq:AdS3qPeninq}\eeq
where $M_{\text{cas}}=-1/8\mathcal{G}_{3}$ is the Casimir energy of backreacting quantum fields. All static and rotating quantum BTZ black holes were found to satisfy (\ref{eq:AdS3qPeninq}) at all orders of backreaction. This implies the existence of weak quantum cosmic censorship\footnote{A notion of weak quantum cosmic censorship may directly follow from `cryptographic censorship' \cite{Engelhardt:2024hpe}, a theorem that states when the time evolution operator of a holographic CFT is approximately pseudorandom on a code subspace, there must be an event horizon in the corresponding bulk dual. Incidentally, certain types of singularities are compatible with approximately pseudorandom time evolution, and thus, by cryptographic censorship, are hidden behind event horizons.} in non-perturbative semi-classical  gravity, for which the quantum Penrose inequality would be a logical input.

\subsection*{Imprints of quantum backreaction beyond thermal equilibrium}

Realistic, astrophysical black holes are generally thought to be dynamical. Such non-stationary black holes lack an equilibrium thermodynamic description, and their horizons change shape and oscillate. One way to explore black hole dynamics is to consider (small) time-dependent perturbations to static or stationary black holes. Such perturbations display characteristic patterns of damped oscillations, dubbed quasi-normal modes (QNMs), that capture deviations away from equilibrium \cite{Berti:2009kk,Konoplya:2011qq}. Further, certain quasi-normal resonances serve as black hole signatures in gravitational wave astronomy. 

Technically, QNMs are derived by examining the fluctuation equations of gauge-invariant perturbations. The simplest example involves the study of probe fields. For example, for a probe scalar field $\phi$, the fluctuation equation is given by the Klein-Gordon equation, 
\beq (\Box+m^2)\phi=0\;,\eeq
where $\phi$ is subsequently decomposed intro Fourier and harmonic modes,  $\phi=e^{i\omega t}Y^m_l(\Omega)\Phi(r)$. The characteristic quasi-normal mode frequencies $\omega$ that solve the fluctuation equation are generally complex, $\omega=\omega_R-i\omega_I$, with a negative imaginary part indicating the decay of perturbations. For holographic AdS black holes, quasi-normal modes appear as poles in relevant retarded two-point correlators of the dual CFT, providing a field-theoretic characterization of black holes out of thermal equilibrium \cite{Horowitz:1999jd}.

It is natural to wonder how backreaction modifies characteristic traits of black hole dynamics. A detailed study of QNMs of the quantum BTZ black hole provides some preliminary insights \cite{QNMstoappear}. Considering probe fields of brane localized matter, dual to CFT operators with conformal dimension $\Delta\in[1,2]$ and spin $s=0,\pm 1/2$, it was found that the QNMs and their overtones exhibit qualitatively different behavior depending on the branch of the qBTZ solution selected. This distinction can be used to differentiate between the types of singularities cloaked by a horizon: dressed conical singularities versus genuine quantum-corrected black holes. Furthermore, the magnitude $\omega_{I}$ of the imaginary part of the leading mode generally \emph{increases} with the strength of backreaction. This implies the thermalization time, $t_{\text{th}}\sim1/\omega_I$, defined by the late-time decay of two-point correlators $\langle\mathcal{O}(0)\mathcal{O}(t)\rangle\sim e^{t/t_{\text{th}}}$, accelerates due to semi-classical effects. This phenomenon can be explained from the perspective of the dual CFT: quantum matter in the bulk with a large central charge corresponds to coupling a large number of light operators in the boundary CFT, thereby increasing the number of channels through which a small perturbation can dissipate.

The pole structure of retarded two-point CFT correlators dual to quantum BTZ QNMs, moreover, exhibits a markedly different behavior than its classical counterpart \cite{QNMstoappear}. This includes, for example, `pole-skipping', i.e., points in momentum space where the retarded Green's function is not unique. Specifically, for the quantum BTZ black hole, the momentum of the pole-skipping points exhibits a non-trivial dependence on the parameter controlling the strength of backreaction. This implies, given the connection between hydrodynamics and chaos in holographic CFTs \cite{Grozdanov:2017ajz,Blake:2018leo}, the chaotic dynamics of black holes are dramatically altered when backreaction is accounted for.\footnote{The connection between hydrodynamics and chaos in 2+1 dimensional Einstein gravity is subtle, because metric fluctuations are pure gauge. This issue is alleviated in braneworld models as they typically contain a massive graviton. Furthermore, a general hydrodynamic framework for chaotic dynamics in 1+1 CFTs has been established \cite{Haehl:2018izb}, linking it to the field theory of soft modes associated with holomorphic and antiholomorphic parameterizations. For further discussion in the context of classical BTZ black holes, see \cite{Cartwright:2024rus}.} Further,  by studying pole collisions in complex momentum space, quantum corrections have a significant impact on the analytic structure of the poles of retarded Green's functions. In particular, the quantum corrections intertwine the tower of modes in a series of level-crossing events noticeably distinct from the level-touching events observed in the classical BTZ geometry \cite{Grozdanov:2019uhi}.

\vspace{3mm} 

\noindent\section*{Acknowledgements}

This review is the culmination of numerous conversations with a plethora of colleagues and collaborators, particularly: Jos\'e Barb\'on, Elena Cáceres, Rafael Carrasco, Casey Cartwright, Ana Climent, Roberto Emparan, Antonia Micol Frassino, Jaume Garriga, Ruth Gregory, Umut G\"ursoy, Robie Hennigar, Seyed Hosseini Mansoori, Clifford Johnson, Eleni Kontou, David Kubiz{\v n}{\' a}k,  Quim Llorens, José Navarro-Salas, Dominik Neuenfeld, Ayan Kumar Patra, Guim Planella Planas, Morteza Rafiee, Watse Sybesma, Marija Toma\v{s}evi\'{c} and Manus Visser. We especially thank Roberto Emparan for encouraging us to write this review. EP is supported by the Cosmoparticle Initiative at UCL. JFP is supported by the `Atracci\'on de Talento' program grant 2020-T1/TIC-20495 and by the Spanish Research Agency through the grants CEX2020-001007-S and PID2021-123017NB-I00, funded by MCIN/AEI/10.13039/501100011033 and by ERDF A way of making Europe. AS is supported by STFC consolidated grant ST/X000753/1.

\appendix 

%%%%%%%%%%%%%%%%%%%%%%%%%%%%%%%%%%%%%%%%%%%%%%%%%%%%%%%%%%%%%%%%%%%%%%%%%%%%%%

\section{Conventions} \label{app:conventions}

Here we summarize our geomemtric conventions and describe the variational principle with Neumann boundary conditions, leading to the brane equations of motion. 

\subsection*{Background geometry}

Let $\mathcal{M}$ be a $d+1$ dimensional spacetime endowed with metric $\hat{g}_{ab}$ with coordinates $x^{a}$ on $\mathcal{M}$. We take a `mostly plus' convention for Lorentzian signature. The Riemann curvature tensor with respect to $\hat{g}_{ab}$ is
\beq R^{c}_{\;dab}=\partial_{a}\Gamma^{c}_{\;bd}-\partial_{b}\Gamma^{c}_{\;ad}+\Gamma^{c}_{\;af}\Gamma^{f}_{\;bd}-\Gamma^{c}_{\;bf}\Gamma^{f}_{\;ad}\;.\label{eq:Riemanngencon}\eeq
%Equivalently, in terms of the Levi-Civita connection $\nabla_{\mu}$, $R^{\rho}_{\;\sigma\mu\nu}=-[\nabla_{\mu},\nabla_{\nu}]V_{\nu}$ and $R^{\rho}_{\;\sigma\mu\nu}V^{\sigma}=[\nabla_{\mu},\nabla_{\nu}]V^{\rho}$ for vectors $V$. Our convention is such that in Euclidean space the intrinsic Ricci scalar of the 2-sphere is positive. 

\subsection*{Hypersurface geometry} 

 Let $\Sigma$ denote a timelike or spacelike $d$-dimensional hypersurface embedded in $\mathcal{M}$. The hypersurface is defined by  a restriction on coordinates $x^{a}$, i.e., introduce a scalar function $\Phi(x^{a})$ which obeys the constraint $\Phi(x^{a})=0$.
 %\footnote{When $\Phi(x^{a})$ is spacelike, $\Phi$ is taken to increase toward the future across the hypersurface, while when $\Sigma$ is timelike $\Phi$ increases outward.}
 The  unit normal  $n_{a}$ to this hypersurface  is $n_{a}=\epsilon \mathcal{N}\partial_{a}\Phi$,
with $\epsilon\equiv n^{2}=n_{a}n^{a}=\pm1$, where $\epsilon=+1$ indicates the hypersurface is timelike, $\epsilon=-1$ has $\Sigma$ spacelike, and $\mathcal{N}$ is a normalization, $\mathcal{N}=|\hat{g}^{ab}\partial_{a}\Phi\partial_{b}\Phi|^{-1/2}$. 
%Further, we take $n^{\alpha}$ to be outwards pointing.
%\footnote{That is, $n^{\alpha}$ points in the direction of increasing $\Phi$, i.e., $n^{\alpha}\partial_{\alpha}\Phi>0$. \DN{Is this just rephrasing the previous footnote?}}
Denote the induced metric $h_{ij}$ on $\Sigma$ and its inverse as $h^{ij}$, which are defined by
\beq h_{ij}\equiv \hat{g}_{ab}e^{a}_{i}e^{b}_{j}\;,\quad  e^{i}_{a}\equiv h^{ij}G_{ab}e^{b}_{j}\;,\eeq 
for vectors  $e^{a}_{i}\equiv\frac{dx^{a}}{d y^{i}}$ tangent to curves contained in $\Sigma$ and coordinates $y^{i}$ intrinsic to $\Sigma$.
%which are tangent to curves contained in $\Sigma$.
By definition, $n_{a}e^{a}_{i}=0$. In terms of the background metric $\hat{g}_{ab}$,
\beq \hat{g}_{ab}=\epsilon n_{a}n_{b}+h_{ij}e^{i}_{a}e^{j}_{b}=\epsilon n_{a}n_{b}+h_{ab}\;,\quad h_{ab}\equiv h_{ij}e^{i}_{a}e^{j}_{b}\;,\label{eq:completeness2}\eeq
where $h_{ab}$ is the projector onto hypersurfaces orthogonal to $n_{a}$. Similarly, $h^{ab}=\hat{g}^{ab}-\epsilon n^{a}n^{b}$.
%Similarly, the inverse metric is 
%\beq g^{\alpha\beta}=\epsilon n^{\alpha}n^{\beta}+h^{ab}e^{\alpha}_{a}e^{\beta}_{b}=\epsilon n^{\alpha}n^{\beta}+h^{\alpha\beta}\;.\label{eq:metdecomp1}\eeq
Note $h^{ab}n_{a}=h^{ab}n_{b}=0$.

Define the extrinsic curvature $K_{ij}$ as
\beq K_{ij}\equiv (\nabla_{b}n_{a})e^{a}_{i}e^{b}_{j}\;.\eeq
The trace of the extrinsic curvature is 
\beq K=h^{ij}K_{ij}=\nabla_{a}n^{a}\;.\eeq
 Equivalently, the extrinsic curvature with respect to bulk coordinates is 
\beq K_{ab}=e^{i}_{a}e^{j}_{b}K_{ij}=-h_{a}^{c}h_{b}^{d}\nabla_{c}n_{d}\;.\eeq
Using the decomposition  (\ref{eq:completeness2}), it follows 
\beq K_{ab}=\nabla_{a}n_{b}-\epsilon n_{a}a_{b}\;,\label{eq:Kmunuacc}\eeq
where $a_{b}$ is the acceleration for the integral curves of the unit normal $n_{b}$, $a_{b}=n^{a}\nabla_{a}n_{b}$.
Since $n^{b}a_{b}=0$, then $K_{ab}n^{b}=n^{b}\nabla_{a}n_{b}=0$. Further $a_{b}\nabla_{a}n^{b}=-n^{b}\nabla_{a}a_{b}$, from which it follows $K=h^{ab}K_{ab}=\hat{g}^{ab}K_{ab}$.

\subsection*{Neumann boundary conditions and the brane equations of motion}

With these conventions, the Einstein-Hilbert action supplemented with a Gibbons-Hawking-York (GHY) boundary term is 
\beq I=\frac{1}{16\pi G_{N}}\int_{\mathcal{M}}d^{d+1}x\sqrt{-\hat{g}}(R-2\Lambda)+\frac{1}{8\pi G_{N}}\int_{\partial\mathcal{M}}d^{d}\xi\sqrt{|h|}\epsilon K\;,\eeq
where here $G_{N}$ is Newton's constant in $d+1$-dimensions. 
%assuming the boundary hypersurface $\partial\mathcal{M}$ obeys Dirichlet boundary conditions. 
The GHY term makes the variational problem well-posed assuming the metric obeys Dirichlet or Neumann (see below) boundary conditions at boundary $\partial\mathcal{M}$. Indeed, a standard computation shows the metric variation of the Einstein-Hilbert term is (upon imposing the Einstein equations of motion)
\beq
\begin{split}
16\pi G_{N}\delta I_{\text{EH}}&=
%\oint_{\partial\mathcal{M}} d^{d-1}y\sqrt{|h|}\epsilon(n_{c}\hat{g}^{db}\delta \Gamma^{c}_{\;bd}-n^{c}\delta\Gamma^{b}_{\;bc})\\
%&=
%\oint_{\partial\mathcal{M}}d^{d-1}\xi\sqrt{|h|}\epsilon D_{\mu}\delta w^{\mu}
\oint_{\partial\mathcal{M}}d^{d}y\sqrt{|h|}\epsilon(K_{ab}-Kh_{ab})\delta h^{ab}-2\delta\left(\oint_{\partial\mathcal{M}}d^{d}y\sqrt{|h|}\epsilon K\right)\;,
\end{split}
\eeq
where we have assumed the manifold $\mathcal{M}$ is void of codimension-2 corners. Thus, 
\beq \delta I=\frac{1}{16\pi G_{N}}\oint_{\partial\mathcal{M}}d^{d}y\sqrt{|h|}\epsilon(K_{ab}-Kh_{ab})\delta h^{ab}\;.\eeq
The action is stationary when either Dirichlet boundary conditions are imposed, $\delta h^{ab}|_{\partial\mathcal{M}}=0$, or Neumann boundary conditions are imposed
\beq (K_{ab}-K h_{ab})|_{\partial\mathcal{M}}=0\;.\label{eq:NeumannBC}\eeq

Next, consider introducing a brane of tension $T$ at $\partial\mathcal{M}$, characterized by a brane action,
\beq I_{\text{brane}}=-\frac{T}{8\pi G_{N}}\int_{\partial\mathcal{M}}d^{d}y\sqrt{|h|}\;.\eeq
 The total action $I+I_{\text{brane}}$ will be stationary provided the bulk Einstein equations hold and the Neumann boundary condition (\ref{eq:NeumannBC}) is modified to
\beq \epsilon K_{ab}|_{\partial\mathcal{M}}=(\epsilon K-T)h_{ab}|_{\partial\mathcal{M}}\;,\label{eq:braneeomNBC}\eeq
 as is referred to as the brane equations of motion. Taking the trace, the tension is
\beq T=\frac{\epsilon(d-1)}{d}K\;,\eeq
such that the brane equation of motion (\ref{eq:braneeomNBC}) becomes
\beq \epsilon K_{ab}|_{\partial\mathcal{M}}=\frac{T}{(d-1)}h_{ab}|_{\partial\mathcal{M}}\;.\eeq

%%%%%%%%%%%%%%%%%%%%%%%%%%%%%%%%%%%%%%%%%%%%%%%%%%%%%%%%%%%%%%%%%%%%
\section{Holographic regularization: a detailed summary} \label{app:holoren}

Here we review holographic renormalization \'a la \cite{deHaro:2000vlm}. For a more algorithmic approach of deriving the effective action, see, e.g., \cite{Kraus:1999di,Papadimitriou:2004ap,Elvang:2016tzz,Bueno:2022log}.

\subsection{Fefferman-Graham expansion and Einstein's equations}

Following \cite{deHaro:2000vlm}, express the asymptotically $\text{AdS}_{d+1}$ bulk metric $\hat{g}_{ab}$ in a Fefferman-Graham (FG) expansion  \cite{Fefferman1985,Fefferman:2007rka} near the asymptotic boundary
\beq ds^{2}_{d+1}=\frac{L^{2}}{4\rho^{2}}d\rho^{2}+\frac{L^{2}}{\rho}g_{ij}(\rho,x)dx^{i}dx^{j}\;,\label{eq:FGmetmain}\eeq
where $L$ is the $\text{AdS}_{d+1}$ curvature scale of the $d+1$-dimensional bulk, and the asymptotic boundary is located at $\rho=0$.\footnote{The asymptotic boundary also lives at $\rho\to\infty$, however we restrict ourselves to the region near $\rho=0$.} Here we use the hat notation to denote the bulk quantities, e.g., the bulk metric is represented by $\hat{g}_{ab}$. The $d$-dimensional boundary submanifold has metric $\hat{g}_{ij}(\rho,x)\equiv h_{ij}(\rho,x)=(L^{2}/\rho) g_{ij}(\rho,x)$.

The non-vanishing Christoffel symbols with respect to metric (\ref{eq:FGmetmain}) are
\beq
\begin{split}
&\hat{\Gamma}^{\rho}_{\;\rho\rho}=-\frac{1}{\rho}\;,\quad \hat{\Gamma}^{\rho}_{\;ij}=2\left(g_{ij}-\rho \partial_{\rho}g_{ij}\right)\;,\\
&\hat{\Gamma}^{k}_{\;\rho i}=\frac{1}{2\rho}[-\delta^{k}_{\;i}+g^{jk}\rho\partial_{\rho}g_{ij}]\;,\quad \hat{\Gamma}^{k}_{\;ij}=\Gamma^{k}_{\;ij}[g]\;.
\end{split}
\label{eq:FGchrisbasicmain}\eeq
 %With these Christoffel symbols, let us now work out the components of the Riemann curvature tensor, which are all we will need to carry out the holographic renormalization procedure. Specifically, 
%\beq \hat{R}^{\rho}_{\;j\rho k}=\partial_{\rho}\Gamma^{\rho}_{\;jk}+\Gamma^{\rho}_{\;\rho\rho}\Gamma^{\rho}_{\;jk}-\Gamma^{\rho}_{\;ik}\Gamma^{i}_{\;j\rho}\;,\eeq
%\beq \hat{R}^{\rho}_{\;jkl}=\partial_{k}\Gamma^{\rho}_{\;jl}-\partial_{l}\Gamma^{\rho}_{\;jk}+\Gamma^{\rho}_{\;ik}\Gamma^{i}_{\;lj}-\Gamma^{\rho}_{\;il}\Gamma^{i}_{\;jk}\;,\eeq
%\beq \hat{R}^{i}_{\;jkl}=R^{i}_{\;jkl}[h]+\Gamma^{i}_{\;k\rho}\Gamma^{\rho}_{\;jl}-\Gamma^{i}_{\;l\rho}\Gamma^{\rho}_{\;jk}\;,\eeq
%where we have dropped the hat notation for the Christoffel symbols as it is apparent via the index structure which quantities are bulk and which are not. Substituting in the Christoffel symbols (\ref{eq:FGchrisbasicmain}) we find explicitly, 
Consequently, the non-zero components of the Riemann curvature tensor are\footnote{We work in the convention where the Riemann curvature tensor is $R^{\rho}_{\;\sigma\mu\nu}=\partial_{\mu}\Gamma^{\rho}_{\;\nu\sigma}+\Gamma^{\rho}_{\;\mu\lambda}\Gamma^{\lambda}_{\;\nu\sigma}-(\mu\leftrightarrow\nu)$.
Equivalently, $R^{\rho}_{\;\sigma\mu\nu}=-[\nabla_{\mu},\nabla_{\nu}]V_{\nu}$ and $R^{\rho}_{\;\sigma\mu\nu}V^{\sigma}=[\nabla_{\mu},\nabla_{\nu}]V^{\rho}$ for vectors $V$. This convention differs from the one used in \cite{deHaro:2000vlm} by an overall minus sign. In our convention, AdS curvature is negative while in \cite{deHaro:2000vlm} AdS curvature is positive.}
\beq 
\begin{split}
&\hat{R}^{\rho}_{\;j\rho k}=-2\rho\partial^{2}_{\rho}g_{jk}-\frac{1}{\rho}g_{jk}+\rho g^{il}(\partial_{\rho}g_{jl})(\partial_{\rho}g_{ik})\;,\\
%\label{eq:RiemFG1main}\eeq
%\beq 
%\begin{split}
&\hat{R}^{\rho}_{\;jkl}=2\rho(\nabla_{l}\partial_{\rho}g_{jk}-\nabla_{k}\partial_{\rho}g_{jl})\;,\\
%\end{split}
%\label{eq:RiemFG2main}\eeq
%\beq
%\begin{split}
&\hat{R}^{i}_{\;jkl}=R^{i}_{jkl}[g]+\frac{1}{\rho}[-\delta^{i}_{k}g_{jl}+\rho\delta^{i}_{k}\partial_{\rho}g_{jl}+\rho g^{im}g_{jl}\partial_{\rho}g_{km}-\rho^{2}g^{im}(\partial_{\rho}g_{jl})(\partial_{\rho}g_{km})]\\
&-\frac{1}{\rho}[-\delta^{i}_{l}g_{jk}+\rho\delta^{i}_{l}\partial_{\rho}g_{jk}+\rho g^{im}g_{jk}\partial_{\rho}g_{lm}-\rho^{2}g^{im}(\partial_{\rho}g_{jk})(\partial_{\rho}g_{lm})]\;,
\end{split}
\eeq
where $\nabla_{i}$ refers to the covariant derivative compatible with $g_{ij}$. 
It is also useful to know
\beq
\begin{split}
\hspace{-4mm}\hat{R}_{ijkl}&=\frac{L^{2}}{\rho}R_{ijkl}[g]-\frac{L^{2}}{\rho^{2}}(g_{ik}g_{jl}-g_{il}g_{jk})+\frac{L^{2}}{\rho}[g_{ik}\partial_{\rho}g_{jl}+g_{jl}\partial_{\rho}g_{ik}-g_{il}\partial_{\rho}g_{jk}-g_{jk}\partial_{\rho}g_{il}]\\
&+L^{2}[(\partial_{\rho}g_{jk})(\partial_{\rho}g_{il})-(\partial_{\rho}g_{jl})(\partial_{\rho}g_{ik})]\;.
\end{split}
\label{eq:RiemFG3main}\eeq
%Note that in (\ref{eq:RiemFG2main}), the covariant derivative $\nabla_{i}$ corresponds to the derivative with respect to the metric $g_{ij}$. 

Let us now work out the form of Einstein's equations near the conformal boundary. One route is to explicitly compute the Ricci tensor and scalar.
%Let's now work out the Einstein equations. We could go and compute the components of the Ricci tensor and Ricci scalar (which have been worked out in Appendix \ref{sec:curvFGdets}), however,
Alternatively, we can use the fact that the bulk spacetime, asymptotically, has a vanishing Weyl tensor such that
\beq \hat{R}_{abcd}=-\frac{1}{L^{2}}[\hat{g}_{ac}\hat{g}_{bd}-\hat{g}_{ad}\hat{g}_{bc}]\;\label{eq:Weylvanish}\eeq
near the boundary. 
In summary, we find
\beq 
\begin{split}
&R_{ijkl}[g]=g_{il}g'_{jk}+g_{jk}g'_{il}-g_{ik}g'_{jl}-g_{jl}g'_{ik}+\rho(g'_{jl}g'_{ik}-g'_{jk}g'_{il})\;,\\
&g''_{jk}-\frac{1}{2}g^{il}g'_{jl}g'_{ik}=0\;,\\
%\left(g''-\frac{1}{2}g'g^{-1}g'\right)_{\hspace{-1mm}jk}=0\;,\\
&\nabla_{l}g'_{jk}-\nabla_{k}g'_{jl}=0\;,
\end{split}
\label{eq:EinEOMmain}\eeq
where we introduced the notation $\partial_{\rho}g_{ij}\equiv g'_{ij}$. Our expressions match Eqs. (7) -- (9) of \cite{Skenderis:1999nb}.\footnote{Note that Eq. (7) of \cite{Skenderis:1999nb} differs by an overall sign and a factor of $L^{2}$ from our expression (\ref{eq:RiemFG3main}). The difference in sign comes from a different convention for the Riemann tensor, one where the cosmological constant for AdS is `positive', and the $L^{2}$ comes from the form of the metric where $h_{ij}^{there}=L^{-2}h_{ij}^{here}$.} Equivalently, by contracting (\ref{eq:EinEOMmain}) with $g^{ij}$, one often expresses Einstein's equations as 
\beq
\begin{split}
&\biggr(Ric[g]+(d-2)g'+\text{Tr}(g^{-1}g')g+\rho(2g'g^{-1}g'-2g''-\text{Tr}(g^{-1}g')g')\biggr)_{\hspace{-1mm} jl}=0\;,\\
&\text{Tr}(g^{-1}g'')-\frac{1}{2}\text{Tr}(g^{-1}g'g^{-1}g')=0\;,\\
&\nabla_{l}\text{Tr}(g^{-1}g')-\nabla^{j}g'_{jl}=0\;,
\end{split}
\label{eq:Einsteineqnsag}\eeq
 matching Eq. (2.5) of \cite{deHaro:2000vlm}.\footnote{To aid the reader, here, for example, $g^{jk}g'_{jk}=\text{Tr}(g^{-1}g')$. Further, to find the first expression in (\ref{eq:Einsteineqnsag}), contract the first expression in (\ref{eq:EinEOMmain}) by $g^{jk}$. Then add zero to the term proportional to $\rho$ as $(g^{ik}g'_{jk}g'_{il}-g^{ik}g'_{jk}g'_{il})$ and use the second expression in (\ref{eq:EinEOMmain}) to recast $g^{ik}g'_{jk}g'_{il}=2g''_{jl}$.} Deriving the Einstein equations in this relied on using (\ref{eq:Weylvanish}), i.e., that the Weyl tensor vanishes near the boundary. Note that for $d=2$ (\ref{eq:Weylvanish}) holds everywhere.

%%%%%%%%%%%%%%%%%%%%%%%%%%%%%%%%%%%%%%%%
\subsection{Perturbatively solving Einstein's equations}

With Einstein's equations in hand, we now want to solve them perturbatively near the boundary $\rho=0$. To this end, we expand the metric $g_{ij}(x,\rho)$ as
\beq
\begin{split}
&g(x,\rho)=g_{(0)}(x)+\rho g_{(2)}(x)+\rho^{2}g_{(4)}(x)+...\quad (d\;\text{odd})\;,\\
&g(x,\rho)=g_{(0)}(x)+\rho g_{(2)}(x)+...+\rho^{d/2}g_{(d)}(x)+\rho^{d/2}(\log\rho) h_{(d)}(x)+\mathcal{O}(\rho^{d/2+1})\quad (d\;\text{even})\;.
\end{split}
\label{eq:pertexpng}\eeq
As we will see, the tensors $g_{(k)}$ are given by some covariant expression with respect to the boundary metric $g_{(0)}$, its Riemann tensor, and their derivatives. Additionally, the subscript or superscript in $g_{(k)}$ indicates the number of derivatives with respect to coordinates $x^{i}$, e.g., $g_{(2)}$ contains two derivatives, $g_{(4)}$ contains four-derivatives, and so forth. 
%We can think of the perturbative expansion (\ref{eq:pertexpng}) in $\rho$ as corresponding to a low energy expansion. 
The basic algorithm for solving Einstein's equations order by order in $\rho$ is to
%The aim then is to solve Einstein's equations order by order in $\rho$.  the basic algorithm is 
differentiate Einstein's equations (\ref{eq:EinEOMmain}) with respect to $\rho$ and then take the limit when $\rho=0$, recasting the coefficients $g_{(k\neq0)}$ in terms of $g_{(0)}$. Notably, for even $d$, this procedure would have broken down at order $d/2$ had the logarithm proportional to $h_{(d)}$ not been introduced (\ref{eq:pertexpng}). Further, the computation differs for $d=2$ and $d>2$ in that the expansion truncates in $d=2$. Below we will focus on the procedure for $d>2$ (see \cite{Henningson:1998gx,Skenderis:1999nb,deHaro:2000vlm} for the $d=2$ case).

%\subsection{Summary of algorithm: $d>2$}

\vspace{2mm}

\noindent \textbf{Coefficients $g_{(k)}$ for $k\neq d$:} We can use the first of Einstein's equations in (\ref{eq:EinEOMmain}) to cast coefficients $g_{(k)}$ for $k\neq d$ solely in terms of covariant expressions of $g_{(0)}$. Begin with $g_{(2)jk}$. Substitute $g(\rho,x)=g_{(0)}+\rho g_{(2)}+\rho^{2}g_{(4)}+...$ into the first expression in (\ref{eq:EinEOMmain}), giving
\beq R_{ijkl}=g^{(0)}_{il}g^{(2)}_{jk}+g^{(0)}_{jk}g^{(2)}_{il}-g^{(0)}_{ik}g_{jl}^{(2)}-g_{jl}^{(0)}g_{ik}^{(2)}+\rho(g^{(2)}_{il}g^{(2)}_{jk}-g^{(2)}_{jl}g^{(2)}_{ik})+\mathcal{O}(\rho^{2})+...\;.\label{eq:Rijklg2}\eeq
Now recall that in the limit $\rho\to0$, the Weyl tensor $W_{ijkl}$ on the boundary vanishes. Using the following useful expression for the Weyl tensor in $d$-dimensions, 
\beq W_{ijkl}=R_{ijkl}+(P_{jk}g_{il}+P_{il}g_{jk}-P_{ik}g_{jl}-P_{jl}g_{ik})\;,\eeq
where $P_{jk}$ is the Schouten tensor,
\beq P_{jk}[g]=\frac{1}{(d-2)}\left(R_{jk}[g]-\frac{1}{2(d-1)}R[g]g_{jk}\right)\;,\eeq
it follows $W_{ijkl}=0$ implies
\beq R_{ijkl}=-(P_{jk}g_{il}+P_{il}g_{jk}-P_{ik}g_{jl}-P_{jl}g_{ik})\;.\eeq
Thus, in the limit $\rho\to0$,
\beq \lim_{\rho\to0}R_{ijkl}=-(P_{jk}[g_{(0)}]g_{il}^{(0)}+P_{il}[g_{(0)}]g_{jk}^{(0)}-P_{ik}[g_{(0)}]g_{jl}^{(0)}-P_{jl}[g_{(0)}]g_{ik}^{(0)})\;,\label{eq:limRijkl}\eeq
with
\beq P_{jk}[g_{(0)}]=\frac{1}{(d-2)}\left(R_{jk}[g_{(0)}]-\frac{1}{2(d-1)}R[g_{(0)}]g_{jk}^{(0)}\right)\;.\eeq
Here we used that $\lim_{\rho\to0}R_{ij}[g]=R_{ij}[g_{(0)}]$, which is easy to show since $R_{ij}$, using the expansion in terms of $g(\rho,x)$ will vanish for any term proportional to $g_{(k)}$ for $k\neq0$.  

Substituting $\lim_{\rho\to0}R_{ijkl}$ (\ref{eq:limRijkl}) into (\ref{eq:Rijklg2}), gives
\beq g^{(0)}_{il}g^{(2)}_{jk}+g^{(0)}_{jk}g^{(2)}_{il}-g^{(0)}_{ik}g_{jl}^{(2)}-g_{jl}^{(0)}g_{ik}^{(2)}=-(P_{jk}[g_{(0)}]g_{il}^{(0)}+P_{il}[g_{(0)}]g_{jk}^{(0)}-P_{ik}[g_{(0)}]g_{jl}^{(0)}-P_{jl}[g_{(0)}]g_{ik}^{(0)})\;.\eeq
Matching like terms, we find
\beq g_{jk}^{(2)}=-P_{jk}[g_{(0)}]=-\frac{1}{(d-2)}\left(R_{jk}[g_{(0)}]-\frac{1}{2(d-1)}R[g_{(0)}]g_{jk}^{(0)}\right)\;,\label{eq:g2ij}\eeq
recovering Eq. (A.1) of \cite{deHaro:2000vlm} (up to an overall sign due to our convention). Note that this same argument also reveals $h_{(2)jk}=0$. 
%We observe that already at this level when $d=2$ the expression diverges.\footnote{Note, however, one could perform the limit more carefully by recalling that in $d=2$ the Ricci tensor is given by $R_{ij}=\frac{1}{2}Rg_{ij}$ (as the Einstein tensor exactly vanishes). Substituting this into the expression for $P_{ij}$ we would seem to find $g_{(2)ij}=-\frac{1}{2}R_{ij}[g_{(0)}]g_{ij}^{(0)}$. However, this is not quite the whole story as we will see later.}
To go to higher order, e.g., $g^{(4)}_{ij}$ one takes $\rho$ derivatives of $R_{ijkl}$ and then takes the limit $\rho\to0$. 

\vspace{2mm}

\noindent \textbf{Trace of $g_{(n)}$:} The third Einstein equation in (\ref{eq:EinEOMmain}) yields the trace of $g_{(n)}$ for any $n$. For example, substituting in the expansion (\ref{eq:pertexpng}) up to $g_{(4)}$ and taking the limit $\rho\to0$ gives
\beq 2g^{(4)}_{jk}-\frac{1}{2}g^{(2)}_{jl}g^{ml}g^{(2)}_{km}=0\Rightarrow g^{(4)}_{jk}=\frac{1}{4}g^{(2)}_{jl}g^{ml}g^{(2)}_{km}\;.\eeq
Taking the trace with by contracting with $g^{jk}$ we are led to
\beq \text{Tr}g_{(4)}=\frac{1}{4}\text{Tr}(g_{(2)}^{2})\;,\eeq
where here $\text{Tr}g_{(4)}=g^{ij}_{(0)}g_{(4)ij}$ and matches Eq. (A.2) in \cite{deHaro:2000vlm}. 

To get to the next order, write $g(\rho,x)=g_{(0)}+\rho g_{(2)}+\rho^{2}g_{(4)}+\rho^{3}g_{(6)}$, and take the derivative of the second Einstein equation (\ref{eq:EinEOMmain}) with respect to $\rho$ and then set $\rho\to0$, giving 
\beq 6g_{jk}^{(6)}-\frac{1}{2}\left[4g^{(4)}_{ij}g^{il}g^{(2)}_{lk}-g^{(2)}_{ij}g^{(2)}_{lk}g^{(2)}_{mn}g^{im}g^{nl}\right]=0\;.\eeq
Taking the trace and simplifying  gives
\beq \text{Tr}g_{(6)}=\frac{2}{3}\text{Tr}(g_{(4)}g_{(2)})-\frac{1}{6}\text{Tr}(g_{(2)}^{3})\;.\eeq
Following this procedure, it is also straightforward to show that $h_{(d)}$ is traceless, i.e.,  $g^{ij}_{(0)}h_{(0)ij}=0$. Likewise, $g^{ij}_{(0)}g_{(3)ij}=g^{ij}_{(0)}g_{(5)ij}=0$. 

\vspace{2mm}

\noindent \textbf{Covariant divergence of $g_{(k)}$:} The second of Einstein's equations (\ref{eq:EinEOMmain}) fixes the covariant divergence of $g_{(k)}$ for any $k$. For example, substituting expansion (\ref{eq:pertexpng}) up to order $\mathcal{O}(\rho)$ and taking 
%\beq \nabla_{l}(g^{jk}g'_{jk})=\nabla^{j}g'_{jl}\;,\eeq
% its contraction with $g^{jk}$ is $\nabla^{j}g'_{jl}=\nabla_{l}(g^{jk}g'_{jk})$. Then, to taking
 the $\rho\to0$ limit of both sides gives
\beq
  \lim_{\rho\to0} \nabla^{j}g'_{jl}=\nabla^{i}g^{(2)}_{ij}=\lim_{\rho\to0}\nabla_{j}(g^{lk}g'_{lk})=\nabla_{j}(g^{kl}_{(0)}g^{(2)}_{kl})=\nabla^{i}(g^{(0)}_{ij}g^{kl}_{(0)}g^{(2)}_{kl})\;.\eeq
Once the limit has been taken, it is understood that the covariant derivative is compatible with $g_{(0)}$.
%and  to arrive to the final equality we raised the index on $\nabla_{i}$ via $g_{(0)}$. 
More compactly,
\beq \nabla^{i}g^{(2)}_{ij}=\nabla^{i}(g_{(0)ij}\text{Tr}g_{(2)})\;,\eeq
matching the first expression in Eq. (A.4) of \cite{deHaro:2000vlm}. To get to the next order we take a $\rho$ derivative of the equation of motion, keeping in mind that the covariant derivative itself has $\rho$ dependence inside, and then take the limit $\rho\to0$ \cite{deHaro:2000vlm}.

\subsection{Regulated bulk action}

Thus far we have shown, given a conformal structure at infinity, an the metric can be determined asymptotically up to order $\rho^{d/2}$, i.e., the coefficient $g_{(d)}$. The aim now is to study the IR divergent structure of the bulk theory gravity, characterized by the action
\beq I_{\text{bulk}}=\frac{1}{16\pi G}\left[\int_{\mathcal{M}}d^{d+1}x\sqrt{-\hat{g}}\left(\hat{R}+\frac{d(d-1)}{L^{2}}\right)+2\int_{\partial\mathcal{M}}d^{d}x\sqrt{h}K\right]\;,\eeq
where $K$ is the trace of the extrinsic curvature of the boundary submanifold $\partial\mathcal{M}$. Here it is understood that $G$ and $L$ refer to the bulk Newton's constant and AdS curvature scale, respectively. As stated in the main text, the on-shell bulk gravity action will have IR divergences near the boundary $\rho=0$.  To regulate IR divergences of the theory, introduce an IR cutoff surface at $\rho=\epsilon$ near the asymptotic boundary, where the preceding analysis has taken place. Thus, in the bulk contribution we consider the integration region $\epsilon\leq\rho\leq\rho_{c}$, where $\rho_{c}$ is some value such that $\rho_{c}\gg\epsilon$.\footnote{Recall that for general dimension $d$, the analysis only applies near $\rho=0$, where the Weyl tensor of the bulk spacetime vanishes. Consequently, $\rho$-integration is performed around $\rho=\epsilon$. Alternatively, when $d=2$, since the three-dimensional Weyl tensor is identically zero everywhere, the perturbative expansion $g_{ij}(\rho,x)$ truncates and the $\rho$-integration can be carried out explicitly (see, e.g., \cite{deHaro:2000vlm,Skenderis:1999nb}).}   The GHY boundary term is simply evaluated at $\rho=\epsilon$. Thence, the IR regulated bulk action is
\beq I_{\text{bulk}}^{\text{reg}}=\frac{1}{16\pi G}\left[\int_{\rho\geq\epsilon}d^{d+1}x\sqrt{-\hat{g}}\left(\hat{R}+\frac{d(d-1)}{L^{2}}\right)+2\int_{\partial\mathcal{M}}d^{d}x\sqrt{h}K|_{\rho=\epsilon}\right]\;,\eeq
To evaluate the bulk contribution, we use that, on-shell, $\hat{R}=-d(d+1)/L^{2}$ near the boundary. Consequently, the bulk contribution becomes	
\beq \int d^{d}x\int_{\rho\geq\epsilon}d\rho \frac{L^{d+1}}{2\rho^{d/2+1}}\left(-\frac{2d}{L^{2}}\right)\sqrt{g(\rho,x)}=\int d^{d}x\int_{\rho\geq\epsilon}d\rho \frac{dL^{d-1}}{\rho^{d/2+1}}\sqrt{g(\rho,x)}\;.\label{eq:bulktermgen}\eeq

The form of the boundary action requires a little more work. We have a hypersurface equation defined by $\Phi\equiv\rho-\epsilon=0$, with unit normal  $n_{a}=[\hat{g}^{ab}(\partial_{a}\Phi)(\partial_{b}\Phi)]^{-1/2}\partial_{a}\Phi$, whose only non-zero component is $n_{\rho}=\frac{2\rho}{L}$.
%\beq n_{\rho}=\frac{1}{\sqrt{(L^{2}/4\rho^{2})(\partial_{\rho}\Phi)^{2}}}\partial_{\rho}\Phi=\frac{2\rho}{L}\;.\eeq
The components of the extrinsic curvature $K_{ab}=\hat{\nabla}_{a}n_{b}$ are easily worked out to be
\beq
\begin{split}
K_{\rho\rho}=K_{i\rho}=K_{\rho i}=0\;,\quad K_{ij}=-\frac{L}{\rho}(g_{ij}-\rho\partial_{\rho}g_{ij})\;,
\end{split}
\eeq
%\beq K_{\rho\rho}=\partial_{\rho}n_{\rho}-\Gamma^{\rho}_{\;\rho\rho}n_{\rho}=-\frac{L}{2\rho^{2}}-\left(-\frac{1}{\rho}\right)\frac{L}{2\rho}=0\;,\eeq
%\beq K_{ij}=-\Gamma^{\rho}_{\;ij}n_{\rho}=-\frac{L}{\rho}(g_{ij}-\rho\partial_{\rho}g_{ij})\;,\eeq
where we used the Christoffel symbols (\ref{eq:FGchrisbasicmain}). The trace $K$ is 
\beq K=\hat{g}_{ij}K_{ij}=
%\frac{\rho}{L^{2}}\left[-\frac{L}{\rho}g^{ij}(g_{ij}-\rho\partial_{\rho}g_{ij})\right]=
-\frac{1}{L}\left(d-\rho g^{ij}\partial_{\rho}g_{ij}\right)\;.\eeq
Also note $\sqrt{h}=\frac{L^{d}}{\rho^{d/2}}\sqrt{g(\rho,x)}$.
%where $g(\rho,x)$ stands for the full metric determinant. 
Therefore, the GHY term becomes
\beq 
\begin{split}
2\int_{\partial\mathcal{M}}d^{d}x\sqrt{h}K&=\int_{\partial\mathcal{M}}d^{d}x\left[-\frac{2L^{d-1}}{\rho^{d/2}}\sqrt{g(x,\rho)}\left(d-\rho g^{ij}\partial_{\rho}g_{ij}\right)\right]_{\rho=\epsilon}\\
&=\int_{\partial\mathcal{M}}d^{d}x\frac{L^{d-1}}{\rho^{d/2}}\left(-2d\sqrt{g(x,\rho)}+2\rho\sqrt{g(x,\rho)}g^{ij}\partial_{\rho}g_{ij}\right)_{\rho=\epsilon}\\
&=\int_{\partial\mathcal{M}}d^{d}x\frac{L^{d-1}}{\rho^{d/2}}\left(-2d\sqrt{g(x,\rho)}+4\rho\partial_{\rho}\sqrt{g(x,\rho)}\right)_{\rho=\epsilon}\;,
\end{split}
\label{eq:GHYgen}\eeq
where to arrive to the final line we used that $\delta\sqrt{g}=\frac{1}{2}\sqrt{g}g^{ij}\delta g_{ij}$.

Combining (\ref{eq:bulktermgen}) and (\ref{eq:GHYgen})  the (on-shell) regulated bulk action to evaluate is
\beq I^{\text{reg}}_{\text{bulk}}=\frac{L^{d-1}}{16\pi G}\int d^{d}x\left[\int_{\rho\geq\epsilon}d\rho \frac{d}{\rho^{d/2+1}}\sqrt{g(\rho,x)}+\frac{1}{\rho^{d/2}}\left(-2d\sqrt{g(x,\rho)}+4\rho\partial_{\rho}\sqrt{g(x,\rho)}\right)_{\rho=\epsilon}\right]\;.\label{eq:regbulkact}\eeq
All that remains is an evaluation of the metric determinant $\sqrt{g(\rho,x)}$, and performing the bulk integral. To evaluate the determinant we apply perturbation theory, which says for $g_{ij}=g^{(0)}_{ij}+q_{ij}$, where $q_{ij}$ is small compared to $g^{(0)}_{ij}$, then
%\beq \sqrt{g}=\sqrt{g_{(0)}}\left(1+\frac{1}{2}q^{i}_{\;i}+\frac{1}{8}q^{i}_{\;i}q^{j}_{\;j}-\frac{1}{4}q^{i}_{\;j}q^{j}_{\;i}\right)\;.\eeq
%In our case, we take $q_{ij}=\rho g_{ij}^{(2)}+\rho^{2}g_{ij}^{(4)}+...$, treating $\rho$ as small. We summarize this procedure below and derive the regulated and renormalized gravity actions both for $d>2$ and $d=2$. 
\beq \sqrt{g}=\sqrt{g_{(0)}}\left(1+\frac{1}{2}q^{i}_{\;i}+\frac{1}{8}q^{i}_{\;i}q^{j}_{\;j}-\frac{1}{4}q^{i}_{\;j}q^{j}_{\;i}\right)\;.\label{eq:metricdetpert}\eeq
For $d>2$, $q_{ij}=\rho g_{ij}^{(2)}+\rho^{2}g_{ij}^{(4)}+...$, treating $\rho$ small. Explicitly, 
\beq q^{i}_{\;i}=g^{ij}_{(0)}q_{ij}=g^{ij}_{(0)}\left(\rho g^{(2)}_{ij}+\rho^{2}g^{(4)}_{ij}+...\right)=\rho\text{Tr}(g_{(2)})+\rho^{2}\text{Tr}(g_{(4)})+...\;,\eeq
such that
\beq q^{i}_{\;i}q^{j}_{\;j}=\rho^{2}(\text{Tr}g_{(2)})^{2}+\rho^{3}(\text{Tr}g_{(2)})(\text{Tr}g_{(4)})+...\;,\eeq
%Next,
%\beq q^{i}_{\;j}=g^{ik}_{(0)}q_{kj}\;,\eeq
and
\beq q^{i}_{\;j}q^{j}_{\;i}=g^{ik}_{(0)}g^{jl}_{(0)}\left[\rho^{2}g^{(2)}_{kj}g^{(2)}_{li}+\rho^{3}(g^{(4)}_{kj}g^{(2)}_{li}+g^{(2)}_{kj}g^{(4)}_{li})+...\right]\;.\eeq
Collecting terms,
\beq
\begin{split}
\sqrt{g}&\approx \sqrt{g_{(0)}}\biggr(1+\frac{1}{2}\rho\text{Tr}g_{(2)}+\rho^{2}\biggr[\frac{1}{2}\text{Tr}g_{(4)}+\frac{1}{8}(\text{Tr}g_{(2)})^{2}-\frac{1}{4}\text{Tr}(g_{(2)}^{2})\biggr]+O(\rho^{3})\biggr)\;.
\end{split}
\eeq
%where we used $g^{ik}_{(0)}g^{jl}_{(0)}g_{kj}^{(2)}g_{li}^{(2)}=\text{Tr}(g_{(2)}^{2})$
%and
%\beq \partial_{\rho}\sqrt{g}=\sqrt{g_{(0)}}\biggr(\frac{1}{2}\text{Tr}g_{(2)}+2\rho\biggr[\frac{1}{2}\text{Tr}g_{(4)}+\frac{1}{8}(\text{Tr}g_{(2)})^{2}-\frac{1}{4}\text{Tr}(g_{(2)}^{2})\biggr]+O(\rho^{2})\biggr)\;.\eeq

Consequently, the term to evaluate on the boundary is, to order $O(\rho^{2})$,
%\beq
%\begin{split}
%&\left(-2d\sqrt{g}+4\rho\partial_{\rho}\sqrt{g}\right)=\sqrt{g_{(0)}}\biggr\{-2d\biggr(1+\frac{1}{2}\rho\text{Tr}g_{(2)}+\rho^{2}\biggr[\frac{1}{2}\text{Tr}g_{(4)}+\frac{1}{8}(\text{Tr}g_{(2)})^{2}-\frac{1}{4}\text{Tr}(g_{(2)}^{2})\biggr]\biggr)\\
%&+2\rho \text{Tr}g_{(2)}+8\rho^{2}\biggr[\frac{1}{2}\text{Tr}g_{(4)}+\frac{1}{8}(\text{Tr}g_{(2)})^{2}-\frac{1}{4}\text{Tr}(g_{(2)}^{2})\biggr]\\
%&=-\sqrt{g_{(0)}}\biggr\{2d+\rho[(d-2)\text{Tr}g_{(2)}]+\rho^{2}\biggr[2(d-4)\biggr(\frac{1}{2}\text{Tr}g_{(4)}+\frac{1}{8}(\text{Tr}g_{(2)})^{2}-\frac{1}{4}\text{Tr}(g_{(2)}^{2})\biggr)\biggr]\biggr\}\;.
%\end{split}
%\eeq
%And thus, 
\beq 
\begin{split}
&\frac{1}{\rho^{d/2}}\left(-2d\sqrt{g}+4\rho\partial_{\rho}\sqrt{g}\right)_{\rho=\epsilon}=-\sqrt{g_{(0)}}\left[2d\epsilon^{-d/2}+(d-2)(\text{Tr}g_{(2)})\epsilon^{-d/2+1}+O(\epsilon^{-d/2+2})\right]\;.
\end{split}
\eeq

%Next consider the bulk action, where we must integrate
%\beq 
%\begin{split}
%\sqrt{g}\frac{d}{\rho^{d/2+1}}&=\sqrt{g_{(0)}}\frac{d}{\rho^{d/2+1}}\left(1+\frac{1}{2}\rho\text{Tr}g_{(2)}+\rho^{2}\biggr[\frac{1}{2}\text{Tr}g_{(4)}+\frac{1}{8}(\text{Tr}g_{(2)})^{2}-\frac{1}{4}\text{Tr}(g_{(2)}^{2})\biggr]\right)\;.
%\end{split}
%\eeq
Meanwhile, the bulk term is
\beq
\begin{split}
&\int_{\epsilon}^{\rho_{c}}\hspace{-2mm}d\rho\sqrt{g}\frac{d}{\rho^{d/2+1}}
%=\sqrt{g_{(0)}}\biggr\{d\left(-\frac{2}{d}\rho^{-d/2}\right)\biggr|^{\rho_{c}}_{\epsilon}+\frac{d}{2}\text{Tr}g_{(2)}\left(-\frac{2}{(d-2)}\rho^{-d/2+1}\right)\biggr|^{\rho_{c}}_{\epsilon}+O(\rho^{-d/2+2})\biggr|^{\rho_{c}}_{\epsilon}\biggr\}\\
=\sqrt{g_{(0)}}\biggr\{-2\rho_{c}^{-d/2}-\frac{d\text{Tr}g_{(2)}}{(d-2)}\rho_{c}^{-d/2+1}+O(\rho_{c}^{-d/2+2})+2\epsilon^{-d/2}\\
&+\frac{d\text{Tr}(g_{(2)})}{(d-2)}\epsilon^{-d/2+1}+O(\epsilon^{-d/2+2})\biggr\}\;.
\end{split}
\eeq
Together, the regulated bulk action may be cast as
\beq I^{\text{reg}}_{\text{bulk}}=I_{\text{div}}+I_{\text{fin}}\;,\label{eq:bulkregactiongenapp}\eeq
the sum of an IR divergent term
\beq I_{\text{div}}=\frac{L}{16\pi G}\int d^{d}x\sqrt{g_{(0)}}\left[\epsilon^{-d/2}a_{(0)}+\epsilon^{-d/2+1}a_{(2)}+\epsilon^{-d/2+2}a_{(4)}+...\right]\;,\eeq
with coefficients\footnote{Note that, naively, $a_{(2)}$ is not valid when $d=2$. The correct term can be found by performing the analysis explicitly when $d=2$, yielding $a_{(2)}^{d=2}=\text{Tr}g_{(2)}$. Similarly, $a_{(4)}$ (which we did not explicitly compute; see \cite{deHaro:2000vlm}) has the coefficient in front replaced with $1/2$ when $d=4$.}
\beq 
\begin{split}
&a_{(0)}=2(1-d)L^{d-2}\;,\quad a_{(2)}=-\frac{(d-1)(d-4)}{(d-2)}L^{d-2}\text{Tr}g_{(2)}\;,\\
&a_{(4)}=\left(\frac{-d^{2}+9d-16}{4(d-4)}\right)L^{d-2}[(\text{Tr}g_{(2)})^{2}-\text{Tr}(g_{(2)}^{2})]\;,
\end{split}
\label{eq:coeffsad}\eeq
and a finite contribution $I_{\text{fin}}$, whose explicit form we will not need. 
%\beq I_{fin}=\frac{L^{d-1}}{16\pi G}\int d^{d}x\sqrt{g_{(0)}}\left[-2\rho_{c}^{-d/2}-\frac{d\text{Tr}g_{(2)}}{(d-2)}\rho_{c}^{-d/2+1}+O(\rho_{c}^{-d/2+2})\right]\;.\eeq

This is not the full story, as the terms $\rho^{d/2}g_{(d)}+\rho^{d/2}\log\rho h_{(d)}$ in the expansion (\ref{eq:pertexpng}) must be accounted for. These will enter the metric determinant at a higher order in $\rho$, however, generally they lead to a logarithmic divergence as well as adding to the finite contributions. For example, reconsider the term
\beq q^{i}_{\;i}=\rho\text{Tr}(g_{(2)})+\rho^{2}\text{Tr}(g_{(4)})+...+\rho^{d/2}\text{Tr}(g_{(d)})+\rho^{d/2}\log\rho\text{Tr}(h_{(d)})\;.\eeq
But, $\text{Tr}(h_{(d)})=0$ \cite{deHaro:2000vlm}. So, including the $\rho^{d/2}\text{Tr}(g_{(d)})$ contribution to the metric determinant, adding to $I_{\text{fin}}$
%we would acquire a finite contribution from the boundary term, namely, 
%\beq \frac{1}{\rho^{d/2}}\sqrt{g_{(0)}}\frac{1}{2}\rho^{d/2}\text{Tr}(g_{(d)})=\sqrt{g_{(0)}}\frac{1}{2}\text{Tr}(g_{(d)})\;,\eeq
%which would be added to $I_{fin}$. Meanwhile, this same term would lead to a logarithmic divergence coming from the bulk term, namely,
and a logarithmic divergence to $I_{\text{div}}$, i.e., 
\beq \int_{\epsilon}^{\rho_{c}}\frac{d}{\rho^{d/2+1}}\left(\frac{1}{2}\rho^{d/2}\text{Tr}g_{(d)}\right)=\frac{d}{2}\text{Tr}g_{(d)}(\log\rho_{c}-\log\epsilon)\;.\eeq
Thus, taking into account this term, the divergent contribution to $I_{\text{bulk}}^{\text{reg}}$ is
\hspace{-10mm} \beq I_{\text{div}}=\frac{L}{16\pi G}\int d^{d}x\sqrt{g_{(0)}}\left[\epsilon^{-d/2}a_{(0)}+\epsilon^{-d/2+1}a_{(2)}+\epsilon^{-d/2+2}a_{(4)}+...+\epsilon^{-1}a_{(d-2)}-\log(\epsilon) a_{(d)}\right]\;,\label{eq:divergentact}\eeq
with $a_{(0)}$, $a_{(2)}$, and $a_{(4)}$ as in (\ref{eq:coeffsad}) and, at this order,
\beq a_{(d)}=L^{d-2}\frac{d}{2}\text{Tr}g_{(d)}\;.\eeq
Note that $a_{(d)}$ here is only including the first term and in general is more complicated (see Eq. (B.1) of \cite{deHaro:2000vlm}).

%As described in the main text, to renormalize the theory, one invokes minimal subtraction by introducing a local counterterm Lagrangian such that $I_{ct}=-I_{div}$, where the renormalized action is then $I_{ren}=I_{reg}+I_{ct}$. 

\subsubsection{Regulated action on boundary}

Thus far, the regulated action is expressed in terms of the metric $g_{(0)}$. It is preferable to express the action in terms of the induced metric $h_{ij}(\epsilon,x)=\frac{L^{2}}{\epsilon}g_{ij}(\epsilon,x)$.
Then, 
\beq \frac{\epsilon^{d/2}}{L^{d}}\sqrt{h(\epsilon,x)}=\sqrt{g(\epsilon,x)}=\sqrt{g_{(0)}(x)}\biggr(1+\frac{1}{2}\epsilon\text{Tr}g_{(2)}+\frac{1}{8}\epsilon^{2}\left[(\text{Tr}g_{(2)})^{2}-\text{Tr}(g_{(2)}^{2})\right]+O(\rho^{3})\biggr)\;,\eeq
where we used $\text{Tr}(g_{(4)})=\frac{1}{4}\text{Tr}(g_{(2)}^{2})$.
Rearranging and performing a power series expansion in $\epsilon$
%Dividing this over and using a standard power series expansion $(1+\epsilon)^{-1}$ since $\epsilon$ is small, we find
yields
\beq \sqrt{g_{(0)}}=\frac{\epsilon^{d/2}}{L^{d}}\sqrt{h(\epsilon,x)}\left(1-\frac{1}{2}\epsilon\text{Tr}g_{(2)}+\frac{1}{8}\epsilon^{2}\left[(\text{Tr}g_{(2)})^{2}+\text{Tr}(g_{(2)}^{2})\right]+O(\epsilon^{3})\right)\;,\label{eq:pertmetg0app}\eeq
matching the first expression in Eq. (B.3) of \cite{deHaro:2000vlm}.

The aim now is to evaluate quantities $\text{Tr} g_{(2)}$ and $\text{Tr}(g_{(2)}^{2})$. To this end, we first recall some relations from metric perturbation. 
%At this point, it is useful to recall some useful relations in perturbation theory.
It is sufficient for our purposes to consider $g_{ij}=g_{(0)ij}+\epsilon g_{(2)ij}$, with $g^{ij}=g_{(0)}^{ij}-\epsilon g_{(2)}^{ij}$. Then, the linear variation of the Christoffel symbol is
\beq
\begin{split}
\label{eq:var_gamm}
   \delta \Gamma^k_{ij}&\equiv \Gamma^k_{ij}[g]-\Gamma^k_{ij}[g_{(0)}]
  % &= \frac{1}{2}\delta g^{kl} \left(\partial_i g_{(0)jl}+\partial_j g_{(0)il}-\partial_l g_{(0)ij}\right)+\frac{1}{2}g_{(0)}^{kl}\left(\partial_i \delta g_{jl}+\partial_j \delta g_{il}-\partial_l \delta g_{ij}\right) \\
  % &=\frac{\epsilon}{2}\left[-g_{(2)}^{kl} \left(\partial_i g_{(0)jl}+\partial_j g_{(0)il}-\partial_l g_{(0)ij}\right)+g_{(0)}^{kl}\left(\partial_i g_{(2)jl}+\partial_j  g_{(2)il}-\partial_l  g_{(2)ij}\right) \right] \\
  %&= \frac{\epsilon}{2}g^{(0)kl}\left[-2\Gamma^n_{ij}[g_{(0)}]g_{(2)nl}+\partial_i g_{(2)jl}+\partial_jg_{(2)il}-\partial_l g_{(2)ij} \right] \\
=\frac{\epsilon}{2}\left[\nabla_i g_{(2)j}^{\ \ k}+ \nabla_j g_{(2)i}^{\ \ k}-\nabla^k  g_{(2)ij} \right] \;,
\end{split}
\eeq
where here $\nabla_{i}$ refers to the Levi-Civita connection compatible with $g^{(0)}_{ij}$. Consequently,
%Note that $\delta \Gamma^k_{ij}$ is a tensor, contrary to the Christoffel symbol itself. To obtain the final line, we added zero in the form of the terms we needed to complete the covariant derivatives. We can play a similar trick to evaluate the variation of the Riemann tensor:
\beq
\begin{split}
    \delta R^a_{bcd}
    %\equiv R^a_{bcd}[g]-R^a_{bcd}[g_{(0)}] \\
    %&= \partial_c \delta\Gamma^a_{bd}-\partial_d \delta \Gamma^a_{bc} +\delta \Gamma^a_{ce}\Gamma^e_{bd}[g_{(0)}]+\Gamma^a_{ce}[g_{(0)}]\Gamma^e_{bd} - \delta \Gamma^a_{de}\Gamma^e_{bc}[g_{(0)}]-\Gamma^a_{de}[g_{(0)}]\Gamma^e_{bc} \\
    &=\nabla_c \delta\Gamma^a_{bd}-\nabla_d \delta\Gamma^a_{bc} \\
    &=\frac{\epsilon}{2}\left[\nabla_c\nabla_b g_{(2)d}^{\ \ a} + \nabla_c\nabla_d g_{(2)b}^{\ \ a} - \nabla_c\nabla^a g_{(2)bd} - \nabla_d\nabla_c g_{(2)b}^{\ \ a} - \nabla_d\nabla_b g_{(2)c}^{\ \ a}+ \nabla_d \nabla^a g_{(2)cb} \right] \ ,\\
\end{split}
\eeq
The leading order change to the Ricci tensor is then
\beq
\begin{split}
    \delta R_{ij}&= \frac{\epsilon}{2}\left[\nabla_a\nabla_i g_{(2)j}^{\ \ a} + \nabla_a\nabla_j g_{(2)i}^{\ \ a} - \Box g_{(2)ij}- \nabla_j\nabla_i g_{(2)a}^{\ \ a} \right] \\
    &= - \frac{\epsilon}{2} \biggr[\nabla_a \left( \nabla_i R_{j}^a[g_{(0)}] +\nabla_j R_{i}^a[g_{(0)}] \right) -\frac{d}{2(d-1)} \nabla_i\nabla_j R[g_{(0)}] \\
    &- \Box R_{ij}[g_{(0)}] + \frac{g_{(0)ij}}{2(d-1)}\Box R[g_{(0)}] \biggr] \ .
\end{split}
\eeq
where the second equality follows from substituting the expression for $g^{(2)}_{ij}$, (\ref{eq:g2ij}).
%where in the last line we have used Equation \ref{eq:var_gamm}. 
%To evaluate the variation in the Ricci tensor, note that, to leading order in $\epsilon$:
%\beq
%\begin{split}
% R_{ij}[g]=R^a_{ibj}[g]g^{kb}g_{ak} &=\left(R^a_{ibj}[g_{(0)}]+\delta R^a_{ibj} \right) \left(g_{(0)}^{kb}-\epsilon g_{(2)}^{kb} \right)\left(g_{(0)ak}+\epsilon g_{(2)ak} \right)  \\
% &= R_{ij}[g_{(0)}] + \delta R_{ij}
%\end{split}
%\eeq
%where we have used the fact that $\delta R^a_{ibj}$ is of order $\epsilon$ and we have defined $\delta R_{ij}=\delta R^a_{ibj}g_{(0)ak}g_{(0)}^{kb}$. 
%It is useful to rewrite the first term as:
Using\footnote{To see this, write $\nabla_a  \nabla_i R_{j}^a = \nabla_i \nabla_a R_{j}^{a}+ g^{ab}[\nabla_a,\nabla_i]R_{bj}$, and then use the contracted Bianchi identity $\nabla_{a}R^{a}_{\;j}=\frac{1}{2}\nabla_{j}R$ and $[\nabla_{a},\nabla_{i}]R_{bj}=-R^{k}_{\;bai}R_{kj}-R^{k}_{\;jai}R_{bk}$.}
\beq \nabla_{a}\nabla_{i}R^{a}_{\;j}=\nabla_{a}\nabla_{j}R^{a}_{\;i}=\frac{1}{2}\nabla_{i}\nabla_{j}R+(R_{ik}R^{k}_{\;j}-R^{k}_{\;jai}R^{a}_{\;k})\;,\eeq
%\beq
%\begin{split}
%    \nabla_a  \nabla_i R_{j}^a[g_{(0)}]  &= \nabla_i \nabla_a R_{j}^a[g_{(0)}]+ [\nabla_a,\nabla_i]R_{j}^a[g_{(0)}] \\
 %   & = \frac{1}{2}\nabla_i\nabla_j R[g_{(0)}] + g_{(0)}^{ab} [\nabla_a,\nabla_i]R_{jb}[g_{(0)}] \\
%    &= \frac{1}{2}\nabla_i\nabla_j R[g_{(0)}] + g_{(0)}^{ab} \left(-R^k_{bai}[g_{(0)}]R_{kj}[g_{(0)}]-R^k_{jai}[g_{(0)}]R_{bk}[g_{(0)}] \right) \ ,
%\end{split}
%\eeq
%where we have used the identity relating the commutator of the covariant derivatives and the Riemann tensor toghether with the contracted Bianchi identity $\nabla_a R^a_j=\frac{1}{2}\nabla_j R$. We can therefore put everything together to find:
we find,
\beq
\begin{split}
 \delta R_{ij} = & -\frac{\epsilon}{2(d-2)} \left[ \frac{d-2}{2(d-1)}\nabla_i \nabla_j R[g_{(0)}] - \Box R_{ij}[g_{(0)}] + \frac{g_{(0)ij}}{2(d-1)} \Box R[g_{(0)}] \right.\\
 & \left. + 2 R_{ki}[g_{(0)}] R^k_j[g_{(0)}] - 2 R_{ljki}[g_{(0)}]R^{lk}[g_{(0)}] \right] \ .
\end{split}
\label{eq:deltaRij}\eeq
Lastly, the Ricci scalar $R[g]=g^{ij}R_{ij}[g]=g^{ij}(R_{ij}[g_{(0)}]+\delta R_{ij})$, to leading order in an $\epsilon$-expansion is
%It is useful to know what the Ricci tensor is expressed as a power series in $\epsilon$. First, note that $\delta R_{ij}$ is traceless with respect to $g_{(0)}$. Then, to leading order in $\epsilon$:
\beq
\begin{split}
    R[g]
    %=g^{ij}R_{ij}[g] &= \left(g_{(0)}^{ij}-\epsilon g_{(2)}^{ij} \right) \left[R_{ij}[g_0]+ \delta R_{ij} \right] \\
    %&= R[g_{(0)}] -\epsilon g_{(2)}^{ij}R_{ij}[g_{(0)}] \\
    %&= R[g_{(0)}] +\frac{\epsilon}{d-2}R_{ij}[g_{(0)}]\left(R^{ij}[g_{(0)}]-\frac{1}{2(d-1)}R[g_{(0)}] g_{(0)}^{ij} \right) \\
    &=R[g_{(0)}] +\frac{\epsilon}{d-2}\left(R^{2}_{ij}[g_{(0)}]-\frac{1}{2(d-1)}R^2[g_{(0)}]\right)\;,
\end{split}
\label{eq:ricciscalar}\eeq
where we used (\ref{eq:g2ij}) and that $g^{ij}_{(0)}\delta R_{ij}=0$.
%Inverting the relationship to linear order in $\epsilon$ is trivial:
%\beq
%   R[g_{(0)}] =R[g] -\frac{\epsilon}{d-2}\left(R^{ij}[g]R_{ij}[g]-\frac{1}{2(d-1)}R^2[g] \right)
%\eeq

With these expressions we have
\beq
\begin{split}
 g^{ij}_{(0)}g_{(2)ij}&=-\frac{1}{2(d-1)}R[g_{(0)}]\\
&\approx -\frac{1}{2(d-1)} \left[R[g] -\frac{\epsilon}{d-2}\left(R^{ij}[g]R_{ij}[g]-\frac{1}{2(d-1)}R^2[g] \right) \right] \\
& \approx - \frac{L^2}{\epsilon}\frac{1}{2(d-1)}  \left[R[h] -\frac{L^2}{d-2}\left(R^{ij}[h]R_{ij}[h]-\frac{1}{2(d-1)}R^2[h] \right) \right] \;,
\end{split}
\eeq
where in the second line we inverted (\ref{eq:ricciscalar}), and in the last line we used $h_{ij}=\frac{L^{2}}{\epsilon}g_{ij}$, such that $R_{ij}[g]=R_{ij}[h]$.
%Finally, using the property that for $\tilde{g}_{ij}=\Omega^2 g_{ij}$:
%\beq
% R_{ij}[\tilde{g}]=R_{ij}[g] \ ,
%\eeq
%we can re-express $\text{Tr}(g_{(2)})$ in terms of curvature tensors for $h_{ij}$ instead: 
Meanwhile, 
\beq
\begin{split}
 g^{ij}_{(2)}g_{(2)ij}&=\frac{1}{(d-2)^{2}}\left(R^{ij}[g_{(0)}]R_{ij}[g_{(0)}]-\frac{(3d-4)}{4(d-1)^{2}}R^{2}[g_{(0)}]\right)\\
&\approx\frac{L^{4}}{\epsilon^{2}}\frac{1}{(d-2)^{2}}\left(R^{ij}[h]R_{ij}[h]-\frac{(3d-4)}{4(d-1)^{2}}R^{2}[h]\right)\;.
\end{split}
\eeq
Consequently, 
\beq
\begin{split}
&\text{Tr}g_{(2)}\approx - \frac{L^2}{\epsilon}\frac{1}{2(d-1)}  \left[R[h] -\frac{L^2}{d-2}\left(R^{2}_{ij}[h]-\frac{1}{2(d-1)}R^2[h] \right) \right]\;,\\
&\text{Tr}(g_{(2)}^{2})\approx\frac{L^{4}}{\epsilon^{2}}\frac{1}{(d-2)^{2}}\left[R^{2}_{ij}[h]-\frac{(3d-4)}{4(d-1)^{2}}R^{2}[h]\right]\;,\\ 
&(\text{Tr}g_{(2)})^{2}+\text{Tr}g_{(2)}^{2}
\approx\frac{L^{4}}{\epsilon^{2}}\biggr[\frac{1}{(d-2)^{2}}R_{ij}^{2}[h]+\frac{R^{2}[h]}{4(d-1)^{2}}\left(\frac{d^{2}-7d+8}{(d-2)^{2}}\right)+O(\mathcal{R}^{3})+...\biggr]\;,\\
&(\text{Tr}g_{(2)})^{2}-\text{Tr}g_{(2)}^{2}\approx-\frac{L^{4}}{\epsilon^{2}}\left[\frac{1}{(d-2)^{2}}R^{2}_{ij}[h] - \frac{dR^{2}}{4(d-2)^{2}(d-1)}+O(\mathcal{R}^{3})+...\right]\;,
\end{split}
\eeq
where $O(\mathcal{R}^{3})$ schematically refers to terms in cubic powers of curvature and the $+...$ corresponds to higher powers in curvature. The first two expressions match Eq. (B.3) of \cite{deHaro:2000vlm}. Substituting these expressions into our perturbative expression for $\sqrt{g_{(0)}}$ (\ref{eq:pertmetg0app}), we find
\beq \sqrt{g_{(0)}}=\sqrt{h}\frac{\epsilon^{d/2}}{L^{d}}\biggr\{1+\frac{L^{2}}{4(d-1)}\left(R-\frac{L^{2}(d-3)}{2(d-2)^{2}}\left[R_{ij}^{2}-\frac{d}{4(d-1)}R^{2}\right]\right)+...\biggr\}\;.
\eeq
Substituting this into the divergent action (\ref{eq:divergentact}), along with coefficients (\ref{eq:coeffsad}) gives,
\beq
\begin{split}
I_{\text{div}}=-\frac{L}{16\pi G(d-2)}\int d^{d}x\sqrt{h}\biggr[&\frac{2(d-1)(d-2)}{L^{2}}+R\\
&+\frac{L^{2}}{(d-2)(d-4)}\left(R_{ij}^{2}-\frac{d}{4(d-1)}R^{2}\right)+...\biggr]\;.
\end{split}
\label{eq:Idivappcom}\eeq
With the divergent action in hand, holographic renormalization is completed via minimal subtraction by adding to the regulated action (\ref{eq:bulkregactiongenapp}) a local counterterm action of the form $I_{\text{ct}}=-I_{\text{div}}$, rendering the action finite in as $\epsilon\to0$. Alternatively, in braneworld holography, where $\epsilon\neq0$, a local counterterm need not be introduced.

%\beq
%\begin{split}
%\sqrt{g_{(0)}}&=\sqrt{h}\frac{\epsilon^{d/2}}{L^{d}}\biggr\{1+\frac{L^{2}}{4(d-1)}\left[-R+\frac{1}{2(d-2)^2}\left((d-3)R_{ij}^{2}-\frac{R^{2}}{4(d-2)}\right)\right]\\
%+...\biggr\}\;.
%\end{split}
%\eeq
%Putting everything together we find:
%\beq
%\begin{split}
%    S_{ct}=-\frac{1}{16\pi G_N}\int d^4x\sqrt{h}\left[\frac{2}{L^2}(1-d)-\frac{R[h]}{d-2}+\frac{L^2}{(d-2)^2(d-4)}\left(R_{ij}[h]R^{ij}[h] + \frac{d}{4(d-1)}R[h]^2\right) + \cdots\right]
%\end{split}
%\eeq

\section{Braneworld basics} \label{app:braneworldbasics}

String theory is a candidate model of quantum gravity which requires the existence of extra dimensions. The observable universe, however, is well described by a four-dimensional spacetime, whose local gravitational dynamics is governed by Einstein's general relativity on macroscopic scales and the background on which Standard Model of particle physics lives. Thus, it is of interest to come up with a mechanism which reduces the number of extra dimensions predicted by string theory such that our observable world emerges as an effective theory in some limit. Historically, the simplest way to reduce the number of extra dimensions is via Kaluza-Klein (KK) dimensional reduction, where the extra dimensions are compactified into a small but finite-dimensional internal manifold. Alternatively, braneworlds provide a mechanism in which a lower-dimensional `brane' (or domain wall) lives in a higher-dimensional `bulk' spacetime, where the brane effectively describes the observable world.

Broadly, braneworlds are in part motivated by the membrane-like solutions in string theory and higher-dimensional supergravity; indeed, some braneworld models have string theoretic realizations (though in this review we will be agnostic to the stringy origins of any particular model). More specifically, braneworlds come in three types: (i) the Arkani-Hamed-Dimopoulos-Dvali (ADD) model \cite{Arkani-Hamed:1998jmv}, (ii) Randall-Sundrum (RS) \cite{Randall:1999ee,Randall:1999vf} and Karch-Randall (KR) models \cite{Karch:2000ct}, and the Dvali-Gabadadze-Porrati (DGP) model \cite{Dvali:2000hr}. As we will briefly review below, the braneworld models distinguish themselves by the type of bulk spacetime the brane is embedded and the way in which gravity is localized on the brane. In particular, the RS and KR models embed a brane inside a bulk anti-de Sitter spacetime -- such that the AdS/CFT correspondence applies -- and gravity localizes either in a way similar to KK reduction or via warped compactification.
%Of course, there exist other braneworld models which mix elements of the three types. For our purposes we are primarily interested in a particular subclass of models belonging to  (ii), and will thus only say a few words of the (i) and (iii) braneworld constructions. 

\subsection*{Braneworld bestiary}

\noindent \textbf{ADD model.} Motivated by resolving the hierarchy problem in particle physics, the original braneworld construction, the ADD model consists of a tensionless brane inside a large bulk spacetime where the extra dimensions are compactified in a manner similar to KK reduction. The bulk spacetime is assumed to be governed by Einstein gravity with zero cosmological constant. Massive KK modes in the extra dimensions are largely ignored such that, upon integrating out the extra dimensions, the effective action is four-dimensional Einstein gravity with scales induced by the higher-dimensional parent theory.  Notably, the case in which the bulk spacetime is five-dimensional is ruled out because the one-dimensional volume of the compactified directions is on astrophysical distance scales; ADD models with at least three extra dimensions are in principle observationally viable. 

\vspace{2mm}

\noindent \textbf{RS models.} There are two distinct Randall-Sundrum models, known as RS-I \cite{Randall:1999ee} and RS-II \cite{Randall:1999vf}. In both models, the bulk is taken to be asymptotically AdS spacetime.\footnote{Historically, the RS models assumed a five-dimensional bulk and four-dimensional brane. More generally the bulk need not be restricted to five dimensions, e.g., in this review we consider a four-dimensional bulk AdS spacetime and three-dimensional branes.} The brane geometry is either asymptotically Minkowski or de Sitter. In the RS-I model, the bulk is bounded two codimension-1 branes, one with positive tension $\tau_{+}$ and the other with negative tension $\tau_{-}$. In the RS-II model the negative tension brane is sent to asymptotic infinity. 

\begin{figure}[t!]
\centering
\includegraphics[width=10cm]{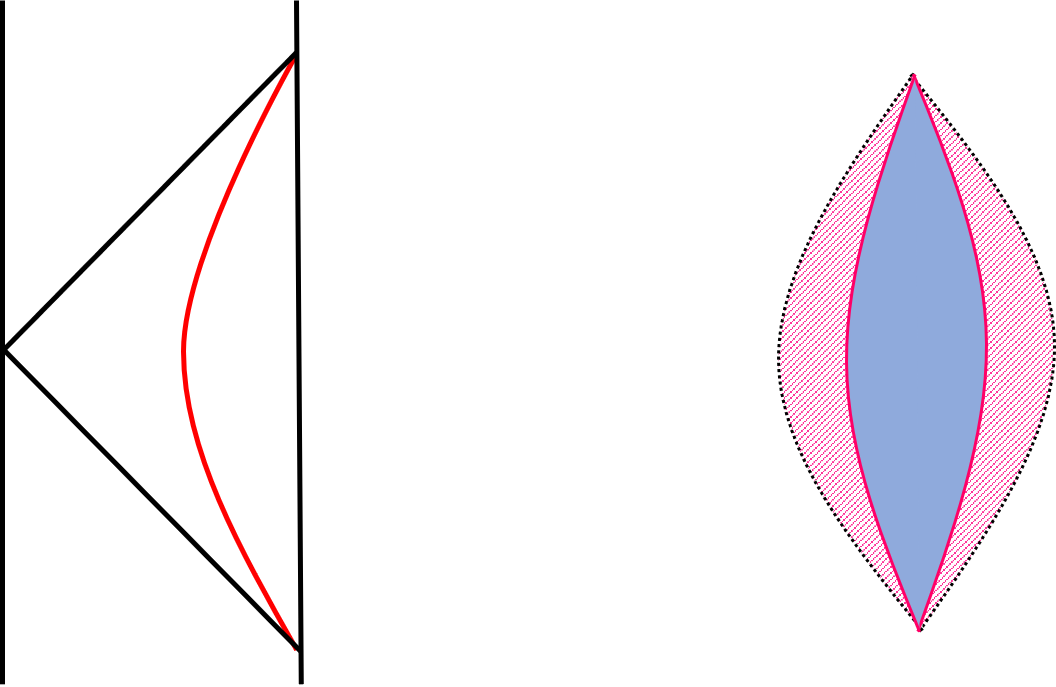}
%\put(-36,80){\Large{$\partial\mathcal{M}$}}
\caption{\small \textbf{RS-I warped solution.} \emph{Left:} Bulk AdS with a `horosphere' (red curve). \emph{Right:} A positive and negative tension brane in bulk AdS, with the shaded (blue) region being retained. The dotted (magenta) regions are excluded and are identified.}
\label{fig:RSIhoro}\end{figure}

More explicitly, the bulk physics of the RS-I model is characterized by five-dimensional Einstein gravity with a negative cosmological constant $\Lambda_{5}=-6/L_{5}^{2}$, plus an action characterizing the two branes:
\beq
\begin{split}
 I_{\text{RS-I}}&=\frac{1}{16\pi G_{5}}\int d^{5}x\sqrt{-\hat{g}}(\hat{R}-2\Lambda_{5})\\
&-\int dy d^{4}x\sqrt{-h}[(\tau_{+}+\mathcal{L}^{(+)}_{\text{mat}})\delta(y)+(\tau_{-}+\mathcal{L}_{\text{mat}}^{(-)})\delta(y-y_{-})]\;.
\end{split}
\eeq
Here, $y$ denotes a bulk spatial coordinate to be integrated over, $h_{ij}$ is the four-dimensional induced metric of the constant-$y$ hypersurfaces, and $\mathcal{L}_{\text{mat}}^{(\pm)}$ refer to matter Lagrangians localized on their respective branes. A negative tension brane is positioned at $y=y_{-}>0$ while a positive tension brane is at $y=0$, and only the region $0<y<y_{-}$ is retained. To complete the space, a $\mathbb{Z}_{2}$ reflection symmetry is assumed across each boundary plane, leading to a jump discontinuity in the extrinsic curvature of each brane. Thus, aside from the bulk Einstein equations, the bulk must obey brane boundary conditions, e.g., the Israel junction conditions which relate the discontinuity in the extrinsic curvature across the branes to the brane tension and the stress tensor $S^{(\pm)}_{ij}$,
\beq \Delta K_{ij}^{\pm}-h_{ij}\Delta K^{\pm}=8\pi G_{5}(\tau_{\pm}h_{ij}+S^{(\pm)}_{ij})\;,\eeq
where $\Delta K_{ij}=K^{+}_{ij}-K^{-}_{ij}$ denotes the difference between the extrinsic curvature across the branes. Imposing the junction conditions sets the branes to have equal and opposite tensions. 
%Solutions to the bulk thus must obey the bulk Einstein equations and the brane boundary conditions.
The simplest such solution to the bulk Einstein equations plus Israel junction conditions is the warped geometry (in horospherical coordinates)
\beq ds^{2}_{5}=dy^{2}+e^{-2|y|/L_{5}}\eta_{ij}dx^{i}dx^{j}\;,\label{eq:warped5RSI}\eeq
where $\eta_{ij}$ is the four-dimensional Minkowski spacetime, though may be replaced by any four-dimensional metric which solves the four-dimensional brane theory. The AdS$_{5}$ horizon is located at $y=\infty$ (which includes a point at infinity) where the coordinate system (\ref{eq:warped5RSI}) breaks down (see Figure \ref{fig:RSIhoro} for an illustration). Mass scales on the negative tension brane will be exponentially suppressed with respect to the positive tension brane. Consequently, beings confined to the negative tension brane would experience weak gravity, while beings on the positive tension (`Planck') brane would experience strong gravity. The RS-I model thus provides a possible solution to the hierarchy problem of the standard model.
%The effective four-dimensional theory on an asymptotically flat brane will have its four-dimensional Planck mass $M^{(4)}_{\text{P}}=G_{4}^{-1/2}$ (setting $c=\hbar=1$) induced from the five-dimensional Planck mass $M^{(5)}_{\text{P}}=G_{5}^{-1/3}$ via $(M^{(4)}_{\text{P}})^{2}=L_{5}(M_{\text{P}}^{(5)})^{3}(e^{2y_{-}/L_{5}}-1)$. 

The starting point for the RS-II model is the same except that the negative tension brane of the RS-I model is sent off to infinity ($y_{-}\to\infty$), such that the five-dimensional warped geometry (\ref{eq:warped5RSI}) takes the form
\beq 
\begin{split}
ds^{2}_{5}&\to dy^{2}+e^{-2|y|/L_{5}}\eta_{ij}dx^{i}dx^{j}\\
&=\frac{L_{5}^{2}}{(L_{5}+|z|)^{2}}(dz^{2}+\eta_{ij}dx^{i}dx^{j})\;,
\end{split}
\eeq
where in the second line the conformal coordinate $z\equiv \text{sign}(y)L_{5}(e^{|y|/L_{5}}-1)$ was introduced. Unlike the RS-I model, the RS-II model is not used to solve the hierarchy problem. Further, although the extra bulk spatial dimension becomes infinite in extent, the higher-dimensional gravity becomes localized on the brane nonetheless.\footnote{Simply, KK excitations are light, have non-vanishing momentum along the extra dimension, and become suppressed near the brane, essentially decoupling from matter fields on the brane. Gravitational interactions between the matter fields are mediated by the `zero mode'.} One way to see this is through a perturbative analysis of the weak gravitational created by isolated matter sources on the brane \cite{Sasaki:1999mi,Garriga:1999yh,Csaki:2000fc}.\footnote{Garriga and Tanaka \cite{Garriga:1999yh} applied a similar perturbative analysis to the RS-I scenario, and found the effective linearized gravity on either brane to be Brans-Dicke theory (with different Brans-Dicke parameters), where the Brans-Dicke scalar, i.e,. the `radion' captures the displacement between the branes.} Define a bulk metric perturbation by $\gamma_{ab}\equiv \hat{g}_{ab}-\hat{g}_{ab}^{(0)}$ (with $\hat{g}^{(0)}_{ab}$ being the unperturbed five-dimensional background). Working in the Randall-Sundrum gauge, 
\beq h_{zz}=h_{iz}=0\;,\quad \partial_{i}h^{i}_{\;j}=0\;,\quad h^{i}_{\;i}=0\;,\eeq
the linearized equations of motion for the metric perturbation are
\beq [-\partial_{z}^{2}+V(z)]\psi_{ij}=\eta^{kl}\partial_{k}\partial_{l}\psi_{ij}\;,\label{eq:lineomsRSII}\eeq
for $\psi_{ij}\equiv \sqrt{|z|+L_{5}}h_{ij}$ and `Volcano' potential 
\beq V(z)=\frac{15}{4(|z|+L_{5})^{2}}-3L_{5}^{2}\delta(z)\;.\eeq
Separating variables as $\psi_{ij}\sim u_{m}(z)e^{ik_{j}x^{j}}$, one obtains an eigenvalue equation for eigenfunctions $u_{m}(z)$ characterizing an effective four-dimensional massive mode with mass $m^{2}=-k_{i}k^{i}$. The general solution for $u_{m}(z)$ is given in terms of Bessel functions. Alongside the massive modes is a discrete set of massless `zero modes'. From the mode functions, the Green's function for $-\partial_{z}^{2}+V(z)$ from which the induced brane metric for a spherical source of mass $M$ will have components (in the RS gauge)
\beq h_{tt}=\frac{2G_{4}M}{r}\left(1+\frac{2L_{5}^{2}}{3r^{2}}\right)\;,\quad h_{ij}=\frac{2G_{4}M}{r}\left(1+\frac{L_{5}^{2}}{3r^{2}}\right)\delta_{ij}\;.\eeq
The $h_{tt}$ metic component is simply a modified Newtonian potential for a source $M$ where the $1/r^{3}$ correction can be shown to be precisely of the form of the corrections to the Newtonian due to 1-loop quantum effects induced by the four-dimensional graviton propagator \cite{Duff:1974ud,Duff:2000mt}. 

\vspace{2mm}

\noindent \textbf{KR model.} An important extension of the RS-II model is the Karch-Randall braneworld model \cite{Karch:2000ct}. In this set-up, the positive tension brane is detuned such that the four-dimensional brane geometry is asymptotically AdS$_{4}$.\footnote{Stringy realizations of the KR construction were uncovered in \cite{Karch:2001cw,Karch:2000gx,DeWolfe:2001pq,Aharony:2003qf} (and more recently, \cite{Karch:2022rvr}).} To this, consider the RS-II model with a two-sided brane with a decreased tension
\beq \tau=\frac{3(1+\delta)}{4\pi L_{5}G_{5}}\;,\eeq
for $\delta<0$. The (unperturbed) five-dimensional warped geometry takes the form
\beq ds^{2}=\frac{L_{5}^{2}}{L_{4}^{2}\sin^{2}[(|z|+z_{0})/L_{4}]}(h_{ij}dx^{i}dx^{j}+dz^{2})\;,\eeq
with
\beq h_{ij}dx^{i}dx^{j}=-\left(1+\frac{r^{2}}{L_{4}^{2}}\right)dt^{2}+\left(1+\frac{r^{2}}{L_{4}^{2}}\right)^{-1}dr^{2}+r^{2}d\Omega_{2}^{2}\;,\eeq
and $z_{0}=L_{4}\text{arcsin}(L_{5}/L_{4})$. The brane is located at $z=0$ and has an effective negative cosmological constant $\Lambda_{4}=-3/L_{4}^{2}\equiv3(2\delta+\delta^{2})/L_{5}^{2}$. A perturbative analysis similar to the RS-II case yields a linearized equation (\ref{eq:lineomsRSII}) though with a different potential $V(z)$. While we will not go into the details, the shape of the potential in the KR model leads to a totally discrete mass $m^{2}$ spectrum. Further, the  $m^{2}=0$ zero-mode are not normalizable. Imposing a normalizability condition shifts the zero mode to an `almost zero mode' with mass $m\sim\mathcal{O}(L_{5}/L_{4}^{2})$. Hence, localized four-dimensional gravity has a massive graviton. Massive gravity theories have the behavior that gravitational attraction becomes short-ranged as opposed to long range, and where gravity waves propagate at subluminal speeds. 
%Thus, massive gravity seemingly does not describe our universe. 
Note, however, for regions where $L_{5}/L_{4}^{2}\ll1$ the graviton becomes essentially massless. 

\vspace{2mm}

\noindent \textbf{DGP model.} In the DGP model, an Einstein-Hilbert term is added to the brane action. The original incarnation of the model consisted of higher-dimensional Einstein gavity with vanishing cosmological constant with the bulk being described by the action 
\beq I_{\text{DGP}}=\frac{1}{16\pi G_{5}}\int d^{5}x\sqrt{-\hat{g}}\hat{R}-\int d^{4}x\sqrt{-h}\left(\frac{1}{16\pi G_{4}}R_{\text{4D}}+\mathcal{L}_{\text{mat}}\right)\;.\eeq
A distinguishing feature of this model is that the graviton propagator is such that the effective four-dimensional theory is left unmodified at short distances.

\subsection*{Stringy connections}

The codimension-1 branes described above are not the same types of branes which appear in string theory, however, there is a stringy connection via AdS/CFT duality. Recall that in string theory Dp-branes are extended objects $p$ (spatial)-dimensional objects open strings with endpoints obeying Dirichlet boundary conditions end on. D-branes are to be viewed as physical, dynamical objects with their own mass and (Ramond-Ramond) charges, charges carried by the string endpoints \cite{Polchinski:1995mt}. In the case of $N$ parallel Dp-branes, each open string endpoint can lie on one of $N$ different branes, yielding a total od $N^{2}$ choices. When these $N$ parallel $Dp$-branes are coincident at the same location, the resulting string theory has a particle spectrum which in part describes a $U(N)$ gauge theory. 

Since string theories have a low energy effective description in terms of an appropriate theory of supergravity, a D-brane or a stack of coincident Dp-branes have a realization as a supergravity solution with of a corresponding mass and charge. A relevant example is a stack of $N$ coincident D3-branes in ten-dimensional spacetime. Such a set D-brane configuration exists in type IIB string theory with string coupling $g_{s}$ and string length scale $\ell_{s}=\sqrt{\alpha'}$. In the decoupling limit, where $\alpha'\to0$ and the ratio of the distance between the non-coincident configuration of parallel branes and coupling, $u\equiv r/\alpha'$, is kept fixed, the gauge theory on the brane is that of four-dimensional $\mathcal{N}=4$ SU(N) super-Yang-Mills theory (a superconformal field theory). The corresponding supergravity solution has line element \cite{Horowitz:1991cd} 
\beq ds^{2}_{10}=f^{-1/2}dx_{\parallel}+f^{1/2}dx_{\perp}^{2}\;,\quad f=1+\frac{4\pi g_{s}\alpha'^{2}}{r^{4}}\;.\label{eq:D3branemet}\eeq
Here $dx_{\parallel}^{2}$ denotes the set of coordinates along the four-dimensional worldvolume of the D3-brane configuration, while $dx_{\perp}^{2}=(dr^{2}+r^{2}d\Omega_{5}^{2})$ is the collection of coordinates of the six-dimensional space perpendicular to the stack of branes where the radial coordinate $r$ characterizes the distance between the branes (before taking the coincident limit) and $d\Omega_{5}^{2}$ denotes the metric of the unit 5-sphere. 

Introduce coordinate $u\equiv r/\alpha'$ such that $f=1+\frac{4\pi g_{s}N}{\alpha'^{2}u^{4}}$. The supergravity geometry (\ref{eq:D3branemet}) of the  aforementioned decoupling limit ($\alpha'\to0$, $u$ fixed) is valid when $g_{s}N\gg1$, i.e., at large 't Hooft coupling $\lambda\equiv g_{s}N$. In this  `near horizon' limit, the 10-dimensional geometry is approximately 
\beq ds^{2}_{10}\approx \ell_{s}^{2}\left[\frac{u^{2}}{\sqrt{4\pi g_{s}N}}dx_{\parallel}^{2}+\sqrt{4\pi g_{s}N}\frac{du^{2}}{u^{2}}+\sqrt{4\pi g_{s}N}d\Omega_{5}^{2}\right]\;.\eeq
This line element is simply AdS$_{5}\times S^{5}$, with AdS$_{5}$ length  $L_{5}=\sqrt{4\pi g_{s}N}\ell_{s}^{2}$ and a 5-sphere of the same radius.\footnote{A coordinate rescaling brings us to the $L_{5}=L_{\text{P}}^{(10)}(g_{s}N)^{1/4}$ in Section \ref{app:evapbhshighd}.} In other words, the supergravity solution has a dual description in terms of the four-dimensional superconformal field theory, a sharp realization of the AdS/CFT correspondence \cite{Maldacena:1997re}.

Now return to the RS-II model describing a single codimension-1 brane sitting inside of an asymptotically AdS$_{5}$ bulk spacetime. There are two comments worth mentioning here. First, Gubser observed when the bulk is the AdS$_{5}$-Schwarzschild black hole, the induced geometry on the brane is that of radiation dominated FRW cosmology \cite{Gubser:1999vj}. Via AdS/CFT, the thermal nature of the bulk black hole has a dual description in terms of a thermal CFT at strong coupling with temperature given by the Hawking temperature of the bulk black hole. This CFT can be viewed as living on the brane giving rise to the (`dark') radiation cosmology. Second, in taking the near-horizon limit of the stack of D3-branes, the RS model may be viewed as effectively cutting off the spacetime outside of the stack of D-branes, such that the RS-brane serves as an end-of-the-world brane. 

The precise connection to AdS/CFT is different for the Karch-Randall construction \cite{Karch:2000ct}. This is because in such a set-up one is no longer in a near-horizon limit of a stack of D3-branes. Rather, one realization is that of a D5-probe brane intersecting a stack of D3 branes \cite{Karch:2000gx,DeWolfe:2001pq}, leading to a doubly holographic interpretation (Section \ref{sec:holoren}).

%%%%%%%%%%%%%%%%%%%%%%%%%%%%%%%%%%%%%%%%%%%%%%%%%%%%%%%%%%%%%%%%%%%%
\section{Geometric elements of the C-metric} \label{app:AdSCmetprops}

The C-metric, as noted in the main text, can be interpreted as an accelerating or a pair of accelerating black holes.\footnote{Historically, the original C-metric  belonged to a classification of types of black hole solutions to Einstein-Maxwell theory owed to Levi-Civita in 1918 \cite{Levicivita1918}. These solutions we rediscovered in the 1960's and further classified, particularly by Ehlers and Kundt \cite{Ehlers:1962zz}, giving the naming scheme of black holes of A,B and C-type metrics. It was not until 1970 that Kinnersley and Walker understood the C-metric as an accelerated black hole \cite{Kinnersley:1970zw}. In 1976, Plebanski and Demianski \cite{Plebanski:1976gy} showed how the C-metric is embedded in a larger family of algebraic type-D solutions. For more on the history and aspects of the C-metric, see \cite{Appels:2018jcs}. } In this Appendix we briefly review the origins of the C-metric and then describe some of its geometric features relevant for the main text. 

\subsection{C-metric from Plebanski-Demianski}

In order to better understand the origins of the C-metric, it is useful to analyse its parent metric first, the Plebanski-Demianski (PD) geometry \cite{Plebanski:1976gy}.\footnote{Refer to Eq. (2.1) \cite{Plebanski:1976gy} with metric functions in Eq. (3.25). Here, however, we take $p,q$ and $\sigma$ to have opposite signs. For further details of the PD spacetimes and its various limits, see, e.g., \cite{Griffiths:2005qp,Podolsky:2006px,Griffiths:2006tk,Podolsky:2021zwr}.} The PD metric describes the most general algebraic type-D spacetime solving Einstein-Maxwell theory in vacuum with an aligned, non-zero electromagnetic field and a cosmological constant $\Lambda$. The line element is
\beq 
\begin{split}
 ds^{2}&=\frac{1}{(p+q)^{2}}\biggr\{-\frac{Q(q)}{1+(pq)^{2}}(d\tau+p^{2}d\sigma)^{2}+\frac{1+(pq)^{2}}{Q(q)}dq^{2}+\frac{1+(pq)^{2}}{P(p)}dp^{2}\\
 &+\frac{P(p)}{1+(pq)^{2}}(d\sigma-q^{2}d\tau)^{2}\biggr\}\;,   
\end{split}
\eeq
with metric functions 
\beq 
\begin{split}
 &Q=-\frac{\Lambda}{6}+g^{2}-\gamma-2nq+\epsilon q^{2}-2mq^{3}+\left(-\frac{\Lambda}{6}+e^{2}+\gamma\right)q^{4}\;,\\
 &P=-\frac{\Lambda}{6}-g^{2}+\gamma-2np-\epsilon p^{2}-2mp^{3}+\left(-\frac{\Lambda}{6}-e^{2}+\gamma\right)p^{4}\;.
\end{split}
\label{eq:metfuncsPD1}\eeq
The solution is completely characterized by seven parameters, namely, the cosmological constant $\Lambda$, and the real parameters $g,e,n,m,\epsilon,\gamma$. Of these, only $e$ and $g$ have a clear physical interpretation, corresponding to electric and magnetic charge, respectively. Parameters $m,n,\epsilon$ and $\gamma$ have a less clear interpretation in general, however, often $m$ and $n$ play the role of the mass and NUT parameters, respectively, while in certain instances $\gamma$ and $\epsilon$ are interpreted as angular momentum and acceleration. %Note further $m,n,e$, and $g$ are said to be `dynamical parameters' in that they feature in the Weyl spinor and hence generate curvature, while $\gamma$ and $\epsilon$ do not and are dubbed `kinematical parameters'.
For convenience, one typically shifts $\gamma\to\gamma+g^{2}+\frac{\Lambda}{6}$, such that the metric functions (\ref{eq:metfuncsPD1}) become
\beq 
\begin{split}
&Q=-\frac{\Lambda}{3}-\gamma-2nq+\epsilon q^{2}-2mq^{3}+(\gamma+e^{2}+g^{2})q^{4}\;,\\
&P=\gamma-2np-\epsilon p^{2}-2mp^{3}-\left(\gamma+e^{2}+g^{2}+\frac{\Lambda}{3}\right)p^{4}\;.
\end{split}
\eeq
Further, while $\Lambda$ here is generic, we will be primarily interested in the $\text{AdS}_{4}$ C-metric, where $\Lambda=-3/L_{4}^{2}$, with $L_{4}$ being the $\text{AdS}_{4}$ curvature scale. 

By appropriate coordinate and parameter rescalings, the C-metric emerges from the PD metric. To see this, first introduce two real parameters $a$ and $A$ and perform the following coordinate rescalings
\beq p\to\sqrt{aA}p\;,\quad q\to\sqrt{aA}q\;,\quad \tau\to\sqrt{\frac{a}{A^{3}}}\tau\;,\quad \sigma\to\sqrt{\frac{a}{A^{3}}}\sigma\;.\eeq
Additionally, rescale the parameters as
\beq 
\label{eq:c-metric1}
\begin{split}
&m\to\left(\frac{A}{a}\right)^{3/2}m\;,\quad n\to\left(\frac{A}{a}\right)^{1/2}n\;,\quad e\to\frac{A}{a}e\;,\quad g\to\frac{A}{a}g\;,\\
&\epsilon\to\frac{A}{a}\epsilon\;,\quad \gamma \to A^{2}\gamma\;,\quad P\to A^{2}P\;,\quad Q\to A^{2}Q\;.
\end{split}
\eeq
The resulting metric is the $\text{AdS}_{4}$ (spinning) C-metric 
\beq 
\begin{split}
ds^{2}&=\frac{1}{A^{2}(p+q)^{2}}\biggr\{-\frac{Q(q)}{1+(aApq)^{2}}(d\tau+aAp^{2}d\sigma)^{2}+\frac{1+(aApq)^{2}}{Q(q)}dq^{2}\\
&+\frac{1+(aApq)^{2}}{P(p)}dp^{2}+\frac{P(p)}{1+(aApq)^{2}}(d\sigma-aAq^{2}d\tau)^{2}\biggr\}\;,
\end{split}
\label{eq:spinningCmetric}\eeq
together with
\beq 
\begin{split}
 &Q=\frac{1}{A^{2}L_{4}^{2}}-\gamma-\frac{2nq}{A}+\epsilon q^{2}-2mAq^{3}+A^{2}(\gamma a^{2}+e^{2}+g^{2})q^{4}\;,\\
 &P=\gamma-\frac{2np}{A}-\epsilon p^{2}-2mAp^{3}+\left(\frac{a^{2}}{L_{4}^{2}}-A^{2}(\gamma a^{2}+e^{2}+g^{2})\right)p^{4}\;.
\end{split}
\label{eq:metfuncsCmetv1}\eeq
Notably, the parameters $\epsilon$ and $\gamma$ do not affect the geometry of the solution, as they do not appear in any curvature invariants \cite{Plebanski:1976gy}, and sometimes dubbed `kinematical parameters'. Consequently, these are parameters are gauge choices and can be chosen to take specific values without affecting the local geometry. Alternatively,  $m,n,e,g$ are said to be `dynamical parameters' in that they generate curvature; setting $m=n=e=g=0$ results in a maximally symmetric background with cosmological constant $\Lambda$.

\subsection*{C-metric used for braneworld black hole}

As written, the C-metric line element in (\ref{eq:c-metric1}) is cosmetically quite different from the one used for studying braneworld black holes in \cite{Emparan:1999wa,Emparan:1999fd}. To clarify the relation, first rename coordinates and metric functions as $q\to-y$, $p\to x$, $Q\to-H$, $P\to G$, $\tau\to t$, $\sigma\to\phi$, and $\epsilon\to-k$. Then, the metric functions (\ref{eq:metfuncsCmetv1}) become
\beq 
\begin{split}
 &H=-\frac{1}{A^{2}L^{2}_{4}}+\gamma-\frac{2n}{A}y+ky^{2}-2mAy^{3}-A^{2}(\gamma a^{2}+e^{2}+g^{2})y^{4}\;,\\
 &G=\gamma-\frac{2n}{A}x+kx^{2}-2mAx^{3}+\left[\frac{a^{2}}{L_{4}^{2}}-A^{2}(\gamma a^{2}+e^{2}+g^{2})\right]x^{4}\;.
\end{split}
\eeq
Now, making use of the aforementioned gauge freedom we set $\gamma=1$, and  further rescale $A^{2}a^{2}\equiv a^{2}$ and $A^{2}(e^{2}+g^{2})=q^{2}$. Then, 
\beq 
\begin{split}
&H=-\frac{1}{A^{2}L_{4}^{2}}+1-\frac{2n}{A}y+ky^{2}-2mAy^{3}-(a^{2}+q^{2})y^{4}\;,\\
&G=1-\frac{2n}{A}x+kx^{2}-2mAx^{3}+\left[\left(\frac{1}{A^{2}L_{4}^{2}}-1\right)a^{2}-q^{2}\right]x^{4}\;.
\end{split}
\eeq
Further, introduce $\lambda\equiv \frac{1}{A^{2}L_{4}^{2}}-1$, yielding
\beq 
\begin{split}
&H=-\lambda-\frac{2n}{A}y+k y^{2}-2mAy^{3}-(a^{2}+q^{2})y^{4}\;,\\
&G=1-\frac{2n}{A}x+kx^{2}-2mAx^{3}+(\lambda a^{2}-q^{2})x^{4}\;.
\end{split}
\label{eq:metfuncscmetricv2}\eeq
The line element, meanwhile, is
\beq 
\begin{split}
ds^{2}&=\frac{1}{A^{2}(x-y)^{2}}\biggr[\frac{H(y)}{\Sigma(x,y)}(dt+ax^{2}d\phi)^{2}-\frac{\Sigma(x,y)}{H(y)}dy^{2}\\
&+\frac{\Sigma(x,y)}{G(x)}dx^{2}+\frac{G(x)}{\Sigma(x,y)}(d\phi-ay^{2}dt)^{2}\biggr]\;,
\end{split}
\label{eq:rotCmetapp2}\eeq
with 
\beq \Sigma(x,y)=1+a^{2}x^{2}y^{2}\;.\eeq
In this context, $q$ represents both electric and magnetic charge, $A$ is the `acceleration' parameter, $a$ the angular rotation, and $n$ is the NUT parameter. Setting $n=q=0$ in the metric functions (\ref{eq:metfuncscmetricv2}) recovers the form of the C-metric used in \cite{Emparan:1999wa,Emparan:1999fd} (cf. Eq. (5.1) of \cite{Emparan:1999fd}). We will explore further details of this metric momentarily.

\vspace{2mm}

\noindent \textbf{Rindler form.} It is easiest to see the accelerating nature of the C-metric by setting $a=n=q=m=0$, $k=-1$ and $L_{4}\to\infty$ such that $\lambda=-1$. Further rescale $t\to At$. Then the line element (\ref{eq:rotCmetapp2}) with metric functions (\ref{eq:metfuncscmetricv2}) becomes
\beq ds^{2}=\frac{1}{A^{2}(x-y)^{2}}\left[-(y^{2}-1)A^{2}dt^{2}+\frac{dy^{2}}{(y^{2}-1)}+\frac{dx^{2}}{(1-x^{2})}+(1-x^{2})d\phi^{2}\right]\;.\eeq
Next introduce the coordinate transformations
\beq \xi=\frac{\sqrt{y^{2}-1}}{(x-y)}\;,\quad \rho=\frac{1}{A}\frac{\sqrt{1-x^{2}}}{(x-y)}\;,\eeq
resulting in
\beq ds^{2}=-\xi^{2}dt^{2}+A^{-2}d\xi^{2}+d\rho^{2}+\rho^{2}d\phi^{2}\;,\eeq
which we recognize as the four-dimensional Rindler metric in cylindrical coordinates. An acceleration horizon with acceleration $A$ occurs at $\xi=0$, or, equivalently, $y=1$.

\vspace{2mm}

\noindent \textbf{Boyer-Lindquist-from.} It is also useful to express the metric in a Boyer-Lindquist-like form. Specifically, to describe an $\text{AdS}_{3}$ Karch-Randall brane, make the identifications
\beq \lambda=\frac{\ell^{2}}{\ell_{3}^{2}}\;,\quad A=\frac{1}{\ell}\;,\quad k=-\kappa\;,\quad 2mA=\mu\;,\quad a\to\frac{a}{\sqrt{\lambda}\ell_{3}}\;,\eeq
together with the coordinate rescaling and change
\beq t\to\frac{t}{\ell}\;,\quad y=-\frac{\ell}{r}\;.\label{eq:coordrescalBL}\eeq
Then the metric (\ref{eq:rotCmetapp2}) becomes
\beq 
\begin{split}
 ds^{2}&=\frac{\ell^{2}}{(\ell+xr)^{2}}\biggr\{-\frac{H(r)}{\Sigma(x,r)}(dt+ax^{2}d\phi)^{2}+\frac{\Sigma(x,r)}{H(r)}dr^{2}\\
 &+r^{2}\left[\frac{\Sigma(x,r)}{G(x)}dx^{2}+\frac{G(x)}{\Sigma(x,r)}\left(d\phi-\frac{a}{r^{2}}dt\right)^{2}\right]\biggr\}
\end{split}
\label{eq:BLformv1Cmet}\eeq
with metric functions
\beq 
\begin{split} 
&H(r)\equiv-\frac{r^{2}}{\ell^{2}}H(-\ell/r)=\frac{r^{2}}{\ell_{3}^{2}}+\kappa-2nr-\frac{\mu\ell}{r}+\frac{(a^{2}+q^{2}\ell^{2})}{r^{2}}\;,\\
&G(x)=1-2n\ell x-\kappa x^{2}-\mu x^{3}+\left(\frac{a^{2}}{\ell_{3}^{2}}-q^{2}\right)x^{4}\;,\\
&\Sigma(x,r)=1+\frac{a^{2}x^{2}}{r^{2}}\;.
\end{split}
\eeq
Note, the $n$ parameter may always be rescaled by $\ell$ or $\ell^{-1}$.  Setting $n=0$ and taking the zero acceleration limit $\ell\to\infty$ (upon setting $\kappa=+1$, and rescaling $\mu\ell\to 2M$ and similarly for charge $q^{2}$), we recover the Kerr-Newman-AdS black hole in Boyer-Lindquist coordinates, with $x=\cos\theta$. Alternatively, the parameter space giving positive curvature on the brane is easily reached via the replacements $\ell_{3}\to iR_{3}$ and set $a\to-a$ and $\kappa=+1$.

\subsection*{Factorized C-metric}

Thus far, the metric factors $P,Q$ or $H,G$ are quadratic functions of a single coordinate and are not readily factorizable. This makes finding the roots of the metric functions complicated. Further, for the rotating C-metric, even when $n=0$, the solution still has a non-zero NUT charge. In 2003 and 2004 Hong and Teo \cite{Hong:2003gx,Hong:2004dm} found a way to express the C-metric in a factorized form, which not only simplified some calculations, but also showed the factorized version of the rotating C-metric is physically distinct from its original non-factorized form.  

To arrive at a factorized form of the rotating C-metric, recall $\text{AdS}_{4}$ spinning C-metric (\ref{eq:spinningCmetric}) with metric functions $Q$ and $P$ (\ref{eq:metfuncsCmetv1}). Then, following Hong and Teo \cite{Hong:2004dm}, use gauge freedom to set the kinematical parameters $\gamma$ and $\epsilon$ to 
\beq \gamma=1\;,\quad \epsilon=1+\frac{a^{2}}{L_{4}^{2}}-A^{2}(a^{2}+e^{2}+g^{2})\;.\eeq
Then, in order to place the metric functions (\ref{eq:metfuncsCmetv1}) in a factorized form, further set $n=-mA^{2}$, leading to the C-metric (\ref{eq:spinningCmetric}), now with factorized metric functions, 
\beq 
\begin{split}
&Q=(q^{2}-1)\left(1-2mAq+A^{2}(a^{2}+e^{2}+g^{2})q^{2}\right)+\frac{1}{A^{2}L^{2}_{4}}(1+a^{2}A^{2}q^{2})\;,\\
&P=(1-p^{2})\left(1+2mAp+p^{2}\left[A^{2}(a^{2}+e^{2}+g^{2})-\frac{a^{2}}{L^{2}_{4}}\right]\right)\;.
\end{split}
\label{eq:factorizedPQ}\eeq
Incidentally, the choice $n=-mA^{2}$ which ultimately factorizes the metric functions, leads to a solution with vanishing NUT charge. Consequently, the factorized and non-factorized forms of the C-metric are physically distinct, and are not related by a coordinate transformation. Note, however, in the non-rotating case ($a=0$) with $\Lambda=0$, the factorized and non-factorized C-metrics are related by a coordinate transformation are thus physically equivalent \cite{Hong:2003gx} (in the limit $a=0$, $n$ acts as a kinematical parameter). The difference between factorized and non-factorized rotating C-metrics arises because the latter has `torsion singularities', conical singularities which have a non-zero angular velocity, while the former do not have torsion singularities. The effect is that the non-factorized rotating C-metric will possess closed timelike curves in a neighborhood of the torsion singularities \cite{Bonnor:2002fk}.\footnote{Hong and Teo note that the only way to remove the closed timelike curves is when the angular velocity of the conical singularities have the same constant value along the entire axis of symmetry \cite{Hong:2004dm}.} 

Further, the factorized C-metric, line element (\ref{eq:spinningCmetric}) with functions (\ref{eq:factorizedPQ}), can be brought to a more standard Boyer-Lindquist form (cf. Eq. (2.31) of \cite{Appels:2018jcs}). The (extended) thermodynamics of this form of the factorized C-metric was analyzed in \cite{Appels:2016uha,Appels:2017xoe,Anabalon:2018ydc}.

We can similarly bring the C-metric used to study braneworld black holes, line element (\ref{eq:metfuncscmetricv2}) with metric functions (\ref{eq:rotCmetapp2}), into a factorized form. All that is required is to use gauge freedom to set 
\beq k= -1-\frac{a^{2}}{A^{2}L_{4}^{2}}+(a^{2}+q^{2})\;,\eeq
and remove the NUT charge via $n=-mA^{2}$. The result is the rotating C-metric (\ref{eq:metfuncscmetricv2}), now with metric functions
\beq 
\begin{split}
H&=-(\lambda+1)(1+a^{2}y^{2})+(1-y^{2})\left(1+2mAy+(a^{2}+q^{2})y^{2}\right)\;,\\
G&=(1-x^{2})\left(1+2Amx-(\lambda a^{2}-q^{2})x^{2}\right)\;,
\end{split}
\eeq
with $\lambda+1=(A^{2}L_{4}^{2})^{-1}$. We see the metric function $H$ is completely factorized when $\Lambda=0$. Making identifications $\lambda=\ell^{2}/\ell_{3}^{2}$, $A=\ell^{-1}$, $a\to a/\ell$, $2mA=\mu$, and using the coordinate change (\ref{eq:coordrescalBL}), puts the factorized C-metric in a more Boyer-Lindquist form (\ref{eq:BLformv1Cmet}), with metric functions 
\beq 
\begin{split}
&H(r)=(r^{2}+a^{2})\left(\frac{1}{\ell_{3}^{2}}+\frac{1}{\ell^{2}}\right)+\left(1-\frac{r^{2}}{\ell^{2}}\right)\left(1-\frac{\mu\ell}{r}+\frac{(a^{2}+q^{2}\ell^{2})}{r^{2}}\right)\;,\\
%H(r)=\frac{r^{2}}{\ell^{2}}\biggr\{\left(1+\frac{\ell^{2}}{\ell_{3}^{2}}\right)\left(1+\frac{a^{2}}{r^{2}}\right)-\left(1-\frac{\ell^{2}}{r^{2}}\right)\left(1-\frac{\mu\ell}{r}+\frac{(a^{2}+q^{2}\ell^{2})}{r^{2}}\right)\biggr\}\;,\\
&G(x)=(1-x^{2})\left(1+\mu x-\left(\frac{a^{2}}{\ell_{3}^{2}}-q^{2}\right)x^{2}\right)\;,\quad \Sigma(x,r)=1+\frac{a^{2}x^{2}}{r^{2}}\;.
\end{split}
\eeq

\subsection{Some properties of the C-Metric}\label{app:propsofCmet}

Let us now detail some of the properties of the C-metric used in the main text. Following the conventions of \cite{Emparan:1999fd},  begin with the uncharged, non-rotating $\text{AdS}_{4}$ C-metric
\beq ds^{2}=\frac{1}{A^{2}(x-y)^{2}}\left[H(y)dt^{2}-\frac{dy^{2}}{H(y)}+\frac{dx^{2}}{G(x)}+G(x)d\phi^{2}\right]\;,\label{eq:genAdSC}\eeq
with
\beq H(y)=-\lambda+ky^{2}-2mAy^{3}\;,\quad G(x)=1+kx^{2}-2mAx^{3}\;,\eeq
obeying $H(\chi)=G(\chi)-(1+\lambda)$. Here $k=+1,0,-1$, which will determine the black hole horizon topology. The bulk Ricci tensor satisfies $\hat{R}_{AB}=-(3/L_{4}^{2})\hat{g}_{AB}$ where $L_{4}\equiv (A\sqrt{\lambda+1})^{-1}$ sets the scale for the bulk cosmological constant. Maintaining a negative cosmological constant in the bulk thus requires $\lambda>-1$, while $G(x)\geq0$ to maintain a Lorentzian signature.
%however, as summarized below various ranges of $\lambda$ describe different asymptotic brane geometries.
The overall factor $(x-y)^{-2}$ in (\ref{eq:genAdSC}) implies the point $y=x$ is infinitely far away from points $y\neq x$ (the point $y=x$ corresponds to the asymptotic $\text{AdS}_{4}$ boundary). The curvature invariants
\beq
\begin{split} 
&\hat{R}^{ABCD}\hat{R}_{ABCD}=24 A^{4}[2A^{2}m^{2}(x-y)^{6}+(1+\lambda)^{2}]\;,\\
&\hat{C}^{ABCD}\hat{C}_{ABCD}=48A^{6}m^{2}(x-y)^{2}\;,
\end{split}
\eeq
reveal curvature singularities at $y\pm\infty$ and $x=\pm\infty$, while the Weyl tensor vanishes asymptotically at $y=x$. To avoid naked singularities, one must restrict coordinate ranges for a given set of parameters. To maintain a `mostly plus' Lorentzian signature requires $G(x)\geq0$, restricting the range of $x$. 

To gain further intuition for the C-metric, it is instructive to consider the simplifying case when $mA=0$. Then consider perform the coordinate transformation
%One can then move to a coordinate frame showing the geometry is locally $\text{AdS}_{4}$ with a three-dimensional submanifold with cosmological constant $\Lambda_{3}=-\lambda$ \cite{Emparan:1999fd}. Specifically, via the coordinate transformation
\beq \tilde{r}=\frac{\sqrt{y^{2}+\lambda x^{2}}}{A(x-y)}\;,\quad \rho=\sqrt{\frac{1+kx^{2}}{y^{2}+\lambda x^{2}}}\;,\eeq
such that the metric (\ref{eq:genAdSC}) becomes
\beq ds^{2}=\frac{d\tilde{r}^{2}}{\frac{\tilde{r}^{2}}{\ell_{4}^{2}}-\lambda}+\tilde{r}^{2}\left[-(\lambda\rho^{2}-k)dt^{2}+\frac{d\rho^{2}}{\lambda\rho^{2}-k}+\rho^{2}d\phi^{2}\right]\;.\label{eq:localAdS4m0}\eeq
Locally the geometry is $\text{AdS}_{4}$, where surfaces of constant $\tilde{r}$ have constant Riemann curvature  with a three-dimensional cosmological constant $\Lambda_{3}=-\lambda$. 
Thus, the sign of $\lambda$ denotes different constant curvature slicings of $\text{AdS}_{4}$. There are three distinct cases: \textbf{(1)} $\lambda=0$, a flat slicing. In this case one must choose $k=\pm1$, where for $k=-1$ the coordinate $t$ is timelike everywhere; \textbf{(2)} $-1<\lambda<0$, leads to a three-dimensional de Sitter slicing. One must select $k=-1$ to have $\text{dS}_{3}$ in static patch coordinates and cosmological horizons, and \textbf{(3)} $\lambda>0$, an $\text{AdS}_{3}$ slicing where the three different values of $k$ distinguish three slicings of $\text{AdS}_{3}$: global coordinates ($k=-1$), the massive BTZ black hole ($k=+1$), and the massless BTZ black hole ($k=0$). 
%The flat ($\lambda=0$) solution was studied in \cite{Emparan:1999wa} while the $\text{AdS}_{3}$ slicings were analyzed in \cite{Emparan:1999fd}.

\vspace{2mm}

\noindent \textbf{Conical singularities.} Each zero of $H(y)$ corresponds to a Killing horizon associated with the time translation Killing vector $\partial_{t}$. Meanwhile, the zeros of $G(x)$ corresponds to an axis for the rotation symmetry $\partial_{\phi}$, i.e., for $\xi^{a}=\partial_{\phi}^{a}$, then $\xi^{2}\sim G(x)$, vanishing at a zero of $G(x)$. For a range of values of $mA$ and $k$, there will be three distinct real zeros\footnote{The cubic $G(x)=-2mA x^{3}+kx^{2}+1=0$ can be solved by introducing $x=z-\frac{k}{3}$ and express in depressed form, $z^{3}+pz+q=0$, with $p=-\frac{k^{2}}{12(mA)^{2}}$, $q=-\frac{[2k^{3}+27(4mA)^{2}]}{27(2mA)^{3}}$, and discriminant $\Delta\equiv-(4p^{3}+27q^{2})=-\frac{k^{3}+27(mA)^{2}}{4(mA)^{4}}$. For $\Delta>0$, $G(x)$ has three distinct real roots. For example, for $k=-1$, three real roots $x_{0}<x_{2}<0<x_{1}$ exist when $0<mA<\frac{1}{3\sqrt{3}}$ \cite{Emparan:1999wa}. For $\Delta<0$, $G(x)$ will have one real and two complex roots.}
 to $G(x)$, $\{x_{0},x_{1},x_{2}\}$, with each zero leading to a distinct conical singularity. One singularity can be removed via\footnote{To see this, introduce $\tilde{x}^{2}=4(x-x_{i})/G'(x_{i})$. Expand the $(x,\phi)$ sector of (\ref{eq:genAdSC}) about a zero of $G(x)$,
 $$G^{-1}(x)dx^{2}+G(x)d\phi^{2}\approx[G'(x_{i})(x-x_{i})]^{-1}dx^{2}+G'(x_{i})(x-x_{i})d\phi^{2}=\tilde{x}^{2}(G'(x_{i})/2)^{2}d\phi^{2}+d\tilde{x}^{2}\;.$$ Periodicity (\ref{eq:periodicityphiapp}) then follows from imposing regularity at $\tilde{x}=0$.\label{fn:consingperiod}}
\beq \phi\sim \phi+\Delta\phi(x_{i})\;,\quad \Delta\phi(x_{i})=\frac{4\pi}{|G'(x_{i})|} \;,\label{eq:periodicityphiapp}\eeq
where $x_{i}$ is one of the zeros. Once the period of $\phi$ has been fixed in this way (say at $x=x_{1}$), the coordinate $\phi$ cannot be readjusted to eliminate the remaining conical singularities at $x=x_{0},x_{2}$. Thus, in general there will be a conical singularity along the axis $x=x_{i}\neq x_{1}$ with angular deficit $\delta=\Delta\phi(x_{1})-\frac{4\pi}{G'(x_{i})}.$\footnote{This relation follows from having set the periodicity (\ref{eq:periodicityphiapp}), such that the coordinate range for $\phi$ is to $0\leq\phi\leq \Delta\phi(x_{1})-\delta$, where $\delta$ is the angular deficit associated with the other conical singularities at $x_{i}\neq x_{1}$.}
This may be interpreted as a cosmic string with tension $\tau_{cs}=\delta/8\pi$. It is this feature which leads one to interpret the C-metric as a a single or pair of accelerating black holes. In the case of a single black hole, a cosmic string attaches at one pole in the background and the black hole, suspending it away from the center of the spacetime, thus inducing its acceleration. 
%This leads to an acceleration horizon with its own thermodynamic description.  
When the cosmic string hits the boundary it produces a conical defect in the boundary geometry, which may or may not be hidden by a horizon on the boundary. 

It is possible to for the metric function $G(x)$ has degenerate roots. For example, when $k=-1$ and $mA=1/3\sqrt{3}$, $G(x)$ has a double root $x_{d}$. Near $x=x_{d}$, the $(x,\phi)$-sector of the line element takes the form a Euclidean hyperboloid
\beq ds^{2}_{x_{d}}=\frac{dx^{2}}{(x-x_{d})^{2}}+(x-x_{d})^{2}d\phi^{2}\;,\eeq
 with a spatial divergence at $x=x_{d}$.

%Once the location of the cosmic string is determined, it is possible to
%(see also \cite{Appels:2016uha,Ball:2020vzo}). 

\vspace{2mm}

\noindent \textbf{Boundary geometry.} The asymptotic $\text{AdS}_{4}$ boundary is located at $y=x$. Stripping off the conformal factor via an appropriate conformal transformation, the boundary metric is 
\beq ds^{2}_{\text{bdry}}=H(x)dt^{2}-\frac{dx^{2}}{H(x)G(x)}+G(x)d\phi^{2}\;.\eeq
This line element describes a black hole with a Killing horizon located at values of $x$ which coincide with the zeros $y_{i}$ of $H(y)$, i.e., $x=y_{i}$. Proper distances between points on the boundary are given using this boundary metric, e.g., $x_{\text{proper}}=\int dx(-H(x)G(x))^{1/2}$. While beyond the scope of this review, it is worth highlighting that, in the context of the AdS/CFT correspondence, specific limits of the $\text{AdS}_{4}$ C-metric are dual to holographic $\text{CFT}$s living on fixed three-dimensional black hole backgrounds \cite{Hubeny:2009ru,Hubeny:2009kz}. Namely, two classes of asymptotically $\text{AdS}_{4}$ solutions include: (i) \emph{black funnels}, i.e., solutions that have a single connected but non-compact horizon, and (ii) \emph{black droplets}, i.e., solutions with two disconnected horizons. In either case, the $\text{AdS}_{4}$ solution is attached to the boundary black hole horizon, where in (ii) the (compact) horizon connected to the boundary horizon is the `droplet' and is suspended above a (deformed) planar black hole. Note that the $A\to\infty$ limit of the C-metric with a de Sitter slicing on the $x=0$ hypersurface is a double-Wick rotation of the hyperbolic $\text{AdS}_{4}$ metric (see Appendix C of \cite{Emparan:2022ijy} for details).

%We are interested in introducing a brane into the $\text{AdS}_{4}$ spacetime. Generally there will be a discontinuity in the extrinsic curvature
%\footnote{The brane $\mathcal{B}$ is a timelike 3-surface with spacelike unit normal $n_{i}$ and extrinsic curvature $K_{ij}=\nabla_{i}n_{j}$.} 
%$K_{ij}[h]$ across the brane which, via the Israel junction conditions \cite{Israel:1966rt} (equations of motion for the brane), is related to the stress-tensor $S_{ij}$ introduced by the brane. In the four-dimensional case at hand, where the brane action is purely tensional, the junction conditions are
%\beq \Delta K_{ij}-h_{ij}\Delta K^{k}_{\;k}=8\pi G_{4}\tau h_{ij}\;,\eeq
%where $\Delta K_{ij}=K_{ij}^{+}-K_{ij}^{-}=2K_{ij}$ and $S_{ij}=-\tau h_{ij}$. Therefore, the tension can be seen as a parameter which fixes the location of the brane $\mathcal{B}$. 

\vspace{2mm}

\noindent \textbf{Umbilic surfaces.} The C-metric has the nice property that its $x=0$ and $y=0$ hypersurfaces are umbilic, i.e., when the extrinsic curvature of the hypersurface is proportional to the induced metric. 
%a natural choice for the location of $\mathcal{B}$ is the surface $x=0$ since it is \emph{umbilic}. 
To see this, the outward pointing unit normal to the $x=0$ hypersurface is $n_{x}^{i}=-A\epsilon(x-y)\sqrt{G(x)}\partial^{i}_{x}$, where $\epsilon=\pm1$ (we will take $\epsilon=+1$ since $x=0$ is a timelike hypersurface). The non-vanishing components of the extrinsic curvature $K_{ij}=\nabla_{i}n_{j}$ obey 
\beq K^{(x)}_{ij}=A \epsilon h_{ij}^{(x)}\;,\label{eq:Kxij}\eeq
%\beq
%\begin{split}
%&K_{tt}|_{x=0}=-\frac{1}{Ay^{2}}H(y)=-A h^{(x)}_{tt}|_{x=0}\;,\\
%&K_{yy}|_{x=0}=\frac{1}{Ay^{2}H(y)}=-A h^{(x)}_{yy}|_{x=0}\;,\\
%&K_{\phi\phi}|_{x=0}=-\frac{1}{Ay^{2}}=-A h^{(x)}_{\phi\phi}|_{x=0}\;,
%\end{split}
%\eeq
with $h^{(x)}_{ij}$ being the induced metric along the $x=0$ surface. 
%Comparing to the Israel junction conditions we identify the brane tension (\ref{eq:branetens}).  
%We then perform surgery on the bulk spacetime by cutting $\text{AdS}_{4}$ at this surface, for which on one of the sides of $\mathcal{B}$ there will be no conical singularities. We then take two copies of the conical singularity free side of the spacetime and glue them together along $\mathcal{B}$ such that the resulting spacetime is free of conical singularities. 
Similarly, the $y=0$ hypersurface is umbilic: with outward pointing unit normal $n_{y}^{i}=-A\epsilon(x-y)\sqrt{H(y)}\partial^{i}_{y}$, then
\beq K^{(y)}_{ij}=A\epsilon\sqrt{-\lambda}h^{(y)}_{ij}\;.\label{eq:Kyij}\eeq
%\beq
%\begin{split}
%&K_{tt}|_{y=0}=-\frac{\epsilon}{Ax^{2}}(-\lambda)^{3/2}=-A\epsilon\sqrt{-\lambda}h^{(y)}_{tt}|_{y=0}\;,\\
%&K_{xx}|_{y=0}=-\frac{\epsilon}{A x^{2} G(x)}\sqrt{-\lambda}=-A\epsilon\sqrt{-\lambda}h^{(y)}_{xx}|_{y=0}\;,\\
%&K_{\phi\phi}|_{y=0}=\frac{-\epsilon G(x)}{A x^{2}}\sqrt{-\lambda}=-A\epsilon\sqrt{-\lambda}h^{(y)}_{\phi\phi}|_{y=0}\;,
%\end{split}
%\eeq
where now $h^{(y)}_{ij}$ is the induced metric along $y=0$.
%such that $K_{ij}=(-A\epsilon\sqrt{-\lambda})h^{(y)}_{ij}$.

\subsubsection*{With rotation}

Now consider the neutral rotating $\text{AdS}_{4}$ C-metric, following the conventions of \cite{Emparan:1999fd},
%\footnote{To recover the form of the metric used in \cite{Emparan:2020znc} one makes the identifications in Footnote \ref{fn:convsapp}, together with $\sqrt{\lambda}a\to a/\ell_{3}$, while to recover the metric used in the main text (\ref{eq:AdS4Ccoord}), one identifies $\sqrt{\lambda}a\to -a/iR_{3}$.} 
%The line element is
\beq 
\begin{split}
ds^{2}&=\frac{1}{A^{2}(x-y)^{2}}\biggr[\frac{H(y)}{\Sigma(x,y)}(dt+ax^{2}d\phi)^{2}-\frac{\Sigma(x,y)}{H(y)}dy^{2}+\frac{\Sigma(x,y)}{G(x)}dx^{2}+\frac{G(x)}{\Sigma(x,y)}(d\phi-ay^{2}dt)^{2}\biggr]\;,
\end{split}
\label{eq:rotCmetapp}\eeq
with metric functions
\beq
\begin{split}
&H(y)=-\lambda+ky^{2}-2mAy^{3}-a^{2}y^{4}\;,\quad \Sigma(x,y)=1+a^{2}x^{2}y^{2}\\
&G(x)=1+kx^{2}-2mAx^{3}+a^{2}\lambda x^{4}\;.
\end{split}
\eeq
%The quartic order in $y^{4}$ and $x^{4}$ also appears for the charged C-metric, and it is therefore straightforward to include charge, however, we will not do so here. 
%As in the static case, this spacetime is obeys $\hat{R}_{AB}=-3/L_{4}^{2}\hat{g}_{AB}$ with the  same scale $L_{4}$.
When $m\neq0$, there is a curvature singularity when $1/y^{2}\Sigma(x,y)=0$, i.e., when both $y\to-\infty$ and $x=0$, which may be understood as a ring singularity familiar to Kerr black holes.

The zeros $x_{i}$ of $G(x)$ now correspond to fixed orbits of the rotational Killing vector 
\beq \xi=\partial_{\phi}-ax_{i}^{2}\partial_{t}\;,\eeq
instead of the Killing vector $\partial_{\phi}^{\mu}$ (which no longer has vanishing norm at $x=x_{i}$). Avoiding a conical defect at $x=x_{1}$ requires one identify points along the integral curves of $\xi$ with an appropriate period, amounting to a coordinate transformation $\tilde{t}=t+ax_{1}^{2}\phi$, where $\phi$ has the same period (\ref{eq:periodicityphiapp}). To see this, expand the metric (\ref{eq:rotCmetapp}) near a zero of $G(x)$. Without loss of generality, the $y=0$ slice is, up to the conformal factor
\beq ds^{2}_{y=0}\approx -\lambda(dt+ax_{i}^{2}d\phi)^{2}+\frac{dx^{2}}{G'(x_{i})(x-x_{i})}+G'(x_{i})(x-x_{i})d\phi^{2}\;.\eeq
Aside from the first term, the $(x,\phi)$ sector has the same form as in the non-rotating case, from which the periodicity of $\phi$ is (\ref{eq:periodicityphiapp}). Including rotation, however, this would not be the correct periodicity for $\phi$. The situation is remedied via the coordinate transformation $\tilde{t}=t+ax_{i}^{2}\phi$, such that, at $x=x_{i}$, $d\tilde{t}=(dt+ax_{i}^{2}d\phi)$. Similarly, at the roots $y_{i}$ of $H(y)$, the Killing vector $\zeta=\partial_{t}+ay_{i}^{2}\partial_{\phi}$
becomes null, defining horizons with angular velocity $\Omega=ay_{i}^{2}$. 

 The asymptotic $\text{AdS}_{4}$ boundary is again at $x=y$ and again gives rise to boundary black holes.  Notably, the $A\to\infty$ limit of the rotating C-metric (see below), with $k=+1$, is a double-Wick rotation of Kerr-$\text{AdS}_{4}$ metric considered in \cite{Hubeny:2009ru} (see Appendix B of \cite{Emparan:2020znc}).

Lastly, as in the static case, the $x=0$ and $y=0$ hypersurfaces are umbilic. Indeed, for spacelike unit normal $n_{x}^{i}=A(x-y)\sqrt{G(x)/\Sigma(x,y)}\partial_{x}^{i}$, the extrinsic curvature satisfies $K_{ij}=-Ah^{(x)}_{ij}$ at $x=0$. Similarly, the $y=0$ hypersurface, with unit normal $n^{i}_{y}=A\epsilon(x-y)\sqrt{H(y)/\Sigma(x,y)}\partial_{y}^{i}$, obeys $K_{ij}=(-A\epsilon\sqrt{-\lambda})h^{(y)}_{ij}$.

\subsection{AdS C-metric on the Poincar\'e disk} \label{appsec:poincdisk}

Here we review how to project the C-metric on the two-dimensional Poincar\'e disk of unit radius to produce the plots shown in Figures \ref{fig:KRbrane}, \ref{fig:dSbranebhs} and \ref{fig:RSflatbrane}.  
%As usual, we'll first discuss the case with AdS geometry on the brane, highlighting only the main differences in the other cases.

\subsection*{Hyperbolic disk}

First, recall two-dimensional hyperbolic space $\mathbb{H}^{2}$ can be embedded in three-dimensional Minkowski space coordinatized by $x^\mu = \{x^0,x^1,x^2\}$ and obeying $\eta_{\mu\nu}x^{\mu}x^{\nu}=-1$. Different choices of coordinates $y^i$ on $\mathbb{H}_2$ correspond to different ways of embedding $\mathbb{H}^{2}$. Particularly relevant for our purposes include (see, e.g.,  \cite{Costa:2001ki}):
%some elementary facts about two-dimensional hyperbolic space $\mathbb{H}^{2}$ (following \cite{Costa:2001ki}).
% $\mathbb{H}_2$ can be embedded in $\mathbb{M}_3$ with coordinates 
% $x^\mu = \{x^0,x^1,x^2\}$ as the hypersurface satisfying:
% \beq
% \eta_{\mu \nu} x^\mu x^\nu = -1 \ .
% \eeq
%Then, different choices of coordinates $y^i$ on $\mathbb{H}_2$ correspond to different parametrisations to such a hypersurface. The ones we will find useful are:
\begin{itemize}

 \item \textbf{Hyperbolic:} Let $\rho \in [0,\infty)$ be a radial coordinate and $\omega \in (-\infty,\infty)$ a hyperbolic angle. Then the coordinate parametrization 
    \begin{equation}
    \begin{cases} 
    \label{eq:hypercoord}
      x^0 = \cosh{\rho}\cosh{\omega}  \\
      x^1 = \sinh{\rho} \\
      x^2 = \cosh{\rho} \sinh{\omega}  
   \end{cases}
    \end{equation}
    leads to the induced two-dimensional geometry for $\mathbb{H}^{2}$
    \beq\label{eq:hyperbolicmetr}
    ds^2 =\eta_{\mu\nu}dx^{\mu}dx^{\nu}= d\rho^2 + \cosh^{2}(\rho) d\omega^2\;.
    \eeq

    \item \textbf{Polar:} Let $\chi \in (-\infty,\infty)$ be a radial coordinate and $\varphi \in [0,2\pi)$ a polar angle. Then
    \begin{equation}
    \begin{cases} 
    \label{eq:polcoord}
      x^0 = \cosh{\chi}  \\
      x^1 = \sinh{\chi} \cos{\varphi} \\
      x^2 = \sinh{\chi} \sin{\varphi}  
   \end{cases}
    \end{equation}
    yields the two-dimensional induced geometry
    \beq\label{eq:polarmetr}
    ds^2 = d\chi^2 + \sinh^{2}(\chi) d\varphi^2\;.
    \eeq

    \item \textbf{Exponential:} Let $\zeta \in (-\infty,\infty)$ be a radial coordinate and $\Psi \in (-\infty,\infty)$ a hyperbolic angle. With coordinates
    \begin{equation}
    \begin{cases}
    \label{eq:expcoord}
      x^0 = \cosh(\zeta) + e^{-\zeta} \Psi^2/2 \\
      x^1 = e^{-\zeta} \Psi \\
      x^2 = \sinh(\zeta) + e^{-\zeta} \Psi^2/2  
   \end{cases}
    \end{equation}
    the induced two-dimensional metric is 
    \beq\label{eq:exponentialmetr}
    ds^2 = d\zeta^2 + e^{-2\zeta} d\Psi^2 \ .
    \eeq
\end{itemize}
Moreover, note it is possible to project the infinite $\mathbb{H}_d$ hyperbolic space onto the $d$-dimensional unit disk using the Poincar\'e projection 
\beq
    x_\text{P}^i \equiv \frac{x^i}{1+x^0} \ .
\label{eq:Poincproj}\eeq
%We use this projection below. 

\subsection*{AdS$_{3}$ foliations}

As discussed in section \ref{ssec:bulkgeomcmet}, it is possible to bring the static AdS$_{4}$ C-metric with vanishing mass parameter $\mu=0$ into the form of empty AdS$_{4}$ (\ref{eq:AdS4empty}), foliated by slices of AdS$_{3}$ using the coordinate transformation (\ref{eq:coordinatetransflat}). 
% it is possible to bring the static C-metric with vanishing mass into a form in which empty AdS$_4$ is manifestly foliated by AdS$_3$ slices by performing the transformation given in Equation \eqref{coordinatetransflat}.
\beq 
\label{eq:coordAdsapp}\cosh(\sigma)=\frac{\ell_{3}}{L_{4}}\frac{1}{|1+\frac{rx}{\ell}|}\sqrt{1+\frac{r^{2}x^{2}}{\ell_{3}^{2}}}\;,\qquad \hat{r}=r\sqrt{\frac{1-\kappa x^{2}}{1+\frac{r^{2}x^{2}}{\ell_{3}^{2}}}}\;.
\eeq
% bring the  C-metric with $\mu=0$ to empty AdS$_{4}$ form 
% \beq ds^{2}=L_{4}^{2}d\sigma^{2}+\frac{L_{4}^{2}}{\ell_{3}^{2}}\cosh^{2}(\sigma)\left[-\left(\kappa+\frac{\hat{r}^{2}}{\ell_{3}^{2}}\right)dt^{2}+\left(\kappa+\frac{\hat{r}^{2}}{\ell_{3}^{2}}\right)^{-1}d\hat{r}^{2}+\hat{r}^{2}d\phi^{2}\right]\;.\eeq
Constant $t$ and $\phi$ slices of empty AdS$_{4}$ (\ref{eq:AdS4empty})  have the induced geometry 
\beq
ds^{2}=L_{4}^{2}d\sigma^{2}+\frac{L_{4}^{2}}{\ell_{3}^{2}}\cosh^{2}(\sigma)\left(\kappa+\frac{\hat{r}^{2}}{\ell_{3}^{2}}\right)^{-1}d\hat{r}^{2}\;.\eeq
This line element is reminiscent of two-dimensional $\mathbb{H}^{2}$ geometry \eqref{eq:hyperbolicmetr}, where $\sigma$ plays the role of $\rho$ and $\hat{r}$ the hyperbolic angle. Indeed,  the coordinate transformation
\beq
 \psi = \ell_3 \ \text{arcsinh}{\frac{\hat{r}}{\ell_3}} \ ,
\eeq
%yields the hyperbolic angle that plays the role of $\omega$. Indeed, this coordinate transformation
brings the metric on the constant $t$ and $\phi$ slice to
\beq
ds^2 = L_4^2 \left(d\sigma^2 + \cosh^2(\sigma)d\psi^2 \right) \ ,
\eeq
which is conformally equivalent to \eqref{eq:hyperbolicmetr}. One can subsequently use coordinates \eqref{eq:hypercoord} and the Poincar\'e projection (\ref{eq:Poincproj}) to represent lines of constant $x$ and $r$ on the two-dimensional disk, as depicted in Figure \ref{fig:KRbrane}. 

A word of caution on coordinate ranges. To cover the whole disk, $\sigma$ should range over all $\mathbb{R}$. In fact, for constant $t$ and $\phi$ the C-metric only covers half of the disk (here we have chosen the $\sigma >0$ range). Nonetheless, the rest of the disk is recovered by rotation of the $\phi$ coordinate. Moreover, the coordinate transformation \eqref{eq:coordAdsapp} is not one-to-one across the whole range of $(x,r)$ coordinates. To make it so, we restrict the range of $x$ for a given $r$ to
\beq
\begin{cases}
    - \ell/r < x \leq 1 \quad & r>0 \\
    x>1 \quad & r<0 \;.
\end{cases}
\eeq
for inverse acceleration $\ell$.

\subsection*{dS$_{3}$ foliations }

The coordinate transformation (\ref{eq:coordtransdsslice}) 
\beq \sinh(\sigma)=\frac{R_{3}}{L_{4}}\frac{1}{|1+\frac{rx}{\ell}|}\sqrt{1-\frac{x^{2}r^{2}}{R_{3}^{2}}}\;,\quad \hat{r}=r\sqrt{\frac{1-x^{2}}{1-\frac{x^{2}r^{2}}{R_{3}^{2}}}}\;,\label{eq:dsbranecoordapp}\eeq
brings the C-metric (with $\mu=0$) into empty AdS$_{4}$ form (\ref{eq:dS3fols}) foliated by dS$_{3}$ slices (after the $\ell_3 \to iR_3$ Wick rotation). 
% A similar coordinate transformation can bring the C-metric (after the $\ell_3 \to iR_3$ Wick rotation) into a form in which AdS$_4$ is foliated by dS$_3$ slices:
% \beq \sinh(\sigma)=\frac{R_{3}}{L_{4}}\frac{1}{|1+\frac{rx}{\ell}|}\sqrt{1-\frac{x^{2}r^{2}}{R_{3}^{2}}}\;,\quad \hat{r}=r\sqrt{\frac{1-x^{2}}{1-\frac{x^{2}r^{2}}{R_{3}^{2}}}}\;,\eeq
At constant $t$ and $\phi$ slices, the two-dimensional metric is
\beq ds^{2}=L_{4}^{2}d\sigma^{2}+\frac{L_{4}^{2}}{R_{3}^{2}}\sinh^{2}(\sigma)\left(1-\frac{\hat{r}^{2}}{R_{3}^{2}}\right)^{-1}d\hat{r}^{2}\;,\eeq
having the form of \eqref{eq:hyperbolicmetr} where $\sigma$ plays the role of $\chi$ and $\hat{r}$ is related to the angle $\varphi$, with $\hat{r} \in [0,R_3)$. 
%is bounded in the coordinate system, making it natural to relate it to a standard angle rather than a hyperbolic one. 
The coordinate transformation
\beq
    \psi = 2 R_3 \arctan{\left(\frac{\hat{r}}{-R_3 + \sqrt{R_3^2-\hat{r}^2}}    \right)} \ ,
\eeq
brings the spatial metric to
\beq
 ds^2 = L_4^2 \left( d\sigma^2 + \sinh^2{(\sigma)} d\psi^2  \right) \ ,
\eeq
which is conformally equivalent to \eqref{eq:polcoord}. 

Note that $\sigma$ is always positive in the coordinate transformation (\ref{eq:dsbranecoordapp}). Consequently, the parametrization only covers the $\sigma\geq 0$ region of the disk, and cannot be recovered by a simple rotation by  $\phi$ coordinate (as in the AdS case). The other half of the disk is covered by analytic continuation of the coordinate system, yielding Figure \ref{fig:dSbranebhs}.

\subsection*{Flat foliations}
Empty AdS$_{4}$ can also be foliated by Mink$_{3}$, and can be understood as a limiting case of either the AdS$_{3}$ and dS$_{3}$ foliations as, respectively, $\ell_3$ and $R_3$ tend to infinity. In particular, the coordinate transformation
\beq e^\sigma = \ell+xr\;,\quad \hat{r} =  r\sqrt{1-x^{2}}\;,\eeq 
produces the constant $t$-$\phi$ geometry
\beq
ds^2 = \ell^2 \left(d\sigma^2 + e^{-2\sigma} d\hat{r}^2 \right) \ ,
\eeq
coinciding with line element \eqref{eq:exponentialmetr}.
Contrary to the previous examples, here $\hat{r}$  plays the role of the hyperbolic angle. 
%The full disk is covered by these coordinates (no need for rotation).
For the coordinate transformation to be meaningful we restrict the values of $x$ such that
\beq 
- \ell/r \leq x <1 \ ,
\eeq
where, recall, $r\geq 0$. In principle, these coordinates cover only half of the disk, while the other half is recovered by a rotation of the angle $\phi$, leading to Figure \ref{fig:RSflatbrane}.

%%%%%%%%%%%%%%%%%%%%%%%%%%%%%%%%%%%%%%%%%%%%%%%%%%%%%%%%%%%%%%%%%%%%%%%%
\section{On-shell Euclidean action of AdS C-metric} \label{app:onshellaction}

Here we review the derivation of the thermodynamics of the bulk system of a regular (non-rotating) $\text{AdS}_{4}$ C-metric with a Karch-Randall or Randall-Sundrum brane using the on-shell Euclidean gravitational action \cite{Kudoh:2004ub}. 

\subsection*{Geometry} 

In Lorentzian signature, we work with the $\text{AdS}_{4}$ C-metric in the form
\beq ds^{2}=\frac{\ell^{2}}{(x-y)^{2}}\left(-H(y)dt^{2}+\frac{dy^{2}}{H(y)}+\frac{dx^{2}}{G(x)}+G(x)d\phi^{2}\right)\;,\label{eq:metkudoh}\eeq
where
\beq H(y)=\lambda-k y^{2}+\mu y^{3}\;,\quad G(x)=1+k x^{2}-\mu x^{3}\;,\label{eq:Hfuncapp}\eeq
with $\lambda\equiv \ell^{2}/\ell_{3}^{2}$ and $\mu=2mA$. To avoid a conical singularity at the zero $x=x_{1}$ of $G(x)$, the period of angular variable $\phi$ is fixed to be $\Delta\phi=\frac{4\pi}{|G(x_{1})|}$, and for a black hole localized on the brane at $x=0$, the $x,y$ coordinate ranges are restricted to be $0\leq x\leq x_{1}$, and $-\infty\leq y\leq x$, where  $y=-\infty$ corresponds to a curvature singularity hidden behind the bulk horizon, located at $y=y_{+}$, the smallest root of $H(y)$. The region $x,y\to0$ corresponds to an asymptotic region far from the black hole (this is apparent in Boyer-Lindquist form).

Recall that the surfaces $x=0$ and $y=0$ are umbilic, satisfying $K^{(x)}_{ij}=-A\epsilon h_{ij}^{(x)}$ (\ref{eq:Kxij}) and $K_{ij}^{(y)}=A\epsilon\sqrt{\lambda}h_{ij}^{(y)}$ (\ref{eq:Kyij}), respectively.\footnote{Since here we work with metric (\ref{eq:metkudoh}), which differs from metric (\ref{eq:genAdSC}) by $H(y)\to -H(y)$, the $\sqrt{-\lambda}$ coefficient in (\ref{eq:Kyij}) is replaced with $-\sqrt{\lambda}$.} Boundary conditions on the $x=0$ and $y=0$ hypersurfaces are governed by Israel's junction conditions \cite{Israel:1966rt}. For purely tensional branes, the junction conditions fix the tensions $\tau_{x}$ and $\tau_{y}$ by relating the discontinuity in the extrinsic curvature across the $x=0$ and $y=0$ surfaces to their respective brane stress-tensors $S_{ij}^{(x)}$ and $S_{ij}^{(y)}$. Specifically, for a $\mathbb{Z}_{2}$-symmetric brane configuration as utilized in the main text, the junction conditions give, 
\beq 2[K_{ij}^{(x,y)}-h_{ij}^{(x,y)}K^{(x,y)}]=8\pi G_{4}S_{ij}^{(x,y)}=-8\pi G_{4}\tau_{x,y}h_{ij}^{(x,y)}\;,\eeq
giving tensions
\beq \tau_{x}=-\frac{\epsilon A}{2\pi G_{4}}\;,\qquad \tau_{y}=-\frac{\epsilon A\sqrt{\lambda}}{2\pi G_{4}}\;.\label{eq:branetensionsxy}\eeq

The Euclidean $\text{AdS}_{4}$ C-metric is found by Wick rotating $t_{E}=it$, for Euclidean time $t_{E}$,
\beq ds^{2}=\frac{\ell^{2}}{(x-y)^{2}}\left(H(y)dt_{E}^{2}+\frac{dy^{2}}{H(y)}+\frac{dx^{2}}{G(x)}+G(x)d\phi^{2}\right)\;,\label{eq:AdS4Euc}\eeq
using $A=\ell^{-1}$.
In Euclidean signature, there will be a conical singularity at the location of the black hole horizon, $y=y_{+}$. To have a regular Euclidean section, the Euclidean time $t_{E}\sim t_{E}+\Delta t_{E}$ is periodicially identified with period
\beq \Delta t_{E}=\frac{4\pi}{|H'(y_{+})|}\;.\eeq
In Euclidean signature the brane and boundary satisfy the same umbilic conditions as before, however, $\epsilon=-1$ in the tensions (\ref{eq:branetensionsxy}).

 \subsection*{On-shell Euclidean action}

 We now follow Gibbons and Hawking \cite{Gibbons:1976ue} to evaluate the quantum gravitational canonical partition function $Z(\beta)$ in the semi-classical limit via the on-shell Euclidean action. Before we evaluate the on-shell action, two comments are in order. First, as an accelerating black hole, the $\text{AdS}_{4}$ C-metric has at least two horizons, a black hole horizon and an acceleration horizon. Due to the system having two horizons, generally with different surface gravities, the two horizon system is not generically in thermodynamic equilibrium. One way to circumvent this problem is to work in a regime in which the black hole is slowly accelerating, where $A<L_{4}^{-1}$ \cite{Podolsky:2002nk}. This translates to $A<A\sqrt{1+\lambda}$, i.e., $\lambda>0$, or, an AdS$_{3}$ slicing on the brane. Below we work in this regime such that we effectively have a single black hole with a single unique temperature given by $\beta^{-1}$. Meanwhile, a dS$_{3}$ slicing, where $-1<\lambda<0$, obeys $A>L_{4}^{-1}$, such that the bulk solution is interpreted as two black holes separated by an acceleration horizon \cite{Dias:2002mi}, and the system is not in thermal equilibrium. This is consistent with the fact this scenario describes a de Sitter black hole localized on the brane, for which the black hole and cosmological horizons are not generally in equilibrium.

 Second, in the evaluation of the on-shell action, it is common to encounter infrared divergences as a boundary is approached, thereby requiring some regularization scheme. Traditionally this accomplished either by the method of background subtraction, or, in the case of asymptotically AdS spacetimes including a local counterterm action \cite{Balasubramanian:1999re,Emparan:1999pm}. The key insight of \cite{Kudoh:2004ub} is that, to recover the thermodynamics of the bulk black hole and, hence, the black hole localized on the brane, no background subtraction or local counterterms are needed. Rather, the potential IR divergences in the total on-shell action are exactly cancelled when branes at $x=0$ and $y=0$ are included.

To this end, the total action $I$ characterizing the bulk Riemannian spacetime $\mathcal{M}$ endowed with Euclidean metric $g$, and branes $\mathcal{B}_{x}$ and $\mathcal{B}_{x}$ embedded at $x,y=0$ is
\beq I=I_{\text{EH}}+I_{\text{GHY}}^{(x)}+I_{\text{GHY}}^{(y)}+I_{\mathcal{B}_{x}}+I_{\mathcal{B}_{y}}\;,\eeq
where the bulk Einstein-Hilbert action 
\beq I_{\text{EH}}=-\frac{1}{16\pi G_{4}}\int_{\mathcal{M}}d^{4}x\sqrt{g}(R-2\Lambda)\;,\label{eq:EHbulkeucapp}\eeq
with $\Lambda=-3/L_{4}^{2}$. To have a well-posed variational problem, Gibbons-Hawking-York boundary (GHY) terms are needed for each brane, 
\beq I_{\text{GHY}}^{(x)}= \frac{1}{8\pi G_{4}}\int_{\mathcal{B}_{x}}d^{3}x\sqrt{h_{(x)}}K^{(x)}\;,\qquad I_{\text{GHY}}^{(y)}=-\frac{1}{8\pi G_{4}}\int_{\mathcal{B}_{y}}d^{3}x\sqrt{h_{(y)}}K^{(y)}\;.\label{eq:GHYtermsapp}\eeq
The brane actions are purely tensional and take the form
\beq I_{\mathcal{B}_{x}}=-\tau_{x}\int_{\mathcal{B}_{x}}d^{3}x\sqrt{h_{(x)}}\;,\qquad I_{\mathcal{B}_{y}}=\tau_{y}\int_{\mathcal{B}_{y}}d^{3}x\sqrt{h_{(y)}}\;.\label{eq:BxByact}\eeq
Each action has IR divergences at $x,y=0$. To remedy this, introduce cutoffs at $x=\epsilon_{x}$ and $y=\epsilon_{y}$, and at the end of the computation take the limit $\epsilon_{x,y}\to0$. 

Let us now evaluate actions (\ref{eq:EHbulkeucapp}) -- (\ref{eq:BxByact}) in the Euclidean background (\ref{eq:AdS4Euc}). The Einstein-Hilbert term evaluates to  
\beq
\begin{split}
I_{\text{EH}}&
%=-\frac{1}{16\pi G_{4}}\int_{\mathcal{M}}d t_{E} d\phi dx dy\sqrt{g}(R-2\Lambda)\\
=\frac{6\ell^{4}}{16\pi G_{4}L_{4}^{2}}\int_{0}^{\Delta t_{E}}d t_{E} \int_{0}^{\Delta\phi}d\phi \int_{\epsilon_{x}}^{x_{1}} dx \int_{y_{+}}^{\epsilon_{y}} dy\frac{1}{(x-y)^{4}}\\
&=\frac{6\ell^{4}}{16\pi G_{4}L_{4}^{2}}\frac{\Delta t_{E}\Delta\phi}{3}\int_{\epsilon_{x}}^{x_{1}}dx\left(\frac{1}{(x-\epsilon_{y})^{3}}-\frac{1}{(x-y_{+})^{3}}\right)\\
&=\frac{\ell^{4}}{16\pi G_{4}L_{4}^{2}}\Delta t_{E}\Delta\phi\left[\frac{1}{(x_{1}-\epsilon_{y})^{2}}-\frac{1}{(x_{1}-y_{+})^{2}}+\frac{1}{(y_{+}-\epsilon_{x})^{2}}-\frac{1}{(\epsilon_{x}-\epsilon_{y})^{2}}\right]\;.
\end{split}
\eeq
where we used $R=-12/L_{4}^{2}$, and $\Lambda=-3/L_{4}^{2}$. The GHY term (\ref{eq:GHYtermsapp}) at $x=\epsilon_{x}$ is
\beq
\begin{split} I_{\text{GHY}}^{(x)} &= \frac{1}{8\pi G_{4}}\int_{\mathcal{B}_{x}}d^{3}x\sqrt{h_{(x)}}K^{(x)}=-\frac{3\ell^{2}}{8\pi G_{4}}\int_{\mathcal{B}_{x}}dt_{E}d\phi dy\frac{1}{(\epsilon_{x}-y)^{3}}\\
&=-\frac{3\ell^{2}}{16\pi G_{4}}\Delta t_{E}\Delta\phi\left[\frac{1}{(\epsilon_{x}-y_{+})^{2}}-\frac{1}{(\epsilon_{x}-\epsilon_{y})^{2}}\right]\;,
\end{split}
\eeq
where we used $K^{(x)}=-\frac{3}{\ell}$ and $\sqrt{G(\epsilon_{x})}\approx 1$ for $\epsilon_{x}\ll1$. Similarly, at $y=\epsilon_{y}\ll1$, 
\beq 
\begin{split}
I_{\text{GHY}}^{(y)}&=-\frac{1}{8\pi G_{4}}\int_{\mathcal{B}_{y}}d^{3}x\sqrt{h_{(y)}}K^{(y)}=-\frac{3\lambda\ell^{2}}{8\pi G_{4}}\int_{\mathcal{B}_{y}}dt_{E}d\phi dx\frac{1}{(x-\epsilon_{y})^{3}}\\
&=-\frac{3\lambda\ell^{2}}{16\pi G_{4}}\Delta t_{E}\Delta\phi\left[\frac{1}{(\epsilon_{x}-\epsilon_{y})^{2}}-\frac{1}{(x_{1}-\epsilon_{y})^{2}}\right]\;,
\end{split}
\eeq
where we used $K^{(y)}=\frac{3\sqrt{\lambda}}{\ell}$ and $\sqrt{H(\epsilon_{y})}\approx \sqrt{\lambda}$. Lastly, the brane actions (\ref{eq:BxByact}) give
\beq I_{\mathcal{B}_{x}}=-\tau_{x}\int_{\mathcal{B}_{x}}d^{3}x\sqrt{h_{(x)}}=\frac{\ell^{2}\Delta t_{E}\Delta\phi}{4\pi G_{4}}\left[\frac{1}{(\epsilon_{x}-\epsilon_{y})^{2}}-\frac{1}{(y_{+}-\epsilon_{x})^{2}}\right]\;,\eeq
\beq I_{\mathcal{B}_{y}}=\tau_{y}\int_{\mathcal{B}_{y}}d^{3}x\sqrt{h_{(y)}}=\frac{\lambda\ell^{2}\Delta t_{E}\Delta\phi}{4\pi G_{4}}\left[\frac{1}{(\epsilon_{x}-\epsilon_{y})^{2}}-\frac{1}{(x_{1}-\epsilon_{y})^{2}}\right]\;,\eeq
with tensions $\tau_{x}=\frac{1}{2\pi G_{4}\ell}$ and  $\tau_{y}=\frac{\sqrt{\lambda}}{2\pi G_{4}\ell}$. Observe that $I_{\text{GHY}}=-\frac{3}{4}I_{\mathcal{B}}$ for either brane. Adding together the Einstein-Hilbert, GHY, and brane actions, and accounting for the $\mathbb{Z}_{2}$ symmetry, the total on-shell Euclidean action is \cite{Kudoh:2004ub}
\beq I_{\text{on-shell}}=\frac{\ell^{2}}{8\pi G_{4}}\Delta t_{E}\Delta\phi\left[\frac{(1+\lambda)}{(x_{1}-y_{+})^{2}}-\frac{1}{x_{1}^{2}}-\frac{\lambda}{y_{+}^{2}}\right]\;,\label{eq:bulkonshellact}\eeq
where we implemented $L^{-2}_{4}=\ell^{-2}(1+\lambda)$, and since all IR divergences cancel, we safely take the limit $\epsilon_{x},\epsilon_{y}\to0$.

It proves useful to introduce parameter $z=-y_{+}/x_{1}$, from which we find
\beq 
\begin{split} 
&x_{1}^{2}=\frac{1}{z^{2}}\frac{\nu^{2}-z^{3}}{k(1+z)}\;,\\
&y_{+}^{2}=\frac{(\nu^{2}-z^{3})}{k(1+z)}\;,\\
&\mu=z(z^{2}+\nu^{2})\sqrt{1+z}\biggr|\frac{k}{(\nu^{2}-z^{3})}\biggr|^{3/2}\;,
\end{split}
\label{eq:paramskudoh}\eeq
where we used $G(x_{1})=H(y_{+})=0$ to solve $x_{1}^{2}$ and $y_{+}^{2}$ and $\mu=(1+kx_{1}^{2})/x_{1}^{3}$. Further, we set $\lambda=\nu^{2}$. Given the range of coordinates $x,y$, the parameter $z$ ranges between $0$ and $\infty$, depending on the value of $k$.\footnote{In particular, for $k=+1$, then $0\leq z<\nu^{2/3}$, while for $k=-1$, then $\nu^{2/3}<z\leq \infty$. Meanwhile, for $k=0$, one has $\mu=1/x_{1}^{1/3}$, and $y_{+}=(\nu^{2}/\mu)^{1/3}$, such that $z=\nu^{2/3}$.} With the parameters (\ref{eq:paramskudoh}), note
\beq \Delta t_{E}\Delta \phi=\frac{16\pi^{2}x_{1}^{2} z^{3}(1+z)^{2}}{[z^{3}+2\nu^{2}+3z\nu^{2}][2z^{3}+\nu^{2}+3z^{2}]}\;,\label{eq:DeltatEphi}\eeq
where we used $H'(y_{+})=-2ky_{+}+3\mu y_{+}^{2}=2kx_{1}z+3\mu x_{1}^{2}z^{2}$, and $\mu x_{1}=\frac{1+kx_{1}^{2}}{x_{1}^{2}}$.
Then,  
\beq 
\begin{split}
I_{\text{on-shell}}&=-\frac{\ell^{2}}{8\pi G_{4}}\frac{\Delta t_{E}\Delta\phi}{x_{1}^{2}}\left[\frac{\nu^{2}(1+2z)+z^{3}(2+z)}{z^{2}(1+z)^{2}}\right]\\
&=-\frac{16\pi^{2}\ell^{2}}{8\pi G_{4}}\frac{z(\nu^{2}+2z\nu^{2}+2z^{3}+z^{4})}{[z^{3}+2\nu^{2}+3z\nu^{2}][2z^{3}+\nu^{2}+3z^{2}]}\\
&=-\frac{8\pi^{2}\ell^{2}z}{8\pi G_{4}}\left(\frac{1}{(2 z^{3}+\nu^{2}+3z^{2})}+\frac{z}{(z^{3}+2\nu^{2}+3\nu^{2}z)}\right)\;.
\end{split}
\eeq
where in the first equality we replaced $y_{+}=-zx_{1}$, and in the second we substituted in (\ref{eq:DeltatEphi}). Notice the parameter $k$ has dropped out of the final expression. 

\subsection*{Thermodynamics in the canonical ensemble}

Following Gibbons and Hawking \cite{Gibbons:1976ue}, the gravitational canonical partition function is given by a Euclidean path integral, which to leading order in a stationary phase approximation is
\beq Z(\beta)=\text{tr}(e^{-\beta \mathcal{H}})\approx e^{-I_{\text{on-shell}}}\;,\eeq
where $\beta$ is the (inverse) temperature $T$ of the system. Unlike, say, the Schwarzschild black hole, the period of Euclidean time $\Delta t_{E}$ is not equal to $\beta$. This is because the coordinates $(t_{E},\phi,y)$ are not canonically normalized in that $\phi\sim \phi+\Delta\phi$ instead of $\phi\sim\phi+2\pi$. Thus, one should instead consider rescaled coordinates $\bar{\phi}=\eta^{-1}\phi$, $\bar{t}_{E}=\eta^{-1}\ell t_{E}$, and $\bar{y}=\eta^{-1}y$ with $\eta\equiv \Delta/2\pi$. In these canonically normalized coordinates, the periodicity of $\bar{t}_{E}$ is such that 
\beq \bar{t}_{E}\sim \bar{t}_{E}+\beta\;,\quad \beta=2\pi\ell\frac{\Delta t_{E}}{\Delta\phi}\;.\eeq
Using the parameters (\ref{eq:paramskudoh}), the inverse temperature may be cast as\footnote{It is also useful to $\Delta\phi=4\pi x_{1}/(3+kx_{1}^{2})$ and $\Delta t_{E}=4\pi x_{1}/z(2kx_{1}^{2}+3z(1+kx_{1}^{2}))$.}
\beq \beta=\frac{2\pi\ell}{z}\left(\frac{\nu^{2}+3z^{2}+2z^{3}}{2\nu^{2}+3z\nu^{2}+z^{3}}\right)\;.\eeq
Since $\log Z(\beta)=-\beta F$, for free energy $F$, then
\beq F=\beta^{-1}I_{\text{on-shell}}=-\frac{\ell z^{2}}{G_{4}}\frac{(z^{3}(2+z)+\nu^{2}(1+2z))}{(\nu^{2}+3z^{2}+2z^{3})^{2}}\;.\eeq
Further, the canonical energy $E$ and entropy $S$ are defined via 
\beq E\equiv-\partial_{\beta}\log Z\;,\qquad S=\beta E+\log Z\;,\eeq
yielding (where we keep $\nu,\ell$ and $G_{4}$ fixed)
\beq E=\frac{\ell z^{2}}{G_{4}}\frac{(1+z)(\nu^{2}-z^{3})}{(\nu^{2}+3z^{2}+2z^{3})^{2}}\;,\eeq
\beq S=\frac{2\pi \ell^{2}z}{G_{4}(\nu^{2}+3z^{2}+2z^{3})}\;.\eeq
From here it is easy to verify $F=E-\beta^{-1}S$. 

To recover the thermodynamic relations of the static qBTZ black hole stated in the main text, simply rescale $z\to \nu z$ in the above quantities:
\beq 
\begin{split}
&I_{\text{on-shell}}=-\frac{2\pi \ell^{2}z}{G_{4}\nu}\frac{[1+2\nu z+\nu z^{3}(2+\nu z)]}{(2+3\nu z+\nu z^{3})(1+3z^{2}+2\nu z^{3})}\;,\\
&\beta=\frac{2\pi \ell}{\nu z}\frac{(1+3z^{2}+2\nu z^{3})}{(2+3\nu z+2\nu z^{3})}\;,\\
&F=-\frac{\ell z^{2}}{G_{4}}\frac{[1+2\nu z+\nu z^{3}(2+\nu z)]}{(1+3z^{2}+2\nu z^{3})^{2}}\;,\\
&E=\frac{\ell z^{2}}{G_{4}}\frac{(1+\nu z)(1-\nu z^{3})}{(1+3z^{2}+2\nu z^{3})^{2}}\;,\\
&S=\frac{2\pi \ell_{3}^{2}}{G_{4}}\frac{\nu z}{(1+3z^{2}+2\nu z^{3})}\;,
\end{split}
\eeq
and subsequently use $\mathcal{G}_{3}=G_{4}/2\ell$ with $G_{3}=\mathcal{G}_{3}\sqrt{1+\nu^{2}}$.

%%%%%%%%%%%%%%%%%%%%%%%%%%%%%%%%%%%%%%%%%%%%%%%%%%%%%%%%%%%%%%%%%%%%%

\bibliography{qbhrefs}

\providecommand{\href}[2]{#2}\begingroup\raggedright\begin{thebibliography}{100}

\bibitem{Birrell:1982ix}
N.~D. Birrell and P.~C.~W. Davies, \emph{{Quantum Fields in Curved Space}}.
\newblock Cambridge Monographs on Mathematical Physics. Cambridge Univ. Press,
  Cambridge, UK, 2, 1984.

\bibitem{Wald:1995yp}
R.~M. Wald, \emph{{Quantum Field Theory in Curved Space-Time and Black Hole
  Thermodynamics}}.
\newblock Chicago Lectures in Physics. University of Chicago Press, Chicago,
  IL, 1995.

\bibitem{Page:1981aj}
D.~N. Page and C.~D. Geilker, \emph{{Indirect Evidence for Quantum Gravity}},
  \href{http://dx.doi.org/10.1103/PhysRevLett.47.979}{\emph{Phys. Rev. Lett.}
  {\bf 47} (1981) 979--982}.

\bibitem{Polyakov:1981rd}
A.~M. Polyakov, \emph{{Quantum Geometry of Bosonic Strings}},
  \href{http://dx.doi.org/10.1016/0370-2693(81)90743-7}{\emph{Phys. Lett. B}
  {\bf 103} (1981) 207--210}.

\bibitem{Christensen:1977jc}
S.~M. Christensen and S.~A. Fulling, \emph{{Trace Anomalies and the Hawking
  Effect}}, \href{http://dx.doi.org/10.1103/PhysRevD.15.2088}{\emph{Phys. Rev.
  D} {\bf 15} (1977) 2088--2104}.

\bibitem{Maldacena:1997re}
J.~M. Maldacena, \emph{{The Large N limit of superconformal field theories and
  supergravity}},
  \href{http://dx.doi.org/10.4310/ATMP.1998.v2.n2.a1}{\emph{Adv. Theor. Math.
  Phys.} {\bf 2} (1998) 231--252},
  [\href{https://arxiv.org/abs/hep-th/9711200}{{\tt hep-th/9711200}}].

\bibitem{tHooft:1993dmi}
G.~'t~Hooft, \emph{{Dimensional reduction in quantum gravity}}, {\emph{Conf.
  Proc. C} {\bf 930308} (1993) 284--296},
  [\href{https://arxiv.org/abs/gr-qc/9310026}{{\tt gr-qc/9310026}}].

\bibitem{Susskind:1994vu}
L.~Susskind, \emph{{The World as a hologram}},
  \href{http://dx.doi.org/10.1063/1.531249}{\emph{J. Math. Phys.} {\bf 36}
  (1995) 6377--6396}, [\href{https://arxiv.org/abs/hep-th/9409089}{{\tt
  hep-th/9409089}}].

\bibitem{Hubeny:2009ru}
V.~E. Hubeny, D.~Marolf and M.~Rangamani, \emph{{Hawking radiation in large N
  strongly-coupled field theories}},
  \href{http://dx.doi.org/10.1088/0264-9381/27/9/095015}{\emph{Class. Quant.
  Grav.} {\bf 27} (2010) 095015}, [\href{https://arxiv.org/abs/0908.2270}{{\tt
  0908.2270}}].

\bibitem{Compere:2008us}
G.~Compere and D.~Marolf, \emph{{Setting the boundary free in AdS/CFT}},
  \href{http://dx.doi.org/10.1088/0264-9381/25/19/195014}{\emph{Class. Quant.
  Grav.} {\bf 25} (2008) 195014}, [\href{https://arxiv.org/abs/0805.1902}{{\tt
  0805.1902}}].

\bibitem{deHaro:2000wj}
S.~de~Haro, K.~Skenderis and S.~N. Solodukhin, \emph{{Gravity in warped
  compactifications and the holographic stress tensor}},
  \href{http://dx.doi.org/10.1088/0264-9381/18/16/307}{\emph{Class. Quant.
  Grav.} {\bf 18} (2001) 3171--3180},
  [\href{https://arxiv.org/abs/hep-th/0011230}{{\tt hep-th/0011230}}].

\bibitem{Arkani-Hamed:1998jmv}
N.~Arkani-Hamed, S.~Dimopoulos and G.~R. Dvali, \emph{{The Hierarchy problem
  and new dimensions at a millimeter}},
  \href{http://dx.doi.org/10.1016/S0370-2693(98)00466-3}{\emph{Phys. Lett. B}
  {\bf 429} (1998) 263--272}, [\href{https://arxiv.org/abs/hep-ph/9803315}{{\tt
  hep-ph/9803315}}].

\bibitem{Randall:1999ee}
L.~Randall and R.~Sundrum, \emph{{A Large mass hierarchy from a small extra
  dimension}}, \href{http://dx.doi.org/10.1103/PhysRevLett.83.3370}{\emph{Phys.
  Rev. Lett.} {\bf 83} (1999) 3370--3373},
  [\href{https://arxiv.org/abs/hep-ph/9905221}{{\tt hep-ph/9905221}}].

\bibitem{Randall:1999vf}
L.~Randall and R.~Sundrum, \emph{{An Alternative to compactification}},
  \href{http://dx.doi.org/10.1103/PhysRevLett.83.4690}{\emph{Phys. Rev. Lett.}
  {\bf 83} (1999) 4690--4693},
  [\href{https://arxiv.org/abs/hep-th/9906064}{{\tt hep-th/9906064}}].

\bibitem{Karch:2000ct}
A.~Karch and L.~Randall, \emph{{Locally localized gravity}},
  \href{http://dx.doi.org/10.1088/1126-6708/2001/05/008}{\emph{JHEP} {\bf 05}
  (2001) 008}, [\href{https://arxiv.org/abs/hep-th/0011156}{{\tt
  hep-th/0011156}}].

\bibitem{Karch:2000gx}
A.~Karch and L.~Randall, \emph{{Open and closed string interpretation of SUSY
  CFT's on branes with boundaries}},
  \href{http://dx.doi.org/10.1088/1126-6708/2001/06/063}{\emph{JHEP} {\bf 06}
  (2001) 063}, [\href{https://arxiv.org/abs/hep-th/0105132}{{\tt
  hep-th/0105132}}].

\bibitem{Kraus:1999di}
P.~Kraus, F.~Larsen and R.~Siebelink, \emph{{The gravitational action in
  asymptotically AdS and flat space-times}},
  \href{http://dx.doi.org/10.1016/S0550-3213(99)00549-0}{\emph{Nucl. Phys. B}
  {\bf 563} (1999) 259--278}, [\href{https://arxiv.org/abs/hep-th/9906127}{{\tt
  hep-th/9906127}}].

\bibitem{Emparan:1999pm}
R.~Emparan, C.~V. Johnson and R.~C. Myers, \emph{{Surface terms as counterterms
  in the AdS / CFT correspondence}},
  \href{http://dx.doi.org/10.1103/PhysRevD.60.104001}{\emph{Phys. Rev. D} {\bf
  60} (1999) 104001}, [\href{https://arxiv.org/abs/hep-th/9903238}{{\tt
  hep-th/9903238}}].

\bibitem{deHaro:2000vlm}
S.~de~Haro, S.~N. Solodukhin and K.~Skenderis, \emph{{Holographic
  reconstruction of space-time and renormalization in the AdS / CFT
  correspondence}},
  \href{http://dx.doi.org/10.1007/s002200100381}{\emph{Commun. Math. Phys.}
  {\bf 217} (2001) 595--622}, [\href{https://arxiv.org/abs/hep-th/0002230}{{\tt
  hep-th/0002230}}].

\bibitem{Skenderis:2002wp}
K.~Skenderis, \emph{{Lecture notes on holographic renormalization}},
  \href{http://dx.doi.org/10.1088/0264-9381/19/22/306}{\emph{Class. Quant.
  Grav.} {\bf 19} (2002) 5849--5876},
  [\href{https://arxiv.org/abs/hep-th/0209067}{{\tt hep-th/0209067}}].

\bibitem{Papadimitriou:2004ap}
I.~Papadimitriou and K.~Skenderis, \emph{{AdS / CFT correspondence and
  geometry}}, \href{http://dx.doi.org/10.4171/013-1/4}{\emph{IRMA Lect. Math.
  Theor. Phys.} {\bf 8} (2005) 73--101},
  [\href{https://arxiv.org/abs/hep-th/0404176}{{\tt hep-th/0404176}}].

\bibitem{Emparan:2002px}
R.~Emparan, A.~Fabbri and N.~Kaloper, \emph{{Quantum black holes as holograms
  in AdS brane worlds}},
  \href{http://dx.doi.org/10.1088/1126-6708/2002/08/043}{\emph{JHEP} {\bf 08}
  (2002) 043}, [\href{https://arxiv.org/abs/hep-th/0206155}{{\tt
  hep-th/0206155}}].

\bibitem{Emparan:1999wa}
R.~Emparan, G.~T. Horowitz and R.~C. Myers, \emph{{Exact description of black
  holes on branes}},
  \href{http://dx.doi.org/10.1088/1126-6708/2000/01/007}{\emph{JHEP} {\bf 01}
  (2000) 007}, [\href{https://arxiv.org/abs/hep-th/9911043}{{\tt
  hep-th/9911043}}].

\bibitem{Emparan:1999fd}
R.~Emparan, G.~T. Horowitz and R.~C. Myers, \emph{{Exact description of black
  holes on branes. 2. Comparison with BTZ black holes and black strings}},
  \href{http://dx.doi.org/10.1088/1126-6708/2000/01/021}{\emph{JHEP} {\bf 01}
  (2000) 021}, [\href{https://arxiv.org/abs/hep-th/9912135}{{\tt
  hep-th/9912135}}].

\bibitem{Emparan:2020znc}
R.~Emparan, A.~M. Frassino and B.~Way, \emph{{Quantum BTZ black hole}},
  \href{http://dx.doi.org/10.1007/JHEP11(2020)137}{\emph{JHEP} {\bf 11} (2020)
  137}, [\href{https://arxiv.org/abs/2007.15999}{{\tt 2007.15999}}].

\bibitem{Emparan:2022ijy}
R.~Emparan, J.~F. Pedraza, A.~Svesko, M.~Toma\v{s}evi\'c and M.~R. Visser,
  \emph{{Black holes in dS$_{3}$}},
  \href{http://dx.doi.org/10.1007/JHEP11(2022)073}{\emph{JHEP} {\bf 11} (2022)
  073}, [\href{https://arxiv.org/abs/2207.03302}{{\tt 2207.03302}}].

\bibitem{Panella:2023lsi}
E.~Panella and A.~Svesko, \emph{{Quantum Kerr-de Sitter black holes in three
  dimensions}}, \href{http://dx.doi.org/10.1007/JHEP06(2023)127}{\emph{JHEP}
  {\bf 06} (2023) 127}, [\href{https://arxiv.org/abs/2303.08845}{{\tt
  2303.08845}}].

\bibitem{Deser:1983tn}
S.~Deser, R.~Jackiw and G.~'t~Hooft, \emph{{Three-Dimensional Einstein Gravity:
  Dynamics of Flat Space}},
  \href{http://dx.doi.org/10.1016/0003-4916(84)90085-X}{\emph{Annals Phys.}
  {\bf 152} (1984) 220}.

\bibitem{Deser:1983nh}
S.~Deser and R.~Jackiw, \emph{{Three-Dimensional Cosmological Gravity: Dynamics
  of Constant Curvature}},
  \href{http://dx.doi.org/10.1016/0003-4916(84)90025-3}{\emph{Annals Phys.}
  {\bf 153} (1984) 405--416}.

\bibitem{Banados:1992wn}
M.~Banados, C.~Teitelboim and J.~Zanelli, \emph{{The Black hole in
  three-dimensional space-time}},
  \href{http://dx.doi.org/10.1103/PhysRevLett.69.1849}{\emph{Phys. Rev. Lett.}
  {\bf 69} (1992) 1849--1851},
  [\href{https://arxiv.org/abs/hep-th/9204099}{{\tt hep-th/9204099}}].

\bibitem{Banados:1992gq}
M.~Banados, M.~Henneaux, C.~Teitelboim and J.~Zanelli, \emph{{Geometry of the
  (2+1) black hole}},
  \href{http://dx.doi.org/10.1103/PhysRevD.48.1506}{\emph{Phys. Rev. D} {\bf
  48} (1993) 1506--1525}, [\href{https://arxiv.org/abs/gr-qc/9302012}{{\tt
  gr-qc/9302012}}].

\bibitem{Souradeep:1992ia}
T.~Souradeep and V.~Sahni, \emph{{Quantum effects near a point mass in
  (2+1)-Dimensional gravity}},
  \href{http://dx.doi.org/10.1103/PhysRevD.46.1616}{\emph{Phys. Rev. D} {\bf
  46} (1992) 1616--1633}, [\href{https://arxiv.org/abs/hep-ph/9208219}{{\tt
  hep-ph/9208219}}].

\bibitem{Soleng:1993yh}
H.~H. Soleng, \emph{{Inverse square law of gravitation in (2+1) dimensional
  space-time as a consequence of Casimir energy}},
  \href{http://dx.doi.org/10.1088/0031-8949/48/6/002}{\emph{Phys. Scripta} {\bf
  48} (1993) 649--652}, [\href{https://arxiv.org/abs/gr-qc/9310007}{{\tt
  gr-qc/9310007}}].

\bibitem{Tanaka:2002rb}
T.~Tanaka, \emph{{Classical black hole evaporation in Randall-Sundrum infinite
  brane world}}, \href{http://dx.doi.org/10.1143/PTPS.148.307}{\emph{Prog.
  Theor. Phys. Suppl.} {\bf 148} (2003) 307--316},
  [\href{https://arxiv.org/abs/gr-qc/0203082}{{\tt gr-qc/0203082}}].

\bibitem{Ross:1992ba}
S.~F. Ross and R.~B. Mann, \emph{{Gravitationally collapsing dust in
  (2+1)-dimensions}},
  \href{http://dx.doi.org/10.1103/PhysRevD.47.3319}{\emph{Phys. Rev. D} {\bf
  47} (1993) 3319--3322}, [\href{https://arxiv.org/abs/hep-th/9208036}{{\tt
  hep-th/9208036}}].

\bibitem{deBuyl:2013ega}
S.~de~Buyl, S.~Detournay, G.~Giribet and G.~S. Ng, \emph{{Baby de Sitter black
  holes and dS$_3$/CFT$_2$}},
  \href{http://dx.doi.org/10.1007/JHEP02(2014)020}{\emph{JHEP} {\bf 02} (2014)
  020}, [\href{https://arxiv.org/abs/1308.5569}{{\tt 1308.5569}}].

\bibitem{Nutku:1993eb}
Y.~Nutku, \emph{{Exact solutions of topologically massive gravity with a
  cosmological constant}},
  \href{http://dx.doi.org/10.1088/0264-9381/10/12/022}{\emph{Class. Quant.
  Grav.} {\bf 10} (1993) 2657--2661}.

\bibitem{Anninos:2009jt}
D.~Anninos, \emph{{Sailing from Warped AdS(3) to Warped dS(3) in Topologically
  Massive Gravity}},
  \href{http://dx.doi.org/10.1007/JHEP02(2010)046}{\emph{JHEP} {\bf 02} (2010)
  046}, [\href{https://arxiv.org/abs/0906.1819}{{\tt 0906.1819}}].

\bibitem{Bousso:2001mw}
R.~Bousso, A.~Maloney and A.~Strominger, \emph{{Conformal vacua and entropy in
  de Sitter space}},
  \href{http://dx.doi.org/10.1103/PhysRevD.65.104039}{\emph{Phys. Rev. D} {\bf
  65} (2002) 104039}, [\href{https://arxiv.org/abs/hep-th/0112218}{{\tt
  hep-th/0112218}}].

\bibitem{Steif:1993zv}
A.~R. Steif, \emph{{The Quantum stress tensor in the three-dimensional black
  hole}}, \href{http://dx.doi.org/10.1103/PhysRevD.49.R585}{\emph{Phys. Rev. D}
  {\bf 49} (1994) 585--589}, [\href{https://arxiv.org/abs/gr-qc/9308032}{{\tt
  gr-qc/9308032}}].

\bibitem{Shiraishi:1993qnr}
K.~Shiraishi and T.~Maki, \emph{{Quantum fluctuation of stress tensor and black
  holes in three dimensions}},
  \href{http://dx.doi.org/10.1103/PhysRevD.49.5286}{\emph{Phys. Rev. D} {\bf
  49} (1994) 5286--5294}, [\href{https://arxiv.org/abs/1804.07872}{{\tt
  1804.07872}}].

\bibitem{Lifschytz:1993eb}
G.~Lifschytz and M.~Ortiz, \emph{{Scalar field quantization on the
  (2+1)-dimensional black hole background}},
  \href{http://dx.doi.org/10.1103/PhysRevD.49.1929}{\emph{Phys. Rev. D} {\bf
  49} (1994) 1929--1943}, [\href{https://arxiv.org/abs/gr-qc/9310008}{{\tt
  gr-qc/9310008}}].

\bibitem{Martinez:1996uv}
C.~Martinez and J.~Zanelli, \emph{{Back reaction of a conformal field on a
  three-dimensional black hole}},
  \href{http://dx.doi.org/10.1103/PhysRevD.55.3642}{\emph{Phys. Rev. D} {\bf
  55} (1997) 3642--3646}, [\href{https://arxiv.org/abs/gr-qc/9610050}{{\tt
  gr-qc/9610050}}].

\bibitem{Casals:2016ioo}
M.~Casals, A.~Fabbri, C.~Mart\'\i{}nez and J.~Zanelli, \emph{{Quantum dress for
  a naked singularity}},
  \href{http://dx.doi.org/10.1016/j.physletb.2016.06.044}{\emph{Phys. Lett. B}
  {\bf 760} (2016) 244--248}, [\href{https://arxiv.org/abs/1605.06078}{{\tt
  1605.06078}}].

\bibitem{Casals:2019jfo}
M.~Casals, A.~Fabbri, C.~Mart\'\i{}nez and J.~Zanelli, \emph{{Quantum-corrected
  rotating black holes and naked singularities in ( 2+1 ) dimensions}},
  \href{http://dx.doi.org/10.1103/PhysRevD.99.104023}{\emph{Phys. Rev. D} {\bf
  99} (2019) 104023}, [\href{https://arxiv.org/abs/1902.01583}{{\tt
  1902.01583}}].

\bibitem{Gubser:1998bc}
S.~S. Gubser, I.~R. Klebanov and A.~M. Polyakov, \emph{{Gauge theory
  correlators from noncritical string theory}},
  \href{http://dx.doi.org/10.1016/S0370-2693(98)00377-3}{\emph{Phys. Lett. B}
  {\bf 428} (1998) 105--114}, [\href{https://arxiv.org/abs/hep-th/9802109}{{\tt
  hep-th/9802109}}].

\bibitem{Witten:1998qj}
E.~Witten, \emph{{Anti-de Sitter space and holography}},
  \href{http://dx.doi.org/10.4310/ATMP.1998.v2.n2.a2}{\emph{Adv. Theor. Math.
  Phys.} {\bf 2} (1998) 253--291},
  [\href{https://arxiv.org/abs/hep-th/9802150}{{\tt hep-th/9802150}}].

\bibitem{Fefferman1985}
C.~Fefferman and C.~R. Graham, \emph{Conformal invariants},  in \emph{\'Elie
  Cartan et les math\'ematiques d'aujourd'hui - Lyon, 25-29 juin 1984},
  no.~S131 in Ast\'erisque.
\newblock Soci\'et\'e math\'ematique de France, 1985.

\bibitem{Fefferman:2007rka}
C.~Fefferman and C.~R. Graham, \emph{{The ambient metric}}, {\emph{Ann. Math.
  Stud.} {\bf 178} (2011) 1--128}, [\href{https://arxiv.org/abs/0710.0919}{{\tt
  0710.0919}}].

\bibitem{Elvang:2016tzz}
H.~Elvang and M.~Hadjiantonis, \emph{{A Practical Approach to the
  Hamilton-Jacobi Formulation of Holographic Renormalization}},
  \href{http://dx.doi.org/10.1007/JHEP06(2016)046}{\emph{JHEP} {\bf 06} (2016)
  046}, [\href{https://arxiv.org/abs/1603.04485}{{\tt 1603.04485}}].

\bibitem{Bueno:2022log}
P.~Bueno, R.~Emparan and Q.~Llorens, \emph{{Higher-curvature Gravities from
  Braneworlds and the Holographic c-theorem}},
  \href{https://arxiv.org/abs/2204.13421}{{\tt 2204.13421}}.

\bibitem{Israel:1966rt}
W.~Israel, \emph{{Singular hypersurfaces and thin shells in general
  relativity}}, \href{http://dx.doi.org/10.1007/BF02710419}{\emph{Nuovo Cim. B}
  {\bf 44S10} (1966) 1}.

\bibitem{Chen:2020uac}
H.~Z. Chen, R.~C. Myers, D.~Neuenfeld, I.~A. Reyes and J.~Sandor,
  \emph{{Quantum Extremal Islands Made Easy, Part I: Entanglement on the
  Brane}}, \href{http://dx.doi.org/10.1007/JHEP10(2020)166}{\emph{JHEP} {\bf
  10} (2020) 166}, [\href{https://arxiv.org/abs/2006.04851}{{\tt 2006.04851}}].

\bibitem{Aguilar-Gutierrez:2023kfn}
S.~E. Aguilar-Gutierrez, P.~Bueno, P.~A. Cano, R.~A. Hennigar and Q.~Llorens,
  \emph{{Aspects of higher-curvature gravities with covariant derivatives}},
  \href{http://dx.doi.org/10.1103/PhysRevD.108.124075}{\emph{Phys. Rev. D} {\bf
  108} (2023) 124075}, [\href{https://arxiv.org/abs/2310.09333}{{\tt
  2310.09333}}].

\bibitem{Emparan:2006ni}
R.~Emparan, \emph{{Black hole entropy as entanglement entropy: A Holographic
  derivation}},
  \href{http://dx.doi.org/10.1088/1126-6708/2006/06/012}{\emph{JHEP} {\bf 06}
  (2006) 012}, [\href{https://arxiv.org/abs/hep-th/0603081}{{\tt
  hep-th/0603081}}].

\bibitem{Myers:2013lva}
R.~C. Myers, R.~Pourhasan and M.~Smolkin, \emph{{On Spacetime Entanglement}},
  \href{http://dx.doi.org/10.1007/JHEP06(2013)013}{\emph{JHEP} {\bf 06} (2013)
  013}, [\href{https://arxiv.org/abs/1304.2030}{{\tt 1304.2030}}].

\bibitem{Takayanagi:2011zk}
T.~Takayanagi, \emph{{Holographic Dual of BCFT}},
  \href{http://dx.doi.org/10.1103/PhysRevLett.107.101602}{\emph{Phys. Rev.
  Lett.} {\bf 107} (2011) 101602}, [\href{https://arxiv.org/abs/1105.5165}{{\tt
  1105.5165}}].

\bibitem{Fujita:2011fp}
M.~Fujita, T.~Takayanagi and E.~Tonni, \emph{{Aspects of AdS/BCFT}},
  \href{http://dx.doi.org/10.1007/JHEP11(2011)043}{\emph{JHEP} {\bf 11} (2011)
  043}, [\href{https://arxiv.org/abs/1108.5152}{{\tt 1108.5152}}].

\bibitem{Karch:2022rvr}
A.~Karch, H.~Sun and C.~F. Uhlemann, \emph{{Double holography in string
  theory}}, \href{http://dx.doi.org/10.1007/JHEP10(2022)012}{\emph{JHEP} {\bf
  10} (2022) 012}, [\href{https://arxiv.org/abs/2206.11292}{{\tt 2206.11292}}].

\bibitem{Omiya:2021olc}
H.~Omiya and Z.~Wei, \emph{{Causal structures and nonlocality in double
  holography}}, \href{http://dx.doi.org/10.1007/JHEP07(2022)128}{\emph{JHEP}
  {\bf 07} (2022) 128}, [\href{https://arxiv.org/abs/2107.01219}{{\tt
  2107.01219}}].

\bibitem{Neuenfeld:2023svs}
D.~Neuenfeld and M.~Srivastava, \emph{{On the causality paradox and the
  Karch-Randall braneworld as an EFT}},
  \href{http://dx.doi.org/10.1007/JHEP10(2023)164}{\emph{JHEP} {\bf 10} (2023)
  164}, [\href{https://arxiv.org/abs/2307.10392}{{\tt 2307.10392}}].

\bibitem{Duff:2000mt}
M.~J. Duff and J.~T. Liu, \emph{{Complementarity of the Maldacena and
  Randall-Sundrum pictures}},
  \href{http://dx.doi.org/10.1088/0264-9381/18/16/310}{\emph{Phys. Rev. Lett.}
  {\bf 85} (2000) 2052--2055},
  [\href{https://arxiv.org/abs/hep-th/0003237}{{\tt hep-th/0003237}}].

\bibitem{Duff:1974ud}
M.~J. Duff, \emph{{Quantum corrections to the schwarzschild solution}},
  \href{http://dx.doi.org/10.1103/PhysRevD.9.1837}{\emph{Phys. Rev. D} {\bf 9}
  (1974) 1837--1839}.

\bibitem{Tanahashi:2011xx}
N.~Tanahashi and T.~Tanaka, \emph{{Black holes in braneworld models}},
  \href{http://dx.doi.org/10.1143/PTPS.189.227}{\emph{Prog. Theor. Phys.
  Suppl.} {\bf 189} (2011) 227--268},
  [\href{https://arxiv.org/abs/1105.2997}{{\tt 1105.2997}}].

\bibitem{Chen:2020hmv}
H.~Z. Chen, R.~C. Myers, D.~Neuenfeld, I.~A. Reyes and J.~Sandor,
  \emph{{Quantum Extremal Islands Made Easy, Part II: Black Holes on the
  Brane}}, \href{http://dx.doi.org/10.1007/JHEP12(2020)025}{\emph{JHEP} {\bf
  12} (2020) 025}, [\href{https://arxiv.org/abs/2010.00018}{{\tt 2010.00018}}].

\bibitem{Bruni:2001fd}
M.~Bruni, C.~Germani and R.~Maartens, \emph{{Gravitational collapse on the
  brane}}, \href{http://dx.doi.org/10.1103/PhysRevLett.87.231302}{\emph{Phys.
  Rev. Lett.} {\bf 87} (2001) 231302},
  [\href{https://arxiv.org/abs/gr-qc/0108013}{{\tt gr-qc/0108013}}].

\bibitem{Kudoh:2003xz}
H.~Kudoh, T.~Tanaka and T.~Nakamura, \emph{{Small localized black holes in
  brane world: Formulation and numerical method}},
  \href{http://dx.doi.org/10.1103/PhysRevD.68.024035}{\emph{Phys. Rev. D} {\bf
  68} (2003) 024035}, [\href{https://arxiv.org/abs/gr-qc/0301089}{{\tt
  gr-qc/0301089}}].

\bibitem{Kudoh:2004kf}
H.~Kudoh, \emph{{Six-dimensional localized black holes: Numerical solutions}},
  \href{http://dx.doi.org/10.1103/PhysRevD.70.029901}{\emph{Phys. Rev. D} {\bf
  69} (2004) 104019}, [\href{https://arxiv.org/abs/hep-th/0401229}{{\tt
  hep-th/0401229}}].

\bibitem{Fitzpatrick:2006cd}
A.~L. Fitzpatrick, L.~Randall and T.~Wiseman, \emph{{On the existence and
  dynamics of braneworld black holes}},
  \href{http://dx.doi.org/10.1088/1126-6708/2006/11/033}{\emph{JHEP} {\bf 11}
  (2006) 033}, [\href{https://arxiv.org/abs/hep-th/0608208}{{\tt
  hep-th/0608208}}].

\bibitem{Yoshino:2008rx}
H.~Yoshino, \emph{{On the existence of a static black hole on a brane}},
  \href{http://dx.doi.org/10.1088/1126-6708/2009/01/068}{\emph{JHEP} {\bf 01}
  (2009) 068}, [\href{https://arxiv.org/abs/0812.0465}{{\tt 0812.0465}}].

\bibitem{Figueras:2011gd}
P.~Figueras and T.~Wiseman, \emph{{Gravity and large black holes in
  Randall-Sundrum II braneworlds}},
  \href{http://dx.doi.org/10.1103/PhysRevLett.107.081101}{\emph{Phys. Rev.
  Lett.} {\bf 107} (2011) 081101}, [\href{https://arxiv.org/abs/1105.2558}{{\tt
  1105.2558}}].

\bibitem{Emparan:2023dxm}
R.~Emparan, R.~Luna, R.~Suzuki, M.~Toma\v{s}evi\'c and B.~Way,
  \emph{{Holographic duals of evaporating black holes}},
  \href{http://dx.doi.org/10.1007/JHEP05(2023)182}{\emph{JHEP} {\bf 05} (2023)
  182}, [\href{https://arxiv.org/abs/2301.02587}{{\tt 2301.02587}}].

\bibitem{Henningson:1998gx}
M.~Henningson and K.~Skenderis, \emph{{The Holographic Weyl anomaly}},
  \href{http://dx.doi.org/10.1088/1126-6708/1998/07/023}{\emph{JHEP} {\bf 07}
  (1998) 023}, [\href{https://arxiv.org/abs/hep-th/9806087}{{\tt
  hep-th/9806087}}].

\bibitem{Skenderis:1999nb}
K.~Skenderis and S.~N. Solodukhin, \emph{{Quantum effective action from the AdS
  / CFT correspondence}},
  \href{http://dx.doi.org/10.1016/S0370-2693(99)01467-7}{\emph{Phys. Lett. B}
  {\bf 472} (2000) 316--322}, [\href{https://arxiv.org/abs/hep-th/9910023}{{\tt
  hep-th/9910023}}].

\bibitem{Carlip:2005tz}
S.~Carlip, \emph{{Dynamics of asymptotic diffeomorphisms in (2+1)-dimensional
  gravity}}, \href{http://dx.doi.org/10.1088/0264-9381/22/14/014}{\emph{Class.
  Quant. Grav.} {\bf 22} (2005) 3055--3060},
  [\href{https://arxiv.org/abs/gr-qc/0501033}{{\tt gr-qc/0501033}}].

\bibitem{Nguyen:2021pdz}
K.~Nguyen, \emph{{Holographic boundary actions in AdS$_{3}$/CFT$_{2}$
  revisited}}, \href{http://dx.doi.org/10.1007/JHEP10(2021)218}{\emph{JHEP}
  {\bf 10} (2021) 218}, [\href{https://arxiv.org/abs/2108.01095}{{\tt
  2108.01095}}].

\bibitem{Dvali:2007hz}
G.~Dvali, \emph{{Black Holes and Large N Species Solution to the Hierarchy
  Problem}}, \href{http://dx.doi.org/10.1002/prop.201000009}{\emph{Fortsch.
  Phys.} {\bf 58} (2010) 528--536},
  [\href{https://arxiv.org/abs/0706.2050}{{\tt 0706.2050}}].

\bibitem{Jackiw:1984je}
R.~Jackiw, \emph{{Lower Dimensional Gravity}},
  \href{http://dx.doi.org/10.1016/0550-3213(85)90448-1}{\emph{Nucl. Phys. B}
  {\bf 252} (1985) 343--356}.

\bibitem{Teitelboim:1983ux}
C.~Teitelboim, \emph{{Gravitation and Hamiltonian Structure in Two Space-Time
  Dimensions}},
  \href{http://dx.doi.org/10.1016/0370-2693(83)90012-6}{\emph{Phys. Lett. B}
  {\bf 126} (1983) 41--45}.

\bibitem{Geng:2022tfc}
H.~Geng, \emph{{Aspects of AdS$_{2}$ quantum gravity and the Karch-Randall
  braneworld}}, \href{http://dx.doi.org/10.1007/JHEP09(2022)024}{\emph{JHEP}
  {\bf 09} (2022) 024}, [\href{https://arxiv.org/abs/2206.11277}{{\tt
  2206.11277}}].

\bibitem{Geng:2022slq}
H.~Geng, A.~Karch, C.~Perez-Pardavila, S.~Raju, L.~Randall, M.~Riojas et~al.,
  \emph{{Jackiw-Teitelboim Gravity from the Karch-Randall Braneworld}},
  \href{http://dx.doi.org/10.1103/PhysRevLett.129.231601}{\emph{Phys. Rev.
  Lett.} {\bf 129} (2022) 231601},
  [\href{https://arxiv.org/abs/2206.04695}{{\tt 2206.04695}}].

\bibitem{Bhattacharjee:2022pcb}
A.~Bhattacharjee and M.~Saha, \emph{{JT gravity from holographic reduction of
  3D asymptotically flat spacetime}},
  \href{http://dx.doi.org/10.1007/JHEP01(2023)138}{\emph{JHEP} {\bf 01} (2023)
  138}, [\href{https://arxiv.org/abs/2211.13415}{{\tt 2211.13415}}].

\bibitem{Aguilar-Gutierrez:2023tic}
S.~E. Aguilar-Gutierrez, A.~K. Patra and J.~F. Pedraza, \emph{{Entangled
  universes in dS wedge holography}},
  \href{http://dx.doi.org/10.1007/JHEP10(2023)156}{\emph{JHEP} {\bf 10} (2023)
  156}, [\href{https://arxiv.org/abs/2308.05666}{{\tt 2308.05666}}].

\bibitem{Neuenfeld:2024gta}
D.~Neuenfeld, A.~Svesko and W.~Sybesma, \emph{{Liouville gravity at the end of
  the world: Deformed defects in AdS/BCFT}},
  \href{https://arxiv.org/abs/2404.07260}{{\tt 2404.07260}}.

\bibitem{Callan:1992rs}
C.~G. Callan, Jr., S.~B. Giddings, J.~A. Harvey and A.~Strominger,
  \emph{{Evanescent black holes}},
  \href{http://dx.doi.org/10.1103/PhysRevD.45.R1005}{\emph{Phys. Rev. D} {\bf
  45} (1992) R1005}, [\href{https://arxiv.org/abs/hep-th/9111056}{{\tt
  hep-th/9111056}}].

\bibitem{Russo:1992ax}
J.~G. Russo, L.~Susskind and L.~Thorlacius, \emph{{The Endpoint of Hawking
  radiation}}, \href{http://dx.doi.org/10.1103/PhysRevD.46.3444}{\emph{Phys.
  Rev. D} {\bf 46} (1992) 3444--3449},
  [\href{https://arxiv.org/abs/hep-th/9206070}{{\tt hep-th/9206070}}].

\bibitem{Susskind:1993if}
L.~Susskind, L.~Thorlacius and J.~Uglum, \emph{{The Stretched horizon and black
  hole complementarity}},
  \href{http://dx.doi.org/10.1103/PhysRevD.48.3743}{\emph{Phys. Rev. D} {\bf
  48} (1993) 3743--3761}, [\href{https://arxiv.org/abs/hep-th/9306069}{{\tt
  hep-th/9306069}}].

\bibitem{Fiola:1994ir}
T.~M. Fiola, J.~Preskill, A.~Strominger and S.~P. Trivedi, \emph{{Black hole
  thermodynamics and information loss in two-dimensions}},
  \href{http://dx.doi.org/10.1103/PhysRevD.50.3987}{\emph{Phys. Rev. D} {\bf
  50} (1994) 3987--4014}, [\href{https://arxiv.org/abs/hep-th/9403137}{{\tt
  hep-th/9403137}}].

\bibitem{Emparan:2000rs}
R.~Emparan, G.~T. Horowitz and R.~C. Myers, \emph{{Black holes radiate mainly
  on the brane}},
  \href{http://dx.doi.org/10.1103/PhysRevLett.85.499}{\emph{Phys. Rev. Lett.}
  {\bf 85} (2000) 499--502}, [\href{https://arxiv.org/abs/hep-th/0003118}{{\tt
  hep-th/0003118}}].

\bibitem{Plebanski:1976gy}
J.~F. Plebanski and M.~Demianski, \emph{{Rotating, charged, and uniformly
  accelerating mass in general relativity}},
  \href{http://dx.doi.org/10.1016/0003-4916(76)90240-2}{\emph{Annals Phys.}
  {\bf 98} (1976) 98--127}.

\bibitem{Podolsky:2002nk}
J.~Podolsky, \emph{{Accelerating black holes in anti-de Sitter universe}},
  \href{http://dx.doi.org/10.1023/A:1013961411430}{\emph{Czech. J. Phys.} {\bf
  52} (2002) 1--10}, [\href{https://arxiv.org/abs/gr-qc/0202033}{{\tt
  gr-qc/0202033}}].

\bibitem{Emparan:2021hyr}
R.~Emparan, A.~M. Frassino, M.~Sasieta and M.~Toma\v{s}evi\'c,
  \emph{{Holographic complexity of quantum black holes}},
  \href{http://dx.doi.org/10.1007/JHEP02(2022)204}{\emph{JHEP} {\bf 02} (2022)
  204}, [\href{https://arxiv.org/abs/2112.04860}{{\tt 2112.04860}}].

\bibitem{Cremonini:2009ih}
S.~Cremonini, J.~T. Liu and P.~Szepietowski, \emph{{Higher Derivative
  Corrections to R-charged Black Holes: Boundary Counterterms and the
  Mass-Charge Relation}},
  \href{http://dx.doi.org/10.1007/JHEP03(2010)042}{\emph{JHEP} {\bf 03} (2010)
  042}, [\href{https://arxiv.org/abs/0910.5159}{{\tt 0910.5159}}].

\bibitem{Chernicoff:2024dll}
M.~Chernicoff, G.~Giribet, J.~Moreno, J.~Oliva, R.~Rojas and C.~R. d.~A.
  Torres, \emph{{Quantum backreactions in (A)dS3 massive gravity and
  logarithmic asymptotic behavior}},
  \href{https://arxiv.org/abs/2404.10127}{{\tt 2404.10127}}.

\bibitem{Caldarelli:1999xj}
M.~M. Caldarelli, G.~Cognola and D.~Klemm, \emph{{Thermodynamics of
  Kerr-Newman-AdS black holes and conformal field theories}},
  \href{http://dx.doi.org/10.1088/0264-9381/17/2/310}{\emph{Class. Quant.
  Grav.} {\bf 17} (2000) 399--420},
  [\href{https://arxiv.org/abs/hep-th/9908022}{{\tt hep-th/9908022}}].

\bibitem{Gibbons:2004ai}
G.~W. Gibbons, M.~J. Perry and C.~N. Pope, \emph{{The First law of
  thermodynamics for Kerr-anti-de Sitter black holes}},
  \href{http://dx.doi.org/10.1088/0264-9381/22/9/002}{\emph{Class. Quant.
  Grav.} {\bf 22} (2005) 1503--1526},
  [\href{https://arxiv.org/abs/hep-th/0408217}{{\tt hep-th/0408217}}].

\bibitem{Lemos:2021jtm}
J.~P.~S. Lemos and P.~Luz, \emph{{All fundamental electrically charged thin
  shells in general relativity: From star shells to tension shell black holes,
  regular black holes, and beyond}},
  \href{http://dx.doi.org/10.1103/PhysRevD.103.104046}{\emph{Phys. Rev. D} {\bf
  103} (2021) 104046}, [\href{https://arxiv.org/abs/2103.15832}{{\tt
  2103.15832}}].

\bibitem{Taylor:2000xw}
M.~Taylor, \emph{{More on counterterms in the gravitational action and
  anomalies}},  \href{https://arxiv.org/abs/hep-th/0002125}{{\tt
  hep-th/0002125}}.

\bibitem{Climent:2024nuj}
A.~Climent, R.~Emparan and R.~A. Hennigar, \emph{{Chemical Potential and Charge
  in Quantum Black Holes}},  \href{https://arxiv.org/abs/2404.15148}{{\tt
  2404.15148}}.

\bibitem{Feng:2024uia}
Y.~Feng, H.~Ma, R.~B. Mann, Y.~Xue and M.~Zhang, \emph{{Quantum Charged Black
  Holes}},  \href{https://arxiv.org/abs/2404.07192}{{\tt 2404.07192}}.

\bibitem{Martinez:1999qi}
C.~Martinez, C.~Teitelboim and J.~Zanelli, \emph{{Charged rotating black hole
  in three space-time dimensions}},
  \href{http://dx.doi.org/10.1103/PhysRevD.61.104013}{\emph{Phys. Rev. D} {\bf
  61} (2000) 104013}, [\href{https://arxiv.org/abs/hep-th/9912259}{{\tt
  hep-th/9912259}}].

\bibitem{Emparan:2000fn}
R.~Emparan, R.~Gregory and C.~Santos, \emph{{Black holes on thick branes}},
  \href{http://dx.doi.org/10.1103/PhysRevD.63.104022}{\emph{Phys. Rev. D} {\bf
  63} (2001) 104022}, [\href{https://arxiv.org/abs/hep-th/0012100}{{\tt
  hep-th/0012100}}].

\bibitem{Maldacena:2010un}
J.~Maldacena, \emph{{Vacuum decay into Anti de Sitter space}},
  \href{https://arxiv.org/abs/1012.0274}{{\tt 1012.0274}}.

\bibitem{Barbon:2010gn}
J.~L.~F. Barbon and E.~Rabinovici, \emph{{Holography of AdS vacuum bubbles}},
  \href{http://dx.doi.org/10.1007/JHEP04(2010)123}{\emph{JHEP} {\bf 04} (2010)
  123}, [\href{https://arxiv.org/abs/1003.4966}{{\tt 1003.4966}}].

\bibitem{Barbon:2011ta}
J.~L.~F. Barbon and E.~Rabinovici, \emph{{AdS Crunches, CFT Falls And
  Cosmological Complementarity}},
  \href{http://dx.doi.org/10.1007/JHEP04(2011)044}{\emph{JHEP} {\bf 04} (2011)
  044}, [\href{https://arxiv.org/abs/1102.3015}{{\tt 1102.3015}}].

\bibitem{Coleman:1980aw}
S.~R. Coleman and F.~De~Luccia, \emph{{Gravitational Effects on and of Vacuum
  Decay}}, \href{http://dx.doi.org/10.1103/PhysRevD.21.3305}{\emph{Phys. Rev.
  D} {\bf 21} (1980) 3305}.

\bibitem{Nariai99}
H.~Nariai, \emph{On a new cosmological solution of einstein's field equations
  of gravitation},
  \href{http://dx.doi.org/10.1023/A:1026602724948}{\emph{General relativity and
  gravitation} {\bf 31} (06, 1999) 963 -- 971}.

\bibitem{Booth:1998gf}
I.~S. Booth and R.~B. Mann, \emph{{Cosmological pair production of charged and
  rotating black holes}},
  \href{http://dx.doi.org/10.1016/S0550-3213(98)00756-1}{\emph{Nucl. Phys. B}
  {\bf 539} (1999) 267--306}, [\href{https://arxiv.org/abs/gr-qc/9806056}{{\tt
  gr-qc/9806056}}].

\bibitem{Bhattacharya:2017scw}
S.~Bhattacharya, \emph{{Kerr-de Sitter spacetime, Penrose process and the
  generalized area theorem}},
  \href{http://dx.doi.org/10.1103/PhysRevD.97.084049}{\emph{Phys. Rev. D} {\bf
  97} (2018) 084049}, [\href{https://arxiv.org/abs/1710.00997}{{\tt
  1710.00997}}].

\bibitem{Anninos:2011vd}
D.~Anninos, S.~de~Buyl and S.~Detournay, \emph{{Holography For a De
  Sitter-Esque Geometry}},
  \href{http://dx.doi.org/10.1007/JHEP05(2011)003}{\emph{JHEP} {\bf 05} (2011)
  003}, [\href{https://arxiv.org/abs/1102.3178}{{\tt 1102.3178}}].

\bibitem{Bekenstein:1972tm}
J.~D. Bekenstein, \emph{{Black holes and the second law}},
  \href{http://dx.doi.org/10.1007/BF02757029}{\emph{Lett. Nuovo Cim.} {\bf 4}
  (1972) 737--740}.

\bibitem{Bekenstein:1973ur}
J.~D. Bekenstein, \emph{{Black holes and entropy}},
  \href{http://dx.doi.org/10.1103/PhysRevD.7.2333}{\emph{Phys. Rev. D} {\bf 7}
  (1973) 2333--2346}.

\bibitem{Hawking:1974sw}
S.~W. Hawking, \emph{{Particle Creation by Black Holes}},
  \href{http://dx.doi.org/10.1007/BF02345020}{\emph{Commun. Math. Phys.} {\bf
  43} (1975) 199--220}.

\bibitem{Hawking:1976de}
S.~W. Hawking, \emph{{Black Holes and Thermodynamics}},
  \href{http://dx.doi.org/10.1103/PhysRevD.13.191}{\emph{Phys. Rev. D} {\bf 13}
  (1976) 191--197}.

\bibitem{Bardeen:1973gs}
J.~M. Bardeen, B.~Carter and S.~W. Hawking, \emph{{The Four laws of black hole
  mechanics}}, \href{http://dx.doi.org/10.1007/BF01645742}{\emph{Commun. Math.
  Phys.} {\bf 31} (1973) 161--170}.

\bibitem{Hawking:1982dh}
S.~W. Hawking and D.~N. Page, \emph{{Thermodynamics of Black Holes in anti-De
  Sitter Space}}, \href{http://dx.doi.org/10.1007/BF01208266}{\emph{Commun.
  Math. Phys.} {\bf 87} (1983) 577}.

\bibitem{Podolsky:2003gm}
J.~Podolsky, M.~Ortaggio and P.~Krtous, \emph{{Radiation from accelerated black
  holes in an anti-de Sitter universe}},
  \href{http://dx.doi.org/10.1103/PhysRevD.68.124004}{\emph{Phys. Rev. D} {\bf
  68} (2003) 124004}, [\href{https://arxiv.org/abs/gr-qc/0307108}{{\tt
  gr-qc/0307108}}].

\bibitem{Appels:2016uha}
M.~Appels, R.~Gregory and D.~Kubiznak, \emph{{Thermodynamics of Accelerating
  Black Holes}},
  \href{http://dx.doi.org/10.1103/PhysRevLett.117.131303}{\emph{Phys. Rev.
  Lett.} {\bf 117} (2016) 131303},
  [\href{https://arxiv.org/abs/1604.08812}{{\tt 1604.08812}}].

\bibitem{Appels:2017xoe}
M.~Appels, R.~Gregory and D.~Kubiznak, \emph{{Black Hole Thermodynamics with
  Conical Defects}},
  \href{http://dx.doi.org/10.1007/JHEP05(2017)116}{\emph{JHEP} {\bf 05} (2017)
  116}, [\href{https://arxiv.org/abs/1702.00490}{{\tt 1702.00490}}].

\bibitem{Anabalon:2018ydc}
A.~Anabal\'on, M.~Appels, R.~Gregory, D.~Kubiz\v{n}\'ak, R.~B. Mann and
  A.~Ovg\"un, \emph{{Holographic Thermodynamics of Accelerating Black Holes}},
  \href{http://dx.doi.org/10.1103/PhysRevD.98.104038}{\emph{Phys. Rev. D} {\bf
  98} (2018) 104038}, [\href{https://arxiv.org/abs/1805.02687}{{\tt
  1805.02687}}].

\bibitem{Appels:2018jcs}
M.~Appels, \emph{{Thermodynamics of Accelerating Black Holes}}.
\newblock PhD thesis, Durham U., Dept. of Math., 2018.

\bibitem{Ball:2020vzo}
A.~Ball and N.~Miller, \emph{{Accelerating black hole thermodynamics with boost
  time}}, \href{http://dx.doi.org/10.1088/1361-6382/ac0766}{\emph{Class. Quant.
  Grav.} {\bf 38} (2021) 145031}, [\href{https://arxiv.org/abs/2008.03682}{{\tt
  2008.03682}}].

\bibitem{Gibbons:1976ue}
G.~W. Gibbons and S.~W. Hawking, \emph{{Action Integrals and Partition
  Functions in Quantum Gravity}},
  \href{http://dx.doi.org/10.1103/PhysRevD.15.2752}{\emph{Phys. Rev. D} {\bf
  15} (1977) 2752--2756}.

\bibitem{Kudoh:2004ub}
H.~Kudoh and Y.~Kurita, \emph{{Thermodynamics of four-dimensional black objects
  in the warped compactification}},
  \href{http://dx.doi.org/10.1103/PhysRevD.70.084029}{\emph{Phys. Rev. D} {\bf
  70} (2004) 084029}, [\href{https://arxiv.org/abs/gr-qc/0406107}{{\tt
  gr-qc/0406107}}].

\bibitem{Myers:2024zhb}
R.~C. Myers, S.-M. Ruan and T.~Ugajin, \emph{{Double Holography of Entangled
  Universes}},  \href{https://arxiv.org/abs/2403.17483}{{\tt 2403.17483}}.

\bibitem{Bekenstein:1974ax}
J.~D. Bekenstein, \emph{{Generalized second law of thermodynamics in black hole
  physics}}, \href{http://dx.doi.org/10.1103/PhysRevD.9.3292}{\emph{Phys. Rev.
  D} {\bf 9} (1974) 3292--3300}.

\bibitem{Wald:1993nt}
R.~M. Wald, \emph{{Black hole entropy is the Noether charge}},
  \href{http://dx.doi.org/10.1103/PhysRevD.48.R3427}{\emph{Phys. Rev. D} {\bf
  48} (1993) R3427--R3431}, [\href{https://arxiv.org/abs/gr-qc/9307038}{{\tt
  gr-qc/9307038}}].

\bibitem{Iyer:1994ys}
V.~Iyer and R.~M. Wald, \emph{{Some properties of Noether charge and a proposal
  for dynamical black hole entropy}},
  \href{http://dx.doi.org/10.1103/PhysRevD.50.846}{\emph{Phys. Rev. D} {\bf 50}
  (1994) 846--864}, [\href{https://arxiv.org/abs/gr-qc/9403028}{{\tt
  gr-qc/9403028}}].

\bibitem{Jacobson:1993vj}
T.~Jacobson, G.~Kang and R.~C. Myers, \emph{{On black hole entropy}},
  \href{http://dx.doi.org/10.1103/PhysRevD.49.6587}{\emph{Phys. Rev. D} {\bf
  49} (1994) 6587--6598}, [\href{https://arxiv.org/abs/gr-qc/9312023}{{\tt
  gr-qc/9312023}}].

\bibitem{Susskind:1994sm}
L.~Susskind and J.~Uglum, \emph{{Black hole entropy in canonical quantum
  gravity and superstring theory}},
  \href{http://dx.doi.org/10.1103/PhysRevD.50.2700}{\emph{Phys. Rev. D} {\bf
  50} (1994) 2700--2711}, [\href{https://arxiv.org/abs/hep-th/9401070}{{\tt
  hep-th/9401070}}].

\bibitem{Solodukhin:2011gn}
S.~N. Solodukhin, \emph{{Entanglement entropy of black holes}},
  \href{http://dx.doi.org/10.12942/lrr-2011-8}{\emph{Living Rev. Rel.} {\bf 14}
  (2011) 8}, [\href{https://arxiv.org/abs/1104.3712}{{\tt 1104.3712}}].

\bibitem{Cooperman:2013iqr}
J.~H. Cooperman and M.~A. Luty, \emph{{Renormalization of Entanglement Entropy
  and the Gravitational Effective Action}},
  \href{http://dx.doi.org/10.1007/JHEP12(2014)045}{\emph{JHEP} {\bf 12} (2014)
  045}, [\href{https://arxiv.org/abs/1302.1878}{{\tt 1302.1878}}].

\bibitem{Pedraza:2021cvx}
J.~F. Pedraza, A.~Svesko, W.~Sybesma and M.~R. Visser, \emph{{Semi-classical
  thermodynamics of quantum extremal surfaces in Jackiw-Teitelboim gravity}},
  \href{http://dx.doi.org/10.1007/JHEP12(2021)134}{\emph{JHEP} {\bf 12} (2021)
  134}, [\href{https://arxiv.org/abs/2107.10358}{{\tt 2107.10358}}].

\bibitem{Svesko:2022txo}
A.~Svesko, E.~Verheijden, E.~P. Verlinde and M.~R. Visser, \emph{{Quasi-local
  energy and microcanonical entropy in two-dimensional nearly de Sitter
  gravity}}, \href{http://dx.doi.org/10.1007/JHEP08(2022)075}{\emph{JHEP} {\bf
  08} (2022) 075}, [\href{https://arxiv.org/abs/2203.00700}{{\tt 2203.00700}}].

\bibitem{Frassino:2024bjg}
A.~M. Frassino, R.~A. Hennigar, J.~F. Pedraza and A.~Svesko, \emph{{Quantum
  inequalities for quantum black holes}},
  \href{https://arxiv.org/abs/2406.17860}{{\tt 2406.17860}}.

\bibitem{Spradlin:2001pw}
M.~Spradlin, A.~Strominger and A.~Volovich, \emph{{Les Houches lectures on de
  Sitter space}},  in \emph{{Les Houches Summer School: Session 76: Euro Summer
  School on Unity of Fundamental Physics: Gravity, Gauge Theory and Strings}},
  pp.~423--453, 10, 2001.
\newblock \href{https://arxiv.org/abs/hep-th/0110007}{{\tt hep-th/0110007}}.

\bibitem{Morvan:2022ybp}
E.~K. Morvan, J.~P. van~der Schaar and M.~R. Visser, \emph{{On the Euclidean
  Action of de Sitter Black Holes and Constrained Instantons}},
  \href{https://arxiv.org/abs/2203.06155}{{\tt 2203.06155}}.

\bibitem{Morvan:2022aon}
E.~K. Morvan, J.~P. van~der Schaar and M.~R. Visser, \emph{{Action, entropy and
  pair creation rate of charged black holes in de Sitter space}},
  \href{https://arxiv.org/abs/2212.12713}{{\tt 2212.12713}}.

\bibitem{Draper:2022xzl}
P.~Draper and S.~Farkas, \emph{{de Sitter black holes as constrained states in
  the Euclidean path integral}},
  \href{http://dx.doi.org/10.1103/PhysRevD.105.126022}{\emph{Phys. Rev. D} {\bf
  105} (2022) 126022}, [\href{https://arxiv.org/abs/2203.02426}{{\tt
  2203.02426}}].

\bibitem{Bousso:1996au}
R.~Bousso and S.~W. Hawking, \emph{{Pair creation of black holes during
  inflation}}, \href{http://dx.doi.org/10.1103/PhysRevD.54.6312}{\emph{Phys.
  Rev. D} {\bf 54} (1996) 6312--6322},
  [\href{https://arxiv.org/abs/gr-qc/9606052}{{\tt gr-qc/9606052}}].

\bibitem{Kastor:2009wy}
D.~Kastor, S.~Ray and J.~Traschen, \emph{{Enthalpy and the Mechanics of AdS
  Black Holes}},
  \href{http://dx.doi.org/10.1088/0264-9381/26/19/195011}{\emph{Class. Quant.
  Grav.} {\bf 26} (2009) 195011}, [\href{https://arxiv.org/abs/0904.2765}{{\tt
  0904.2765}}].

\bibitem{Dolan:2010ha}
B.~P. Dolan, \emph{{The cosmological constant and the black hole equation of
  state}}, \href{http://dx.doi.org/10.1088/0264-9381/28/12/125020}{\emph{Class.
  Quant. Grav.} {\bf 28} (2011) 125020},
  [\href{https://arxiv.org/abs/1008.5023}{{\tt 1008.5023}}].

\bibitem{Cvetic:2010jb}
M.~Cvetic, G.~W. Gibbons, D.~Kubiznak and C.~N. Pope, \emph{{Black Hole
  Enthalpy and an Entropy Inequality for the Thermodynamic Volume}},
  \href{http://dx.doi.org/10.1103/PhysRevD.84.024037}{\emph{Phys. Rev. D} {\bf
  84} (2011) 024037}, [\href{https://arxiv.org/abs/1012.2888}{{\tt
  1012.2888}}].

\bibitem{Johnson:2014xza}
C.~V. Johnson, \emph{{Thermodynamic Volumes for AdS-Taub-NUT and
  AdS-Taub-Bolt}},
  \href{http://dx.doi.org/10.1088/0264-9381/31/23/235003}{\emph{Class. Quant.
  Grav.} {\bf 31} (2014) 235003}, [\href{https://arxiv.org/abs/1405.5941}{{\tt
  1405.5941}}].

\bibitem{Chamblin:1999tk}
A.~Chamblin, R.~Emparan, C.~V. Johnson and R.~C. Myers, \emph{{Charged AdS
  black holes and catastrophic holography}},
  \href{http://dx.doi.org/10.1103/PhysRevD.60.064018}{\emph{Phys. Rev. D} {\bf
  60} (1999) 064018}, [\href{https://arxiv.org/abs/hep-th/9902170}{{\tt
  hep-th/9902170}}].

\bibitem{Kubiznak:2012wp}
D.~Kubiznak and R.~B. Mann, \emph{{P-V criticality of charged AdS black
  holes}}, \href{http://dx.doi.org/10.1007/JHEP07(2012)033}{\emph{JHEP} {\bf
  07} (2012) 033}, [\href{https://arxiv.org/abs/1205.0559}{{\tt 1205.0559}}].

\bibitem{Kubiznak:2014zwa}
D.~Kubiznak and R.~B. Mann, \emph{{Black hole chemistry}},
  \href{http://dx.doi.org/10.1139/cjp-2014-0465}{\emph{Can. J. Phys.} {\bf 93}
  (2015) 999--1002}, [\href{https://arxiv.org/abs/1404.2126}{{\tt 1404.2126}}].

\bibitem{Johnson:2014yja}
C.~V. Johnson, \emph{{Holographic Heat Engines}},
  \href{http://dx.doi.org/10.1088/0264-9381/31/20/205002}{\emph{Class. Quant.
  Grav.} {\bf 31} (2014) 205002}, [\href{https://arxiv.org/abs/1404.5982}{{\tt
  1404.5982}}].

\bibitem{Kubiznak:2016qmn}
D.~Kubiznak, R.~B. Mann and M.~Teo, \emph{{Black hole chemistry: thermodynamics
  with Lambda}}, \href{http://dx.doi.org/10.1088/1361-6382/aa5c69}{\emph{Class.
  Quant. Grav.} {\bf 34} (2017) 063001},
  [\href{https://arxiv.org/abs/1608.06147}{{\tt 1608.06147}}].

\bibitem{Dolan:2014cja}
B.~P. Dolan, \emph{{Bose condensation and branes}},
  \href{http://dx.doi.org/10.1007/JHEP10(2014)179}{\emph{JHEP} {\bf 10} (2014)
  179}, [\href{https://arxiv.org/abs/1406.7267}{{\tt 1406.7267}}].

\bibitem{Kastor:2014dra}
D.~Kastor, S.~Ray and J.~Traschen, \emph{{Chemical Potential in the First Law
  for Holographic Entanglement Entropy}},
  \href{http://dx.doi.org/10.1007/JHEP11(2014)120}{\emph{JHEP} {\bf 11} (2014)
  120}, [\href{https://arxiv.org/abs/1409.3521}{{\tt 1409.3521}}].

\bibitem{Caceres:2015vsa}
E.~Caceres, P.~H. Nguyen and J.~F. Pedraza, \emph{{Holographic entanglement
  entropy and the extended phase structure of STU black holes}},
  \href{http://dx.doi.org/10.1007/JHEP09(2015)184}{\emph{JHEP} {\bf 09} (2015)
  184}, [\href{https://arxiv.org/abs/1507.06069}{{\tt 1507.06069}}].

\bibitem{Karch:2015rpa}
A.~Karch and B.~Robinson, \emph{{Holographic Black Hole Chemistry}},
  \href{http://dx.doi.org/10.1007/JHEP12(2015)073}{\emph{JHEP} {\bf 12} (2015)
  073}, [\href{https://arxiv.org/abs/1510.02472}{{\tt 1510.02472}}].

\bibitem{Caceres:2016xjz}
E.~Caceres, P.~H. Nguyen and J.~F. Pedraza, \emph{{Holographic entanglement
  chemistry}}, \href{http://dx.doi.org/10.1103/PhysRevD.95.106015}{\emph{Phys.
  Rev. D} {\bf 95} (2017) 106015},
  [\href{https://arxiv.org/abs/1605.00595}{{\tt 1605.00595}}].

\bibitem{Rosso:2020zkk}
F.~Rosso and A.~Svesko, \emph{{Novel aspects of the extended first law of
  entanglement}}, \href{http://dx.doi.org/10.1007/JHEP08(2020)008}{\emph{JHEP}
  {\bf 08} (2020) 008}, [\href{https://arxiv.org/abs/2003.10462}{{\tt
  2003.10462}}].

\bibitem{Frassino:2022zaz}
A.~M. Frassino, J.~F. Pedraza, A.~Svesko and M.~R. Visser,
  \emph{{Higher-Dimensional Origin of Extended Black Hole Thermodynamics}},
  \href{http://dx.doi.org/10.1103/PhysRevLett.130.161501}{\emph{Phys. Rev.
  Lett.} {\bf 130} (2023) 161501},
  [\href{https://arxiv.org/abs/2212.14055}{{\tt 2212.14055}}].

\bibitem{Penrose:1968ar}
R.~Penrose, \emph{{Structure of space-time}},  in \emph{{Battelle Rencontres}},
  pp.~121--235, 1968.

\bibitem{Penrose:1969pc}
R.~Penrose, \emph{{Gravitational collapse: The role of general relativity}},
  \href{http://dx.doi.org/10.1023/A:1016578408204}{\emph{Riv. Nuovo Cim.} {\bf
  1} (1969) 252--276}.

\bibitem{Amo:2023bbo}
M.~Amo, A.~M. Frassino and R.~A. Hennigar, \emph{{Entropy Bounds for Rotating
  AdS Black Holes}},
  \href{http://dx.doi.org/10.1103/PhysRevLett.131.241401}{\emph{Phys. Rev.
  Lett.} {\bf 131} (2023) 241401},
  [\href{https://arxiv.org/abs/2307.03011}{{\tt 2307.03011}}].

\bibitem{Klemm:2014rda}
D.~Klemm, \emph{{Four-dimensional black holes with unusual horizons}},
  \href{http://dx.doi.org/10.1103/PhysRevD.89.084007}{\emph{Phys. Rev. D} {\bf
  89} (2014) 084007}, [\href{https://arxiv.org/abs/1401.3107}{{\tt
  1401.3107}}].

\bibitem{Hennigar:2014cfa}
R.~A. Hennigar, D.~Kubiz\v{n}\'ak and R.~B. Mann, \emph{{Entropy Inequality
  Violations from Ultraspinning Black Holes}},
  \href{http://dx.doi.org/10.1103/PhysRevLett.115.031101}{\emph{Phys. Rev.
  Lett.} {\bf 115} (2015) 031101}, [\href{https://arxiv.org/abs/1411.4309}{{\tt
  1411.4309}}].

\bibitem{Hennigar:2015cja}
R.~A. Hennigar, D.~Kubiz\v{n}\'ak, R.~B. Mann and N.~Musoke,
  \emph{{Ultraspinning limits and super-entropic black holes}},
  \href{http://dx.doi.org/10.1007/JHEP06(2015)096}{\emph{JHEP} {\bf 06} (2015)
  096}, [\href{https://arxiv.org/abs/1504.07529}{{\tt 1504.07529}}].

\bibitem{Appels:2019vow}
M.~Appels, L.~Cuspinera, R.~Gregory, P.~Krtou\v{s} and D.~Kubiz\v{n}\'ak,
  \emph{{Are \textquotedblleft{}Superentropic\textquotedblright{} black holes
  superentropic?}},
  \href{http://dx.doi.org/10.1007/JHEP02(2020)195}{\emph{JHEP} {\bf 02} (2020)
  195}, [\href{https://arxiv.org/abs/1911.12817}{{\tt 1911.12817}}].

\bibitem{Frassino:2015oca}
A.~M. Frassino, R.~B. Mann and J.~R. Mureika, \emph{{Lower-Dimensional Black
  Hole Chemistry}},
  \href{http://dx.doi.org/10.1103/PhysRevD.92.124069}{\emph{Phys. Rev. D} {\bf
  92} (2015) 124069}, [\href{https://arxiv.org/abs/1509.05481}{{\tt
  1509.05481}}].

\bibitem{Johnson:2019mdp}
C.~V. Johnson, \emph{{Instability of super-entropic black holes in extended
  thermodynamics}},
  \href{http://dx.doi.org/10.1142/S0217732320500984}{\emph{Mod. Phys. Lett. A}
  {\bf 35} (2020) 2050098}, [\href{https://arxiv.org/abs/1906.00993}{{\tt
  1906.00993}}].

\bibitem{Johnson:2019wcq}
C.~V. Johnson, V.~L. Martin and A.~Svesko, \emph{{Microscopic description of
  thermodynamic volume in extended black hole thermodynamics}},
  \href{http://dx.doi.org/10.1103/PhysRevD.101.086006}{\emph{Phys. Rev. D} {\bf
  101} (2020) 086006}, [\href{https://arxiv.org/abs/1911.05286}{{\tt
  1911.05286}}].

\bibitem{Maldacena:1998bw}
J.~M. Maldacena and A.~Strominger, \emph{{AdS(3) black holes and a stringy
  exclusion principle}},
  \href{http://dx.doi.org/10.1088/1126-6708/1998/12/005}{\emph{JHEP} {\bf 12}
  (1998) 005}, [\href{https://arxiv.org/abs/hep-th/9804085}{{\tt
  hep-th/9804085}}].

\bibitem{Birmingham:2002ph}
D.~Birmingham, I.~Sachs and S.~N. Solodukhin, \emph{{Relaxation in conformal
  field theory, Hawking-Page transition, and quasinormal normal modes}},
  \href{http://dx.doi.org/10.1103/PhysRevD.67.104026}{\emph{Phys. Rev. D} {\bf
  67} (2003) 104026}, [\href{https://arxiv.org/abs/hep-th/0212308}{{\tt
  hep-th/0212308}}].

\bibitem{Frassino:2023wpc}
A.~M. Frassino, J.~F. Pedraza, A.~Svesko and M.~R. Visser, \emph{{Reentrant
  phase transitions of quantum black holes}},
  \href{http://dx.doi.org/10.1103/PhysRevD.109.124040}{\emph{Phys. Rev. D} {\bf
  109} (2024) 124040}, [\href{https://arxiv.org/abs/2310.12220}{{\tt
  2310.12220}}].

\bibitem{Johnson:2023dtf}
C.~V. Johnson and R.~Nazario, \emph{{Specific Heats for Quantum BTZ Black Holes
  in Extended Thermodynamics}},  \href{https://arxiv.org/abs/2310.12212}{{\tt
  2310.12212}}.

\bibitem{HosseiniMansoori:2024bfi}
S.~A. Hosseini~Mansoori, J.~F. Pedraza and M.~Rafiee, \emph{{Criticality and
  thermodynamic geometry of quantum BTZ black holes}},
  \href{https://arxiv.org/abs/2403.13063}{{\tt 2403.13063}}.

\bibitem{Gunasekaran:2012dq}
S.~Gunasekaran, R.~B. Mann and D.~Kubiznak, \emph{{Extended phase space
  thermodynamics for charged and rotating black holes and Born-Infeld vacuum
  polarization}}, \href{http://dx.doi.org/10.1007/JHEP11(2012)110}{\emph{JHEP}
  {\bf 11} (2012) 110}, [\href{https://arxiv.org/abs/1208.6251}{{\tt
  1208.6251}}].

\bibitem{Altamirano:2013ane}
N.~Altamirano, D.~Kubiznak and R.~B. Mann, \emph{{Reentrant phase transitions
  in rotating anti\textendash{}de Sitter black holes}},
  \href{http://dx.doi.org/10.1103/PhysRevD.88.101502}{\emph{Phys. Rev. D} {\bf
  88} (2013) 101502}, [\href{https://arxiv.org/abs/1306.5756}{{\tt
  1306.5756}}].

\bibitem{Frassino:2014pha}
A.~M. Frassino, D.~Kubiznak, R.~B. Mann and F.~Simovic, \emph{{Multiple
  Reentrant Phase Transitions and Triple Points in Lovelock Thermodynamics}},
  \href{http://dx.doi.org/10.1007/JHEP09(2014)080}{\emph{JHEP} {\bf 09} (2014)
  080}, [\href{https://arxiv.org/abs/1406.7015}{{\tt 1406.7015}}].

\bibitem{Ahmed:2023dnh}
M.~B. Ahmed, W.~Cong, D.~Kubiznak, R.~B. Mann and M.~R. Visser,
  \emph{{Holographic CFT phase transitions and criticality for rotating AdS
  black holes}}, \href{http://dx.doi.org/10.1007/JHEP08(2023)142}{\emph{JHEP}
  {\bf 08} (2023) 142}, [\href{https://arxiv.org/abs/2305.03161}{{\tt
  2305.03161}}].

\bibitem{Frassino:2024fin}
A.~M. Frassino, J.~V. Rocha and A.~P. Sanna, \emph{{Weak cosmic censorship and
  the rotating quantum BTZ black hole}},
  \href{https://arxiv.org/abs/2405.04597}{{\tt 2405.04597}}.

\bibitem{Anderson:1994hg}
P.~R. Anderson, W.~A. Hiscock and D.~A. Samuel, \emph{{Stress - energy tensor
  of quantized scalar fields in static spherically symmetric space-times}},
  \href{http://dx.doi.org/10.1103/PhysRevD.51.4337}{\emph{Phys. Rev. D} {\bf
  51} (1995) 4337--4358}.

\bibitem{Hawking:2000bb}
S.~W. Hawking, T.~Hertog and H.~S. Reall, \emph{{Trace anomaly driven
  inflation}}, \href{http://dx.doi.org/10.1103/PhysRevD.63.083504}{\emph{Phys.
  Rev. D} {\bf 63} (2001) 083504},
  [\href{https://arxiv.org/abs/hep-th/0010232}{{\tt hep-th/0010232}}].

\bibitem{Herzog:2013ed}
C.~P. Herzog and K.-W. Huang, \emph{{Stress Tensors from Trace Anomalies in
  Conformal Field Theories}},
  \href{http://dx.doi.org/10.1103/PhysRevD.87.081901}{\emph{Phys. Rev. D} {\bf
  87} (2013) 081901}, [\href{https://arxiv.org/abs/1301.5002}{{\tt
  1301.5002}}].

\bibitem{Fabbri:2005zn}
A.~Fabbri, S.~Farese, J.~Navarro-Salas, G.~J. Olmo and H.~Sanchis-Alepuz,
  \emph{{Semiclassical zero-temperature corrections to Schwarzschild spacetime
  and holography}},
  \href{http://dx.doi.org/10.1103/PhysRevD.73.104023}{\emph{Phys. Rev. D} {\bf
  73} (2006) 104023}, [\href{https://arxiv.org/abs/hep-th/0512167}{{\tt
  hep-th/0512167}}].

\bibitem{Fabbri:2005nt}
A.~Fabbri, S.~Farese, J.~Navarro-Salas, G.~J. Olmo and H.~Sanchis-Alepuz,
  \emph{{Static quantum corrections to the Schwarzschild spacetime}},
  \href{http://dx.doi.org/10.1088/1742-6596/33/1/059}{\emph{J. Phys. Conf.
  Ser.} {\bf 33} (2006) 457--462},
  [\href{https://arxiv.org/abs/hep-th/0512179}{{\tt hep-th/0512179}}].

\bibitem{Ho:2017joh}
P.-M. Ho and Y.~Matsuo, \emph{{Static Black Holes With Back Reaction From
  Vacuum Energy}},
  \href{http://dx.doi.org/10.1088/1361-6382/aaac8f}{\emph{Class. Quant. Grav.}
  {\bf 35} (2018) 065012}, [\href{https://arxiv.org/abs/1703.08662}{{\tt
  1703.08662}}].

\bibitem{Arrechea:2019jgx}
J.~Arrechea, C.~Barcel\'o, R.~Carballo-Rubio and L.~J. Garay,
  \emph{{Schwarzschild geometry counterpart in semiclassical gravity}},
  \href{http://dx.doi.org/10.1103/PhysRevD.101.064059}{\emph{Phys. Rev. D} {\bf
  101} (2020) 064059}, [\href{https://arxiv.org/abs/1911.03213}{{\tt
  1911.03213}}].

\bibitem{Beltran-Palau:2022nec}
P.~Beltr\'an-Palau, A.~del R\'\i{}o and J.~Navarro-Salas, \emph{{Quantum
  corrections to the Schwarzschild metric from vacuum polarization}},
  \href{http://dx.doi.org/10.1103/PhysRevD.107.085023}{\emph{Phys. Rev. D} {\bf
  107} (2023) 085023}, [\href{https://arxiv.org/abs/2212.08089}{{\tt
  2212.08089}}].

\bibitem{Cai:2009ua}
R.-G. Cai, L.-M. Cao and N.~Ohta, \emph{{Black Holes in Gravity with Conformal
  Anomaly and Logarithmic Term in Black Hole Entropy}},
  \href{http://dx.doi.org/10.1007/JHEP04(2010)082}{\emph{JHEP} {\bf 04} (2010)
  082}, [\href{https://arxiv.org/abs/0911.4379}{{\tt 0911.4379}}].

\bibitem{Fernandes:2023vux}
P.~G.~S. Fernandes, \emph{{Rotating black holes in semiclassical gravity}},
  \href{http://dx.doi.org/10.1103/PhysRevD.108.L061502}{\emph{Phys. Rev. D}
  {\bf 108} (2023) L061502}, [\href{https://arxiv.org/abs/2305.10382}{{\tt
  2305.10382}}].

\bibitem{Astorino:2011mw}
M.~Astorino, \emph{{Accelerating black hole in 2+1 dimensions and 3+1 black
  (st)ring}}, \href{http://dx.doi.org/10.1007/JHEP01(2011)114}{\emph{JHEP} {\bf
  01} (2011) 114}, [\href{https://arxiv.org/abs/1101.2616}{{\tt 1101.2616}}].

\bibitem{Xu:2011vp}
W.~Xu, K.~Meng and L.~Zhao, \emph{{Accelerating BTZ spacetime}},
  \href{http://dx.doi.org/10.1088/0264-9381/29/15/155005}{\emph{Class. Quant.
  Grav.} {\bf 29} (2012) 155005}, [\href{https://arxiv.org/abs/1111.0730}{{\tt
  1111.0730}}].

\bibitem{Arenas-Henriquez:2022www}
G.~Arenas-Henriquez, R.~Gregory and A.~Scoins, \emph{{On acceleration in three
  dimensions}}, \href{http://dx.doi.org/10.1007/JHEP05(2022)063}{\emph{JHEP}
  {\bf 05} (2022) 063}, [\href{https://arxiv.org/abs/2202.08823}{{\tt
  2202.08823}}].

\bibitem{Arenas-Henriquez:2023hur}
G.~Arenas-Henriquez, A.~Cisterna, F.~Diaz and R.~Gregory, \emph{{Accelerating
  Black Holes in $2+1$ dimensions: Holography revisited}},
  \href{http://dx.doi.org/10.1007/JHEP09(2023)122}{\emph{JHEP} {\bf 09} (2023)
  122}, [\href{https://arxiv.org/abs/2308.00613}{{\tt 2308.00613}}].

\bibitem{Tian:2024mew}
J.~Tian and T.~Lai, \emph{{Aspects of three-dimensional C-metric}},
  \href{http://dx.doi.org/10.1007/JHEP03(2024)079}{\emph{JHEP} {\bf 03} (2024)
  079}, [\href{https://arxiv.org/abs/2401.04457}{{\tt 2401.04457}}].

\bibitem{Camps:2010sn}
J.~Camps and R.~Emparan, \emph{{A New Class of Accelerating Black Hole
  Solutions}}, \href{http://dx.doi.org/10.1103/PhysRevD.82.024009}{\emph{Phys.
  Rev. D} {\bf 82} (2010) 024009}, [\href{https://arxiv.org/abs/1005.1175}{{\tt
  1005.1175}}].

\bibitem{Kodama:2008wf}
H.~Kodama, \emph{{Accelerating a Black Hole in Higher Dimensions}},
  \href{http://dx.doi.org/10.1143/PTP.120.371}{\emph{Prog. Theor. Phys.} {\bf
  120} (2008) 371--411}, [\href{https://arxiv.org/abs/0804.3839}{{\tt
  0804.3839}}].

\bibitem{Chamblin:1999by}
A.~Chamblin, S.~W. Hawking and H.~S. Reall, \emph{{Brane world black holes}},
  \href{http://dx.doi.org/10.1103/PhysRevD.61.065007}{\emph{Phys. Rev. D} {\bf
  61} (2000) 065007}, [\href{https://arxiv.org/abs/hep-th/9909205}{{\tt
  hep-th/9909205}}].

\bibitem{Gregory:2000gf}
R.~Gregory, \emph{{Black string instabilities in Anti-de Sitter space}},
  \href{http://dx.doi.org/10.1088/0264-9381/17/18/103}{\emph{Class. Quant.
  Grav.} {\bf 17} (2000) L125--L132},
  [\href{https://arxiv.org/abs/hep-th/0004101}{{\tt hep-th/0004101}}].

\bibitem{Gregory:1993vy}
R.~Gregory and R.~Laflamme, \emph{{Black strings and p-branes are unstable}},
  \href{http://dx.doi.org/10.1103/PhysRevLett.70.2837}{\emph{Phys. Rev. Lett.}
  {\bf 70} (1993) 2837--2840},
  [\href{https://arxiv.org/abs/hep-th/9301052}{{\tt hep-th/9301052}}].

\bibitem{Gregory:1994bj}
R.~Gregory and R.~Laflamme, \emph{{The Instability of charged black strings and
  p-branes}}, \href{http://dx.doi.org/10.1016/0550-3213(94)90206-2}{\emph{Nucl.
  Phys. B} {\bf 428} (1994) 399--434},
  [\href{https://arxiv.org/abs/hep-th/9404071}{{\tt hep-th/9404071}}].

\bibitem{Emparan:2002jp}
R.~Emparan, J.~Garcia-Bellido and N.~Kaloper, \emph{{Black hole astrophysics in
  AdS brane worlds}},
  \href{http://dx.doi.org/10.1088/1126-6708/2003/01/079}{\emph{JHEP} {\bf 01}
  (2003) 079}, [\href{https://arxiv.org/abs/hep-th/0212132}{{\tt
  hep-th/0212132}}].

\bibitem{Chamblin:2004vr}
A.~Chamblin and A.~Karch, \emph{{Hawking and Page on the brane}},
  \href{http://dx.doi.org/10.1103/PhysRevD.72.066011}{\emph{Phys. Rev. D} {\bf
  72} (2005) 066011}, [\href{https://arxiv.org/abs/hep-th/0412017}{{\tt
  hep-th/0412017}}].

\bibitem{Hubeny:2009rc}
V.~E. Hubeny, D.~Marolf and M.~Rangamani, \emph{{Hawking radiation from AdS
  black holes}},
  \href{http://dx.doi.org/10.1088/0264-9381/27/9/095018}{\emph{Class. Quant.
  Grav.} {\bf 27} (2010) 095018}, [\href{https://arxiv.org/abs/0911.4144}{{\tt
  0911.4144}}].

\bibitem{Tanahashi:2007wt}
N.~Tanahashi and T.~Tanaka, \emph{{Time-symmetric initial data of large
  brane-localized black hole in RS-II model}},
  \href{http://dx.doi.org/10.1088/1126-6708/2008/03/041}{\emph{JHEP} {\bf 03}
  (2008) 041}, [\href{https://arxiv.org/abs/0712.3799}{{\tt 0712.3799}}].

\bibitem{Gregory:2008br}
R.~Gregory, S.~F. Ross and R.~Zegers, \emph{{Classical and quantum gravity of
  brane black holes}},
  \href{http://dx.doi.org/10.1088/1126-6708/2008/09/029}{\emph{JHEP} {\bf 09}
  (2008) 029}, [\href{https://arxiv.org/abs/0802.2037}{{\tt 0802.2037}}].

\bibitem{Figueras:2011va}
P.~Figueras, J.~Lucietti and T.~Wiseman, \emph{{Ricci solitons, Ricci flow, and
  strongly coupled CFT in the Schwarzschild Unruh or Boulware vacua}},
  \href{http://dx.doi.org/10.1088/0264-9381/28/21/215018}{\emph{Class. Quant.
  Grav.} {\bf 28} (2011) 215018}, [\href{https://arxiv.org/abs/1104.4489}{{\tt
  1104.4489}}].

\bibitem{Abdolrahimi:2012qi}
S.~Abdolrahimi, C.~Cattoen, D.~N. Page and S.~Yaghoobpour-Tari, \emph{{Large
  Randall-Sundrum II Black Holes}},
  \href{http://dx.doi.org/10.1016/j.physletb.2013.02.034}{\emph{Phys. Lett. B}
  {\bf 720} (2013) 405--409}, [\href{https://arxiv.org/abs/1206.0708}{{\tt
  1206.0708}}].

\bibitem{Abdolrahimi:2012pb}
S.~Abdolrahimi, C.~Catto\"en, D.~N. Page and S.~Yaghoobpour-Tari,
  \emph{{Spectral methods in general relativity and large Randall-Sundrum II
  black holes}},
  \href{http://dx.doi.org/10.1088/1475-7516/2013/06/039}{\emph{JCAP} {\bf 06}
  (2013) 039}, [\href{https://arxiv.org/abs/1212.5623}{{\tt 1212.5623}}].

\bibitem{Figueras:2013jja}
P.~Figueras and S.~Tunyasuvunakool, \emph{{CFTs in rotating black hole
  backgrounds}},
  \href{http://dx.doi.org/10.1088/0264-9381/30/12/125015}{\emph{Class. Quant.
  Grav.} {\bf 30} (2013) 125015}, [\href{https://arxiv.org/abs/1304.1162}{{\tt
  1304.1162}}].

\bibitem{Banerjee:2021qei}
S.~Banerjee, U.~Danielsson and S.~Giri, \emph{{Dark bubbles and black holes}},
  \href{http://dx.doi.org/10.1007/JHEP09(2021)158}{\emph{JHEP} {\bf 09} (2021)
  158}, [\href{https://arxiv.org/abs/2102.02164}{{\tt 2102.02164}}].

\bibitem{Biggs:2021iqw}
W.~D. Biggs and J.~E. Santos, \emph{{Rotating Black Holes in Randall-Sundrum II
  Braneworlds}},
  \href{http://dx.doi.org/10.1103/PhysRevLett.128.021601}{\emph{Phys. Rev.
  Lett.} {\bf 128} (2022) 021601},
  [\href{https://arxiv.org/abs/2108.00016}{{\tt 2108.00016}}].

\bibitem{Fabbri:2007kr}
A.~Fabbri and G.~P. Procopio, \emph{{Quantum effects in black holes from the
  Schwarzschild black string?}},
  \href{http://dx.doi.org/10.1088/0264-9381/24/22/003}{\emph{Class. Quant.
  Grav.} {\bf 24} (2007) 5371--5382},
  [\href{https://arxiv.org/abs/0704.3728}{{\tt 0704.3728}}].

\bibitem{Hirayama:2001bi}
T.~Hirayama and G.~Kang, \emph{{Stable black strings in anti-de Sitter space}},
  \href{http://dx.doi.org/10.1103/PhysRevD.64.064010}{\emph{Phys. Rev. D} {\bf
  64} (2001) 064010}, [\href{https://arxiv.org/abs/hep-th/0104213}{{\tt
  hep-th/0104213}}].

\bibitem{Page:1982fm}
D.~N. Page, \emph{{Thermal Stress Tensors in Static Einstein Spaces}},
  \href{http://dx.doi.org/10.1103/PhysRevD.25.1499}{\emph{Phys. Rev. D} {\bf
  25} (1982) 1499}.

\bibitem{Emparan:2013moa}
R.~Emparan, R.~Suzuki and K.~Tanabe, \emph{{The large D limit of General
  Relativity}}, \href{http://dx.doi.org/10.1007/JHEP06(2013)009}{\emph{JHEP}
  {\bf 06} (2013) 009}, [\href{https://arxiv.org/abs/1302.6382}{{\tt
  1302.6382}}].

\bibitem{Bhattacharyya:2015dva}
S.~Bhattacharyya, A.~De, S.~Minwalla, R.~Mohan and A.~Saha, \emph{{A membrane
  paradigm at large D}},
  \href{http://dx.doi.org/10.1007/JHEP04(2016)076}{\emph{JHEP} {\bf 04} (2016)
  076}, [\href{https://arxiv.org/abs/1504.06613}{{\tt 1504.06613}}].

\bibitem{Emparan:2020inr}
R.~Emparan and C.~P. Herzog, \emph{{Large D limit of Einstein\textquoteright{}s
  equations}}, \href{http://dx.doi.org/10.1103/RevModPhys.92.045005}{\emph{Rev.
  Mod. Phys.} {\bf 92} (2020) 045005},
  [\href{https://arxiv.org/abs/2003.11394}{{\tt 2003.11394}}].

\bibitem{Emparan:2021ewh}
R.~Emparan, D.~Licht, R.~Suzuki, M.~Toma\v{s}evi\'c and B.~Way, \emph{{Black
  tsunamis and naked singularities in AdS}},
  \href{http://dx.doi.org/10.1007/JHEP02(2022)090}{\emph{JHEP} {\bf 02} (2022)
  090}, [\href{https://arxiv.org/abs/2112.07967}{{\tt 2112.07967}}].

\bibitem{Sorkin:1984kjy}
R.~D. Sorkin, \emph{{On the Entropy of the Vacuum outside a Horizon}},  in
  \emph{{10th International Conference on General Relativity and Gravitation}},
  vol.~2, pp.~734--736, 1984.
\newblock \href{https://arxiv.org/abs/1402.3589}{{\tt 1402.3589}}.

\bibitem{Bombelli:1986rw}
L.~Bombelli, R.~K. Koul, J.~Lee and R.~D. Sorkin, \emph{{A Quantum Source of
  Entropy for Black Holes}},
  \href{http://dx.doi.org/10.1103/PhysRevD.34.373}{\emph{Phys. Rev. D} {\bf 34}
  (1986) 373--383}.

\bibitem{Srednicki:1993im}
M.~Srednicki, \emph{{Entropy and area}},
  \href{http://dx.doi.org/10.1103/PhysRevLett.71.666}{\emph{Phys. Rev. Lett.}
  {\bf 71} (1993) 666--669}, [\href{https://arxiv.org/abs/hep-th/9303048}{{\tt
  hep-th/9303048}}].

\bibitem{Frolov:1993ym}
V.~P. Frolov and I.~Novikov, \emph{{Dynamical origin of the entropy of a black
  hole}}, \href{http://dx.doi.org/10.1103/PhysRevD.48.4545}{\emph{Phys. Rev. D}
  {\bf 48} (1993) 4545--4551}, [\href{https://arxiv.org/abs/gr-qc/9309001}{{\tt
  gr-qc/9309001}}].

\bibitem{Callan:1994py}
C.~G. Callan, Jr. and F.~Wilczek, \emph{{On geometric entropy}},
  \href{http://dx.doi.org/10.1016/0370-2693(94)91007-3}{\emph{Phys. Lett. B}
  {\bf 333} (1994) 55--61}, [\href{https://arxiv.org/abs/hep-th/9401072}{{\tt
  hep-th/9401072}}].

\bibitem{Sakharov:1967pk}
A.~D. Sakharov, \emph{{Vacuum quantum fluctuations in curved space and the
  theory of gravitation}},
  \href{http://dx.doi.org/10.1070/PU1991v034n05ABEH002498}{\emph{Dokl. Akad.
  Nauk Ser. Fiz.} {\bf 177} (1967) 70--71}.

\bibitem{Visser:2002ew}
M.~Visser, \emph{{Sakharov's induced gravity: A Modern perspective}},
  \href{http://dx.doi.org/10.1142/S0217732302006886}{\emph{Mod. Phys. Lett. A}
  {\bf 17} (2002) 977--992}, [\href{https://arxiv.org/abs/gr-qc/0204062}{{\tt
  gr-qc/0204062}}].

\bibitem{Jacobson:1994iw}
T.~Jacobson, \emph{{Black hole entropy and induced gravity}},
  \href{https://arxiv.org/abs/gr-qc/9404039}{{\tt gr-qc/9404039}}.

\bibitem{Frolov:1996aj}
V.~P. Frolov, D.~V. Fursaev and A.~I. Zelnikov, \emph{{Statistical origin of
  black hole entropy in induced gravity}},
  \href{http://dx.doi.org/10.1016/S0550-3213(96)00678-5}{\emph{Nucl. Phys. B}
  {\bf 486} (1997) 339--352}, [\href{https://arxiv.org/abs/hep-th/9607104}{{\tt
  hep-th/9607104}}].

\bibitem{Frolov:1997up}
V.~P. Frolov and D.~V. Fursaev, \emph{{Mechanism of generation of black hole
  entropy in Sakharov's induced gravity}},
  \href{http://dx.doi.org/10.1103/PhysRevD.56.2212}{\emph{Phys. Rev. D} {\bf
  56} (1997) 2212--2225}, [\href{https://arxiv.org/abs/hep-th/9703178}{{\tt
  hep-th/9703178}}].

\bibitem{Ryu:2006bv}
S.~Ryu and T.~Takayanagi, \emph{{Holographic derivation of entanglement entropy
  from AdS/CFT}},
  \href{http://dx.doi.org/10.1103/PhysRevLett.96.181602}{\emph{Phys. Rev.
  Lett.} {\bf 96} (2006) 181602},
  [\href{https://arxiv.org/abs/hep-th/0603001}{{\tt hep-th/0603001}}].

\bibitem{Ryu:2006ef}
S.~Ryu and T.~Takayanagi, \emph{{Aspects of Holographic Entanglement Entropy}},
  \href{http://dx.doi.org/10.1088/1126-6708/2006/08/045}{\emph{JHEP} {\bf 08}
  (2006) 045}, [\href{https://arxiv.org/abs/hep-th/0605073}{{\tt
  hep-th/0605073}}].

\bibitem{Lewkowycz:2013nqa}
A.~Lewkowycz and J.~Maldacena, \emph{{Generalized gravitational entropy}},
  \href{http://dx.doi.org/10.1007/JHEP08(2013)090}{\emph{JHEP} {\bf 08} (2013)
  090}, [\href{https://arxiv.org/abs/1304.4926}{{\tt 1304.4926}}].

\bibitem{Takayanagi:2019tvn}
T.~Takayanagi and K.~Tamaoka, \emph{{Gravity Edges Modes and Hayward Term}},
  \href{http://dx.doi.org/10.1007/JHEP02(2020)167}{\emph{JHEP} {\bf 02} (2020)
  167}, [\href{https://arxiv.org/abs/1912.01636}{{\tt 1912.01636}}].

\bibitem{Botta-Cantcheff:2020ywu}
M.~Botta-Cantcheff, P.~J. Martinez and J.~F. Zarate, \emph{{R\'enyi entropies
  and area operator from gravity with Hayward term}},
  \href{http://dx.doi.org/10.1007/JHEP07(2020)227}{\emph{JHEP} {\bf 07} (2020)
  227}, [\href{https://arxiv.org/abs/2005.11338}{{\tt 2005.11338}}].

\bibitem{Kastikainen:2023yyk}
J.~Kastikainen and A.~Svesko, \emph{{Gravitational R\'enyi entropy from corner
  terms}}, \href{http://dx.doi.org/10.1103/PhysRevD.109.126017}{\emph{Phys.
  Rev. D} {\bf 109} (2024) 126017},
  [\href{https://arxiv.org/abs/2312.06765}{{\tt 2312.06765}}].

\bibitem{Kastikainen:2023omj}
J.~Kastikainen and A.~Svesko, \emph{{Cornering gravitational entropy}},
  \href{http://dx.doi.org/10.1007/JHEP06(2024)160}{\emph{JHEP} {\bf 06} (2024)
  160}, [\href{https://arxiv.org/abs/2312.13357}{{\tt 2312.13357}}].

\bibitem{Lashkari:2013koa}
N.~Lashkari, M.~B. McDermott and M.~Van~Raamsdonk, \emph{{Gravitational
  dynamics from entanglement 'thermodynamics'}},
  \href{http://dx.doi.org/10.1007/JHEP04(2014)195}{\emph{JHEP} {\bf 04} (2014)
  195}, [\href{https://arxiv.org/abs/1308.3716}{{\tt 1308.3716}}].

\bibitem{Faulkner:2013ica}
T.~Faulkner, M.~Guica, T.~Hartman, R.~C. Myers and M.~Van~Raamsdonk,
  \emph{{Gravitation from Entanglement in Holographic CFTs}},
  \href{http://dx.doi.org/10.1007/JHEP03(2014)051}{\emph{JHEP} {\bf 03} (2014)
  051}, [\href{https://arxiv.org/abs/1312.7856}{{\tt 1312.7856}}].

\bibitem{Haehl:2017sot}
F.~M. Haehl, E.~Hijano, O.~Parrikar and C.~Rabideau, \emph{{Higher Curvature
  Gravity from Entanglement in Conformal Field Theories}},
  \href{http://dx.doi.org/10.1103/PhysRevLett.120.201602}{\emph{Phys. Rev.
  Lett.} {\bf 120} (2018) 201602},
  [\href{https://arxiv.org/abs/1712.06620}{{\tt 1712.06620}}].

\bibitem{Swingle:2014uza}
B.~Swingle and M.~Van~Raamsdonk, \emph{{Universality of Gravity from
  Entanglement}},  \href{https://arxiv.org/abs/1405.2933}{{\tt 1405.2933}}.

\bibitem{Agon:2020mvu}
C.~A. Ag\'on, E.~C\'aceres and J.~F. Pedraza, \emph{{Bit threads,
  Einstein\textquoteright{}s equations and bulk locality}},
  \href{http://dx.doi.org/10.1007/JHEP01(2021)193}{\emph{JHEP} {\bf 01} (2021)
  193}, [\href{https://arxiv.org/abs/2007.07907}{{\tt 2007.07907}}].

\bibitem{Agon:2021tia}
C.~A. Ag\'on and J.~F. Pedraza, \emph{{Quantum bit threads and holographic
  entanglement}}, \href{http://dx.doi.org/10.1007/JHEP02(2022)180}{\emph{JHEP}
  {\bf 02} (2022) 180}, [\href{https://arxiv.org/abs/2105.08063}{{\tt
  2105.08063}}].

\bibitem{Cooper:2019rwk}
S.~Cooper, D.~Neuenfeld, M.~Rozali and D.~Wakeham, \emph{{Brane dynamics from
  the first law of entanglement}},
  \href{http://dx.doi.org/10.1007/JHEP03(2020)023}{\emph{JHEP} {\bf 03} (2020)
  023}, [\href{https://arxiv.org/abs/1912.05746}{{\tt 1912.05746}}].

\bibitem{Jacobson:1995ab}
T.~Jacobson, \emph{{Thermodynamics of space-time: The Einstein equation of
  state}}, \href{http://dx.doi.org/10.1103/PhysRevLett.75.1260}{\emph{Phys.
  Rev. Lett.} {\bf 75} (1995) 1260--1263},
  [\href{https://arxiv.org/abs/gr-qc/9504004}{{\tt gr-qc/9504004}}].

\bibitem{Parikh:2009qs}
M.~K. Parikh and S.~Sarkar, \emph{{Beyond the Einstein Equation of State: Wald
  Entropy and Thermodynamical Gravity}},
  \href{http://dx.doi.org/10.3390/e18040119}{\emph{Entropy} {\bf 18} (2016)
  119}, [\href{https://arxiv.org/abs/0903.1176}{{\tt 0903.1176}}].

\bibitem{Guedens:2011dy}
R.~Guedens, T.~Jacobson and S.~Sarkar, \emph{{Horizon entropy and higher
  curvature equations of state}},
  \href{http://dx.doi.org/10.1103/PhysRevD.85.064017}{\emph{Phys. Rev. D} {\bf
  85} (2012) 064017}, [\href{https://arxiv.org/abs/1112.6215}{{\tt
  1112.6215}}].

\bibitem{Parikh:2017aas}
M.~Parikh and A.~Svesko, \emph{{Einstein\textquoteright{}s equations from the
  stretched future light cone}},
  \href{http://dx.doi.org/10.1103/PhysRevD.98.026018}{\emph{Phys. Rev. D} {\bf
  98} (2018) 026018}, [\href{https://arxiv.org/abs/1712.08475}{{\tt
  1712.08475}}].

\bibitem{Parikh:2018anm}
M.~Parikh, S.~Sarkar and A.~Svesko, \emph{{Local first law of gravity}},
  \href{http://dx.doi.org/10.1103/PhysRevD.101.104043}{\emph{Phys. Rev. D} {\bf
  101} (2020) 104043}, [\href{https://arxiv.org/abs/1801.07306}{{\tt
  1801.07306}}].

\bibitem{Svesko:2018qim}
A.~Svesko, \emph{{Equilibrium to Einstein: Entanglement, Thermodynamics, and
  Gravity}}, \href{http://dx.doi.org/10.1103/PhysRevD.99.086006}{\emph{Phys.
  Rev. D} {\bf 99} (2019) 086006},
  [\href{https://arxiv.org/abs/1810.12236}{{\tt 1810.12236}}].

\bibitem{Engelhardt:2014gca}
N.~Engelhardt and A.~C. Wall, \emph{{Quantum Extremal Surfaces: Holographic
  Entanglement Entropy beyond the Classical Regime}},
  \href{http://dx.doi.org/10.1007/JHEP01(2015)073}{\emph{JHEP} {\bf 01} (2015)
  073}, [\href{https://arxiv.org/abs/1408.3203}{{\tt 1408.3203}}].

\bibitem{Dong:2013qoa}
X.~Dong, \emph{{Holographic Entanglement Entropy for General Higher Derivative
  Gravity}}, \href{http://dx.doi.org/10.1007/JHEP01(2014)044}{\emph{JHEP} {\bf
  01} (2014) 044}, [\href{https://arxiv.org/abs/1310.5713}{{\tt 1310.5713}}].

\bibitem{Camps:2013zua}
J.~Camps, \emph{{Generalized entropy and higher derivative Gravity}},
  \href{http://dx.doi.org/10.1007/JHEP03(2014)070}{\emph{JHEP} {\bf 03} (2014)
  070}, [\href{https://arxiv.org/abs/1310.6659}{{\tt 1310.6659}}].

\bibitem{Hawking:2000da}
S.~Hawking, J.~M. Maldacena and A.~Strominger, \emph{{de Sitter entropy,
  quantum entanglement and AdS / CFT}},
  \href{http://dx.doi.org/10.1088/1126-6708/2001/05/001}{\emph{JHEP} {\bf 05}
  (2001) 001}, [\href{https://arxiv.org/abs/hep-th/0002145}{{\tt
  hep-th/0002145}}].

\bibitem{Fursaev:2000ym}
D.~V. Fursaev, \emph{{Black hole thermodynamics, induced gravity and gravity in
  brane worlds}},  in \emph{{International Conference on Quantization, Gauge
  Theory, and Strings: Conference Dedicated to the Memory of Professor Efim
  Fradkin}}, pp.~462--470, 6, 2000.
\newblock \href{https://arxiv.org/abs/hep-th/0009164}{{\tt hep-th/0009164}}.

\bibitem{Jacobson:1999mi}
T.~Jacobson, \emph{{On the nature of black hole entropy}},
  \href{http://dx.doi.org/10.1063/1.1301569}{\emph{AIP Conf. Proc.} {\bf 493}
  (1999) 85--97}, [\href{https://arxiv.org/abs/gr-qc/9908031}{{\tt
  gr-qc/9908031}}].

\bibitem{Hawking:1976ra}
S.~W. Hawking, \emph{{Breakdown of Predictability in Gravitational Collapse}},
  \href{http://dx.doi.org/10.1103/PhysRevD.14.2460}{\emph{Phys. Rev. D} {\bf
  14} (1976) 2460--2473}.

\bibitem{Page:1993wv}
D.~N. Page, \emph{{Information in black hole radiation}},
  \href{http://dx.doi.org/10.1103/PhysRevLett.71.3743}{\emph{Phys. Rev. Lett.}
  {\bf 71} (1993) 3743--3746},
  [\href{https://arxiv.org/abs/hep-th/9306083}{{\tt hep-th/9306083}}].

\bibitem{Penington:2019npb}
G.~Penington, \emph{{Entanglement Wedge Reconstruction and the Information
  Paradox}}, \href{http://dx.doi.org/10.1007/JHEP09(2020)002}{\emph{JHEP} {\bf
  09} (2020) 002}, [\href{https://arxiv.org/abs/1905.08255}{{\tt 1905.08255}}].

\bibitem{Almheiri:2019psf}
A.~Almheiri, N.~Engelhardt, D.~Marolf and H.~Maxfield, \emph{{The entropy of
  bulk quantum fields and the entanglement wedge of an evaporating black
  hole}}, \href{http://dx.doi.org/10.1007/JHEP12(2019)063}{\emph{JHEP} {\bf 12}
  (2019) 063}, [\href{https://arxiv.org/abs/1905.08762}{{\tt 1905.08762}}].

\bibitem{Almheiri:2019qdq}
A.~Almheiri, T.~Hartman, J.~Maldacena, E.~Shaghoulian and A.~Tajdini,
  \emph{{Replica Wormholes and the Entropy of Hawking Radiation}},
  \href{http://dx.doi.org/10.1007/JHEP05(2020)013}{\emph{JHEP} {\bf 05} (2020)
  013}, [\href{https://arxiv.org/abs/1911.12333}{{\tt 1911.12333}}].

\bibitem{Penington:2019kki}
G.~Penington, S.~H. Shenker, D.~Stanford and Z.~Yang, \emph{{Replica wormholes
  and the black hole interior}},
  \href{http://dx.doi.org/10.1007/JHEP03(2022)205}{\emph{JHEP} {\bf 03} (2022)
  205}, [\href{https://arxiv.org/abs/1911.11977}{{\tt 1911.11977}}].

\bibitem{Goto:2020wnk}
K.~Goto, T.~Hartman and A.~Tajdini, \emph{{Replica wormholes for an evaporating
  2D black hole}}, \href{http://dx.doi.org/10.1007/JHEP04(2021)289}{\emph{JHEP}
  {\bf 04} (2021) 289}, [\href{https://arxiv.org/abs/2011.09043}{{\tt
  2011.09043}}].

\bibitem{Pedraza:2021ssc}
J.~F. Pedraza, A.~Svesko, W.~Sybesma and M.~R. Visser, \emph{{Microcanonical
  action and the entropy of Hawking radiation}},
  \href{http://dx.doi.org/10.1103/PhysRevD.105.126010}{\emph{Phys. Rev. D} {\bf
  105} (2022) 126010}, [\href{https://arxiv.org/abs/2111.06912}{{\tt
  2111.06912}}].

\bibitem{Almheiri:2019hni}
A.~Almheiri, R.~Mahajan, J.~Maldacena and Y.~Zhao, \emph{{The Page curve of
  Hawking radiation from semiclassical geometry}},
  \href{http://dx.doi.org/10.1007/JHEP03(2020)149}{\emph{JHEP} {\bf 03} (2020)
  149}, [\href{https://arxiv.org/abs/1908.10996}{{\tt 1908.10996}}].

\bibitem{Dvali:2000hr}
G.~R. Dvali, G.~Gabadadze and M.~Porrati, \emph{{4-D gravity on a brane in 5-D
  Minkowski space}},
  \href{http://dx.doi.org/10.1016/S0370-2693(00)00669-9}{\emph{Phys. Lett. B}
  {\bf 485} (2000) 208--214}, [\href{https://arxiv.org/abs/hep-th/0005016}{{\tt
  hep-th/0005016}}].

\bibitem{Neuenfeld:2021bsb}
D.~Neuenfeld, \emph{{Homology conditions for RT surfaces in double
  holography}}, \href{http://dx.doi.org/10.1088/1361-6382/ac51e7}{\emph{Class.
  Quant. Grav.} {\bf 39} (2022) 075009},
  [\href{https://arxiv.org/abs/2105.01130}{{\tt 2105.01130}}].

\bibitem{Lee:2022efh}
J.~H. Lee, D.~Neuenfeld and A.~Shukla, \emph{{Bounds on gravitational brane
  couplings and tomography in AdS$_{3}$ black hole microstates}},
  \href{http://dx.doi.org/10.1007/JHEP10(2022)139}{\emph{JHEP} {\bf 10} (2022)
  139}, [\href{https://arxiv.org/abs/2206.06511}{{\tt 2206.06511}}].

\bibitem{Geng:2021mic}
H.~Geng, A.~Karch, C.~Perez-Pardavila, S.~Raju, L.~Randall, M.~Riojas et~al.,
  \emph{{Entanglement phase structure of a holographic BCFT in a black hole
  background}}, \href{http://dx.doi.org/10.1007/JHEP05(2022)153}{\emph{JHEP}
  {\bf 05} (2022) 153}, [\href{https://arxiv.org/abs/2112.09132}{{\tt
  2112.09132}}].

\bibitem{Karch:2023ekf}
A.~Karch, C.~Perez-Pardavila, M.~Riojas and M.~Youssef, \emph{{Subregion
  entropy for the doubly-holographic global black string}},
  \href{http://dx.doi.org/10.1007/JHEP05(2023)195}{\emph{JHEP} {\bf 05} (2023)
  195}, [\href{https://arxiv.org/abs/2303.09571}{{\tt 2303.09571}}].

\bibitem{Geng:2021wcq}
H.~Geng, Y.~Nomura and H.-Y. Sun, \emph{{Information paradox and its resolution
  in de Sitter holography}},
  \href{http://dx.doi.org/10.1103/PhysRevD.103.126004}{\emph{Phys. Rev. D} {\bf
  103} (2021) 126004}, [\href{https://arxiv.org/abs/2103.07477}{{\tt
  2103.07477}}].

\bibitem{Geng:2020qvw}
H.~Geng and A.~Karch, \emph{{Massive islands}},
  \href{http://dx.doi.org/10.1007/JHEP09(2020)121}{\emph{JHEP} {\bf 09} (2020)
  121}, [\href{https://arxiv.org/abs/2006.02438}{{\tt 2006.02438}}].

\bibitem{Geng:2023qwm}
H.~Geng, \emph{{Revisiting Recent Progress in the Karch-Randall Braneworld}},
  \href{https://arxiv.org/abs/2306.15671}{{\tt 2306.15671}}.

\bibitem{Geng:2021hlu}
H.~Geng, A.~Karch, C.~Perez-Pardavila, S.~Raju, L.~Randall, M.~Riojas et~al.,
  \emph{{Inconsistency of islands in theories with long-range gravity}},
  \href{http://dx.doi.org/10.1007/JHEP01(2022)182}{\emph{JHEP} {\bf 01} (2022)
  182}, [\href{https://arxiv.org/abs/2107.03390}{{\tt 2107.03390}}].

\bibitem{Czech:2017ryf}
B.~Czech, \emph{{Einstein Equations from Varying Complexity}},
  \href{http://dx.doi.org/10.1103/PhysRevLett.120.031601}{\emph{Phys. Rev.
  Lett.} {\bf 120} (2018) 031601},
  [\href{https://arxiv.org/abs/1706.00965}{{\tt 1706.00965}}].

\bibitem{Caputa:2018kdj}
P.~Caputa and J.~M. Magan, \emph{{Quantum Computation as Gravity}},
  \href{http://dx.doi.org/10.1103/PhysRevLett.122.231302}{\emph{Phys. Rev.
  Lett.} {\bf 122} (2019) 231302},
  [\href{https://arxiv.org/abs/1807.04422}{{\tt 1807.04422}}].

\bibitem{Susskind:2019ddc}
L.~Susskind, \emph{{Complexity and Newton's Laws}},
  \href{http://dx.doi.org/10.3389/fphy.2020.00262}{\emph{Front. in Phys.} {\bf
  8} (2020) 262}, [\href{https://arxiv.org/abs/1904.12819}{{\tt 1904.12819}}].

\bibitem{Pedraza:2021mkh}
J.~F. Pedraza, A.~Russo, A.~Svesko and Z.~Weller-Davies, \emph{{Lorentzian
  Threads as Gatelines and Holographic Complexity}},
  \href{http://dx.doi.org/10.1103/PhysRevLett.127.271602}{\emph{Phys. Rev.
  Lett.} {\bf 127} (2021) 271602},
  [\href{https://arxiv.org/abs/2105.12735}{{\tt 2105.12735}}].

\bibitem{Pedraza:2021fgp}
J.~F. Pedraza, A.~Russo, A.~Svesko and Z.~Weller-Davies, \emph{{Sewing
  spacetime with Lorentzian threads: complexity and the emergence of time in
  quantum gravity}},
  \href{http://dx.doi.org/10.1007/JHEP02(2022)093}{\emph{JHEP} {\bf 02} (2022)
  093}, [\href{https://arxiv.org/abs/2106.12585}{{\tt 2106.12585}}].

\bibitem{Pedraza:2022dqi}
J.~F. Pedraza, A.~Russo, A.~Svesko and Z.~Weller-Davies, \emph{{Computing
  spacetime}}, \href{http://dx.doi.org/10.1142/S021827182242010X}{\emph{Int. J.
  Mod. Phys. D} {\bf 31} (2022) 2242010},
  [\href{https://arxiv.org/abs/2205.05705}{{\tt 2205.05705}}].

\bibitem{Carrasco:2023fcj}
R.~Carrasco, J.~F. Pedraza, A.~Svesko and Z.~Weller-Davies, \emph{{Gravitation
  from optimized computation: Einstein and beyond}},
  \href{http://dx.doi.org/10.1007/JHEP09(2023)167}{\emph{JHEP} {\bf 09} (2023)
  167}, [\href{https://arxiv.org/abs/2306.08503}{{\tt 2306.08503}}].

\bibitem{Susskind:2014moa}
L.~Susskind, \emph{{Entanglement is not enough}},
  \href{http://dx.doi.org/10.1002/prop.201500095}{\emph{Fortsch. Phys.} {\bf
  64} (2016) 49--71}, [\href{https://arxiv.org/abs/1411.0690}{{\tt
  1411.0690}}].

\bibitem{Stanford:2014jda}
D.~Stanford and L.~Susskind, \emph{{Complexity and Shock Wave Geometries}},
  \href{http://dx.doi.org/10.1103/PhysRevD.90.126007}{\emph{Phys. Rev. D} {\bf
  90} (2014) 126007}, [\href{https://arxiv.org/abs/1406.2678}{{\tt
  1406.2678}}].

\bibitem{Susskind:2014rva}
L.~Susskind, \emph{{Computational Complexity and Black Hole Horizons}},
  \href{http://dx.doi.org/10.1002/prop.201500092}{\emph{Fortsch. Phys.} {\bf
  64} (2016) 24--43}, [\href{https://arxiv.org/abs/1403.5695}{{\tt
  1403.5695}}].

\bibitem{Brown:2015lvg}
A.~R. Brown, D.~A. Roberts, L.~Susskind, B.~Swingle and Y.~Zhao,
  \emph{{Complexity, action, and black holes}},
  \href{http://dx.doi.org/10.1103/PhysRevD.93.086006}{\emph{Phys. Rev. D} {\bf
  93} (2016) 086006}, [\href{https://arxiv.org/abs/1512.04993}{{\tt
  1512.04993}}].

\bibitem{Brown:2015bva}
A.~R. Brown, D.~A. Roberts, L.~Susskind, B.~Swingle and Y.~Zhao,
  \emph{{Holographic Complexity Equals Bulk Action?}},
  \href{http://dx.doi.org/10.1103/PhysRevLett.116.191301}{\emph{Phys. Rev.
  Lett.} {\bf 116} (2016) 191301},
  [\href{https://arxiv.org/abs/1509.07876}{{\tt 1509.07876}}].

\bibitem{Swingle:2009bg}
B.~Swingle, \emph{{Entanglement Renormalization and Holography}},
  \href{http://dx.doi.org/10.1103/PhysRevD.86.065007}{\emph{Phys. Rev. D} {\bf
  86} (2012) 065007}, [\href{https://arxiv.org/abs/0905.1317}{{\tt
  0905.1317}}].

\bibitem{Swingle:2012wq}
B.~Swingle, \emph{{Constructing holographic spacetimes using entanglement
  renormalization}},  \href{https://arxiv.org/abs/1209.3304}{{\tt 1209.3304}}.

\bibitem{Bao:2018pvs}
N.~Bao, G.~Penington, J.~Sorce and A.~C. Wall, \emph{{Beyond Toy Models:
  Distilling Tensor Networks in Full AdS/CFT}},
  \href{http://dx.doi.org/10.1007/JHEP11(2019)069}{\emph{JHEP} {\bf 11} (2019)
  069}, [\href{https://arxiv.org/abs/1812.01171}{{\tt 1812.01171}}].

\bibitem{Jahn:2021uqr}
A.~Jahn and J.~Eisert, \emph{{Holographic tensor network models and quantum
  error correction: a topical review}},
  \href{http://dx.doi.org/10.1088/2058-9565/ac0293}{\emph{Quantum Sci.
  Technol.} {\bf 6} (2021) 033002},
  [\href{https://arxiv.org/abs/2102.02619}{{\tt 2102.02619}}].

\bibitem{Belin:2021bga}
A.~Belin, R.~C. Myers, S.-M. Ruan, G.~S\'arosi and A.~J. Speranza, \emph{{Does
  Complexity Equal Anything?}},
  \href{http://dx.doi.org/10.1103/PhysRevLett.128.081602}{\emph{Phys. Rev.
  Lett.} {\bf 128} (2022) 081602},
  [\href{https://arxiv.org/abs/2111.02429}{{\tt 2111.02429}}].

\bibitem{Belin:2022xmt}
A.~Belin, R.~C. Myers, S.-M. Ruan, G.~S\'arosi and A.~J. Speranza,
  \emph{{Complexity equals anything II}},
  \href{http://dx.doi.org/10.1007/JHEP01(2023)154}{\emph{JHEP} {\bf 01} (2023)
  154}, [\href{https://arxiv.org/abs/2210.09647}{{\tt 2210.09647}}].

\bibitem{Hernandez:2020nem}
J.~Hernandez, R.~C. Myers and S.-M. Ruan, \emph{{Quantum extremal islands made
  easy. Part III. Complexity on the brane}},
  \href{http://dx.doi.org/10.1007/JHEP02(2021)173}{\emph{JHEP} {\bf 02} (2021)
  173}, [\href{https://arxiv.org/abs/2010.16398}{{\tt 2010.16398}}].

\bibitem{Chen:2023tpi}
B.~Chen, Y.~Liu and B.~Yu, \emph{{Holographic complexity of rotating quantum
  black holes}}, \href{http://dx.doi.org/10.1007/JHEP01(2024)055}{\emph{JHEP}
  {\bf 01} (2024) 055}, [\href{https://arxiv.org/abs/2307.15968}{{\tt
  2307.15968}}].

\bibitem{Aguilar-Gutierrez:2023ccv}
S.~E. Aguilar-Gutierrez, B.~Craps, J.~Hernandez, M.~Khramtsov, M.~Knysh and
  A.~Shukla, \emph{{Holographic complexity: braneworld gravity versus the Lloyd
  bound}}, \href{http://dx.doi.org/10.1007/JHEP03(2024)173}{\emph{JHEP} {\bf
  03} (2024) 173}, [\href{https://arxiv.org/abs/2312.12349}{{\tt 2312.12349}}].

\bibitem{RafaInPrep}
R.~Carrasco, J.~F. Pedraza and A.~Svesko, \emph{Work in progress}.

\bibitem{1974IAUS...64...82P}
R.~{Penrose}, \emph{{Gravitational Collapse}},  in \emph{Gravitational
  Radiation and Gravitational Collapse} (C.~{Dewitt-Morette}, ed.), vol.~64,
  p.~82, Jan., 1974.

\bibitem{Penrose:1980ge}
R.~Penrose, \emph{{Singularities and time asymmetry}}, pp.~581--638.
\newblock 1980.

\bibitem{Simpson:1973ua}
M.~Simpson and R.~Penrose, \emph{{Internal instability in a Reissner-Nordstrom
  black hole}}, \href{http://dx.doi.org/10.1007/BF00792069}{\emph{Int. J.
  Theor. Phys.} {\bf 7} (1973) 183--197}.

\bibitem{Poisson:1990eh}
E.~Poisson and W.~Israel, \emph{{Internal structure of black holes}},
  \href{http://dx.doi.org/10.1103/PhysRevD.41.1796}{\emph{Phys. Rev. D} {\bf
  41} (1990) 1796--1809}.

\bibitem{Lanir:2018vgb}
A.~Lanir, A.~Ori, N.~Zilberman, O.~Sela, A.~Maline and A.~Levi, \emph{{Analysis
  of quantum effects inside spherical charged black holes}},
  \href{http://dx.doi.org/10.1103/PhysRevD.99.061502}{\emph{Phys. Rev. D} {\bf
  99} (2019) 061502}, [\href{https://arxiv.org/abs/1811.03672}{{\tt
  1811.03672}}].

\bibitem{Zilberman:2019buh}
N.~Zilberman, A.~Levi and A.~Ori, \emph{{Quantum fluxes at the inner horizon of
  a spherical charged black hole}},
  \href{http://dx.doi.org/10.1103/PhysRevLett.124.171302}{\emph{Phys. Rev.
  Lett.} {\bf 124} (2020) 171302},
  [\href{https://arxiv.org/abs/1906.11303}{{\tt 1906.11303}}].

\bibitem{Dias:2019ery}
O.~J.~C. Dias, H.~S. Reall and J.~E. Santos, \emph{{The BTZ black hole violates
  strong cosmic censorship}},
  \href{http://dx.doi.org/10.1007/JHEP12(2019)097}{\emph{JHEP} {\bf 12} (2019)
  097}, [\href{https://arxiv.org/abs/1906.08265}{{\tt 1906.08265}}].

\bibitem{Hollands:2019whz}
S.~Hollands, R.~M. Wald and J.~Zahn, \emph{{Quantum instability of the Cauchy
  horizon in Reissner\textendash{}Nordstr\"om\textendash{}deSitter spacetime}},
  \href{http://dx.doi.org/10.1088/1361-6382/ab8052}{\emph{Class. Quant. Grav.}
  {\bf 37} (2020) 115009}, [\href{https://arxiv.org/abs/1912.06047}{{\tt
  1912.06047}}].

\bibitem{Emparan:2020rnp}
R.~Emparan and M.~Toma\v{s}evi\'c, \emph{{Strong cosmic censorship in the BTZ
  black hole}}, \href{http://dx.doi.org/10.1007/JHEP06(2020)038}{\emph{JHEP}
  {\bf 06} (2020) 038}, [\href{https://arxiv.org/abs/2002.02083}{{\tt
  2002.02083}}].

\bibitem{Kolanowski:2023hvh}
M.~Kolanowski and M.~Toma\v{s}evi\'c, \emph{{Singularities in 2D and 3D quantum
  black holes}}, \href{http://dx.doi.org/10.1007/JHEP12(2023)102}{\emph{JHEP}
  {\bf 12} (2023) 102}, [\href{https://arxiv.org/abs/2310.06014}{{\tt
  2310.06014}}].

\bibitem{Wald:1974hkz}
R.~Wald, \emph{{Gedanken experiments to destroy a black hole}},
  \href{http://dx.doi.org/10.1016/0003-4916(74)90125-0}{\emph{Annals Phys.}
  {\bf 82} (1974) 548--556}.

\bibitem{Sorce:2017dst}
J.~Sorce and R.~M. Wald, \emph{{Gedanken experiments to destroy a black hole.
  II. Kerr-Newman black holes cannot be overcharged or overspun}},
  \href{http://dx.doi.org/10.1103/PhysRevD.96.104014}{\emph{Phys. Rev. D} {\bf
  96} (2017) 104014}, [\href{https://arxiv.org/abs/1707.05862}{{\tt
  1707.05862}}].

\bibitem{Penrose:1973um}
R.~Penrose, \emph{{Naked singularities}},
  \href{http://dx.doi.org/10.1111/j.1749-6632.1973.tb41447.x}{\emph{Annals N.
  Y. Acad. Sci.} {\bf 224} (1973) 125--134}.

\bibitem{Huisken01}
G.~Huisken and T.~Ilmanen, \emph{{The Inverse Mean Curvature Flow and the
  Riemannian Penrose Inequality}},
  \href{http://dx.doi.org/10.4310/jdg/1090349447}{\emph{Journal of Differential
  Geometry} {\bf 59} (2001) 353 -- 437}.

\bibitem{Bray01}
H.~L. Bray, \emph{{Proof of the Riemannian Penrose Inequality Using the
  Positive Mass Theorem}},
  \href{http://dx.doi.org/10.4310/jdg/1090349428}{\emph{Journal of Differential
  Geometry} {\bf 59} (2001) 177 -- 267}.

\bibitem{Bray:2007opu}
H.~L. Bray and D.~A. Lee, \emph{{On the Riemannian Penrose inequality in
  dimensions less than 8}},
  \href{http://dx.doi.org/10.1215/00127094-2009-020}{\emph{Duke Math. J.} {\bf
  148} (2009) 81--106}, [\href{https://arxiv.org/abs/0705.1128}{{\tt
  0705.1128}}].

\bibitem{Mars:2009cj}
M.~Mars, \emph{{Present status of the Penrose inequality}},
  \href{http://dx.doi.org/10.1088/0264-9381/26/19/193001}{\emph{Class. Quant.
  Grav.} {\bf 26} (2009) 193001}, [\href{https://arxiv.org/abs/0906.5566}{{\tt
  0906.5566}}].

\bibitem{Itkin:2011ph}
I.~Itkin and Y.~Oz, \emph{{Penrose Inequality for Asymptotically AdS Spaces}},
  \href{http://dx.doi.org/10.1016/j.physletb.2012.01.007}{\emph{Phys. Lett. B}
  {\bf 708} (2012) 307--308}, [\href{https://arxiv.org/abs/1106.2683}{{\tt
  1106.2683}}].

\bibitem{Folkestad:2022dse}
r.~Folkestad, \emph{{Penrose Inequality as a Constraint on the Low Energy Limit
  of Quantum Gravity}},
  \href{http://dx.doi.org/10.1103/PhysRevLett.130.121501}{\emph{Phys. Rev.
  Lett.} {\bf 130} (2023) 121501},
  [\href{https://arxiv.org/abs/2209.00013}{{\tt 2209.00013}}].

\bibitem{Bousso:2019var}
R.~Bousso, A.~Shahbazi-Moghaddam and M.~Tomasevic, \emph{{Quantum Penrose
  Inequality}},
  \href{http://dx.doi.org/10.1103/PhysRevLett.123.241301}{\emph{Phys. Rev.
  Lett.} {\bf 123} (2019) 241301},
  [\href{https://arxiv.org/abs/1908.02755}{{\tt 1908.02755}}].

\bibitem{Bousso:2019bkg}
R.~Bousso, A.~Shahbazi-Moghaddam and M.~Toma\v{s}evi\'c, \emph{{Quantum
  Information Bound on the Energy}},
  \href{http://dx.doi.org/10.1103/PhysRevD.100.126010}{\emph{Phys. Rev. D} {\bf
  100} (2019) 126010}, [\href{https://arxiv.org/abs/1909.02001}{{\tt
  1909.02001}}].

\bibitem{Engelhardt:2024hpe}
N.~Engelhardt, r.~Folkestad, A.~Levine, E.~Verheijden and L.~Yang,
  \emph{{Cryptographic Censorship}},
  \href{https://arxiv.org/abs/2402.03425}{{\tt 2402.03425}}.

\bibitem{Berti:2009kk}
E.~Berti, V.~Cardoso and A.~O. Starinets, \emph{{Quasinormal modes of black
  holes and black branes}},
  \href{http://dx.doi.org/10.1088/0264-9381/26/16/163001}{\emph{Class. Quant.
  Grav.} {\bf 26} (2009) 163001}, [\href{https://arxiv.org/abs/0905.2975}{{\tt
  0905.2975}}].

\bibitem{Konoplya:2011qq}
R.~A. Konoplya and A.~Zhidenko, \emph{{Quasinormal modes of black holes: From
  astrophysics to string theory}},
  \href{http://dx.doi.org/10.1103/RevModPhys.83.793}{\emph{Rev. Mod. Phys.}
  {\bf 83} (2011) 793--836}, [\href{https://arxiv.org/abs/1102.4014}{{\tt
  1102.4014}}].

\bibitem{Horowitz:1999jd}
G.~T. Horowitz and V.~E. Hubeny, \emph{{Quasinormal modes of AdS black holes
  and the approach to thermal equilibrium}},
  \href{http://dx.doi.org/10.1103/PhysRevD.62.024027}{\emph{Phys. Rev. D} {\bf
  62} (2000) 024027}, [\href{https://arxiv.org/abs/hep-th/9909056}{{\tt
  hep-th/9909056}}].

\bibitem{QNMstoappear}
C.~Cartwright, U.~Gursoy, J.~F. Pedraza and G.~Planella~Planas, \emph{To
  appear}.

\bibitem{Grozdanov:2017ajz}
S.~Grozdanov, K.~Schalm and V.~Scopelliti, \emph{{Black hole scrambling from
  hydrodynamics}},
  \href{http://dx.doi.org/10.1103/PhysRevLett.120.231601}{\emph{Phys. Rev.
  Lett.} {\bf 120} (2018) 231601},
  [\href{https://arxiv.org/abs/1710.00921}{{\tt 1710.00921}}].

\bibitem{Blake:2018leo}
M.~Blake, R.~A. Davison, S.~Grozdanov and H.~Liu, \emph{{Many-body chaos and
  energy dynamics in holography}},
  \href{http://dx.doi.org/10.1007/JHEP10(2018)035}{\emph{JHEP} {\bf 10} (2018)
  035}, [\href{https://arxiv.org/abs/1809.01169}{{\tt 1809.01169}}].

\bibitem{Haehl:2018izb}
F.~M. Haehl and M.~Rozali, \emph{{Effective Field Theory for Chaotic CFTs}},
  \href{http://dx.doi.org/10.1007/JHEP10(2018)118}{\emph{JHEP} {\bf 10} (2018)
  118}, [\href{https://arxiv.org/abs/1808.02898}{{\tt 1808.02898}}].

\bibitem{Cartwright:2024rus}
C.~Cartwright, \emph{{An example of the convergence of hydrodynamics in strong
  external fields}},  \href{https://arxiv.org/abs/2403.12638}{{\tt
  2403.12638}}.

\bibitem{Grozdanov:2019uhi}
S.~Grozdanov, P.~K. Kovtun, A.~O. Starinets and P.~Tadi\'c, \emph{{The complex
  life of hydrodynamic modes}},
  \href{http://dx.doi.org/10.1007/JHEP11(2019)097}{\emph{JHEP} {\bf 11} (2019)
  097}, [\href{https://arxiv.org/abs/1904.12862}{{\tt 1904.12862}}].

\bibitem{Sasaki:1999mi}
M.~Sasaki, T.~Shiromizu and K.-i. Maeda, \emph{{Gravity, stability and energy
  conservation on the Randall-Sundrum brane world}},
  \href{http://dx.doi.org/10.1103/PhysRevD.62.024008}{\emph{Phys. Rev. D} {\bf
  62} (2000) 024008}, [\href{https://arxiv.org/abs/hep-th/9912233}{{\tt
  hep-th/9912233}}].

\bibitem{Garriga:1999yh}
J.~Garriga and T.~Tanaka, \emph{{Gravity in the brane world}},
  \href{http://dx.doi.org/10.1103/PhysRevLett.84.2778}{\emph{Phys. Rev. Lett.}
  {\bf 84} (2000) 2778--2781},
  [\href{https://arxiv.org/abs/hep-th/9911055}{{\tt hep-th/9911055}}].

\bibitem{Csaki:2000fc}
C.~Csaki, J.~Erlich, T.~J. Hollowood and Y.~Shirman, \emph{{Universal aspects
  of gravity localized on thick branes}},
  \href{http://dx.doi.org/10.1016/S0550-3213(00)00271-6}{\emph{Nucl. Phys. B}
  {\bf 581} (2000) 309--338}, [\href{https://arxiv.org/abs/hep-th/0001033}{{\tt
  hep-th/0001033}}].

\bibitem{Karch:2001cw}
A.~Karch and L.~Randall, \emph{{Localized gravity in string theory}},
  \href{http://dx.doi.org/10.1103/PhysRevLett.87.061601}{\emph{Phys. Rev.
  Lett.} {\bf 87} (2001) 061601},
  [\href{https://arxiv.org/abs/hep-th/0105108}{{\tt hep-th/0105108}}].

\bibitem{DeWolfe:2001pq}
O.~DeWolfe, D.~Z. Freedman and H.~Ooguri, \emph{{Holography and defect
  conformal field theories}},
  \href{http://dx.doi.org/10.1103/PhysRevD.66.025009}{\emph{Phys. Rev. D} {\bf
  66} (2002) 025009}, [\href{https://arxiv.org/abs/hep-th/0111135}{{\tt
  hep-th/0111135}}].

\bibitem{Aharony:2003qf}
O.~Aharony, O.~DeWolfe, D.~Z. Freedman and A.~Karch, \emph{{Defect conformal
  field theory and locally localized gravity}},
  \href{http://dx.doi.org/10.1088/1126-6708/2003/07/030}{\emph{JHEP} {\bf 07}
  (2003) 030}, [\href{https://arxiv.org/abs/hep-th/0303249}{{\tt
  hep-th/0303249}}].

\bibitem{Polchinski:1995mt}
J.~Polchinski, \emph{{Dirichlet Branes and Ramond-Ramond charges}},
  \href{http://dx.doi.org/10.1103/PhysRevLett.75.4724}{\emph{Phys. Rev. Lett.}
  {\bf 75} (1995) 4724--4727},
  [\href{https://arxiv.org/abs/hep-th/9510017}{{\tt hep-th/9510017}}].

\bibitem{Horowitz:1991cd}
G.~T. Horowitz and A.~Strominger, \emph{{Black strings and P-branes}},
  \href{http://dx.doi.org/10.1016/0550-3213(91)90440-9}{\emph{Nucl. Phys. B}
  {\bf 360} (1991) 197--209}.

\bibitem{Gubser:1999vj}
S.~S. Gubser, \emph{{AdS / CFT and gravity}},
  \href{http://dx.doi.org/10.1103/PhysRevD.63.084017}{\emph{Phys. Rev. D} {\bf
  63} (2001) 084017}, [\href{https://arxiv.org/abs/hep-th/9912001}{{\tt
  hep-th/9912001}}].

\bibitem{Levicivita1918}
T.~Levi-Civita, \emph{{Ds2 einsteiniani in campi newtoniani}}, {\emph{Atti
  Accad. Nazl. Lincei} {\bf 27} (1918) }.

\bibitem{Ehlers:1962zz}
J.~Ehlers and W.~Kundt, \emph{{Exact solutions of the gravitational field
  equations}}, .

\bibitem{Kinnersley:1970zw}
W.~Kinnersley and M.~Walker, \emph{{Uniformly accelerating charged mass in
  general relativity}},
  \href{http://dx.doi.org/10.1103/PhysRevD.2.1359}{\emph{Phys. Rev. D} {\bf 2}
  (1970) 1359--1370}.

\bibitem{Griffiths:2005qp}
J.~B. Griffiths and J.~Podolsky, \emph{{A New look at the Plebanski-Demianski
  family of solutions}},
  \href{http://dx.doi.org/10.1142/S0218271806007742}{\emph{Int. J. Mod. Phys.
  D} {\bf 15} (2006) 335--370},
  [\href{https://arxiv.org/abs/gr-qc/0511091}{{\tt gr-qc/0511091}}].

\bibitem{Podolsky:2006px}
J.~Podolsky and J.~B. Griffiths, \emph{{Accelerating Kerr-Newman black holes in
  (anti-)de Sitter space-time}},
  \href{http://dx.doi.org/10.1103/PhysRevD.73.044018}{\emph{Phys. Rev. D} {\bf
  73} (2006) 044018}, [\href{https://arxiv.org/abs/gr-qc/0601130}{{\tt
  gr-qc/0601130}}].

\bibitem{Griffiths:2006tk}
J.~B. Griffiths, P.~Krtous and J.~Podolsky, \emph{{Interpreting the C-metric}},
  \href{http://dx.doi.org/10.1088/0264-9381/23/23/008}{\emph{Class. Quant.
  Grav.} {\bf 23} (2006) 6745--6766},
  [\href{https://arxiv.org/abs/gr-qc/0609056}{{\tt gr-qc/0609056}}].

\bibitem{Podolsky:2021zwr}
J.~Podolsky and A.~Vratny, \emph{{New improved form of black holes of type D}},
  \href{http://dx.doi.org/10.1103/PhysRevD.104.084078}{\emph{Phys. Rev. D} {\bf
  104} (2021) 084078}, [\href{https://arxiv.org/abs/2108.02239}{{\tt
  2108.02239}}].

\bibitem{Hong:2003gx}
K.~Hong and E.~Teo, \emph{{A New form of the C metric}},
  \href{http://dx.doi.org/10.1088/0264-9381/20/14/321}{\emph{Class. Quant.
  Grav.} {\bf 20} (2003) 3269--3277},
  [\href{https://arxiv.org/abs/gr-qc/0305089}{{\tt gr-qc/0305089}}].

\bibitem{Hong:2004dm}
K.~Hong and E.~Teo, \emph{{A New form of the rotating C-metric}},
  \href{http://dx.doi.org/10.1088/0264-9381/22/1/007}{\emph{Class. Quant.
  Grav.} {\bf 22} (2005) 109--118},
  [\href{https://arxiv.org/abs/gr-qc/0410002}{{\tt gr-qc/0410002}}].

\bibitem{Bonnor:2002fk}
W.~B. Bonnor, \emph{{Closed timelike curves in general relativity}},
  \href{http://dx.doi.org/10.1142/S0218271803004122}{\emph{Int. J. Mod. Phys.
  D} {\bf 12} (2003) 1705--1708},
  [\href{https://arxiv.org/abs/gr-qc/0211051}{{\tt gr-qc/0211051}}].

\bibitem{Hubeny:2009kz}
V.~E. Hubeny, D.~Marolf and M.~Rangamani, \emph{{Black funnels and droplets
  from the AdS C-metrics}},
  \href{http://dx.doi.org/10.1088/0264-9381/27/2/025001}{\emph{Class. Quant.
  Grav.} {\bf 27} (2010) 025001}, [\href{https://arxiv.org/abs/0909.0005}{{\tt
  0909.0005}}].

\bibitem{Costa:2001ki}
S.~S. Costa, \emph{{A Description of several coordinate systems for hyperbolic
  spaces}},  \href{https://arxiv.org/abs/math-ph/0112039}{{\tt
  math-ph/0112039}}.

\bibitem{Dias:2002mi}
O.~J.~C. Dias and J.~P.~S. Lemos, \emph{{Pair of accelerated black holes in
  anti-de Sitter background: AdS C metric}},
  \href{http://dx.doi.org/10.1103/PhysRevD.67.064001}{\emph{Phys. Rev. D} {\bf
  67} (2003) 064001}, [\href{https://arxiv.org/abs/hep-th/0210065}{{\tt
  hep-th/0210065}}].

\bibitem{Balasubramanian:1999re}
V.~Balasubramanian and P.~Kraus, \emph{{A Stress tensor for Anti-de Sitter
  gravity}}, \href{http://dx.doi.org/10.1007/s002200050764}{\emph{Commun. Math.
  Phys.} {\bf 208} (1999) 413--428},
  [\href{https://arxiv.org/abs/hep-th/9902121}{{\tt hep-th/9902121}}].

\end{thebibliography}\endgroup

\end{document}